\documentclass[11pt]{article}

\usepackage{graphicx}
\usepackage{color}
\usepackage{rotating}
\usepackage{amsmath}
\usepackage{amssymb}
\usepackage{amsthm}
\usepackage{float}
 \usepackage{url}
\usepackage{subfigure}

\usepackage{natbib} 

\usepackage{enumerate}
\usepackage{enumitem}

\usepackage{dsfont}
 
 \usepackage{algorithm}
\usepackage{algorithmic}
\usepackage{hyperref}
\usepackage{array,arydshln}

\newtheorem{theorem}{Theorem}[section]

\newtheorem{proposition}[theorem]{Proposition} 
\newtheorem{remark}{Remark}[section]

\newcommand\ba {\mathbf a}
\newcommand\bb {\mathbf b}

\newcommand\bu {\mathbf u}

\newcommand\bx {\mathbf x}

\newcommand\bz {\mathbf z}

\newcommand\bB {\mathbf B}
\newcommand\bC {\mathbf C}

\newcommand\bV {\mathbf V}

\newcommand\indica {\mathbb{I}}

\newcommand\uno {\mathds{1}}
\newcommand\vech {\mbox{vech}}

\newcommand\wa {\widehat{{a}}}
 
\newcommand\wb {\widehat{{b}}}
\newcommand\wbb {\widehat{\bb}}

\newcommand\wu {\widehat{u}}

\newcommand\wz {\widehat{z}}

\newcommand\wx {\widehat{x}}
\newcommand\wbu {\widehat{\bu}}
\newcommand\wbx {\widehat{\bx}}
\newcommand\wbz {\widehat{\bz}}

\newcommand\wD {\widehat{D}}

\newcommand\wX {\widehat{X}}

\newcommand\wta {\widetilde{a}}

\newcommand\itA {{\mathcal{A}}}
\newcommand\itB {{\mathcal{B}}}
\newcommand\itC {{\mathcal{C}}}
\newcommand\itD {{\mathcal{D}}}
\newcommand\itE {{\mathcal{E}}}
\newcommand\itF {{\mathcal{F}}}

\newcommand\itH {{\mathcal{H}}}
\newcommand\itI {{\mathcal{I}}}
\newcommand\itJ {{\mathcal{J}}}

\newcommand\itL {{\mathcal{L}}}

\newcommand\itT {{\mathcal{T}}}

\newcommand\itV {{\mathcal{V}}}
\newcommand\itW {{\mathcal{W}}}

\newcommand\bmu {\mbox{\boldmath $\mu$}}

\newcommand\bnu {\mbox{\boldmath $\nu$}}

\newcommand\bxi {\mbox{\boldmath $\xi$}}

\newcommand\walfa {\widehat{\alpha}}

\newcommand\wbeta {\widehat{\beta}}

\newcommand\wphi {\widehat{\phi}}

\newcommand\wlam {\widehat{\lambda}}

\newcommand\wmu {\widehat{\mu}}
\newcommand\wbmu {\widehat{\bmu}}

\newcommand\wbnu {\widehat{\bnu}}
\newcommand\wsigma {\widehat{\sigma}}

\newcommand\wup {\widehat{\upsilon}}

\newcommand\wxi {\widehat{\xi}}
\newcommand\wbxi  {\widehat{\bxi}}

\newcommand\wGamma {\widehat{\Gamma}}

\newcommand\wUps  {\widehat{\Upsilon}}

\newcommand\wtalfa {\widetilde{\alpha}}

\newcommand\wtbeta {\widetilde{\beta}}

\newcommand\wttheta {\widetilde{\theta}}
\newcommand\wtup {\widetilde{\upsilon}}

\newcommand\wtUpsilon {\widetilde{\Upsilon}}
\def\real{\mathbb{R}}
\def\natu{\mathbb{N}}

\newcommand{\esp}{\mathbb{E}}
\newcommand{\prob}{\mathbb{P}}

\newcommand{\var}{\mbox{\sc Var}}

\newcommand{\trasp}{^{\mbox{\footnotesize \sc t}}}

\newcommand{\traspbis}{{\mbox{\footnotesize \sc t}}}

\newcommand\bcero {{\bf{0}}}

\def\argmin{\mathop{\mbox{argmin}}}

\newcommand\noi{\noindent}

\def\square{\ifmmode\sqr\else{$\sqr$}\fi}
\def\sqr{\vcenter{
         \hrule height.1mm
         \hbox{\vrule width.1mm height2.2mm\kern2.18mm
\vrule width.1mm}
         \hrule height.1mm}}

\newcommand{\eme}{\mbox{\scriptsize \sc m}}

\newcommand{\tuk}{\mbox{\scriptsize \sc t}}

\newcommand{\ini}{\mbox{\footnotesize \sc ini}}

\newcommand{\rob}{\mbox{\scriptsize \sc r}}

\newcommand{\sm}{\mbox{\footnotesize    \sc sm}}

\newcommand\trim {\mbox{\footnotesize  \sc tr}}

\newcommand\ls {\mbox{\scriptsize\sc ls}}

\newcommand{\out}{\mbox{\scriptsize\sc out}}

\newcommand{\cont}{\mbox{\scriptsize\sc co}}

 \usepackage{xcolor}

\usepackage{hyperref} %PARA QUE AL PINCHAR VAYAS A LA ECUACION

\begin{document}

%%%%%%%%%%%%%%%%%% ABSTRACT%%%%%%%%%%%%%%%%%%%%%%%%%%%%%%%%

\title{Robust estimation for  functional quadratic regression models}
\author{Graciela Boente$^1$ and Daniela Parada$^2$\\
$^1$ Universidad de Buenos Aires and CONICET\\
$^2$ Universidad de Buenos Aires}

\date{\today}
 
	\maketitle

%%%%%%%%%%%%%%%%%% ABSTRACT%%%%%%%%%%%%%%%%%%%%%%%%%%%%%%%%
\begin{abstract}
Functional quadratic regression models postulate a polynomial relationship rather than a linear one between a scalar response and a functional covariate. As in functional linear regression, vertical and especially high--leverage outliers may affect the classical estimators. For that reason,  providing reliable estimators in such situations is an important issue. Taking into account that the functional polynomial model is equivalent to  a regression model that is a polynomial of the same order in the functional principal component scores of the predictor processes, our proposal combines  robust estimators of the principal directions  with robust regression estimators based on a bounded loss function
and a preliminary residual scale estimator. Fisher--consistency of the proposed method is derived under mild assumptions. The results of a numerical study show  the benefits of the robust proposal over  the one based on sample principal directions and least  squares for the considered contaminating scenarios. 
The usefulness of the proposed approach is also illustrated through the analysis of a real data set which also reveals  that when the potential outliers are removed the classical method  behave very similarly to the robust one computed with all the data.
\end{abstract}


\normalsize
\newpage
\section{Introduction}\label{sec:intro}

In the last decades, functional explanatory variables have been included in regression models either nonparametrically  or through parametric models. Within the field of functional data analysis, some excellent overviews are provided in \citet{ferraty2006nonparametric} who presents a careful treatment of nonparametric models and also in the books by \citet{ramsay2002applied,ramsay2005functional}, \citet{horvath2012inference} and \citet{hsing2015theoretical} who place emphasis on the functional linear model. Various aspects of this last model  including implementations  and asymptotic theory,  have been studied among others in  \citet{cardot2003spline}, \citet{shen2004ftest}, \citet{cardot2005estimation}, \citet{cai2006prediction}, \citet{hall2007methodology}, \citet{febrero2017functional} and \citet{reiss2017methods}. The functional linear model imposes a structural linear constraint on the regression relationship which may or may not be satisfied. Some procedures to test the goodness of fit in such models have been discussed among others in \citet{garcia2014goodness}, \citet{cuesta2019goodness} and \citet{patilea2020testing}.

The linear constraint circumvents the \textsl{curse of dimensionality} present when considering fully nonparametric models since in the infinite--dimensional function space, the elements of a finite sample of random functions are very far away from each other. However, as pointed out in \citet{yao2010functional} and \citet{horvath2013test}, this linear model imposes a constraint on the regression relationship that may be too restrictive for some applications. To preserve a reasonable structural constraint, but at the same time improving the model flexibility within the class of parametric models, \citet{yao2010functional} defined a functional polynomial model analogous to the extension from simple linear regression to polynomial regression. As in functional linear regression,   regularization   is key step to define the estimators. For that reason, \citet{yao2010functional} and \citet{horvath2013test} project the predictor on the  eigenfunctions basis of the process, which is then truncated at a reasonable number of included components, leading to a parsimonious representation. With this representation, \citet{yao2010functional} have shown that the functional polynomial regression model can be represented as a polynomial regression model in the functional principal component scores of the predictor process. %For that reason, it seems sensible  to implement the model as polynomial regression in the principal components of predictor processes. 

In this paper, we consider independent and identically distributed observations with the same distribution as  $(y,X )$, where the response $y \in \real$ is related to the functional explanatory variable $X\in L^2(\itI)$ according to the quadratic model $y = \alpha_0+ \langle \beta_0, X \rangle + \langle X , \Upsilon_0 X \rangle +\sigma_0\;\epsilon $, where $\langle \cdot, \cdot \rangle$ denotes the usual $L^2(\itI)$ inner product, $\sigma_0 > 0$ is a residual scale parameter, and $\epsilon$ is the error term, independent of $X$. In this model, the regression parameter $\beta_0$ is
assumed to be in $L^2(\itI)$ and $\Upsilon_0: L^2(\itI) \to L^2(\itI)$ is a linear operator, that without loss of generality may be assumed to be self--adjoint, that is, if $\Upsilon_0 \, u(t)=\int_{\itI} \upsilon_0(s,t) u(s) ds$ then $\upsilon_0(s,t) =\upsilon_0(t,s) $. Furthermore, we will also assume that $\Upsilon_0$ is Hilbert--Schmidt, that is,  $\int_{\itI}\int_{\itI} \upsilon_0^2(s,t) ds\,dt<\infty$. The quadratic term $ \langle X  , \Upsilon_0 X  \rangle $ appearing in the model reflects that beyond the effect that the values $X(t)$, $t\in {\itI}$, have on the response, the products $\{X (s)X (t)\}$, for $s,t \in {\itI}$, are also included as additional predictors. 

As it has been extensively described,  small proportions of outliers and other atypical observations can affect seriously the estimators for regression  models and the situation in functional linear or quadratic models is not an exception. Robust proposals for functional linear regression models using $P-$splines or $B-$splines were considered in \citet{maronna2013robust}, \citet{boente2020robust} and \citet{kalogridis2021robust}, while an approach combining robust functional principal components and robust linear regression was studied in \citet{kalogridis2019robust}. 
 
As mentioned   in \citet{Hubert:Rousseeuw:Segaert:2015}, different types of outliers may arise  when considering functional data. These author pointed out that, in the functional setting,  atypical data    might consist of curves that behave differently from the others displaying a persistent behaviour either in shift, amplitude and/or shape making more difficult their detection. Several  detection criteria have been given in the literature based on different notions of depths, dimension reduction and/or  visualization tools. Among others, we can mention the procedures described in  
\citet{Febrero:Galeano:Gonzalez:07, Febrero:Galeano:Gonzalez:08}, \citet{Hyndman:Shang:2010}, \citet{sun2011functional},    \citet{Arribas:Romo:2014}, \citet{Rousseeuw:Raymaekers:Hubert:2018}, \citet{Dai:Genton:2019}. 

However, it should be noticed that when providing robust procedures for linear regression models with covariates in $\real^p$, outliers in the covariates are not automatically eliminated in a first step using some diagnostic method. The main reason is that the    atypical data in the explanatory variables may not always be bad high--leverage observations, since some of them may help in the fitting process.   $MM-$estimators with bounded loss functions  provide an alternative choice for robust regression, in which good leverage points are not discarded.  The same approach should be followed  when dealing with functional covariates and quadratic models, so even when  different outlier detection rules exist, it is better to adapt the best practices of robust estimation to this setting.

In this paper, we adapt the  robust procedures for  multiple linear regression estimators to the functional quadratic regression model.  More precisely, we first compute robust estimators of the  principal directions with the aim of providing    finite--dimensional candidates  for the estimators of both    the functional regression parameter and the quadratic operator. We then  apply $MM-$regression estimators \citep{yohai1987high} that are based on a bounded loss function and a preliminary residual scale estimator to the residuals obtained from these finite--dimensional spaces. The initial scale estimator ensures that the estimators of $\beta_0$ and $\Upsilon_0$ are scale equivariant, while the  bounded loss function and the robust principal directions guarantee that the resulting procedure will be robust against high--leverage outliers.  It is worth mentioning that the presence of outliers in the functional covariates may affect the estimation procedure when the sample  principal components are used to estimate the regression function and quadratic operator, even when $MM-$estimators are used. The main reason is that a distorted estimator of the principal direction will affect the scores of all the observations in that direction, that is why  robust  estimators of the principal direction  are needed. Among others, one may consider  the spherical principal components  introduced in \citet{locantore1999robust} and studied in \citet{gervini2008robust}, \citet{boente2014characterization} and \citet{boente2019spatial} or  the projection--pursuit approach considered in \citet{hyndman2007robust} and \citet{bali2011robust}.

We illustrate our approach with the Tecator data set \cite[see][]{ferraty2006nonparametric}. This food quality--control data contains 215 samples of  finely chopped meat with different percentages of fat, protein and moisture content. For each sample, a spectrometric curve of absorbances was measured using a Tecator Infratec Food and Feed Analyzer. To predict the fat content of a meat sample from  its absorbance spectrum, \citet{yao2010functional} fitted a functional quadratic model, while \citet{horvath2013test} tested the significance of the quadratic term. However,  \citet{boente2017robust} and  \citet{febrero2012statistical}, among others, showed the presence of atypical data in the  spectrometric curves. Thus, a reliable analysis of   the Tecator data set  requires procedures protecting from outliers in the absorbance spectrum.

The rest of the paper is organized as follows. The model and our proposed estimators are described in Section \ref{sec:estimadores}. Fisher--consistency of the procedure is studied in Section  \ref{sec:Fisher} both for finite--dimensional and infinite--dimensional processes. In Section \ref{sec:simu}, the performance and advantages of the proposed methods are illustrated   for finite--samples. Section \ref{sec:tecator} contains the Tecator data set analysis, while final comments are given in Section \ref{sec:concl}.

\section{Model and estimators} \label{sec:estimadores}

As mentioned in the Introduction, the   functional quadratic regression model assumes that the observations $(y_i, X_i)$, $1\le i\le n$, are independent and identically distributed realizations of the random element $(y,X)$, where $y \in \real$ is the response variable, $X$ is a stochastic process on $L^2(\itI)$, the space of square integrable functions on the interval $\itI$. The relationship between the response and the explanatory variable  is given by: 
\begin{equation}\label{eq:cuad}
y = \alpha_0+ \langle \beta_0, X \rangle + \langle X , \Upsilon_0 X \rangle +\sigma_0\;\epsilon \,,
\end{equation}
where $\langle \cdot, \cdot \rangle$ denotes the usual $L^2(\itI)$ inner product, $\epsilon$ is independent of $X$, $\sigma_0 > 0$ is the unknown error scale parameter, $\beta_0 \in L^2(\itI)$ is the regression coefficient  and  $\Upsilon_0: L^2(\itI) \to L^2(\itI)$ is the linear self--adjoint and Hilbert--Schmidt operator corresponding to the quadratic term, where  $\itI$ is a compact interval. For simplicity, we will assume that $\itI=[0,1]$. At this instance, to avoid requiring moments to the errors, but at the same time to identify the regression function, we require that   $\epsilon$ has a symmetric distribution $G(\cdot)$ with scale parameter $1$.  

Just as in \citet{kalogridis2019robust} who considered the functional linear regression model, to obtain a proper finite--dimensional approximation of $X$, we assume that   $\esp \|X\|^2<\infty$, where $\|X\|^2=\langle X, X\rangle$.  From now on, we denote as $\phi_j$, $j\ge 1$, the eigenfunctions of the covariance operator $\Gamma$ of $X$ and as $\lambda_j$, $j\ge 1$, the related eigenvalues ordered such that $\lambda_j\ge \lambda_{j+1}$, for all $j$. In such a case, the Karhunen--Lo\`{e}ve representation of $X$ is $X=\mu + \sum_{j \ge 1} \xi_{j} \, \phi_j$, where $\mu=\esp (X)$ and the scores $\xi_{j} \, = \, \langle X - \mu \, , \phi_j \, \rangle$ are uncorrelated random variables with mean zero and variance $\lambda_j$. 
 
Note that  in \citet{yao2010functional} and \citet{horvath2013test} model \eqref{eq:cuad} is written in terms of the centered process   $X^{(c)}=X-\mu$ as 
 \begin{equation}\label{eq:cuadcent}
y = \alpha_0^{*}+ \langle \beta_0^{*}, X^{(c)} \rangle + \langle X^{(c)} , \Upsilon_0 X^{(c)} \rangle +\sigma_0\;\epsilon \,.
\end{equation}
Clearly, the parameters in both models  \eqref{eq:cuad}  and  \eqref{eq:cuadcent} are related as follows  $\alpha_0^{*}= \alpha_0+ \langle \mu, \beta_0\rangle +\langle \mu, \Upsilon_0 \mu\rangle$ and $\beta_0^{*}= \beta_0+ 2\, \Upsilon_0\mu$. Taking into account the Karhunen--Lo{\'e}ve of the process $X$, \citet{yao2010functional} suggest to estimate  $\alpha_0^{*}$, $\beta_0^{*}$ and $ \Upsilon_0$  using the  scores on the linear space spanned by the first eigenfunctions of the covariance operator. Moreover, \citet{yao2010functional} derived explicit expressions for the   coefficients $b_{0,j}^{*}$, $j\ge 1$ and $v_{0,j\ell} $, $j,\ell\ge 1$ of $\beta_0^{*}$ and $\Upsilon_0$, respectively in \eqref{eq:cuadcent} and suggested to estimate them by plugging--in  the unknown scores by their predicted values and replacing the expectations by their  sample counterparts. To define the final estimators of $\beta_0^{*}$ and $ \Upsilon_0$, they  approximated  their infinite expansions  by a small one that uses only $p$ estimated eigenfunctions.   \citet{horvath2013test}   used   a finite approximation  and a least squares approach to estimate the parameters and to construct a test for significance of the quadratic operator.

Let us consider model \eqref{eq:cuad},  similar expansions than those given below can be obtained when using the centered model \eqref{eq:cuadcent}, replacing  $\alpha_0 $ and $\beta_0 $ by $\alpha_0^{*}$, $\beta_0^{*}$, respectively and $X$ by  $ X^{(c)} $, so that $x_{j}=\langle X , \phi_j\rangle$ needs to be replaced by $\langle X^{(c)} , \phi_j\rangle= \xi_j$ and $\wx_{ij}$ by $ \langle \wX_i-\wmu, \wphi_j \rangle= \wxi_{ij}$, for proper estimators $\wmu$ of $\mu$ and $\wphi_j$ of $\phi_j$. Section \ref{sec:FPC} revisits some well known robust estimators for these quantities.

To motivate the estimators to be used, we begin by expanding $\beta_0$ and $\Upsilon_0$ over the basis of eigenfunctions. Note that  $\{\phi_j\otimes \phi_\ell\}_{j\ge 1, \ell \ge 1}$  is a proper basis on the space of self--adjoint Hilbert--Schmidt operator. Hence, we have the following expansions for $\beta_0$ and $\Upsilon_0$  
\begin{equation*}
\beta_0 =\sum_{j=1}^{\infty} b_{0,j}  \phi_j\qquad \qquad \Upsilon_0=\sum_{j=1}^{\infty}  v_{0,jj} \phi_j\otimes \phi_j+\sum_{j=1}^{\infty} \sum_{\ell=j+1}^{\infty} v_{0,j\ell} \left(\phi_j\otimes \phi_\ell +\phi_\ell\otimes \phi_j \right)\,,
\label{eq:desarrollo-beta-upsilon}
\end{equation*} 
where $b_{0,j} = \langle\beta , \phi_j\rangle$ and  $v_{0,j\ell}= \langle \phi_j, \Upsilon_0 \phi_\ell\rangle = v_{0,\ell j}$ with $\sum_{j\ge 1} b_{0,j} ^2<\infty$ and $\sum_{j\ge 1} \sum_{\ell \ge 1}  v_{0,j\ell}^2<\infty$.
Thus, replacing in \eqref{eq:cuad}, we get that
\begin{align*}
y & = \alpha_0 + \langle \beta_0 , X \rangle + \langle X  , \Upsilon_0 X  \rangle +\sigma_0\;\epsilon\\
 & = \alpha_0 + \sum_{j=1}^{\infty} b_{0,j}  x_j +\sum_{j=1}^{\infty}  v_{0,jj} x_j^2+  \sum_{j=1}^{\infty} \sum_{\ell=j+1}^{\infty} v_{0,j\ell} \langle X  , \left(\phi_j\otimes  \phi_\ell +\phi_\ell\otimes \phi_j \right)\, X  \rangle +\sigma_0\;\epsilon\\
  & = \alpha_0 + \sum_{j=1}^{\infty} b_{0,j}  x_j +\sum_{j=1}^{\infty}  v_{0,jj} x_j^2+ 2\, \sum_{j=1}^{\infty} \sum_{\ell=j+1}^{\infty} v_{0,j\ell} x_j x_{\ell} +\sigma_0\;\epsilon\\
  & = \alpha_0 + \sum_{j=1}^{\infty} b_{0,j} x_j +  \sum_{j=1}^{\infty} \sum_{\ell=j}^{\infty} \left(2- \uno_{j=\ell}\right) v_{0,j\ell} x_j x_{\ell} +\sigma_0\;\epsilon\,,
\end{align*}
where $x_{j}=\langle X , \phi_j\rangle= \xi_j + \langle \mu , \phi_j\rangle$ and  $\uno_{j=\ell}$ equals 1 if $j=\ell$ and $0$, otherwise. 

It is worth mentioning that even when robust estimators of the principal directions are obtained, for atypical trajectories $X_i$ their predicted scores $\widehat{\xi}_{ ij}= \langle X_i-\wmu, \wphi_j \rangle$ may be distorted. Hence, the least squares procedure used in \citet{horvath2013test} will not lead to resistant estimators.  Moreover, vertical outliers which correspond to atypical values only in the responses may also be present in the sample, affecting also these estimators. For that reason, we will follow a different approach combining robust estimators of the principal directions and robust regression estimators.  

More precisely, assume that robust estimators of the location $\mu$ and the eigenfunctions $\phi_j$ are available and denote them $\wmu$ and $\wphi_j$, respectively. In such a case, one may predict $X_i$  using a small number $p$ of principal directions as $\wX_i= \wmu    +  \sum_{j=1}^p \,\wxi_{ ij} \, \wphi_j$, which allows to approximate the regression function $g(X)=\alpha_0 + \langle \beta_0 , X \rangle + \langle X  , \Upsilon_0 X  \rangle $  at $X_i$ as
\begin{equation*}
 g(X_i)\approx  g(\wX_i)= \alpha_0 + \langle \beta_0 ,  \wX_i  \rangle + \langle \wX_i   , \Upsilon_0 \wX_i \rangle  = \alpha_0 + \sum_{j=1}^p   \langle \wphi_j, \beta_0 \rangle \, \wx_{ ij} + \sum_{j=1}^p \sum_{\ell=1}^p    \langle \wphi_j, \Upsilon_0 \wphi_\ell\rangle\, \wx_{ ij} \wx_{ i\ell}\,, 
\label{eq:approx1}
\end{equation*}
where $\wx_{ij}= \langle \wX_i, \wphi_j \rangle= \langle \wmu, \wphi_j \rangle + \wxi_{ij}$. Noticing that $ \langle \wphi_j, \Upsilon_0 \wphi_\ell\rangle=  \langle \wphi_\ell, \Upsilon_0 \wphi_j\rangle$, since $\Upsilon_0$ is self adjoint, we can write 
\begin{align*}
g(\wX_i) & =    \alpha_0 + \sum_{j=1}^p  \langle \wphi_j, \beta_0 \rangle \, \wx_{ ij}  + \sum_{j=1}^p \sum_{\ell=j+1}^p  2 \langle \wphi_j, \Upsilon_0 \wphi_\ell\rangle  \, \wx_{ ij} \wx_{ i\ell} + \sum_{j=1}^p \langle \wphi_j, \Upsilon_0 \wphi_j\rangle  \, \wx_{ ij}^2  \\
 &= \alpha_0 + \sum_{j=1}^p  b_{j}  \, \wx_{ ij}  + \sum_{j=1}^p \sum_{\ell=j}^p u_{j\ell}\, \wx_{ ij} \wx_{ i\ell}  \,. 
%\label{eq:approx}
\end{align*}
As mentioned above, the unknown coefficients $b_j =\langle \wphi_j, \beta_0 \rangle$ and $u_{j\ell}= (2-\uno_{j=\ell}) \langle \wphi_j, \Upsilon_0 \wphi_\ell\rangle$ are estimated in \citet{horvath2013test} using a least squares approach. 

The above expansions suggest that one possible way to estimate $\beta_0$ and $\Upsilon_0$ is to restrict the set of possible candidates to those belonging to the linear spaces spanned by $\wphi_1,\dots, \wphi_p$ and $\{\wphi_j\otimes \wphi_\ell\}_{1\le j,\ell\le p}$, respectively. To ensure resistance to atypical data, including vertical outliers and atypical observations in the covariates, we will use $MM-$estimators combined with robust estimators of the principal directions, as described below in Section \ref{sec:defest}.

\subsection{Some robust principal direction estimators}{\label{sec:FPC}}
As mentioned in the Introduction, several robust estimators for the principal directions have been considered in the literature since  the spherical principal components  introduced in \citet{locantore1999robust}. As it is well known, the spherical principal directions are the eigenfunctions of the sample sign covariance function, which is just the sample covariance function of the centered curves projected on the unit sphere. More precisely,  let $\wmu$ stand for an estimator of the location of $X$ such as  the sample spatial median $\wmu_{\sm} =  \argmin_{\theta \in L^2(\itI)}\sum_{i=1}^n \left( \|X_i - \theta \| - \|X_i \| \right) $ and define the operator $\wGamma^{\mbox{\footnotesize \sc s}}$ as 
$$\wGamma^{\mbox{\footnotesize \sc s}}=\frac{1}{n}\sum_{i=1}^n     \frac{(X_i-\wmu)\otimes (X_i-\wmu)}{\|X_i-\wmu\|^2} \,.$$
\citet{gervini2008robust} and \citet{cardot2013efficient} have shown that $\wmu_{\sm}$ is a consistent estimator of the spatial median $\mu_{\sm}=\argmin_{\theta \in L^2(\itI)}\esp \left( \|X - \theta \| - \|X \| \right)$. Consistency of $\wGamma^{\mbox{\footnotesize \sc s}}$ to  the sign operator defined as
$${\Gamma}^{{\mbox{\footnotesize \sc s}}}= \esp\left\{  \frac{(X-\mu)\otimes (X-\mu)}{\|X-\mu\|^2}\right\} \,, $$
was derived in Theorem 1 in \citet{boente2019spatial}, whenever $\wmu$ is consistent to $\mu$. These authors also obtained the asymptotic distribution of $\wGamma^{\mbox{\footnotesize \sc s}}$ and that of its eigenfunctions, that is, the asymptotic distribution of the spherical principal directions, see   Proposition 1 in \citet{boente2019spatial}. 

It is worth mentioning that when considering  the spherical principal directions, one need to choose a robust estimator for the location $\mu$ of the process $X$ to center the data. As discussed in \citet{boente2019spatial}, several robust location functionals and their related estimators may be considered. Among others, the geometric median or spatial median defined above is the usual choice  when using the spatial operator. However, other choices are possible including the $\alpha-$trimmed mean defined in \citet{fraiman2001trimmed} or the $M-$estimators defined in \citet{sinova2018m} which are consistent under some model assumptions. Also, estimators defined through a suitable depth notion may be used defining the related median as the deepest point. Note that all these procedures provide Fisher--consistent estimators when considering a symmetric process around $\mu \in \itH$, meaning that $X-\mu$ and $\mu-X$ have the same distribution.   

As when estimating $\mu$, one important issue to be considered when defining robust estimators of the directions $\phi_j$ is that they are indeed estimating the target directions, a property which is usually known as Fisher--consistency. Theorem 3 in \citet{gervini2008robust} shows that if the process is finite--dimensional, i.e., $X=\mu + \sum_{k = 1}^q \xi_{k} \, \phi_k$ and the standardized scores  $(\xi_{1}/\sqrt{\lambda_1}, \dots, \xi_{q}/\sqrt{\lambda_q})$, $\lambda_1\ge \dots \ge \lambda_q>0$, have a symmetric distribution with exchangeable marginals, then the eigenfunctions of $\Gamma^{\mbox{\footnotesize \sc s}}$ are $\phi_j$, $1\le j\le q$. Furthermore,  as mentioned in \citet{boente2014characterization}, for infinite--dimensional processes, if $X$ is an elliptical process $\itE(\mu,   \Gamma)$, then $\Gamma^{\mbox{\footnotesize \sc s}}$ has the same eigenfunctions as $\Gamma$ and in the same order. These two properties do not require the existence of second moments, making the procedure adequate when we suspect that atypical curves may arise among the functional covariates. 

Other procedures to  robustly estimate the principal directions include the projection--pursuit approach considered in \citet{hyndman2007robust} and generalized in \citet{bali2011robust} to include a penalization, so as to ensure that the principal direction estimators are smooth. The projection--pursuit estimators provide a Fisher--consistent method at elliptical processes  and  consistent estimators, under mild conditions. A procedure based on projecting the observations over a known basis and performing robust principal components analysis on the coefficients has been proposed in \citet{sawant2012functional}. Robust alternatives based on estimating the eigenspace have been also considered. In this direction, we can mention  the $M-$type smoothing spline estimators proposed in \citet{lee2013m} who proposed a sequential algorithm that robustly fits one--dimensional linear spaces. Other alternatives are the $S-$estimators defined in \citet{boente2015sestimators} or those defined in \citet{cevallos2016methods} who, as in \citet{sawant2012functional}, consider the coefficients of the data over a   finite--dimensional basis and then apply a robust multivariate method to estimate principal subspaces. It is worth mentioning that, even when, for elliptically distributed random processes, these last two procedures are Fisher--consistent methods to estimate the linear space spanned by the first eigenfunctions, they do not give estimators of the principal directions themselves but to the linear space spanned by them, so a proper basis in that space should then be selected.    

From now on, $\wphi_j$, $1\le j\le p$, will stand for the principal direction estimators obtained by one of these methods.

\subsection{The estimators of $\beta_0$ and $\Upsilon_0$}{\label{sec:defest}}

As mentioned above, one way to regularize the problem and avoid the curse of dimensionality imposed by dealing with functional covariates is to restrict the set of possible candidates for the estimators of $\beta_0$ and $\Upsilon_0$ to those belonging to the linear spaces spanned by $\wphi_1,\dots, \wphi_p$ and $\{\wphi_j\otimes \wphi_\ell\}_{1\le j,\ell\le p}$. For that purpose, from now on, we denote as $\vech(\cdot)$ the half--vectorization that  stacks the columns of the lower triangular portion of the
matrix under each other. 

To define our estimators, for any symmetric matrix $\bV$, we define $u_{j\ell}=(2-\uno_{j=\ell}) v_{j\ell} $ and  $\bu=\vech(\{u_{j\ell}\})_{1 \le j\le \ell\le p}=\vech(\{(2-\uno_{j=\ell}) v_{j\ell}\})_{1 \le j\le \ell\le p}\in \real^{p\times(p+1)/2}$. Furthermore, given   $\bb\in \real^{p}$ and symmetric matrix $\bV\in \real^{p \times p}$,
let  $\beta_{\bb}$ and $\Upsilon_{\bu}$ stand for
\begin{align}
\beta_{\bb} &= \sum_{j=1}^{p} b_j \,    \wphi_j \;,\nonumber\\ 
  \Upsilon_{\bu}& = \sum_{j=1}^{\infty} \sum_{\ell=1}^{\infty} v_{j\ell}  \wphi_j\otimes \wphi_\ell  =\sum_{j=1}^{p} \sum_{\ell=j }^{p} (2-\uno_{j=\ell}) v_{j\ell} \,    \wphi_j\otimes \wphi_\ell=\sum_{j=1}^{p} \sum_{\ell=j }^{p}   u_{j\ell} \,    \wphi_j\otimes \wphi_\ell  \,.
  \label{eq:candidatoUPS}
  \end{align}
To define the robust estimators, we use robust regression $MM-$estimators \citep{yohai1987high}, that is, we compute a residual scale estimator using an initial robust regression estimator and then we calculate a regression $M-$estimator using a bounded loss function and standardized residuals. 

In what follows the loss functions $\rho_j : \real \to \real_+$, $j=0,1$ to be used below correspond to  bounded  $\rho-$functions as defined in  \citet{maronna2019robust}. The Tukey's bisquare function $\rho_{\,\tuk,\,c}(t) =\min\left(1 - (1-(t/c)^2)^3, 1\right)$  provides an example of bounded $\rho-$function. The tuning parameter $c>0$ is chosen to balance   the robustness and efficiency properties of the associated estimators. 
 
We define the residuals $r_i(a,\beta_{\bb}, \Upsilon_{\bu})$, $1 \le i \le n$, with respect to the corresponding approximations  $\beta_{\bb}$ and $\Upsilon_{\bu} $ as
\begin{equation*} \label{eq:residuals}
r_i(a,\beta_{\bb}, \Upsilon_{\bu}) \, =  \, y_i - a-\sum_{j=1}^{p} b_j \, \wx_{ij} -  \sum_{j=1}^{p} \sum_{\ell=1}^{p} v_{j\ell} \,  \wx_{ij}\,  \wx_{i\ell} %= y_i - a-\sum_{j=1}^{p} b_j \, \wx_{ij} - \sum_{j=1}^{p}\sum_{\ell=j}^{p} u_{j\ell} \wx_{ij}\,  \wx_{i\ell}
\, = \, y_i - a - \bb\trasp \wbx_i - \bu\trasp \wbz_i \, ,
\end{equation*}
where $\wbx_i= (\wx_{i1}, \dots, \wx_{ip})\trasp$,   $\wbz_i=(\wz_{i,1}, \dots, \wz_{i,q})\trasp$ with
$\wbz_i=\vech(\{(\wx_{ij}\,\wx_{ i\ell}\})_{1 \le j\le \ell\le p}\in \real^q$, $q={p\times(p+1)/2}$, and $\wx_{ij} = \langle X_i , \wphi_j\rangle$.

First, we compute an $S-$estimator of regression and its associated residual scale. Let $\rho_0$ be a bounded $\rho-$function and $s_n(a,\beta_{\bb}, \eta_{\ba} )$ be the $M-$scale estimator of the residuals given as the solution to the following equation:
\begin{equation} \label{eq:s-est}
\frac{1}{n-(p+q)}\sum_{i=1}^n \rho_{0}\left(\frac{r_i(a,\beta_{\bb}, \Upsilon_{\bu})}{s_n(a,\beta_{\bb}, \Upsilon_{\bu})}\right) \, = \, b\, ,
\end{equation}
where $b = E( \rho_{0} (\epsilon) )$. This choice of $b$ ensures that the scale estimators are indeed Fisher--consistent. Note that as in \citet{boente2020robust}, we use $1 / (n-(p +q))$ instead of $1/n$ in \eqref{eq:s-est} above to control the effect of a possibly large number of parameters ($p + q$) relative to the sample size \cite[see][]{maronna2019robust}. Recall that if  $\rho_{0}=\rho_{\,\tuk,\,c_0}$,
the choices $c_0 = 1.54764$ and $b = 1/2$ above yield a scale estimator that is Fisher--consistent when the errors have a normal distribution, and with a 50\% breakdown point in finite--dimensional regression models.
$S-$regression estimators are defined as the minimizers of the $M-$scale above:
\begin{equation} \label{eq:m-scale}
(\wa_{\ini},\wbb_{\ini}, \wbu_{\ini}) \ = \ \argmin_{a,\bb, \bu}  \, s_n(a,\beta_{\bb}, \Upsilon_{\bu}) \, .
\end{equation}
The associated residual scale estimator is 
\begin{equation*} \label{eq:scale_est}
\wsigma  \, =  
s_n(\wa_{\ini},\beta_{\wbb_{\ini}}, \Upsilon_{\wbu_{\ini}}) =\min_{a,\bb, \bu}   s_n(a,\beta_{\bb}, \Upsilon_{\bu})\, .
\end{equation*}
 
Let $\rho_{1}$ be a $\rho-$function such that  $\rho_{1} \le \rho_{0}$ and $\sup_t\rho_1(t)=\sup_t\rho_0(t)$. As it is well known, if $\rho_{0}=\rho_{\,\tuk,\,c_0}$ and $\rho_{1}=\rho_{\,\tuk,\,c_1}$, then $\rho_0\le \rho_1$ when $c_1>c_0$. We now compute an $M-$estimator using the residual scale estimator $\wsigma$ and the loss function  $\rho_{1}$ as
\begin{equation}
(\wa, \wbb, \wbu)  \ = \    \argmin_{a,\bb, \bu} L_n( a,\bb, \bu) \ = \    \argmin_{a,\bb, \bu} \sum_{i=1}^n \rho_{1} \left ( \frac{r_i(a,\beta_{\bb}, \Upsilon_{\bu}) }{\wsigma} \right )\,, 
\label{eq:estfinitos}
\end{equation}
where $L_n( a,\bb, \bu) = \sum_{i=1}^n \rho_{1} \left (  {r_i(a,\beta_{\bb}, \Upsilon_{\bu}) }/{\wsigma} \right )$. Note that $\wa$ provides an estimator of $\alpha_0$, that will be denoted $\walfa$. If we denote as $\wu_{j\ell}$, the elements of a symmetric matrix such that $ \wbu = \vech(\{ \wu_{j\ell}  \}_{1 \le j\le \ell\le p} )$, the resulting estimators of the regression function $\beta_0 $ and the quadratic operator  $\Upsilon_0$ are given by
\begin{eqnarray}
\wbeta = \sum_{j=1}^{p} \wb_j \wphi_j \, , \quad   \mbox{and }  \quad \wUps= \sum_{j=1}^{p} \wu_{jj} \wphi_j\otimes \wphi_j + \sum_{1\le j< \ell\le p}    \frac{1}{2}\wu_{j\ell} \left(\wphi_j\otimes \wphi_\ell +\wphi_\ell\otimes \wphi_j\right) \,.
\label{eq:estimadores}
\end{eqnarray}
It is worth mentioning that if model \eqref{eq:cuadcent} is considered, given a robust consistent estimator $\wmu$ of $\mu$,   estimators of $\alpha_0^{*}$ and  $\beta_0^{*} $ may be constructed from those obtained in   \eqref{eq:estimadores} as
\begin{equation}
\label{eq:estimadores-star1}
 \walfa^{*}=\walfa+ \langle \wmu, \wbeta\rangle+ \langle \wmu, \wUps \wmu\rangle \qquad \mbox{and} \qquad \wbeta^{*}=\wbeta + 2\, \wUps \wmu\,.
 \end{equation}
Note that when defining both $ \walfa^{*}$ and $\wbeta^{*}$ in the above expressions, only the coordinates of $\wmu$ on the finite--dimensional basis $\wphi_1,\dots, \wphi_p$ are used. 

It is worth mentioning that the transformation made in \eqref{eq:estimadores-star1} to construct the estimators $ \walfa^{*}$ and $\wbeta^{*}$ is equivalent to directly obtaining estimators of $\alpha_0^{*}$, $\beta_0^{*} $ and $\Upsilon_0$ using $MM-$estimators with the predicted scores. More precisely, let us denote   $\wbxi_i= (\wxi_{i1}, \dots, \wxi_{ip})\trasp$,   $\wbz_i^{*}=(\wz_{i,1}^{*}, \dots, \wz_{i,q}^{*})\trasp$ with
$\wbz_i^{*}=\vech(\{\wxi_{ij}\,\wxi_{ i\ell}\})_{1 \le j\le \ell\le p}\in \real^q$, $q={p\times(p+1)/2}$, and $\wxi_{ij} = \langle X_i-\wmu , \wphi_j\rangle=x_{ij}- \langle \wmu, \wphi_j \rangle$ and the possible candidates for estimating $\beta_0^{*}$ as 
$$\beta_{\bb^{*}}^{*}= \sum_{j=1}^{p} b_j^{*} \,    \wphi_j \, ,$$
where    $\bb^{*}\in \real^{p}$. The possible  candidates for estimating the quadratic operator are given by  \eqref{eq:candidatoUPS}. The residuals are now defined as 
$$ 
r_i^{*}(a^{*},\beta_{\bb^{*}}^{*}, \Upsilon_{\bu}) \, =  \, y_i - a^{*} -\sum_{j=1}^{p} b_j^{*} \, \wxi_{ij}   
 -  \sum_{j=1}^{p} \sum_{\ell=1}^{p} v_{j\ell} \,  \wxi_{ij}\,  \wxi_{i\ell}  = \, y_i - a^{*} - \bb^{*\,\traspbis} \wbxi_i - \bu\trasp \wbz_i^{*} \, ,
$$
where we use the upper--script $^{*}$ to make clear that we are dealing with the centered model. 
Then, one may consider $S-$regression estimators to obtain estimators of $\sigma_0$, that is, 
\begin{equation*} \label{eq:scale_est-star}
\wsigma^{*}  \, =  
s_n^{*}(\wa_{\ini}^{*},\beta_{\wbb_{\ini}}^{*}, \Upsilon_{\wbu_{\ini}}^{*}) =\min_{a^{*},\bb^{*}, \bu}   s_n^{*}(a^{*},\beta_{\bb^{*}}^{*}, \Upsilon_{\bu})\, ,
\end{equation*}
where $s_n^{*}(a^{*},\beta_{\bb^{*}}^{*}, \Upsilon_{\bu})$  is defined as in \eqref{eq:s-est}, that is, 
\begin{equation*} \label{eq:s-est-star}
\frac{1}{n-(p+q)}\sum_{i=1}^n \rho_{0}\left(\frac{r_i^{*}(a^{*},\beta_{\bb^{*}}^{*}, \Upsilon_{\bu})}{s_n^{*}(a^{*},\beta_{\bb^{*}}^{*}, \Upsilon_{\bu})}\right) \, = \, b\, .
\end{equation*}
 The $M-$estimator of the coefficient is obtained using the residual scale estimator $\wsigma^{*}$ and the loss function  $\rho_{1}$ as
\begin{equation}
(\walfa^{*}, \wbb^{*}, \wbu^{*})  \ = \    \argmin_{a^{*},\bb^{*}, \bu} L_n^{*}( a^{*},\bb^{*}, \bu) \ = \    \argmin_{a^{*},\bb^{*}, \bu} \sum_{i=1}^n \rho_{1} \left ( \frac{r_i^{*}(a^{*},\beta_{\bb^{*}}^{*}, \Upsilon_{\bu}) }{\wsigma^{*}} \right )\,. 
\label{eq:estfinitos*}
\end{equation}
 The resulting estimators of    $\beta_0^{*} $ and    $\Upsilon_0$ are then equal to
\begin{eqnarray}
\wbeta^{*} = \sum_{j=1}^{p} \wb_j^{*} \wphi_j \, , \quad   \mbox{and }  \quad \wUps= \sum_{j=1}^{p} \wu_{jj}^{*} \wphi_j\otimes \wphi_j + \sum_{1\le j< \ell\le p}    \frac{1}{2}\wu_{j\ell}^{*} \left(\wphi_j\otimes \wphi_\ell +\wphi_\ell\otimes \wphi_j\right) \,,
\label{eq:estimadores-star}
\end{eqnarray}
where we used the notation $ \wbu^{*}= \vech(\{ \wu_{j\ell}^{*} \}_{1 \le j\le \ell\le p} )$. Taking into account the relations between $\wbx_{i}$ and $\wbz_{i}$ with  $\wbxi_{i}$ and  $\wbz_i^{*}$, respectively, straightforward calculations allow to see that, for any $a \in \real$, $\bb \in \real^p$ and   symmetric matrix $\bV \in \real^{p\times p}$, we have  
$  r_i(a,\beta_{\bb}, \Upsilon_{\bu})\,=\, r_i^{*}(a^{*},\beta_{\bb^{*}}^{*}, \Upsilon_{\bu}) $, where 
\begin{align*}
a^{*} & = a + \sum_{j=1}^p b_j \wmu_j + \sum_{j=1}^p \sum_{\ell=1}^p  v_{j\ell }\wmu_j \wmu_{\ell}= a + \bb\trasp  \wbmu  + \sum_{j=1}^p \sum_{\ell=j}^p (2-\uno_{j=\ell}) v_{j\ell} \wmu_j \wmu_{\ell} = a + \bb\trasp  \wbmu  + \bu\trasp \wbnu \,,\\
b_j^{*} & = b_j +   \sum_{\ell=1}^p \left(  v_{j\ell}+ v_{ \ell\,j } \right)  \wmu_{\ell} = b_j + 2  \sum_{\ell=1}^p    v_{j\ell}  \wmu_{\ell}  = b_j + 2 \langle \Upsilon_{\bu} \wmu, \wphi_j\rangle= b_j + 2  \sum_{\ell=j}^p    u_{j\ell}  \wmu_{\ell} = b_j + 2      \bu^{(j)\,\traspbis}  \wbmu  \,,
\end{align*}
with $\wmu_j=\langle \wmu, \wphi_j\rangle$, $\wbmu =(\wmu_1,\dots, \wmu_p)\trasp$,  $\wbnu= \vech(\{(\wmu_{j}\,\wmu_{ \ell}\})_{1 \le j\le \ell\le p}\in \real^q$ and $\bu=(\bu^{(1)\,\traspbis}, \dots, \bu^{(p)\,\traspbis})\trasp$ where $\bu^{(j)\,\traspbis}\in \real^{p-j+1}$. Hence, due to the equivariance of $M-$estimators, we have that $\wsigma^{*}=\wsigma$, $\wbu^{*}=\wbu$, $ \walfa^{*}=\walfa+ \langle \wmu, \wbeta\rangle+ \langle \wmu, \wUps \wmu\rangle$ and $ \wb_j^{*}=\wb_j + 2\, \wu_{jj} \,\langle \wphi_j, \wmu\rangle + \sum_{  \ell=j+1}^p \wu_{j\ell}  \,\langle \wphi_\ell, \wmu\rangle$ which leads to the estimators defined in \eqref{eq:estimadores-star1}. Even though either considering model \eqref{eq:cuad} or \eqref{eq:cuadcent} there is one-to-one transformation that relate the  $MM-$estimators in one model to those in the other one, when considering Fisher--consistency it will be better to use the proposed procedure under model \eqref{eq:cuadcent}. The main reason is that, unless the kernel of the covariance operator of $X$ reduces to $\{0\}$, the parameters are not uniquely identified. Under  model \eqref{eq:cuad}, this lack of uniqueness also affects the intercept parameter, while under  model \eqref{eq:cuadcent} it only involves   $\beta_0^{*} $ and    $\Upsilon_0$.

\subsection{Additional remarks}
\paragraph{Smooth estimators} 
The above considered estimators of the regression parameter $\beta_0$ and the quadratic kernel $\upsilon_0$ will be Fisher--consistent under mild regularity conditions (see Section \ref{sec:Fisher}) and our simulation study will show that they also are resistant to high--leverage trajectories. However, if smoothness of the estimated  function and kernel is a requirement, two possibilities arise. On the one hand,  as done in the functional linear model by \citet{kalogridis2019robust}, the practitioner  may add a penalty term in  the loss functions $L_n( a,\bb, \bu)$ or  $L_n^{*}( a^{*},\bb^{*}, \bu) $ defined in \eqref{eq:estfinitos} or \eqref{eq:estfinitos*}, respectively. However, unlike the method proposed by these authors, even when working with the centered trajectories and model \eqref{eq:cuadcent}, the quadratic model does not allow to obtain easily the penalized coefficients  transforming those obtained from \eqref{eq:estfinitos} as in ridge regression, since the predicted squared scores $\wxi_{ij}^2$ are not centered. Instead of adding computational burden to the procedure, we suggest to use   the smoothed robust principal components defined in \citet{bali2011robust} obtained  penalizing the scale. The choice of this basis, which is an orthonormal basis,   guarantees the smoothness of the regression and the quadratic kernel estimators.

\paragraph{Sparse trajectories} 
The estimators defined in Section \ref{sec:defest} only depend on proper predictors of the scores and estimators of the location $\mu$ and the eigenfunctions $\phi_j$, since $\wx_{ij} = \langle X_i , \wphi_j\rangle = \langle \wmu, \wphi_j \rangle + \wxi_{ij}$. Hence, as mentioned in the classical case by \citet{yao2010functional}, they can be implemented when the functional predictors are derived from sparse and irregular measurements.  The non--robust procedure proposed in \citet{yao2005functional} predicts the scores and estimates the principal directions via the so--called Principal Analysis by Conditional Estimation (PACE) algorithm, implemented in \texttt{R} through the package \texttt{fdapace} developed by \citet{chen2020pace}.  

When the practitioner suspects that outliers may be present among the functional covariates, the center function $\mu$ may be estimated by aggregating the available information, for instance, using a robust local $M-$estimator as in \citet{boente2021robust} or the $M-$type smoothing spline one defined in \citet{kalogridis2021robust}. The stepwise procedure proposed in \citet{boente2021robust} may be used to estimate the scatter operator $\Gamma$ and its eigenfunctions. Besides, the scores may be predicted using the conditional distribution of the scores given the observed trajectories. The robust $MM-$estimators defined in \eqref{eq:estimadores-star} can be obtained using the predicted scores $\wxi_{ij}$ and the estimator $\wmu$ of $\mu$.

\paragraph{Semi--functional models} 
The estimators defined above can  be extended to other  functional   models, such as those involving a nonparametric component using $B-$splines to approximate the unknown function. More specifically, consider the model 
$$
y_i \, =  \langle X_i, \beta_0 \rangle +  \langle X_i, \Upsilon_0 X_i\rangle+  \eta_0(z_i) + \sigma_0 \, \epsilon_i \, ,
$$
where   $z_i \in \itJ$, $1\le i \le n$, is another explanatory variable and $\eta_0: \itJ\to \real$ is an unknown smooth function with $\itJ$ a compact interval which, without loss of generality, we will assume equal to $ \itJ=[0,1]$.  Note that in order to identify $\eta_0$ the intercept coefficient is avoided in the above model. To define $MM-$estimators in this setting, we consider $B-$splines estimators of $\eta_0$. More precisely, we fix a desired spline order $\ell$ and consider  $m_n$ knots  to approximate  $\eta_0$. The corresponding $B-$spline  basis has dimension  $k=k_{n}=m_n +\ell$, see Corollary 4.10 in \citet{schumaker1981spline} and will be denoted $\{B_j : 1\leq j\leq k_{n}\}$. In the sequel, 
 $\eta_{\ba}$ stands for the spline $\eta_{\ba} (z)=\sum_{j=1}^{k} a_j \,   B_j(z )$.

The residuals  are now defined as
$$r_i( \eta_\ba,\beta_{\bb}, \Upsilon_{\bu}) \, =  \, y_i -   \sum_{j=1}^{p} b_j \, \wx_{ij} - 
\sum_{j=1}^{p}\sum_{\ell=j}^{p} u_{j\ell} \wz_{ij\ell}-\sum_{j=1}^{k} a_j \,   B_j(z_i )\, = \, y_i - \bb\trasp \wbx_i - \bu\trasp \wbz_i -\ba\trasp \bB_i \,,$$
with  $\bB_i=( B_1 (z_i), \dots,  B_{k} (z_i))\trasp$ and the estimators may be defined as before, but   minimizing over $(\ba,\bb, \bu)\in \real^{ k+ p +q}$ in \eqref{eq:m-scale} and \eqref{eq:estfinitos}.

\section{Fisher--consistency}{\label{sec:Fisher}}
From now on, denote as $P$ the probability measure related to $(y,X)$ where the response $y$ and the functional covariates satisfy the quadratic model \eqref{eq:cuad} and as $P_X$ the probability measure of $X$. Let $\Gamma$ be the covariance operator of $X$ and denote as $\phi_j$, $j\ge 1$, its eigenfunctions with related   eigenvalues  $\lambda_1 \geq \lambda_2 \geq \dots$. In this section, we will study  the functionals associated to the estimators of the centered model \eqref{eq:cuadcent}. For notation simplicity, we omit the upper--script $^{*}$. Two situations will be considered. In the first one, as in \citet{kalogridis2019robust} we will assume that the process is a finite--dimensional one, that is, $\lambda_j=0$, for $j>q$ and some $q\ge 1$. Therefore,  only the components of $\beta_0$ and $\Upsilon_0$ over the linear space induced by the corresponding eigenfunctions will be identifiable. This motivates the definition of the projections given in \eqref{eq:beta0p}, below. In the second case, we will consider an infinite--dimensional process where the kernel of  $\Gamma$, denoted $\mbox{ker}(\Gamma)$, reduces to $\{0\}$, in which case smoothness assumptions will be required to the eigenfunctions of $\Gamma$.

\subsection{Fisher--consistency for finite--dimensional processes}{\label{sec:FCfinito}}
Given $p$, define the projections of $\mu$, $\beta_0$ and $\Upsilon_0$ over the finite--dimensional spaces spanned by $\phi_1, \dots, \phi_p$ and $\{\phi_j\otimes \phi_{\ell}\}_{1\le j,\ell \le p}$ as  
\begin{equation}
\mu_p =\sum_{j=1}^p \langle \mu, \phi_j\rangle \phi_j\,, \qquad \beta_{0,p} = \sum_{j=1}^p \langle \beta_0, \phi_j\rangle \,\phi_j\qquad \mbox{and} \qquad  \Upsilon_{0,p}=  \sum_{j=1}^p\sum_{\ell=1}^p \langle \phi_j, \Upsilon_{0} \phi_\ell \rangle \,\phi_j\otimes \phi_{\ell}\,.
\label{eq:beta0p}
\end{equation}
The functional related to the   estimation procedure described in Section  \ref{sec:defest} can be defined as follows. Let $\mu_{\rob}(P_X)$ be a location functional for the process $X$. For $1\le j\le p$ denote as $\phi_{\rob,j}(P_X)$   any robust  principal component direction  functional  from which the estimators are constructed and as $\lambda_{\rob,j}(P_X)$ the related eigenvalues ordered such that $\lambda_{\rob,j}(P_X)\ge  \lambda_{\rob,j+1}(P_X)$. For each fixed $p$, define $\bx_p(P_X)=(\langle X-\mu_{\rob}(P_X), \phi_{\rob,1}(P_X)\rangle, \dots, \langle X-\mu_{\rob}(P_X), \phi_{\rob,p}(P_X)\rangle)\trasp$ and $\bz_p(P_X)=\vech(\{( x_{j}(P_X)\, x_{  \ell}(P_X)\})_{1 \le j\le \ell\le p}\in \real^q$, $q={p\times(p+1)/2}$, with $x_{j}(P_X) = \langle X -  \mu_{\rob}(P_X), \phi_{\rob,j}(P_X)\rangle$. 

Given a $\rho-$function $\rho_1$ denote as $\beta(P)$ and $\Upsilon(P)$ the functionals, 
\begin{align}
\beta(P) &=  \sum_{j=1}^{p} b_j(P) \phi_{\rob,j}(P_X) \, ,\nonumber\\
\Upsilon(P)&=  \sum_{j=1}^{p} u_{jj}(P) \phi_{\rob,j}(P_X)\otimes \phi_{\rob,j}(P_X) +   \frac{1}{2} \sum_{1\le j< \ell\le p}    u_{j\ell}(P) \left(\phi_{\rob,j}(P_X)\otimes \phi_{\rob,\ell}(P_X) +\phi_{\rob,\ell}(P_X)\otimes \phi_{\rob,j}(P_X)\right) 
\nonumber\\
& = \sum_{1\le j, \ell\le p}    v_{j\ell}(P)  \phi_{\rob,j}(P_X)\otimes \phi_{\rob,\ell}(P_X)\,,
\label{eq:funcional} 
\end{align}
where $ v_{jj}(P)= u_{jj}(P) $, $ v_{j\ell}(P) = v_{ \ell\, j}(P)  =  u_{j\ell}(P) /2$ and $(\alpha(P), \bb(P), \bu(P))$ are such that
\begin{equation}
(\alpha(P), \bb(P), \bu(P))  \ = \    \argmin_{a,\bb, \bu}  \esp\rho_{1} \left ( \frac{  y -a- \bb\trasp \bx_p(P_X)- \bu\trasp \bz_p(P_X)  }{\sigma_0} \right )\,. 
\label{eq:funfinitos} 
\end{equation}
For simplicity, we denote as $\itF$ the space of linear, self--adjoint and Hilbert--Schmidt operators $\Upsilon:L^2(0,1)\to L^2(0,1 )$ and as $\itF_p$ the finite--dimensional linear space of $\itF$ defined as
$$\itF_p= \{\Upsilon \in \itF: \Upsilon= \sum_{1\le j,\ell\le p} v_{j\ell} \phi_j\otimes \phi_{\ell} \mbox{ with } v_{j\ell} =v_{ \ell j} \}\,.$$  
Moreover, $\itH_p$ will stand for the linear space spanned by the first $p$ eigenfunctions of $\Gamma$, that is,  $\itH_p=\{\beta\in L^2(0,1): \beta=\sum_{j=1}^p b_j \phi_j \mbox{ for some } b_j\in\real\}$. Then,  the set of the possible candidates allowing to define $(\alpha(P),\beta(P) ,\Upsilon(P))$ equals
$\itC_p=\real\times \itH_p\times\itF_p$.

Throughout this section, we will consider the following assumptions  
\begin{enumerate}[label=\textbf{C\arabic*}]

\item\label{ass:rho}: The function $\rho : \real \to [0, \infty)$ is bounded, continuous, even, non--decreasing on $[0,+\infty)$, and such that $\rho(0)=0$. Moreover, $\lim_{u\to \infty} \rho(u)\ne 0$ and if $0 \leq u< v$ with $\rho(v) < \sup_u \rho(u)$ then $\rho(u)<\rho(v)$. Furthermore, assume that $ \sup_u \rho(u)=1$.  
 
\item\label{ass:densidad}: The random variable $\epsilon$ has a density function $f_0(t)$ that is even, non-increasing in $|t|$, and strictly decreasing for $|t|$ in a neighbourhood of $0$.  
\item \label{ass:FCphies}: $\mu_{\rob}(\cdot)$ and $\phi_{\rob,j}(\cdot)$,$1\le j\le p$, are Fisher--consistent at $P_X$.
\item \label{ass:finiteKL}: $X$ has a finite--dimensional Karhunen--Lo{\`e}ve decomposition given by $X=\mu+\sum_{j=1}^q \xi_j \phi_j$.
\item \label{ass:probaX}:  $\prob(\langle X-\mu,\beta\rangle + \langle X-\mu,\Upsilon (X-\mu)\rangle=a )<1$, for any $(a, \beta,\Upsilon)\in \itC_q$ such that  $(a, \beta,\Upsilon)\ne 0$, where $q$ is given in \ref{ass:finiteKL}.
%\item \label{ass:probaX1}:  $\prob(\langle X-\mu,\beta\rangle + \langle X-\mu,\Upsilon (X-\mu)\rangle=a ) <1$, for any $a\in \real$, $\beta\in L^2(0,1)$ and any operator $\Upsilon \in \itF$.
 \end{enumerate}

\begin{remark}
Condition \ref{ass:rho} corresponds to the requirements in  \citet{maronna2019robust} for a bounded  $\rho-$function. Condition \ref{ass:densidad} is a usual assumption when considering robust estimators in   linear regression models. Conditions ensuring Fisher--consistency of the principal direction functional and of the location functional required in \ref{ass:FCphies} were discussed in Section \ref{sec:FPC}. 
It is worth mentioning that, when assumption \ref{ass:FCphies} holds, \eqref{eq:funcional} and \eqref{eq:funfinitos} can be written as
 $$(\alpha(P), \beta(P), \Upsilon(P))  \ = \    \argmin_{a\in \real,\beta\in \itH_p, \Upsilon\in \itF_p}  L(a,\beta,\Upsilon, \mu,\sigma_0)\,,$$ 
where 
 $$L(a,\beta,\Upsilon,\nu, \sigma)=\esp \,\rho_1\left(\frac{y -a- \langle X-\nu, \beta\rangle - \langle X-\nu,\Upsilon (X-\nu)\rangle}{\sigma}\right)\,.$$
 Moreover, under  \ref{ass:finiteKL}, we have that 
$$L(\alpha_0,\beta_0,\Upsilon_0,\mu, \sigma)= L(\alpha_0,\beta_{0,q},\Upsilon_{0,q}, \mu,\sigma)= L(\alpha_0,\beta_{0,q},\Upsilon_{0,q}, \mu_q,\sigma)=  \esp\rho_{1} \left ( \frac{  y - \bb\trasp \bx_q - \bu\trasp \bz_q  }{\sigma} \right )\,,$$ 
where $\mu_q$ is defined in \eqref{eq:beta0p}, $\bx_q=(x_1,\dots, x_q)\trasp=(\langle X-\mu, \phi_1\rangle, \dots,\langle X-\mu, \phi_q\rangle)\trasp=(\langle X-\mu_q, \phi_1\rangle, \dots,\langle X-\mu_q, \phi_q\rangle)\trasp$ and $\bz_q=\vech(\{ x_j x_{\ell} \})_{1 \le j\le \ell\le q}$. 

Note that, for any  $\beta=\sum_{j=1}^q b_j \phi_j  \in \itH_q$ and   $\Upsilon= \sum_{1\le j,\ell\le q} v_{j\ell} \phi_j\otimes \phi_{\ell} \in \itF_q$, we have that $\langle X-\mu,\beta\rangle + \langle X-\mu,\Upsilon (X-\mu)\rangle= \bxi\trasp \bb + \bxi\trasp \bV\bxi$, where  $\bxi=(\xi_1,\dots, \xi_q)\trasp$, $\bV\in \real^{q\times q}$ is the symmetric matrix with  $(j,\ell)-$element $v_{j\ell}$ and $\bb=(b_1,\dots, b_q)\trasp$. Hence, assumption  \ref{ass:probaX} holds if $\Lambda_q^{-1/2}\bxi$ is absolutely continuous. If in addition   $\Lambda_q^{-1/2}\bxi$ is an spherically distributed random vector, we obtain  assumption  (C2) in \citet{kalogridis2019robust}. Furthermore,  \ref{ass:probaX}  is valid not only under   \ref{ass:finiteKL}, but also when $X$ is elliptically distributed, i.e., $X \sim \itE(\mu, \Gamma,\phi)$, as defined in \citet{bali2009} and  $\Gamma$ does not have finite rank.
\end{remark}

\begin{remark}
A more restrictive assumption than \ref{ass:probaX}, is to require that  $\prob(\langle X-\mu,\beta\rangle + \langle X-\mu,\Upsilon (X-\mu)\rangle=a ) <1$, for any $(a, \beta, \Upsilon)\in \real\times L^2(0,1)\times \itF$. In such case, the kernel of the covariance operator $\Gamma$,  denoted $\mbox{ker}(\Gamma)$, reduces to $\{0\}$. However, our set of possible candidates is a finite--dimensional space and for that reason, \ref{ass:finiteKL} and \ref{ass:probaX} are required. In particular, note that when   \ref{ass:finiteKL} holds, $\mbox{ker}(\Gamma)$ is the infinite--dimensional linear space orthogonal to the linear space spanned by $\phi_1,\dots, \phi_q$. In this case, as mentioned in \citet{cardot2003spline} for the functional linear regression model,  the parameters $\alpha_0$, $\beta_0$ and $\Upsilon_0$ are not uniquely defined. Effectively,  for any $\gamma\in  \mbox{ker}(\Gamma)$, $\prob(\langle X-\mu, \gamma \rangle = 0)=1$, so that
$\prob(\langle X-\mu, \beta_0 \rangle = \langle X-\mu,\beta_0 +\gamma\rangle)=1$ and the model may be reparametrized using $\beta_0+\gamma$. Similarly, the linear operator $\Upsilon_0+  \gamma\otimes\gamma$ also provides a  valid parametrization for the quadratic parameter in model \eqref{eq:cuadcent}, meaning that the parameters in \eqref{eq:cuadcent} (or  \eqref{eq:cuad}) are not identifiable. Hence, under assumption \ref{ass:finiteKL}, the linear and quadratic parameters should be defined via equivalence class identifying all the parameters $(\beta_0,\Upsilon_0)$ whose projection over $\itH_q\times \itF_q$ is the same.  Taking into account that model  \eqref{eq:cuadcent} remains unchanged if we replace $\beta_0$ and $\Upsilon_0$ by $\beta_{0,q}$ and $\Upsilon_{0,q}$, respectively, it is sensible to define    $\beta_{0,q}$ and $ \Upsilon_{0,q}$ as in \eqref{eq:beta0p}, that is, as the projection of $\beta_0$ and $\Upsilon_0$ over $\itH_q$ and $\itF_q$, respectively. Recall that $\itH_q$  is  the range of the covariance operator of $X$. Thus, our target will be to estimate $\beta_{0,q}$ and $\Upsilon_{0,q}$ and  Fisher--consistency in this case means that the projections of $\beta(P)$ and $\Upsilon(P)$ over $\itH_q$  and $\itF_q$,   are indeed     $\beta_{0,q}$ and $\Upsilon_{0,q}$.
 
\end{remark}

It is worth mentioning that similar arguments to those considered in Lemma S.1.1 in \citet{boente2020robust} allow to show that, if  \ref{ass:densidad}  holds  and   $\rho_1$ satisfies \ref{ass:rho}, then  for any  $a\in \real$, $\beta\in L^2(0,1)$ and any operator $\Upsilon \in \itF$, we have that $L(\alpha,\beta,\Upsilon,  \mu,\sigma)\ge L(\alpha_0,\beta_0,\Upsilon_0,  \mu,\sigma)$, for any $\sigma>0$. Furthermore, if      $\prob(\langle X-\mu,\beta\rangle + \langle X-\mu,\Upsilon (X-\mu)\rangle=a ) <1$, for any $(a, \beta, \Upsilon)\in \real\times L^2(0,1)\times \itF$, $(\alpha_0,\beta_0,\Upsilon_0)$ is the unique minimizer of $L(\alpha,\beta,\Upsilon, \mu, \sigma)$. However, the set of possible candidates for our functionals  is not the space  $\real\times L^2(0,1)\times  \itF$ but the reduced space $\itC_p=\real\times \itH_p\times\itF_p$, for that reason \ref{ass:finiteKL} and \ref{ass:probaX} are required and the following Lemma whose proof is relegated to the Appendix shows that the functionals  $\alpha(P) $, $\beta(P) $ and $\Upsilon(P)$ are indeed Fisher--consistent. 
 
\noi \begin{proposition}{\label{prop:FCKL}}
Assume that \ref{ass:densidad}, \ref{ass:FCphies}, \ref{ass:finiteKL} and \ref{ass:probaX} hold. Let $\rho_1$ be a  function satisfying \ref{ass:rho}. Given $p\ge q$ and  any $\sigma>0$, let $(\alpha(P), \beta(P),\Upsilon(P))$ be defined through \eqref{eq:funfinitos} and \eqref{eq:funcional}. Then, we have that,  $\alpha(P)=\alpha_0$, $\pi(\beta(P), \itH_q)=\beta_{0,q}$ and $\pi(\Upsilon(P), \itF_q)=\Upsilon_{0,q}$, where we have denoted as $\pi(u,\itL)$ the orthogonal projection of $u$ over the finite--dimensional space $\itL$. In particular, when $p=q$,  $L(\alpha,\beta,\Upsilon, \sigma)$ is uniquely minimized  over the set  $\itC_q$. 
\end{proposition}

\subsection{Fisher--consistency for infinite--dimensional processes}
In order to consider the situation of purely infinite--dimensional processes, that is, when   $\Gamma$ does not have finite rank, we will strengthen the dependence of the functionals $(\alpha(P), \beta(P), \Upsilon(P))$ defined through \eqref{eq:funcional} and \eqref{eq:funfinitos} on the dimension $p$ by denoting them as  $(\alpha_p(P), \beta_p(P), \Upsilon_p(P))$. The following result shows that the lowest value of 	$L(\alpha,\beta,\Upsilon, \mu,\sigma_0)$  over $\itC_p$ converges to $L(\alpha_0,\beta_0,\Upsilon_0,\mu, \sigma_0)$ which is the smallest possible value. This is the
infinite--dimensional counterpart of the classical result for robust linear regression.

\noi \begin{proposition}{\label{prop:Lphaciainfty}}
Let $\rho_1$ be a  function satisfying \ref{ass:rho} and let $(\alpha_p(P), \beta_p(P), \Upsilon_p(P))$ be defined through \eqref{eq:funcional} and \eqref{eq:funfinitos}.  Assume that \ref{ass:densidad} and \ref{ass:FCphies}  hold. Then, we have that,  
$$\lim_{p\to \infty}  L(\alpha_p(P), \beta_p(P), \Upsilon_p(P), \mu,\sigma_0) =\lim_{p\to \infty} \argmin_{a\in \real,\beta\in \itH_p, \Upsilon\in \itF_p}  L(\alpha,\beta,\Upsilon, \mu,\sigma_0)= L(\alpha_0,\beta_0,\Upsilon_0,\mu, \sigma_0)\,.$$
\end{proposition}

In order to show that, when $X$ does not have a finite--dimensional expansion as in \ref{ass:finiteKL},  the functional $(\alpha_p(P), \beta_p(P), \Upsilon_p(P))$ is still Fisher--consistent in the sense that it converges to the true parameters, we will need some additional assumptions.
\begin{enumerate}[label=\textbf{C\arabic*}]
\setcounter{enumi}{5} 
\item \label{ass:phiesder}: The eigenfunctions of the covariance operator $\Gamma$ are differentiable and such that $ \phi_j^{\prime}\in L^2(0,1)$, for all $j\ge 1$.
\item \label{ass:probaX1}:  $\prob(\langle X-\mu,\beta\rangle + \langle X-\mu,\Upsilon (X-\mu)\rangle=a ) <c <1$, for any $a\in \real$, $\beta\in L^2(0,1)$ and any operator $\Upsilon \in \itF$ such that  $(a, \beta,\Upsilon)\ne 0$.
 \end{enumerate}
From now on,  $\itW^{1,2}=\itW^{1,2}(0,1)$ stands for the Sobolev space 
$$\itW^{1,2}(0,1)=\{f\in L^2(0,1): \mbox{the weak derivative of $f$ exists and } \int_0^1 \left\{f^{\prime}(t)\right\}^2\, dt<\infty\}\,,$$ 
with norm   given by $\|f\|_{\itW^{1,2}}^2=\|f\|_{L^2(0,1)}^2+ \|f^{\prime}\|_{L^2(0,1)}^2$. Besides, we will also denote as $\itW_2^{1,2}$ the Sobolev space of two--dimensional functions in $ \itT=(0,1)\times(0,1)$, that is,  
\begin{align*}
\itW_2^{1,2} = & \{f\in L^2(\itT): \mbox{ the weak derivatives of $f$ exist for any $\nu=(\nu_1,\nu_2)$ such that $|\nu|\le 1$, and } \\
& D^{(\nu)}f \in L^2(\itT) \}\,, 
\end{align*}
where $ D^{(\nu)}f $ stands for the partial derivative $ D^{(\nu)}f=\partial^{|\nu|} f/\partial t^{\nu_1} \partial s^{\nu_2}$. We consider the norm in $\itW_2^{1,2}$ given by $\|f\|_{\itW_2^{1,2}}^2=\sum_{|\nu|\le 1} \| D^{(\nu)}f\|_{L^2(\itT)}^2$. 

Assumption \ref{ass:phiesder} means that $ \phi_j \in \itW^{1,2}$, while Assumption \ref{ass:probaX1}  is the functional version of assumption (A.3) in  \citet{yohai1987high} adapted to functional quadratic models.  We have the following result that states Fisher--consistency in the sense that  the considered functional provides a finite--dimensional approximation to the true parameters when the process is infinite--dimensional.

\noi \begin{proposition}{\label{prop:FCinfty}}
Assume that $\esp\|X\|^2<\infty$ and that \ref{ass:densidad}, \ref{ass:FCphies} and \ref{ass:phiesder} hold. Let $\rho_1$ be a  function satisfying \ref{ass:rho} and assume that  \ref{ass:probaX1}  holds with $c< 1-b_{\rho_1}$, where $b_{\rho_1}=L(\alpha_0,\beta_0,\Upsilon_0,\mu, \sigma_0) $. Then, if $(\alpha_p(P), \beta_p(P), \Upsilon_p(P))$ are defined through \eqref{eq:funcional} and \eqref{eq:funfinitos}, we have that,  
$$\lim_{p\to \infty}  |\alpha_p(P)-\alpha_0|+ \| \beta_p(P)-\beta_0\|_{L^2(0,1)}+ \|\Upsilon_p(P)-\Upsilon_0\|_{\itF}=0\,.$$
\end{proposition}

Denote as $\upsilon_p(P)$ and  $\upsilon_0$ the kernels related to the operators $\Upsilon_p(P)$ and $\Upsilon_0$, that is,
$$(\upsilon_p(P))(t,s)= \sum_{1\le j, \ell\le p}    v_{j\ell}(P)  \phi_{\rob,j}(P_X)(t) \phi_{\rob,\ell}(P_X)(s)\,,$$
where the coefficients $ v_{j\ell}(P)$ are given in \eqref{eq:funcional} and 
$$\upsilon_0(t,s)= \sum_{j=1}^\infty
\sum_{\ell=1}^\infty     v_{0,j\ell}   \phi_j (t) \phi_{ \ell} (s)\,,$$
with $ v_{0,j\ell}=\langle \phi_j, \Upsilon_{0} \phi_\ell \rangle$. It is worth mentioning that, using the Fisher--consistency of the principal direction functionals we get that 
$\|\Upsilon_p(P)-\Upsilon_0\|_{\itF}=\|\upsilon_p(P)-\upsilon_0\|_{L^2((0,1)\times(0,1))}$. Moreover, under  \ref{ass:FCphies} and \ref{ass:phiesder}, $\beta_p(P)\in \itW^{1,2}$, while $\upsilon_p(P)\in  \itW_2^{1,2}$.

%%%%%%%%%%%%%%%%%%%%%%%%
%MONTE CARLO
%%%%%%%%%%%%%%%%%%%%%%%%%%%%%%%%
\section{Simulation study}{\label{sec:simu}}
In this section, we report the results of a Monte Carlo study designed to investigate the finite--sample properties 
of the robust estimators proposed for the  functional quadratic regression model:
\begin{equation}
  \label{eq:monte-carlo-modelo}
  y_i = \alpha_0+\langle \beta_0, X_i\rangle \, + \, \langle X_i, \Upsilon_0 X_i\rangle \, + \, \sigma_0\,\epsilon_i \, , \quad i=1, \ldots, n \, . 
\end{equation}
We considered several choices for the parameters, including the case where $\Upsilon_0=0$, that is, where the true model is a linear one.
The first setting mimics the one considered in \citet{boente2020robust} for the covariate distribution and some of the contamination schemes. The second one corresponds to the functional quadratic model studied in \citet{yao2010functional}. In both cases, we selected  $\alpha_0=0$ and $\itI=[0,1]$ and for each setting we generated $n_R = 1000$ samples of size $n = 300$.  

We compared two estimators: the classical procedure based on least squares (\textsc{ls}), and the $MM-$estimators (\textsc{mm}) from Section \ref{sec:estimadores}. The $MM-$estimators were computed using   a bounded $\rho-$function  $\rho_0$ to obtain the   residual  scale estimator in \eqref{eq:s-est}  and also a bounded $\rho_1$ for the $M-$step \eqref{eq:estfinitos}. For $j=0,1$, we choose  $\rho_j= \rho_{\,\tuk,\,c_j}$, the bisquare function, with tuning constants $c_0=1.54764$ ($b=1/2$) and  $c_1=3.444$. All calculations were performed in \verb=R=.  

To compute the estimators of the principal directions we use the eigenfunctions of the sample covariance when considering the classical procedure and the spherical principal directions for the robust one, since the former are very sensitive to atypical curves. Denote $\lambda_k^{{\mbox{\footnotesize \sc s}}}$ the   eigenvalues of $\Gamma^{{\mbox{\footnotesize \sc s}}}$. When second moment exists, the values    $\lambda_k^{{\mbox{\footnotesize \sc s}}}$ are  shrinked  with respect to those of the scatter operator as follows $ \lambda_k^{{\mbox{\footnotesize \sc s}}}  = \lambda_k \, \esp\left( {\xi^2_k}/{\sum_{j \geq 1} \lambda_j \xi_j^2}\right)$. To avoid situations in which the eigenvalues related to the sign operator are too close and will not allow to identify easily the order of the estimated eigenfunctions, it is better to order the eigenfunctions $\wphi_j$ according to the values of a robust scale of the projected data, $\langle X_i-\wmu, \wphi_j\rangle$, $1\le i \le n$, which are resistant estimators $\wlam_j$ of the $j-$th eigenvalue of $\Gamma$.

Both for the classical and robust method, we select the dimension $p$ used in the regularization as the smallest number explaining at least 90\% of total variation obtained. In the classical case, the total variation is obtained  using the eigenvalues of the sample covariance, while when considering the robust procedure, the eigenvalue estimators $\{\wlam_j\}$ defined above as a robust scale of the projected data, are taken. We use as robust scale the  $M-$scale computed with the bisquare function.

To evaluate the performance of each estimator, we looked at their integrated squared bias and mean integrated squared error. These were computed on a grid of $M = 100$ equally spaced points on $[0, 1]$ and $[0,1]\times [0,1]$, for $\wbeta$ and $\wup$, respectively. More specifically, if $\wbeta_j$ is the estimate of the function $\beta$ obtained with the $j-$th sample ($1 \le j \le n_R$) and  $\wup_j$ is the estimate of $\upsilon$, we compute approximations for the integrated square bias as
\begin{align*}
 \mbox{Bias}^2(\wbeta)   & =    \frac 1M \sum_{s=1}^M
 \left( \frac{1}{n_R} \sum_{j=1}^{n_R} \wbeta_j(t_s) - \beta_0 (t_s) \right )^2 \,,  \\
  \mbox{Bias}^2(\wup)  &  =    \frac 1{M^2} \sum_{s=1}^M\sum_{\ell=1}^M
 \left( \frac{1}{n_R} \sum_{j=1}^{n_R} \wup_j(t_s, t_\ell) - \upsilon_0 (t_s, t_\ell) \right )^2\,, 
\end{align*}
and of the integrated squared errors as
\begin{align*}
\mbox{MISE}(\wbeta)  &  =    \frac 1M \sum_{s=1}^M  \frac{1}{n_R} \sum_{j=1}^{n_R} \left( \wbeta_j(t_s) - \beta_0 (t_s) \right )^2  \,, 
\\
  \mbox{MISE}(\wup)  &  = \frac 1{M^2} \sum_{s=1}^M\sum_{\ell=1}^M \frac{1}{n_R} \sum_{j=1}^{n_R} 
 \left( \wup_j(t_s, t_\ell) - \upsilon_0 (t_s, t_\ell) \right )^2\,,
 \end{align*}
where $t_1\le \dots \le t_M$ are equispaced points on   $\itI=[0, 1]$. Note that   when $\upsilon_0=0$, $\mbox{Bias}^2(\wup) $ measures the squared bias of the estimated coefficients, while for the other quadratic kernels it also gives a measure of how biased  are the principal direction estimators. This difference is also inherited by the mean integrated squared error.
As mentioned in \citet{he1998monotone} who studied estimators under a nonparametric regression model and in \citet{boente2020robust} who considered estimators of the slope and of the nonparametric component under a functional partial linear model,   the squared $\mbox{Bias}$ and the $\mbox{MISE}$ may be heavily influenced by numerical errors at the boundaries of the grid or near them. For that reason,   we also consider trimmed versions of the above computed without the $q$ first and last points on the grid, that is, 
 \begin{align*}
 \mbox{Bias}_{\trim}^2(\wbeta)  &  =    \frac 1{M-2q} \sum_{s=q+1}^{M-q}
 \left( \frac{1}{n_R} \sum_{j=1}^{n_R} \wbeta_j(t_s) - \beta_0 (t_s) \right )^2 \,,\\
  \mbox{Bias}_{\trim}^2(\wup)   & =    \frac 1{(M-2q)^2} \sum_{s=q+1}^{M-q}\sum_{\ell=q+1}^{M-q}
 \left( \frac{1}{n_R} \sum_{j=1}^{n_R} \wup_j(t_s, t_\ell) - \upsilon_0 (t_s, t_\ell) \right )^2\,, 
 \\
 \mbox{MISE}_{\trim}(\wbeta)  & =    \frac 1{M-2q} \sum_{s=q+1}^{M-q}  
 \frac{1}{n_R} \sum_{j=1}^{n_R} \left( \wbeta_j(t_s) - \beta_0 (t_s) \right )^2 \,, \\
  \mbox{MISE}_{\trim}^2(\wup)  &  =    \frac 1{(M-2q)^2} \sum_{s=q+1}^{M-q}\sum_{\ell=q+1}^{M-q} \frac{1}{n_R} \sum_{j=1}^{n_R}
 \left(  \wup_j(t_s, t_\ell) - \upsilon_0 (t_s, t_\ell) \right )^2\, .
 \end{align*}
We chose $q=[M\times 0.05]$ which uses the central 90\% interior points in the grid, which for both models was an equally spaced grid of points in $[0,1]$ with size $M=100$.
 
\subsection{Model 1}
\begin{figure}[ht!] %[tp]
  \begin{center} 
  \begin{tabular}{ccc}
  \multicolumn{2}{c}{Quadratic kernels} & Regression parameter \\
   $\upsilon_{0,1}$ &  $\upsilon_{0,2}$ & $\beta_0$\\
   \includegraphics[width=0.3\textwidth]{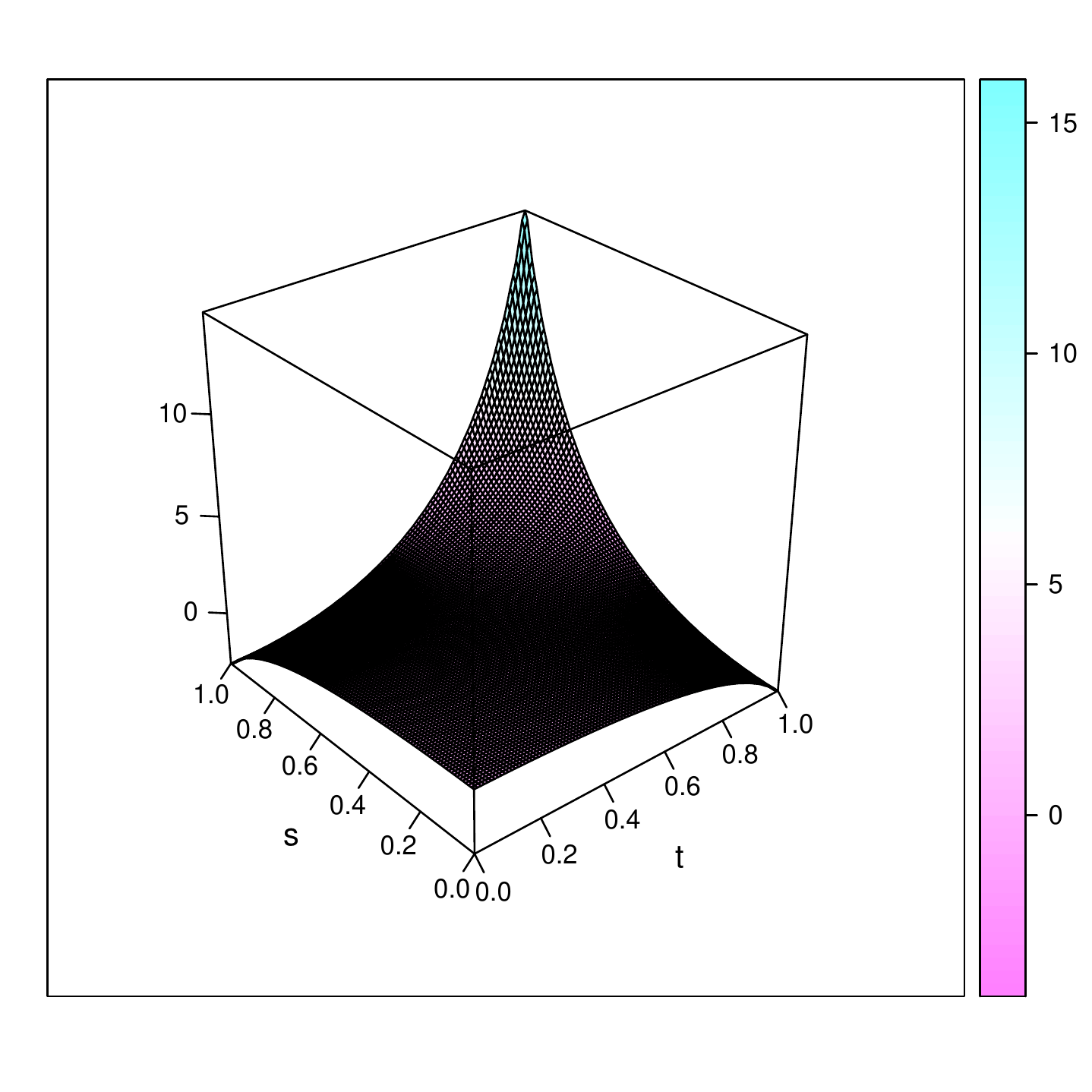} &
   \includegraphics[width=0.3\textwidth]{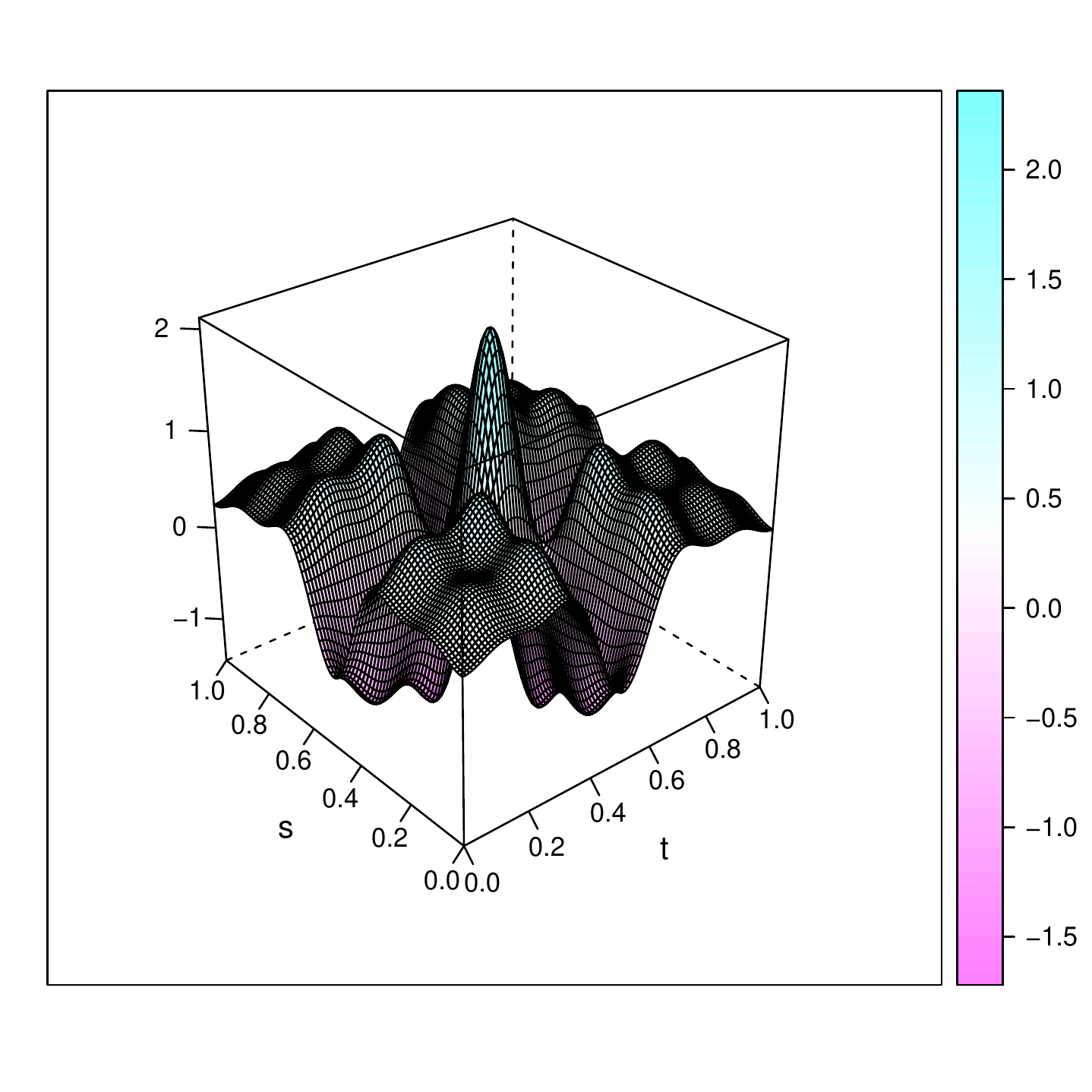} &
    \includegraphics[width=0.29\textwidth]{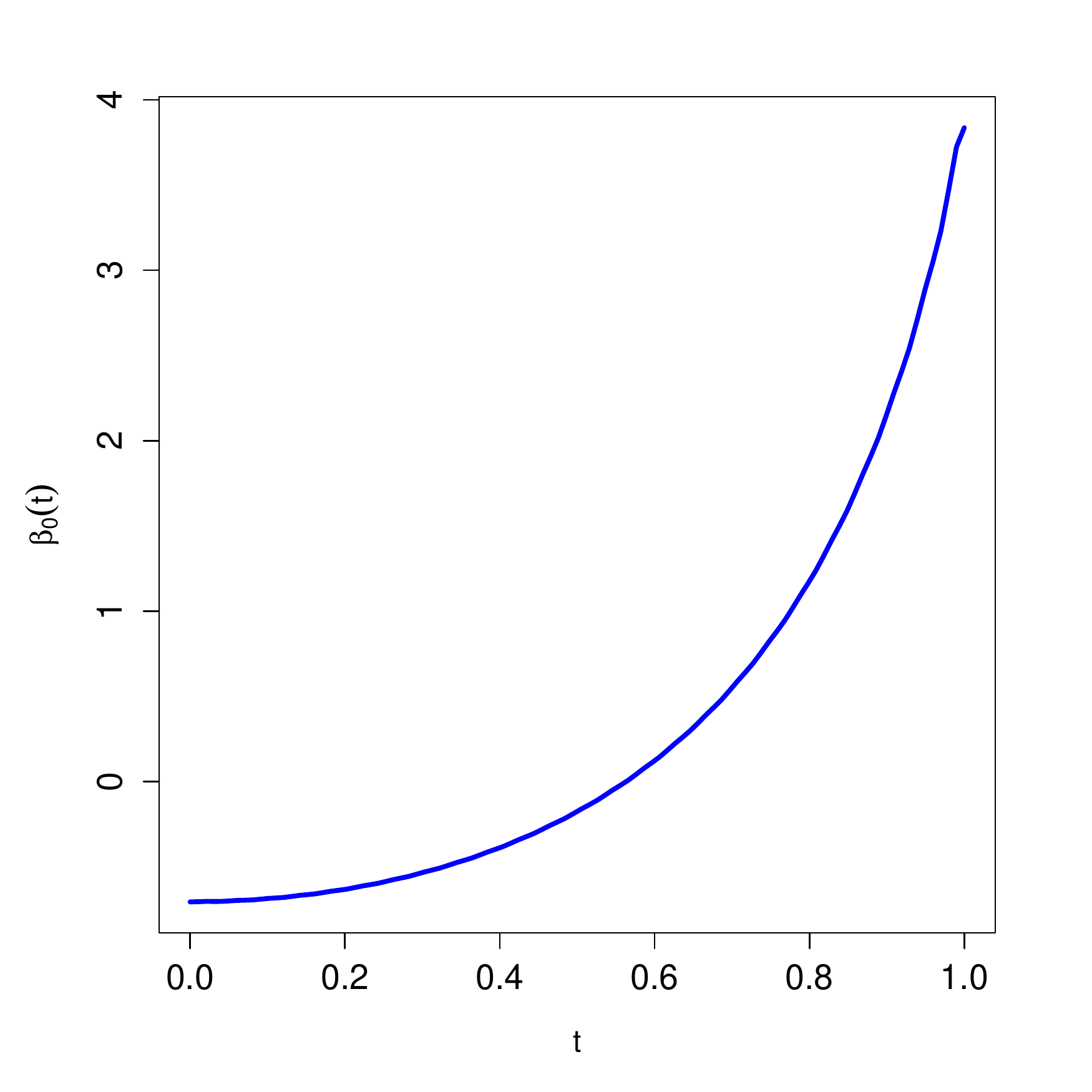} 
\end{tabular}
  \caption{\small \label{fig:parametros-verdaderos} True parameters.}
\end{center}
\end{figure}

{\label{sec:model1}}
In this section, we considered the functional quadratic regression model \eqref{eq:monte-carlo-modelo}, where  $\sigma_0=1$ and the regression parameter equals the one used in \citet{boente2020robust}, that is, $\beta_0(t) = \sum_{j=1}^{50} b_{j,0} \phi_j(t)$ with the basis $\phi_1(t) \equiv 1$, $\phi_j(t) = \sqrt{2} \cos ((j-1)\pi t)$, $j\geq 2$, and  $\bb_0=(b_{1,0}, \dots, b_{50,0})\trasp$ where $b_{1,0} = 0.3$ and $b_{j,0} = 4(-1)^{j+1}j^{-2}$, $j \geq 2$. We label this model as \textbf{Model 1}. Besides, we considered three possible choices for the quadratic operator, that will be labelled \textbf{Model$_{1,0}$} to \textbf{Model$_{1,2}$} and correspond to: 
\begin{itemize}
\item \textbf{Model$_{1,0}$}: $\Upsilon_{0,0}=0$ which means that the true model is a linear one. 
\item \textbf{Model$_{1,1}$}: $\Upsilon_{0,1}= \sum_{j=1}^{50} \sum_{\ell=1}^{50} c_{j\ell, 1} \phi_j\otimes  \phi_\ell$ with $\bC_{ 1}=( c_{j\ell, 1})_{1\le j, \ell\le 50}= 5 \bb_1\, \bb_1\trasp$ and $\bb_1=\bb_0$ with $\bb_0$ defined above.
\item  \textbf{Model$_{1,2}$}: $\Upsilon_{0,2}= \sum_{j=1}^{5} \sum_{\ell=1}^{5} c_{j\ell, 2} \phi_{2(j-1)+1}\otimes  \phi_{2(\ell-1)+1}$ with $\bC_{2}=( c_{j\ell, 2})_{1\le j, \ell\le 5}= 5 \bb_{2}\, \bb_2\trasp$, where $\bb_2=(b_{1,2}, \dots, b_{5,2})\trasp$, $b_{1,2}=b_{2,2}=0.3$, $b_{j,2} =3(-1)^{j+1}j^{-2}$, for $j=3,4,5$.
\end{itemize}
Figure \ref{fig:parametros-verdaderos} shows the function   $\beta_0$ and the surface $\upsilon_{0,j}(s,t)$, $j=1,2$, associated to each choice of a non--null operator $\Upsilon_0$.

The process that generates the functional covariates $X_i(t)$ was Gaussian with mean 0 and covariance operator with eigenfunctions  $\phi_j(t)$. For uncontaminated samples that will be denoted $C_0$, the scores  $\xi_{ij} $ were independent Gaussian random variables $\xi_{ij}\sim N(0,j^{-2})$, and the errors $\epsilon_i \sim N(0,1)$, independent of  $X_i$. Taking into account that $\var(\xi_{ij})\le 1/2500$ when $j > 50$, the process was approximated numerically using the first 50 terms of its Karhunen--Lo\`{e}ve representation. 
 
Table \ref{tab:tabla-1}  reports the  squared bias and   $\mbox{MISE}$ and their trimmed counterparts for samples without outliers. 
  
\begin{table}[ht!]
  \centering
  \small 
  \renewcommand{\arraystretch}{1.4}   
  \setlength{\tabcolsep}{1.8pt}
	\begin{tabular}{r rr :  rr |rr  :  rr |rr  : rr   }
	\hline  
	& \multicolumn{4}{c|}{$\Upsilon_{0,0}$}& \multicolumn{4}{c|}{$\Upsilon_{0,1}$}& \multicolumn{4}{c }{$\Upsilon_{0,2}$}\\
	\hline
	&\multicolumn{2}{c:}{$\wbeta$} &\multicolumn{2}{ c|}{$\wup$}&\multicolumn{2}{c:}{$\wbeta$} &\multicolumn{2}{ c|}{$\wup$}&\multicolumn{2}{c:}{$\wbeta$} &\multicolumn{2}{ c }{$\wup$} \\
	\hline
	& Bias$^2$ & MISE & Bias$^2$ & MISE  & Bias$^2$ & MISE & Bias$^2$ & MISE  & Bias$^2$ & MISE & Bias$^2$ & MISE  \\
	\hline
	
	\textsc{ls} & 0.2958 & 2.7659 & 0.0270 & 58.0046 & 0.2966 & 2.7919 & 23.5860 & 86.4361 & 0.2958 & 2.7636 & 1.2817 & 59.5208\\
	\textsc{mm} & 0.2983 & 3.4331 & 0.0488 & 80.3626 & 0.3006 & 3.4731 & 22.7776 & 109.9980  & 0.2973 & 3.4421 & 1.2584 & 82.2667 \\ \hline  
	& Bias$^2_{\trim}$ & MISE$_{\trim}$ & Bias$^2_{\trim}$ & MISE$_{\trim}$ & Bias$^2_{\trim}$ & MISE$_{\trim}$ & Bias$^2_{\trim}$ & MISE$_{\trim}$ & Bias$^2_{\trim}$ & MISE$_{\trim}$ & Bias$^2_{\trim}$ & MISE$_{\trim}$  \\
	\hline
	\textsc{ls} & 0.0656 & 2.3135 & 0.0214 & 48.8398 & 0.0665 & 2.3384 & 3.4632 & 55.7912  & 0.0654 & 2.3099 & 1.4807 & 50.5278 \\
	  \textsc{mm} & 0.0624 & 2.9323 & 0.0370 & 68.3566 & 0.0622 & 2.9664 & 3.1481 & 76.5740  &0.0625 & 2.9411 & 1.4579 & 70.3716   \\
	\hline  
	\end{tabular}
	\caption{ \small \label{tab:tabla-1} 
    Integrated squared bias  and mean integrated squared errors and their trimmed versions (multiplied by 10) over $n_R = 1000$ clean samples under \textbf{Model 1}. }
    %For each model, the top two rows report the Monte Carlo estimates of these measures using a grid of 100 equispaced points, and the bottom two correspond to their trimmed versions using the 90\% inner grid points. }
\end{table}

We note that the boundary effect is more pronounced for the estimators of $\Upsilon_0$, but it is present for $\wbeta$ as well. In particular, due to the shape near $(1,1)$ of $\upsilon_{0,1}$ the  squared bias is seven times larger than the trimmed one. 
 
Based on this observation, in what follows, we report the trimmed measures in all Tables and Figures.

We considered three contamination scenarios. The first one contains outliers in the response variables and is expected to affect 
mainly the  estimation of $\beta_0$ and $\alpha_0$. The other ones include  high--leverage  outliers in the functional explanatory variables, which typically affect the estimation of the linear regression parameter or the quadratic operator. Specifically, we constructed our samples as follows:
\begin{itemize}
\item Scenario $C_{1,\mu}$: here only the regression errors are contaminated in order to produce ``vertical outliers''. Their distribution $G$ is given by $G(u)=0.9 \, \Phi(u )+0.1\, \Phi\left((u-\mu)/ 0.5 \right)$, with $\Phi$ the standard normal distribution function. 

\item Scenario $C_{2,\mu}$: in these settings, we introduce high--leverage outliers by contaminating the functional covariates $X_i$ and the errors simultaneously. Outliers in the $X_i$'s are generated by perturbing the distribution of the second score 
in the Karhunen--Lo\`{e}ve representation of the process. We denote with the superscript $(\mbox{\textsc{co}})$ the contaminated observations. Specifically, we sample $v_i\sim Bi(1, 0.10)$ and then:
  \begin{itemize}
  \item if $v_i = 0$, let $\epsilon_i^{(\cont)}=\epsilon_i$ and $X_i^{(\cont)}=X_i$;
  \item if $v_i=1$, let $\epsilon_i^{(\cont)} \sim N( \mu, 0.25)$ and
    $X_i^{(\cont)}=\sum_{j=1}^{50} \xi_{ij}^{(\cont)} \phi_j$, with
    $\xi_{ij}^{(\cont)}\sim N(0,j^{-2})$ for $j\ne 2$ and
    $\xi_{i2}^{(\cont)}\sim N(\mu/2, 0.25) $. 
   \end{itemize} 
 The responses are generated as $y_i^{(\cont)} = \langle \beta_0, X_i^{(\cont)}\rangle + \langle X_i^{(\cont)},\Upsilon_0   X_i^{(\cont)}\rangle  +   \epsilon_i^{(\cont)}$. 

 \item Scenario  $C_{3,\mu, \delta }$: high--leverage outliers are introduced contaminating the functional covariates $X_i$ and modifying the responses as in \citet{maronna2013robust}. More precisely, outliers in the covariates are generated by adding a constant to the coefficient process, that is, we sample $v_i\sim Bi(1, 0.10)$ and then:
  \begin{itemize}
  \item if $v_i = 0$, let $y_i^{(\cont)}=y_i$ and $X_i^{(\cont)}=X_i$;
  \item if $v_i=1$, let    $X_i^{(\cont)}= \sum_{j=1}^{50}  \xi_{ij}^{(\cont)} \phi_j$ where $\xi_{ij}^{(\cont)}\sim N(\mu,j^{-2})$, while    $y_i^{(\cont)}=\delta\, y_i$.
 \end{itemize} 
\end{itemize}
 
Both  $C_{1,\mu}$ and  $C_{2,\mu}$ depend on the parameter $\mu \in \real$. In this experiment, we looked at  values of $\mu$ varying between 8 and 20 with a step of 2. They produce a range of contamination scenarios ranging from mild to severe. Besides, in contamination $C_{3,\mu, \delta }$, we combine values of $\mu =2,4,6,8$ with $\delta=0.2,0.4,0.6,0.8$ to affect the quadratic component. When considering the   contamination schemes, Table \ref{tab:tabla-Upsilon} reports the maximum value of the trimmed squared bias and   $\mbox{MISE}$ over the different values of $\mu$ and/or $\delta$ and we simply label the situation as $C_j$, for $j=1,2,3$ to avoid burden notation.

\begin{figure}[ht!]%[tp]
  \small
  \begin{center} 
   \newcolumntype{M}{>{\centering\arraybackslash}m{\dimexpr.05\linewidth-1\tabcolsep}}
   \newcolumntype{G}{>{\centering\arraybackslash}m{\dimexpr.3\linewidth-1\tabcolsep}}
	\begin{tabular}{GGG}
 	 $C_0$ & $C_{2,12}$ & $C_{3,4,0.4}$\\
      \includegraphics[width=0.3\textwidth]{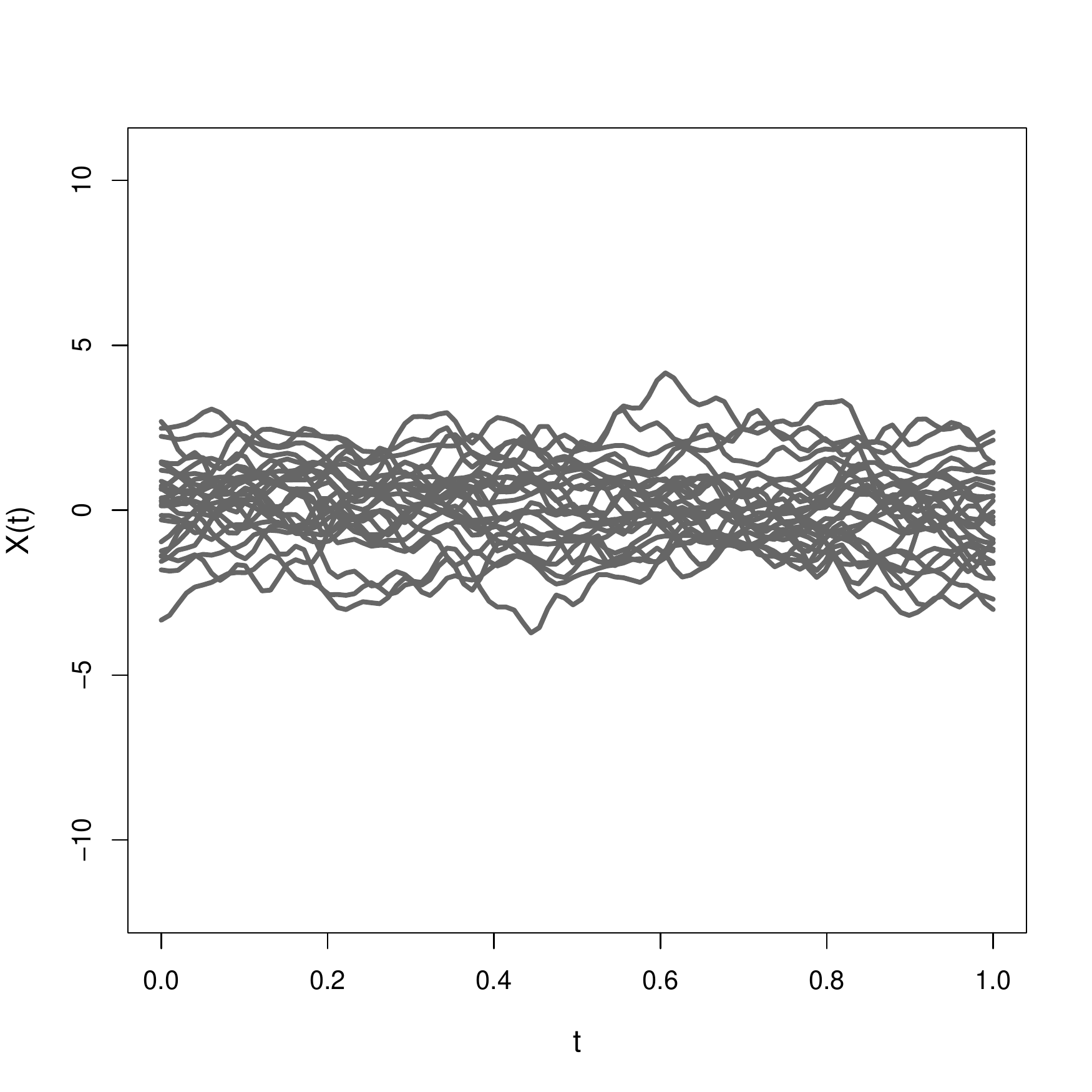} & 
      \includegraphics[width=0.3\textwidth]{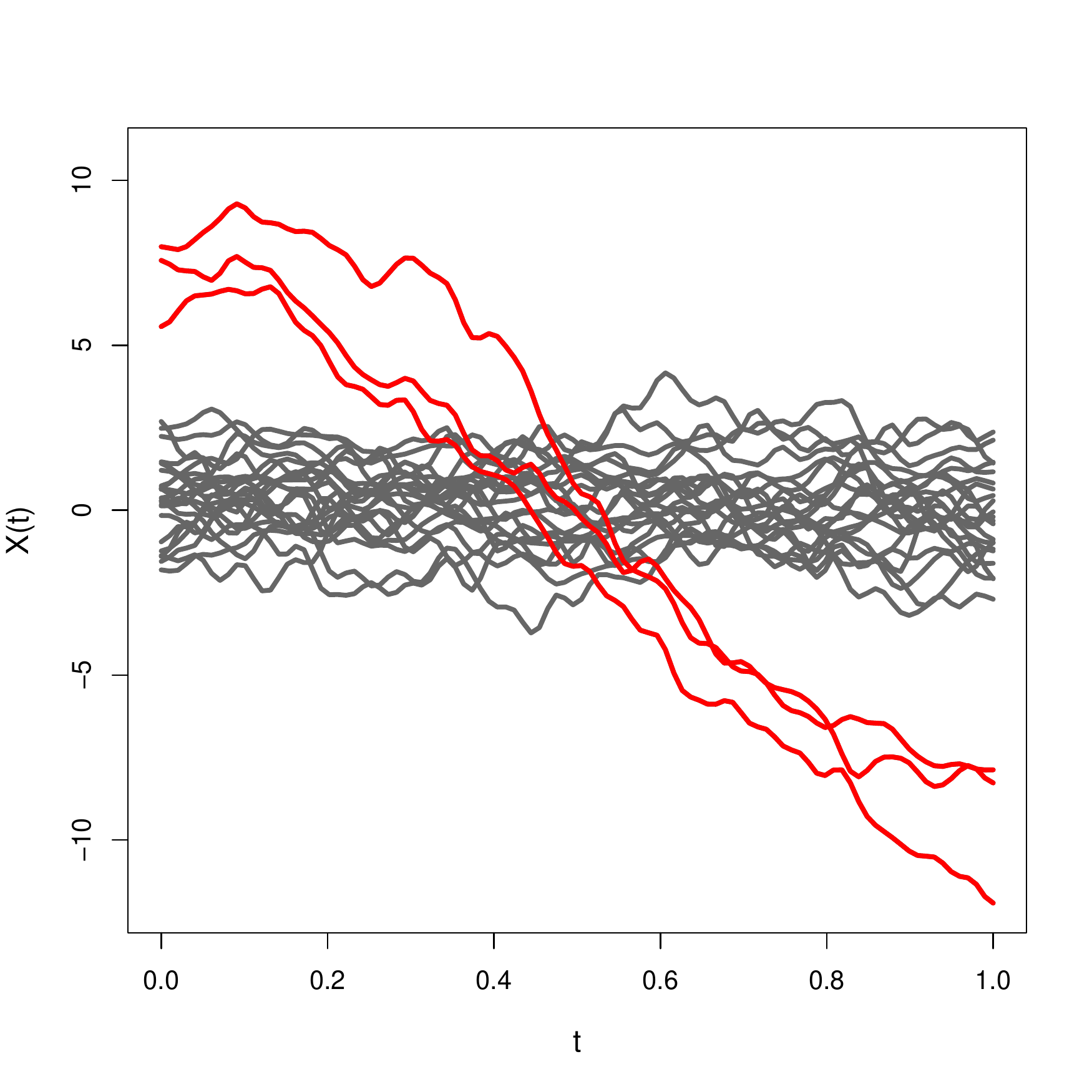} &
      \includegraphics[width=0.3\textwidth]{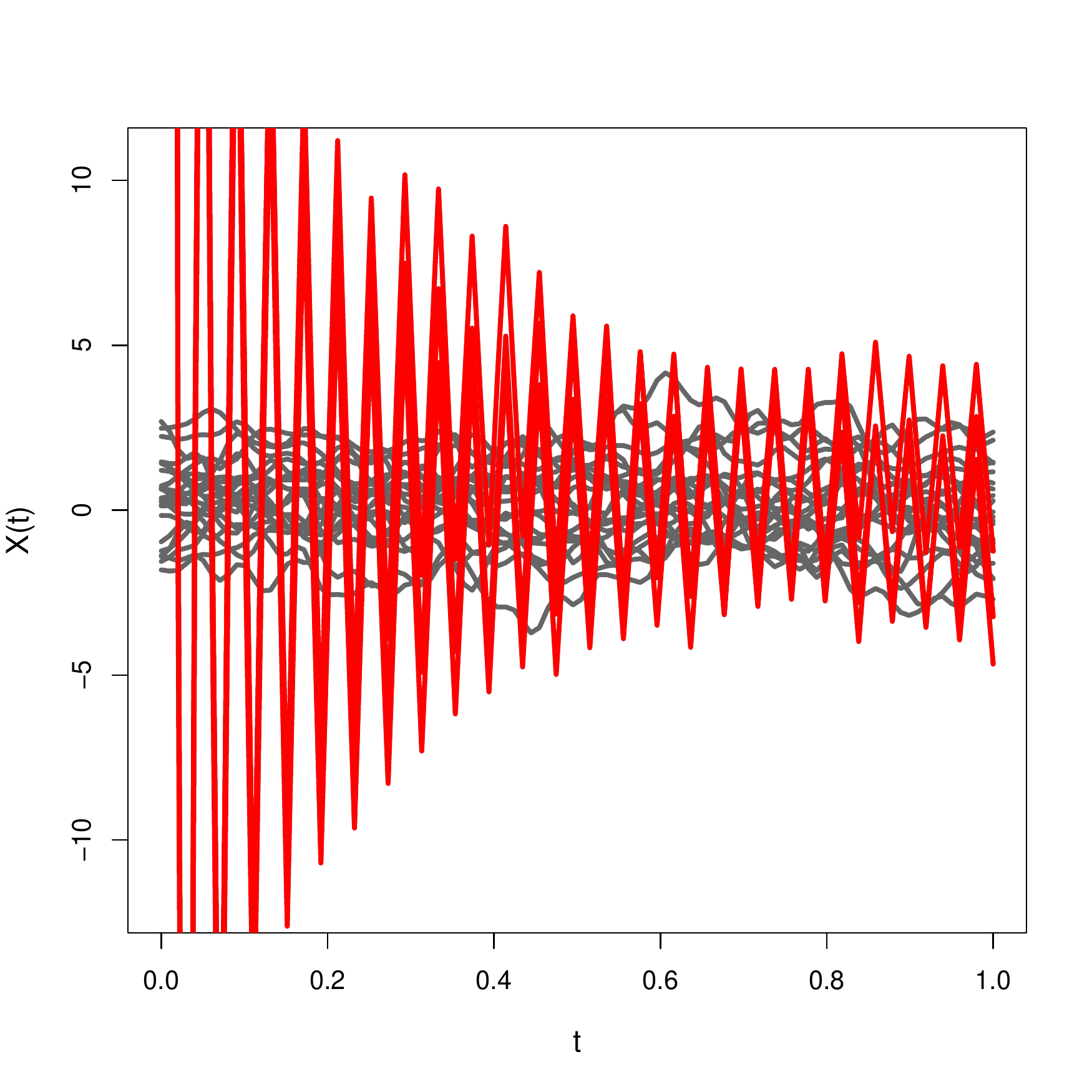}
      \end{tabular}
   \end{center}
  \vskip-0.2in
  \caption{ \label{fig:trayectorias} 25 trajectories $X_i(t)$ with and without contamination.}
\end{figure}

As an illustration of the type of outliers generated, Figure \ref{fig:trayectorias} shows 25 randomly chosen functional covariates $X_i(t)$, for one sample generated under $C_0$ (with no outliers), one obtained under $C_{2,12}$ and the other one under  $C_{3,4,0.4}$.

\begin{table}
\footnotesize
   \renewcommand{\arraystretch}{1.4}   
  \setlength{\tabcolsep}{2pt}
\begin{tabular}{rr rr :  rr |rr  :  rr |rr  : rr   }
\hline  
&& \multicolumn{4}{c|}{$\Upsilon_{0,0}$}& \multicolumn{4}{c|}{$\Upsilon_{0,1}$}& \multicolumn{4}{c }{$\Upsilon_{0,2}$}\\
\hline
&&\multicolumn{2}{c:}{$\wbeta$} &\multicolumn{2}{ c|}{$\wup$}&\multicolumn{2}{c:}{$\wbeta$} &\multicolumn{2}{ c|}{$\wup$}&\multicolumn{2}{c:}{$\wbeta$} &\multicolumn{2}{ c }{$\wup$} \\
\hline

& & Bias$^2_{\trim}$ & MISE$_{\trim}$ & Bias$^2_{\trim}$ & MISE$_{\trim}$   
& Bias$^2_{\trim}$ & MISE$_{\trim}$ & Bias$^2_{\trim}$ & MISE$_{\trim}$ 
& Bias$^2_{\trim}$ & MISE$_{\trim}$ & Bias$^2_{\trim}$ & MISE$_{\trim}$\\
\hline
$C_0$  & \textsc{ls} & 0.0656 & 2.3135 & 0.0214 & 48.8398  
					&  0.0665 & 2.3384 & 3.4632 & 55.7912 
					& 0.0654 & 2.3099 & 1.4807 & 50.5278
					\\  
   
 & \textsc{mm} & 0.0624 & 2.9323 & 0.0370 & 68.3566 
 			& 0.0622 & 2.9664 & 3.1481 & 76.5740
 			& 0.0625 & 2.9411 & 1.4579 & 70.3716
 			\\ 
\hline
 $C_{1}$  & \textsc{ls} & 0.0788 & 82.5051 & 1.6237 & 1867.4839 
 						& 0.0803 & 82.5097 & 5.7713 & 1874.4401
 						& 0.0788 & 82.5052 & 3.0274 & 1869.5741
 						\\ 
 & \textsc{mm} & 0.0591 & 2.9711 & 0.0461 & 69.2086 
 			   & 0.0595 & 3.0037 & 3.1985 & 77.3132
 			   & 0.0590 & 2.9749 & 1.4963 & 70.9218 
 			   \\ 
\hline
 $C_{2}$  & \textsc{ls} & 2.6091 & 2.8690 & 1.0453 & 3.0408 
 						& 2.5548 & 6.2146 & 86.3542 & 87.2354 
 						& 2.6110 & 2.8761 & 5.3909 & 6.8795 
 						\\ 
 & \textsc{mm} & 0.3125 & 3.0785 & 1.5264 & 54.2488 
 			   & 0.2624 & 3.0535 & 7.4495 & 64.6150 
 			   & 0.3111 & 3.0842 & 3.0050 & 56.5554
 			   \\ 
\hline
$C_{3}$ & \textsc{ls} & 9.3426 & 9.3797 & 0.0000 & 0.0000 
					  & 9.3422 & 9.5766 & 218.1688 & 218.1692 
					  & 9.3426 & 9.3879 & 5.2112 & 5.2112
					  \\   %ES LA C4 
 & \textsc{mm} & 0.0844 & 3.9225 & 0.0366 & 55.2039 
 			   & 0.0845 & 4.0217 & 5.5522 & 65.8103
 			   & 0.0854 & 3.9209 & 1.5840 & 57.0941
 			   \\ 
 \hline
\end{tabular}
\caption{ \small \label{tab:tabla-Upsilon} Trimmed versions of the integrated squared bias  and mean integrated squared errors (multiplied by 10) for clean and contaminated samples, \textbf{Model 1}. The reported values under contamination correspond to the worst situation.}
\end{table}

\begin{figure}[ht!]
 \begin{center}
 \newcolumntype{M}{>{\centering\arraybackslash}m{\dimexpr.05\linewidth-1\tabcolsep}}
   \newcolumntype{G}{>{\centering\arraybackslash}m{\dimexpr.4\linewidth-1\tabcolsep}}
\begin{tabular}{M GG}
 &  $\wbeta$  & $\wup$  \\[-4ex]
{\small   $C_{1,\mu}$} &  
\includegraphics[scale=0.38]{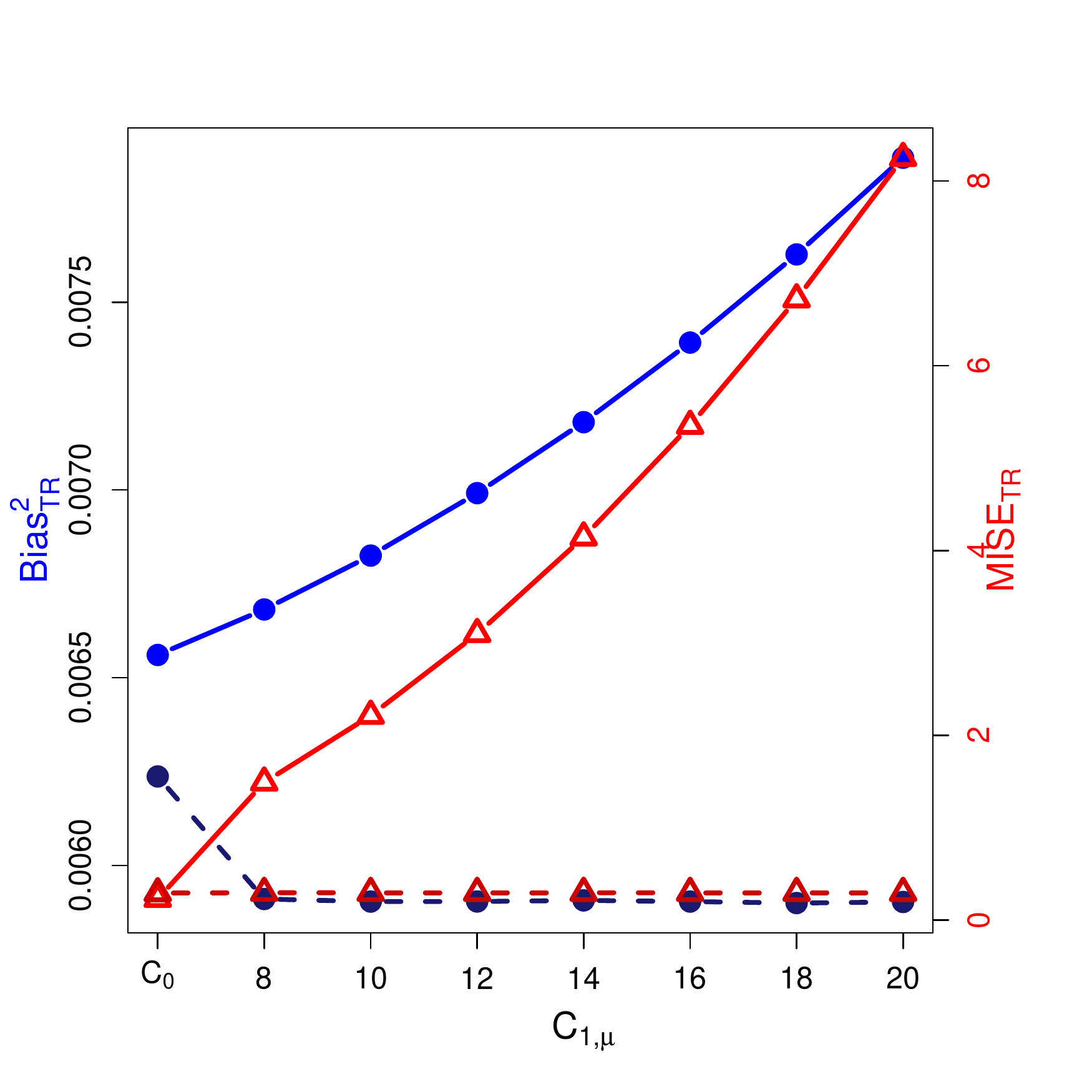} 
&  \includegraphics[scale=0.38]{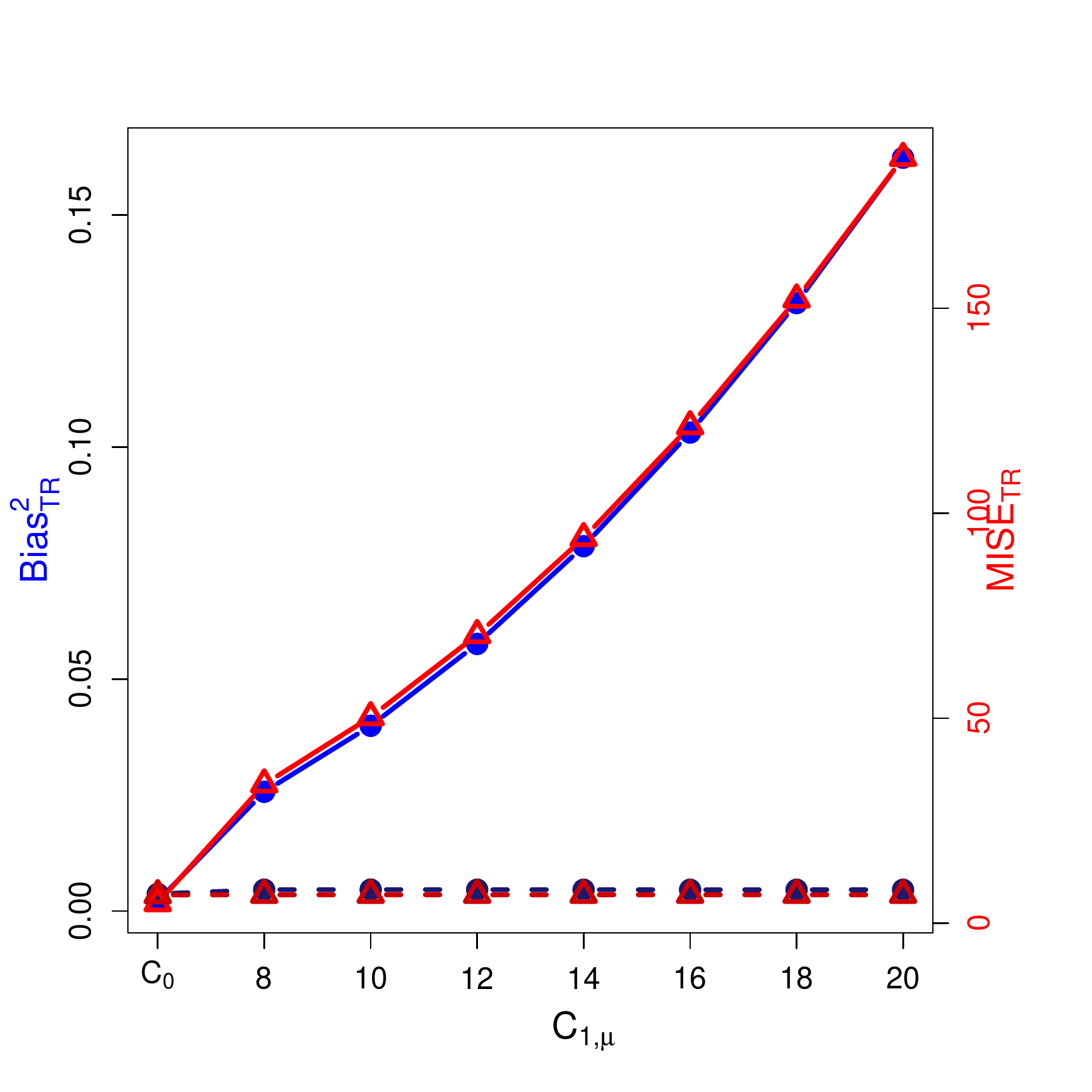}
 \\[-5ex]
 {\small   $C_{2,\mu}$} &  
 \includegraphics[scale=0.38]{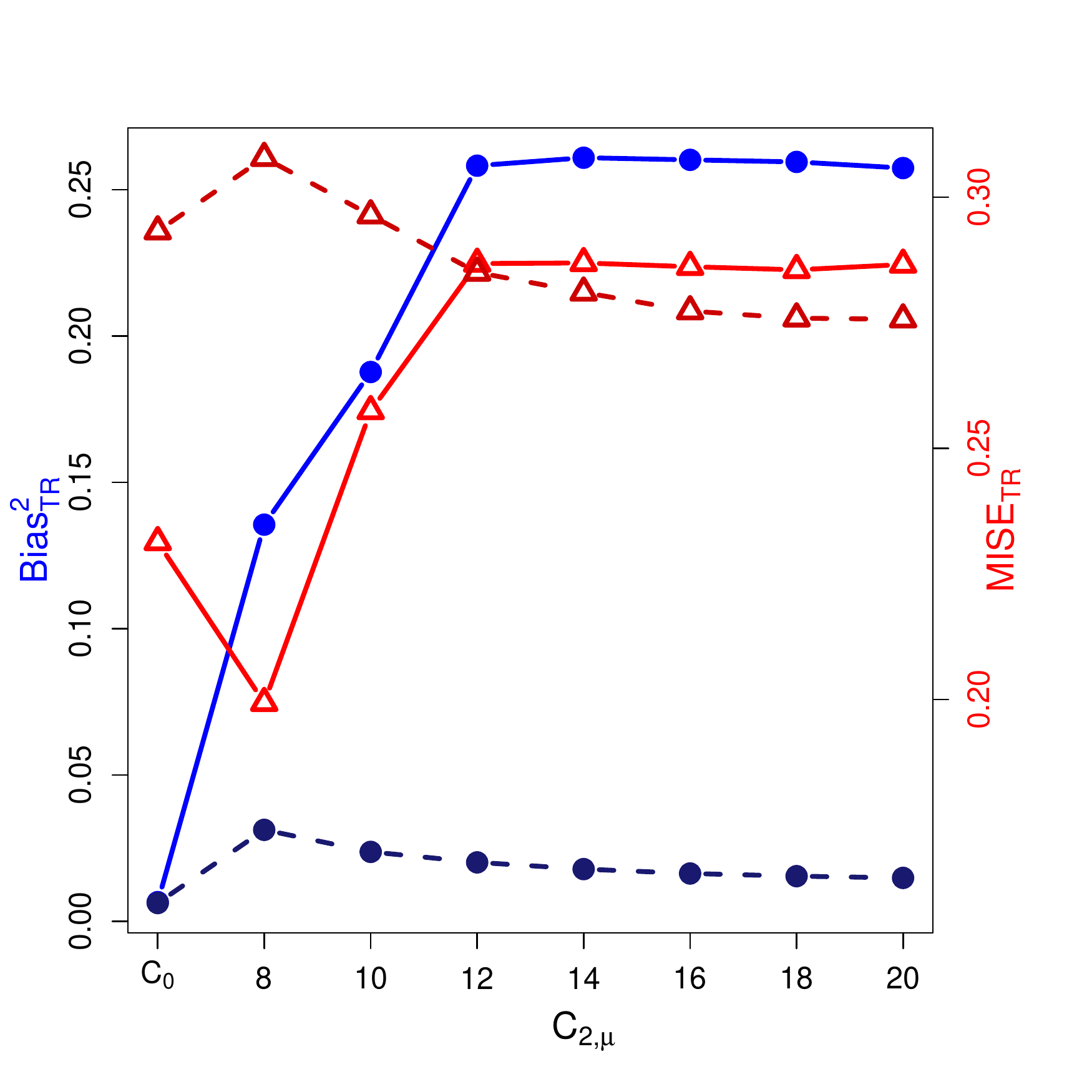} 
&  \includegraphics[scale=0.38]{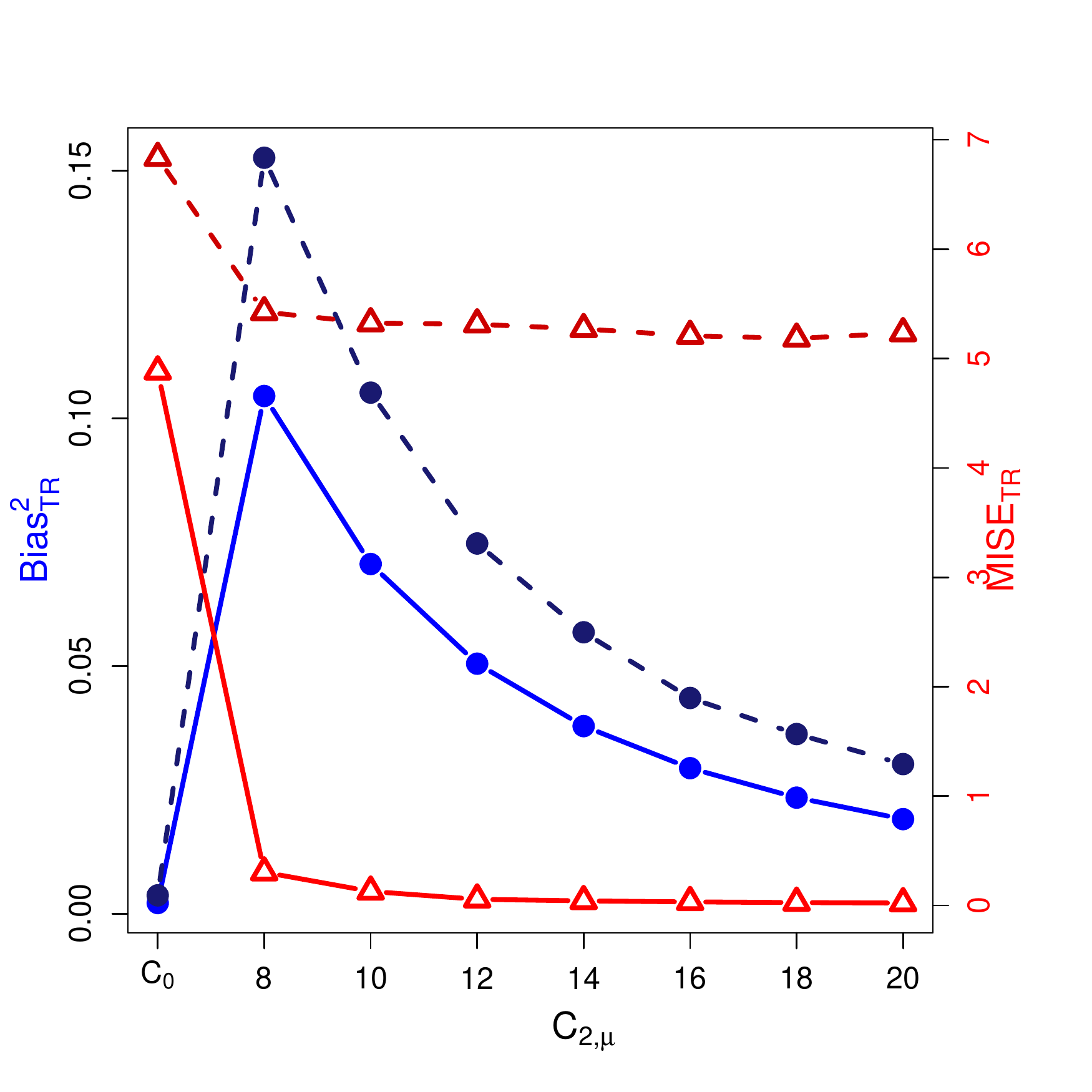}\\[-5ex]
{\small   $C_{3,2,\delta}$} &  
\includegraphics[scale=0.38]{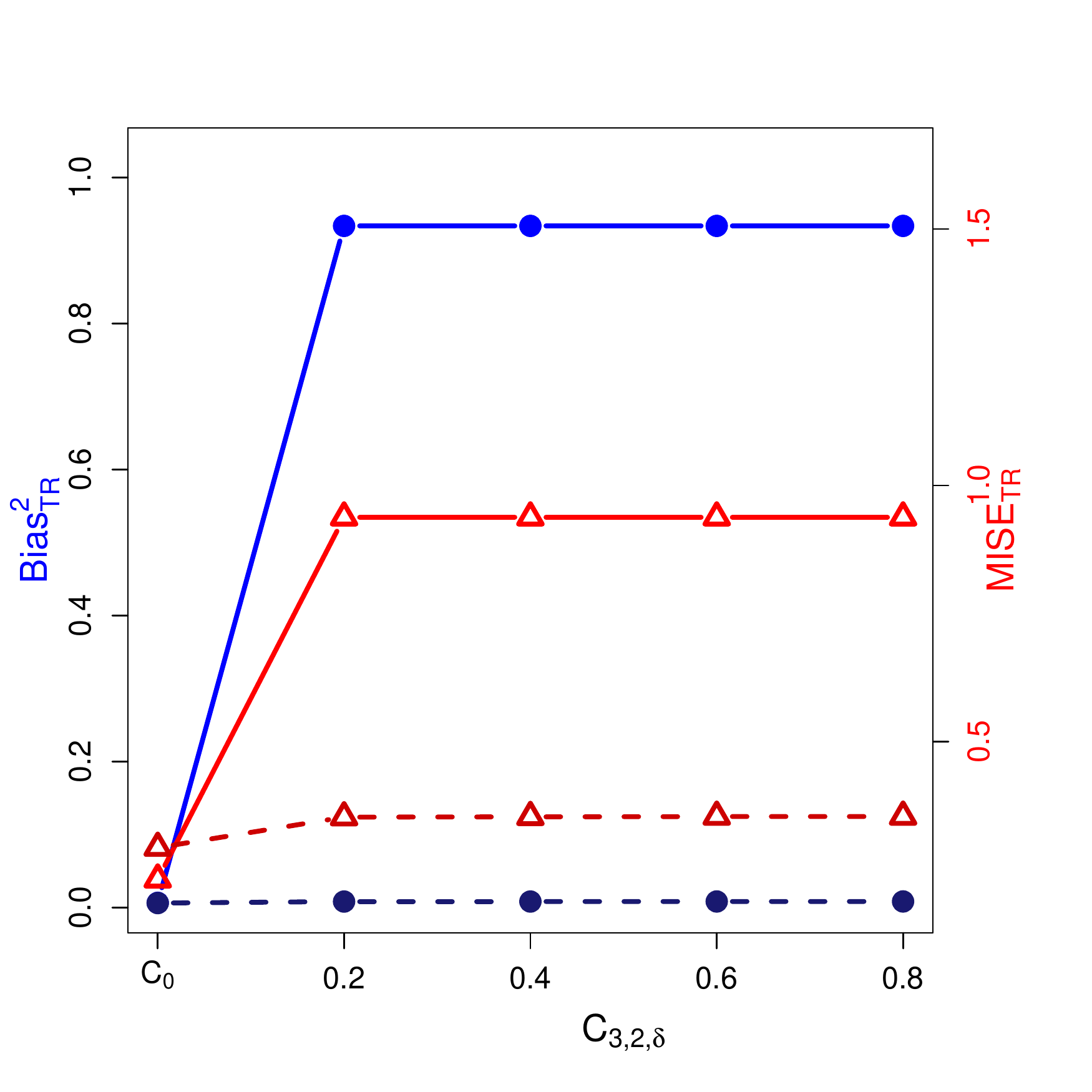} 
&  \includegraphics[scale=0.38]{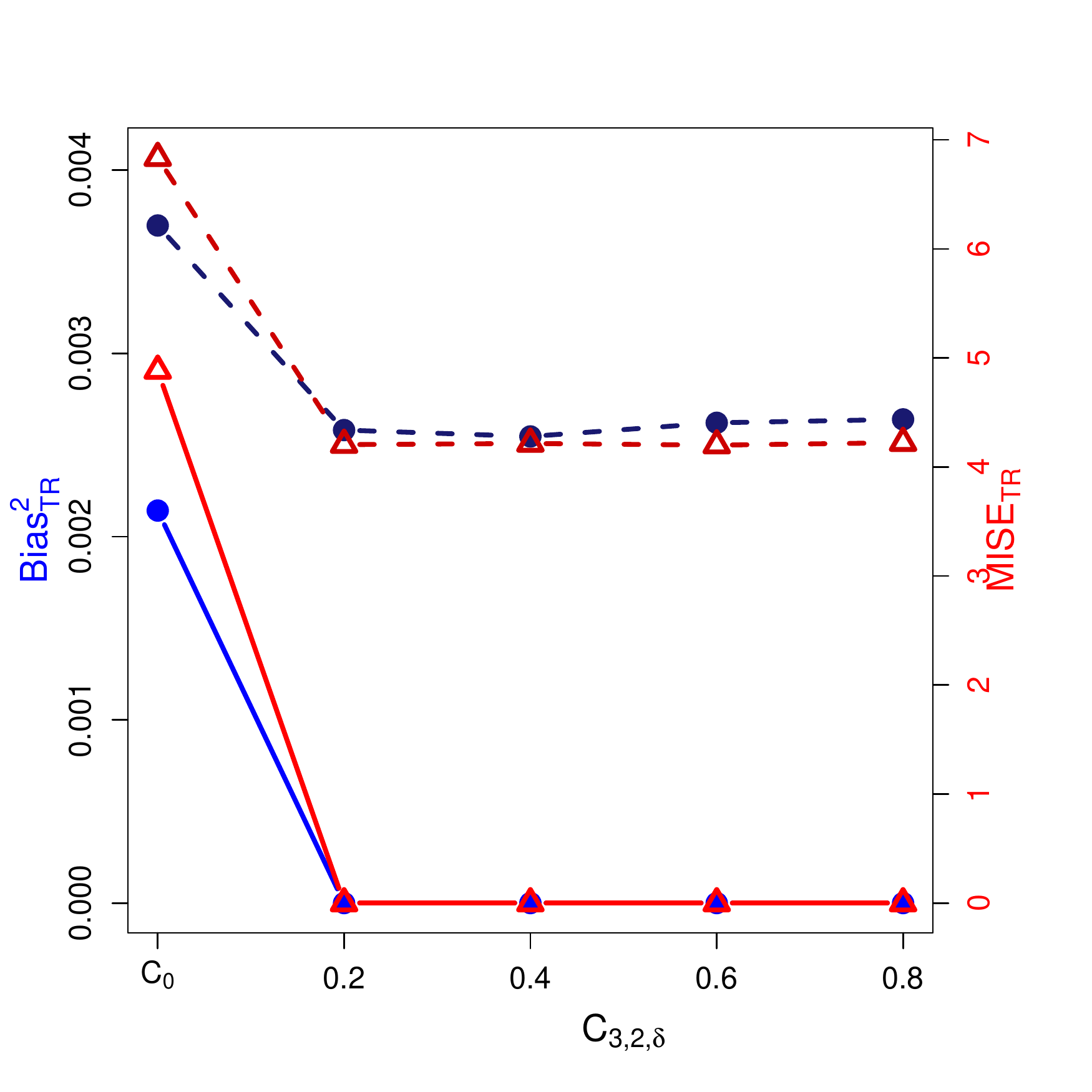}
 \end{tabular}
\caption{\small \label{fig:BIAS-MISE-Upsilon0}  Plots of the trimmed squared bias and MISE of the estimators of $\beta_0$ and $\upsilon_0$ 
as a function of $\mu$ or $\delta$ for each contamination scenario, under \textbf{Model 1} with $\Upsilon_0=0$.  The solid and dashed   lines correspond to the least squares and $MM-$estimators, respectively. The squared bias is indicated with circles, and the MISE with triangles. }
\end{center} 
\end{figure}

\begin{figure}[ht!]
 \begin{center}
 \newcolumntype{M}{>{\centering\arraybackslash}m{\dimexpr.05\linewidth-1\tabcolsep}}
   \newcolumntype{G}{>{\centering\arraybackslash}m{\dimexpr.4\linewidth-1\tabcolsep}}
\begin{tabular}{M GG}
 &  $\wbeta$  & $\wup$  \\[-4ex]
{\small   $C_{1,\mu}$} &  
\includegraphics[scale=0.38]{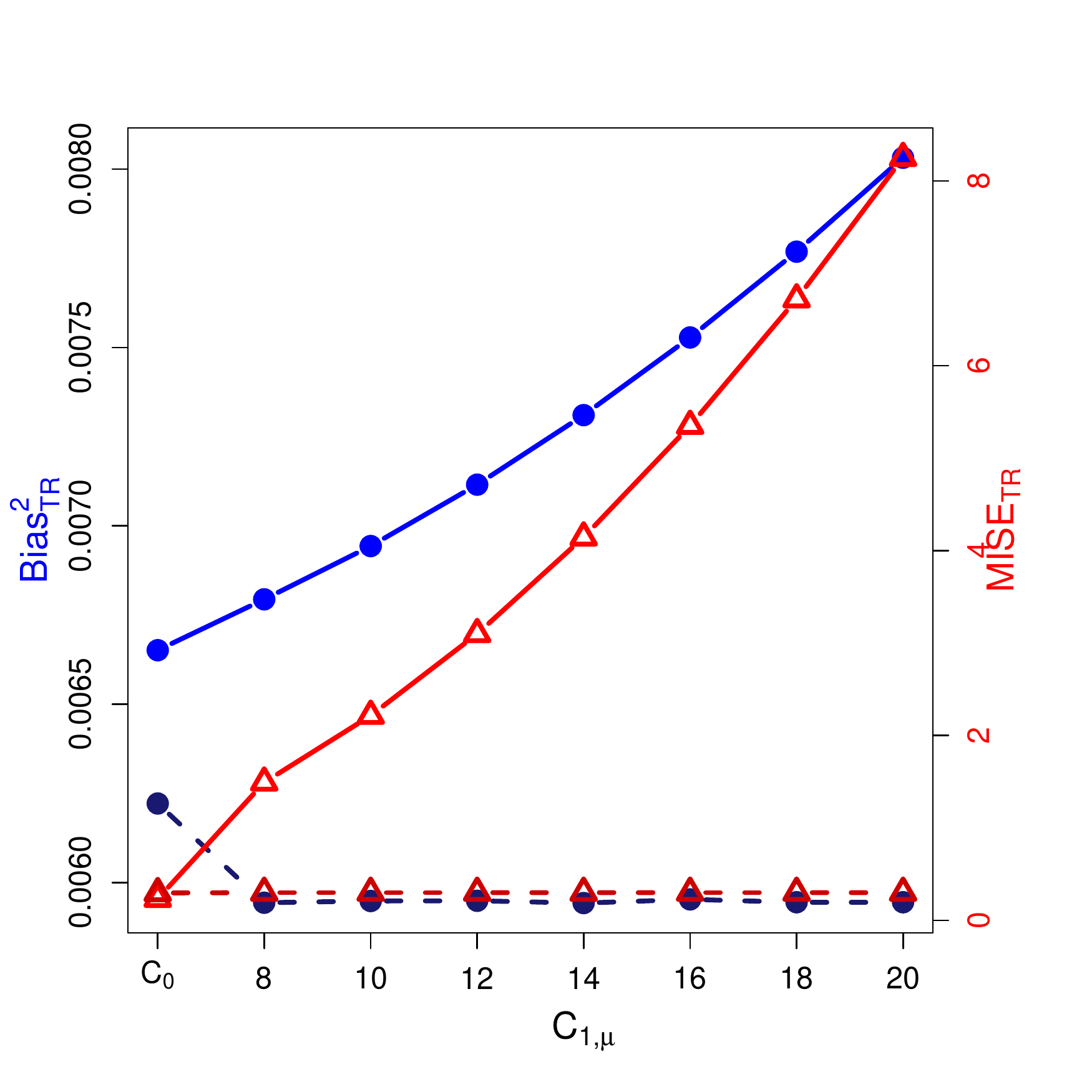} 
&  \includegraphics[scale=0.38]{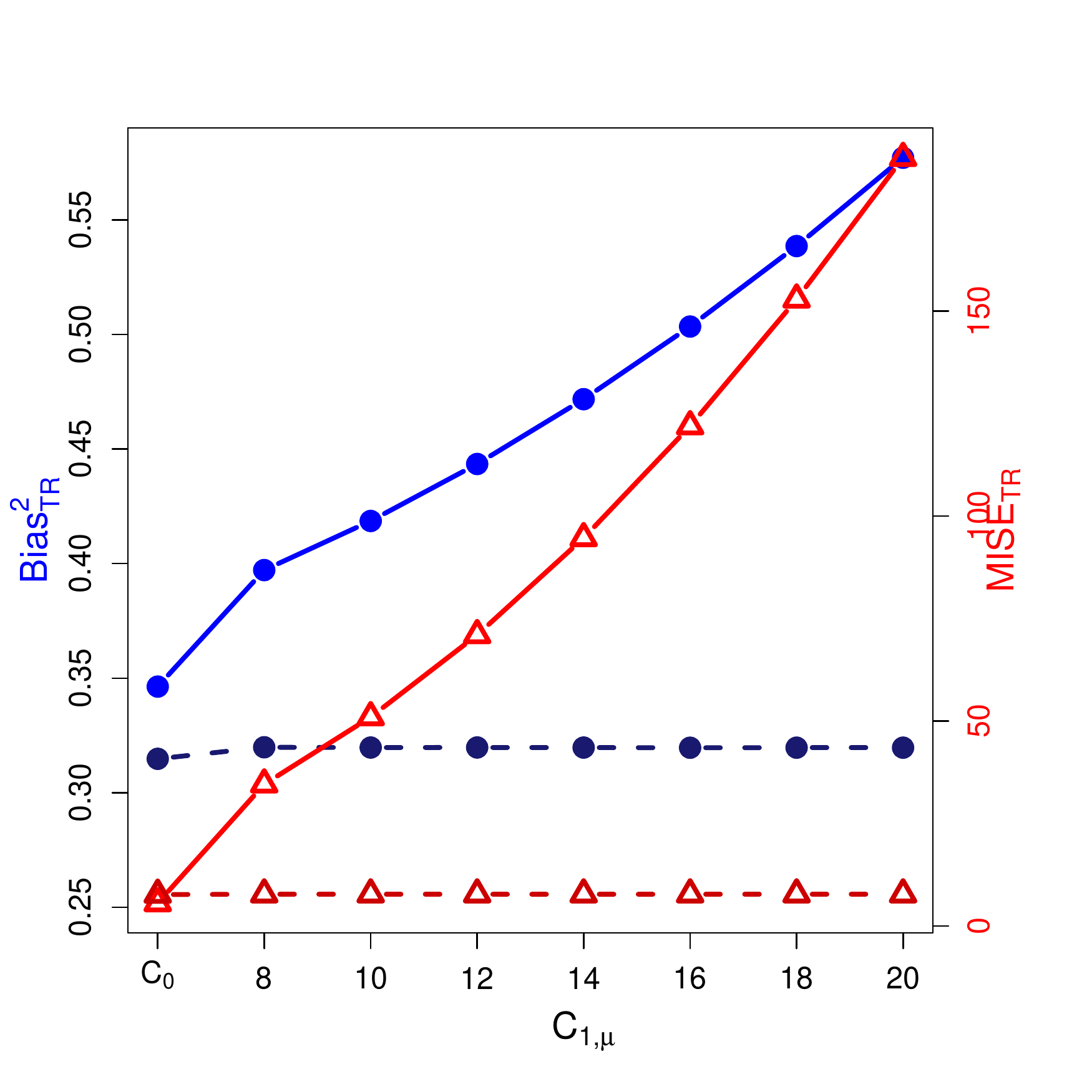}
 \\[-5ex]
 {\small   $C_{2,\mu}$} &  
 \includegraphics[scale=0.38]{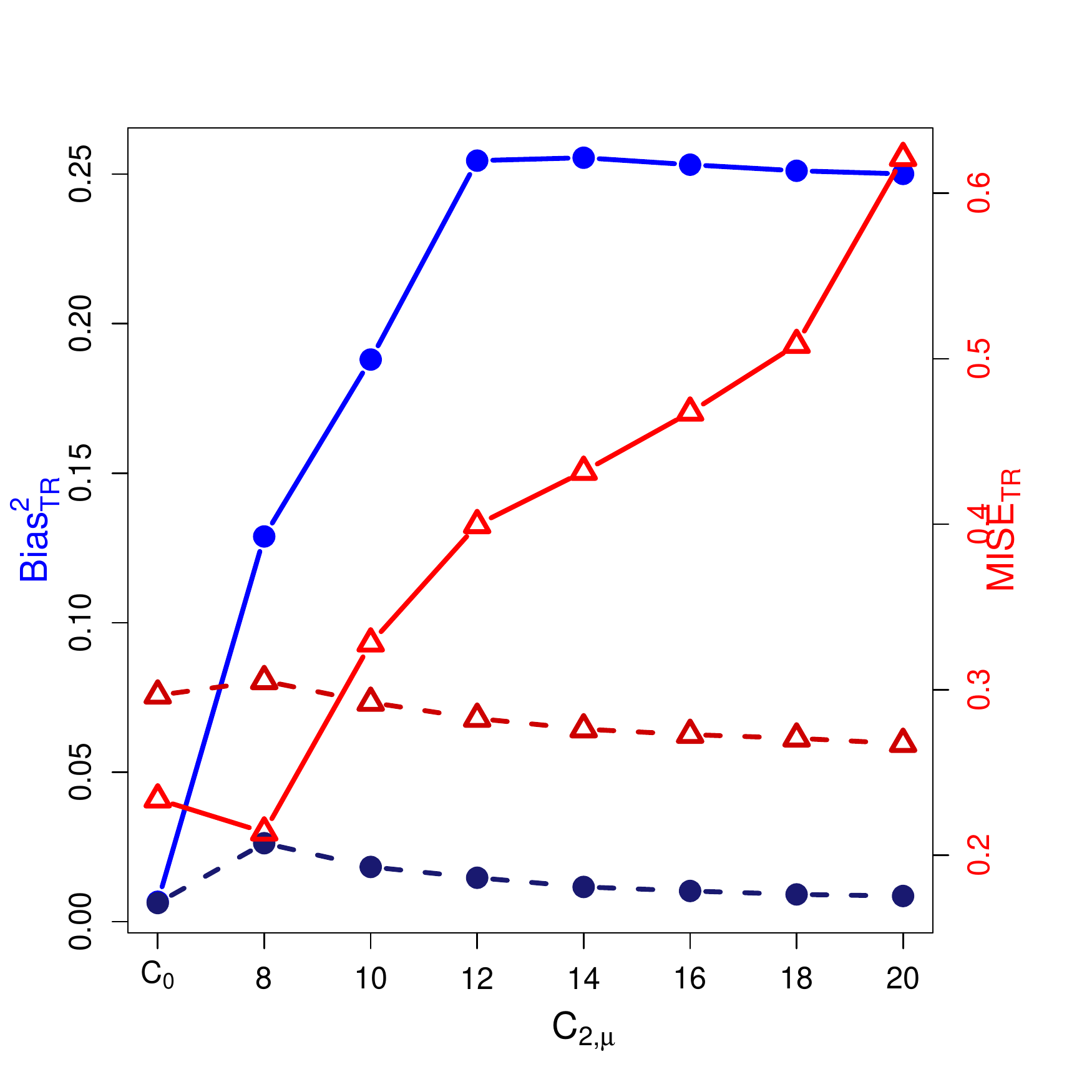} 
&  \includegraphics[scale=0.38]{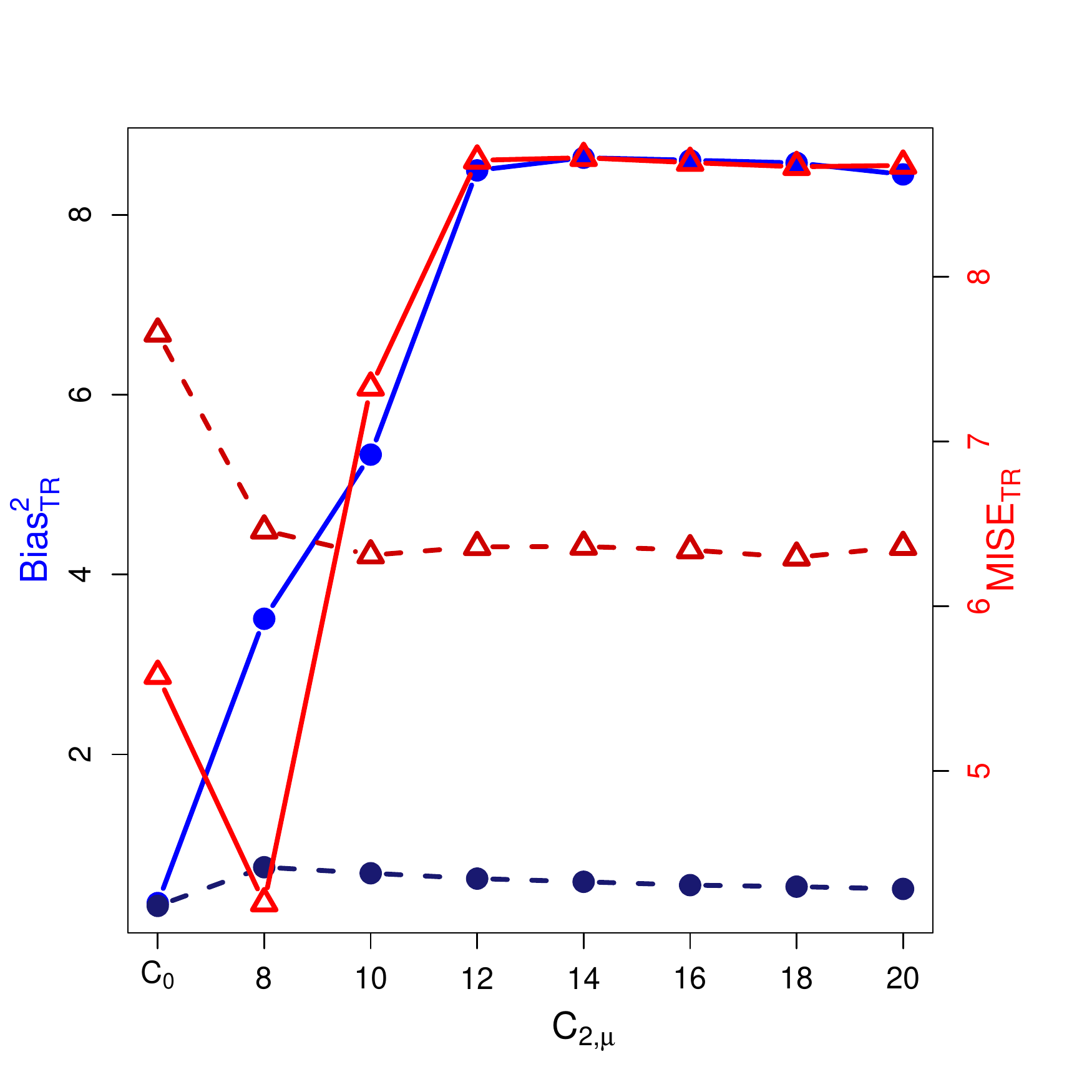}\\[-5ex]
{\small   $C_{3,2,\delta}$} &  
\includegraphics[scale=0.38]{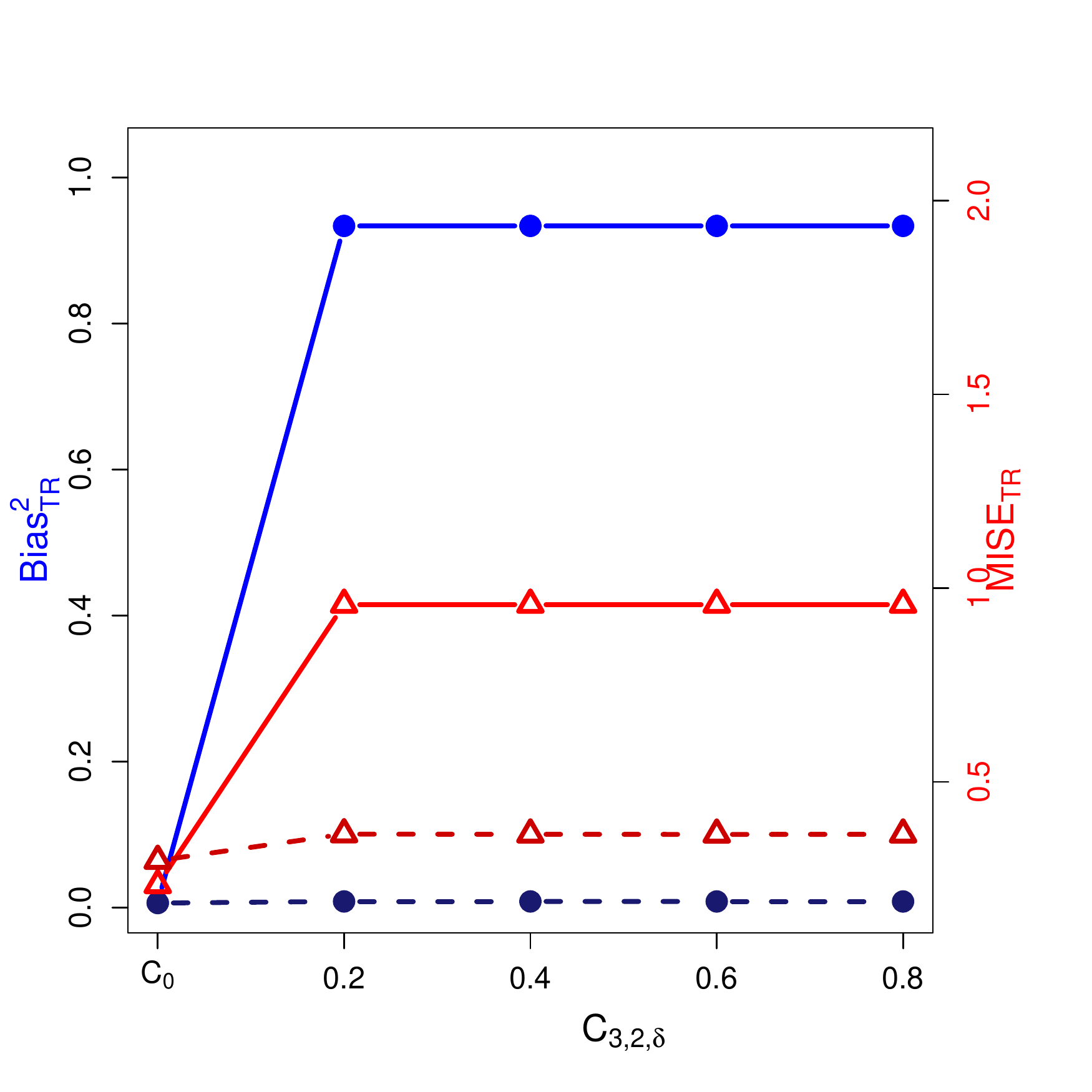} 
&  \includegraphics[scale=0.38]{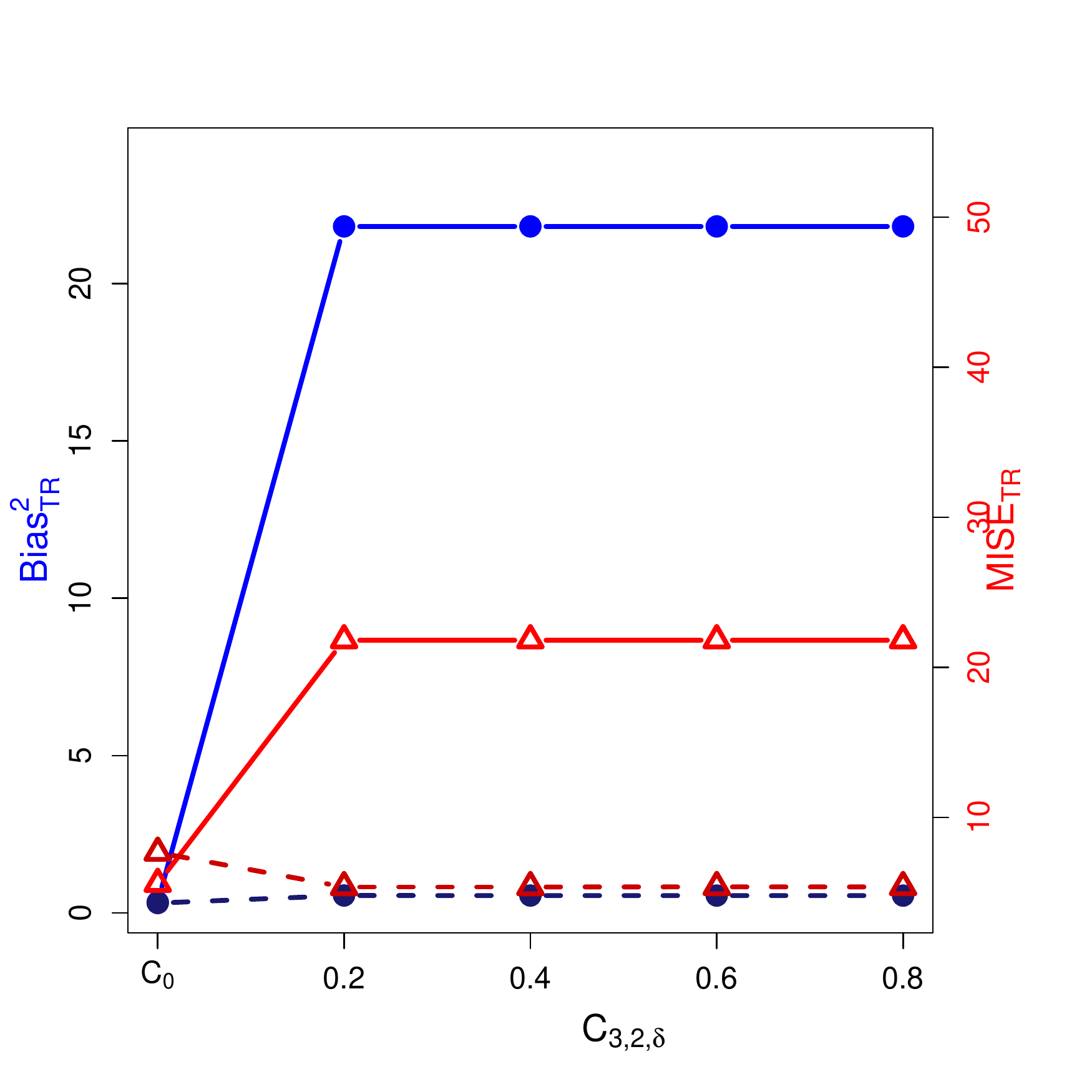}
 \end{tabular}
\caption{\small \label{fig:BIAS-MISE-Upsilon1}  Plots of the trimmed squared bias and MISE of the estimators of $\beta_0$ and $\upsilon_0$ 
as a function of  $\mu$ or $\delta$  for each contamination scenario, under \textbf{Model 1} with  $\Upsilon_0=\Upsilon_{0,1}$.  The solid and dashed   lines correspond to the least squares and $MM-$estimators, respectively. The squared bias is indicated with circles, and the MISE with triangles. }
\end{center} 
\end{figure}

\begin{figure}[ht!]
 \begin{center}
 \newcolumntype{M}{>{\centering\arraybackslash}m{\dimexpr.05\linewidth-1\tabcolsep}}
   \newcolumntype{G}{>{\centering\arraybackslash}m{\dimexpr.4\linewidth-1\tabcolsep}}
\begin{tabular}{M GG}
 &  $\wbeta$  & $\wup$  \\[-4ex]
{\small   $C_{1,\mu}$} &  
\includegraphics[scale=0.38]{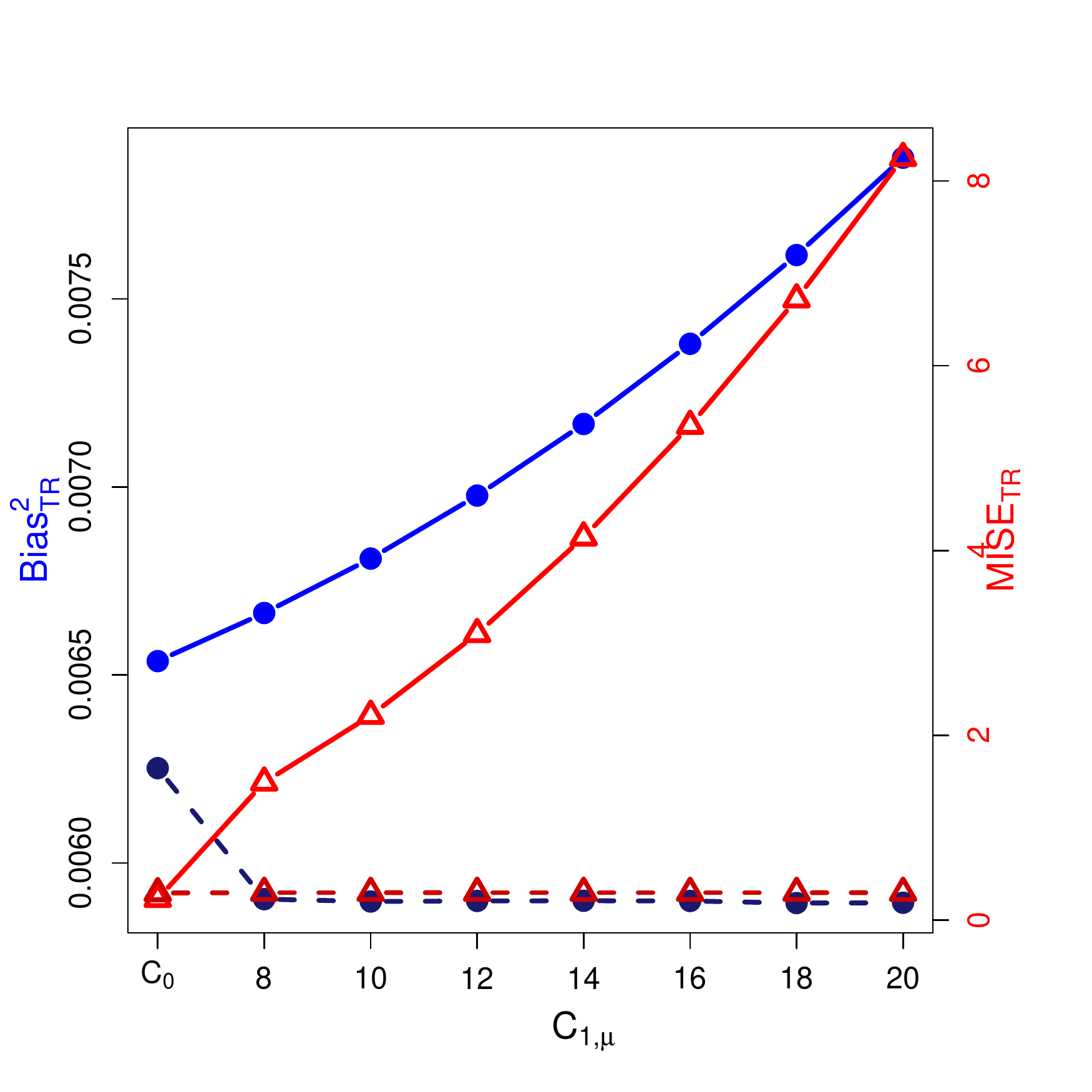} 
&  \includegraphics[scale=0.38]{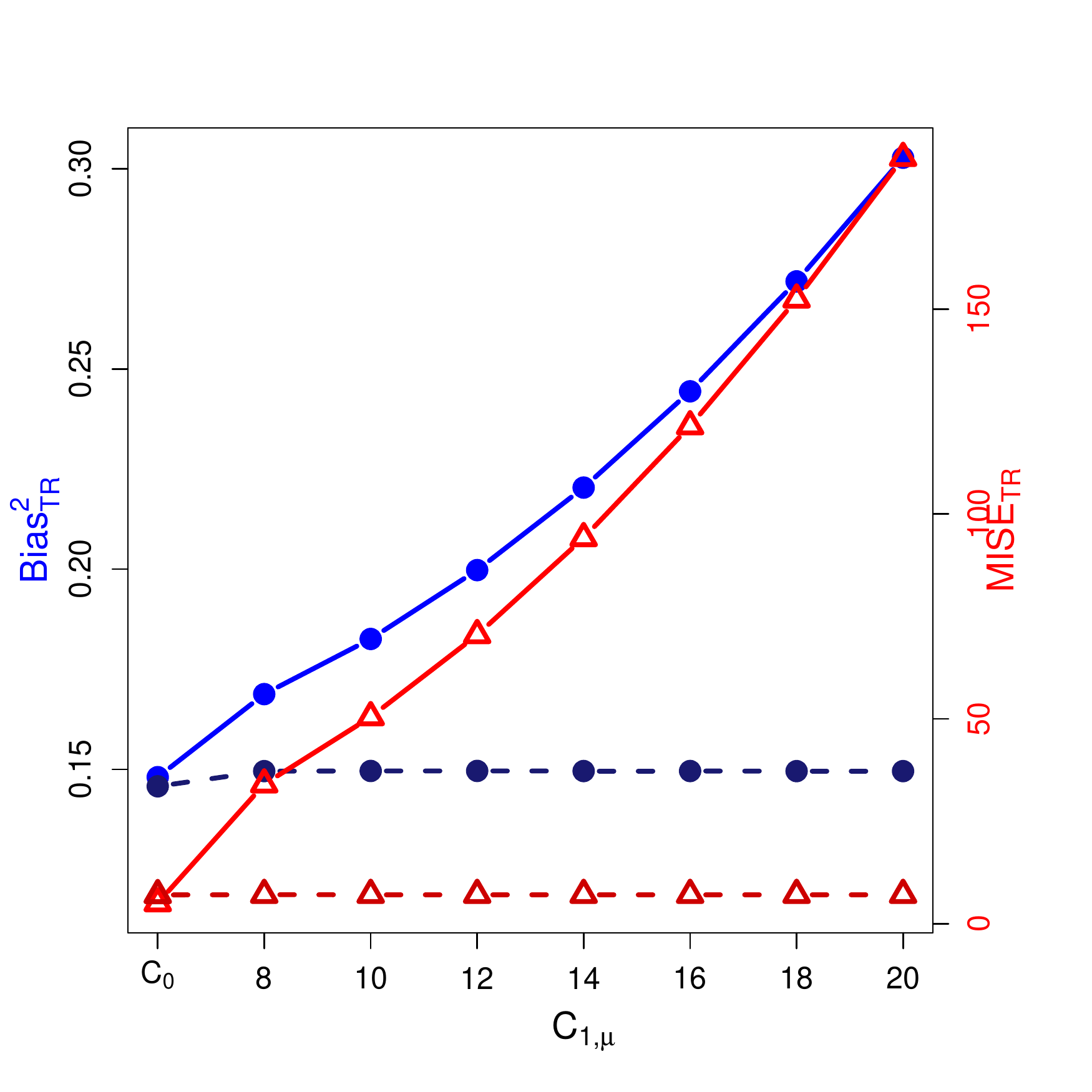}
 \\[-5ex]
 {\small   $C_{2,\mu}$} &  
 \includegraphics[scale=0.38]{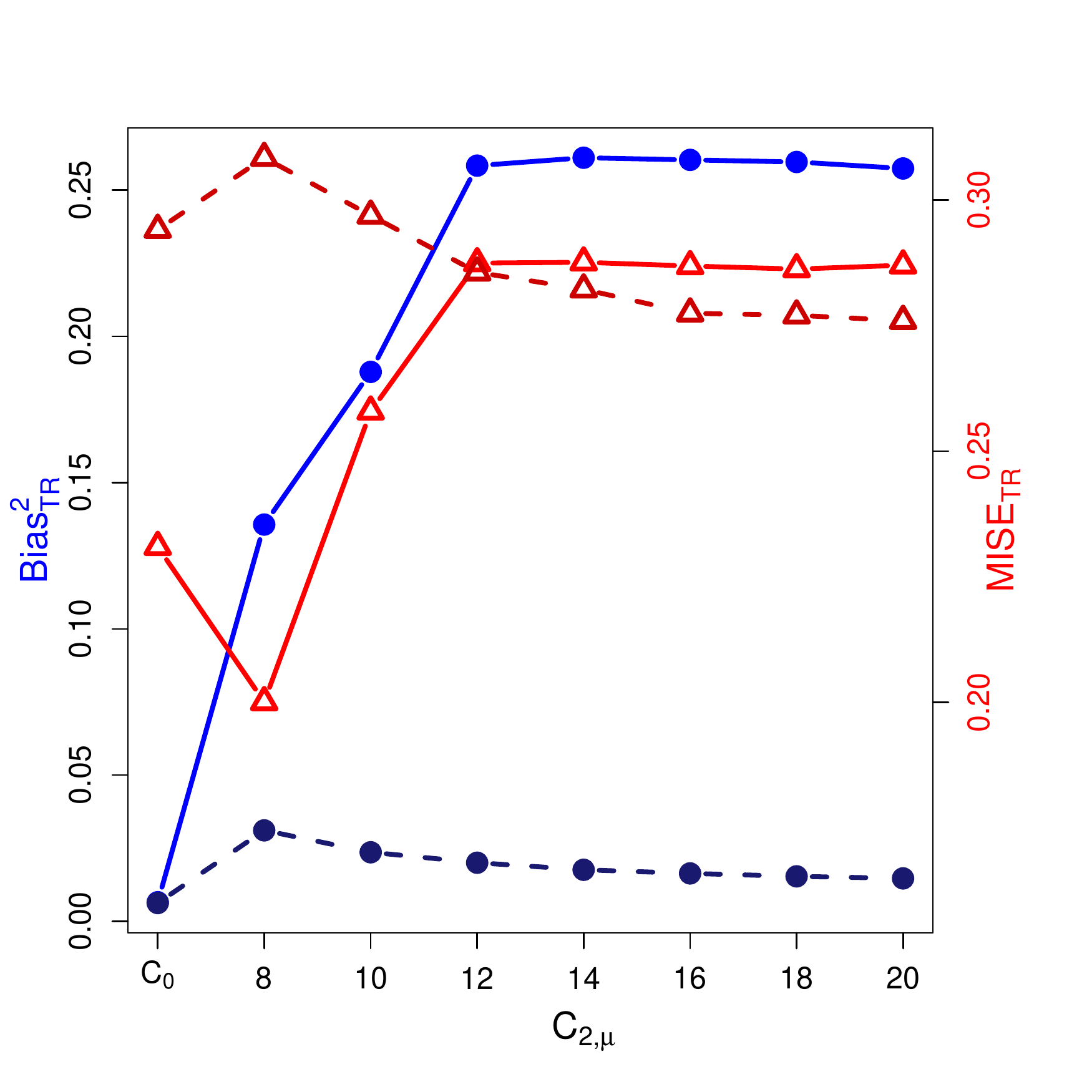} 
&  \includegraphics[scale=0.38]{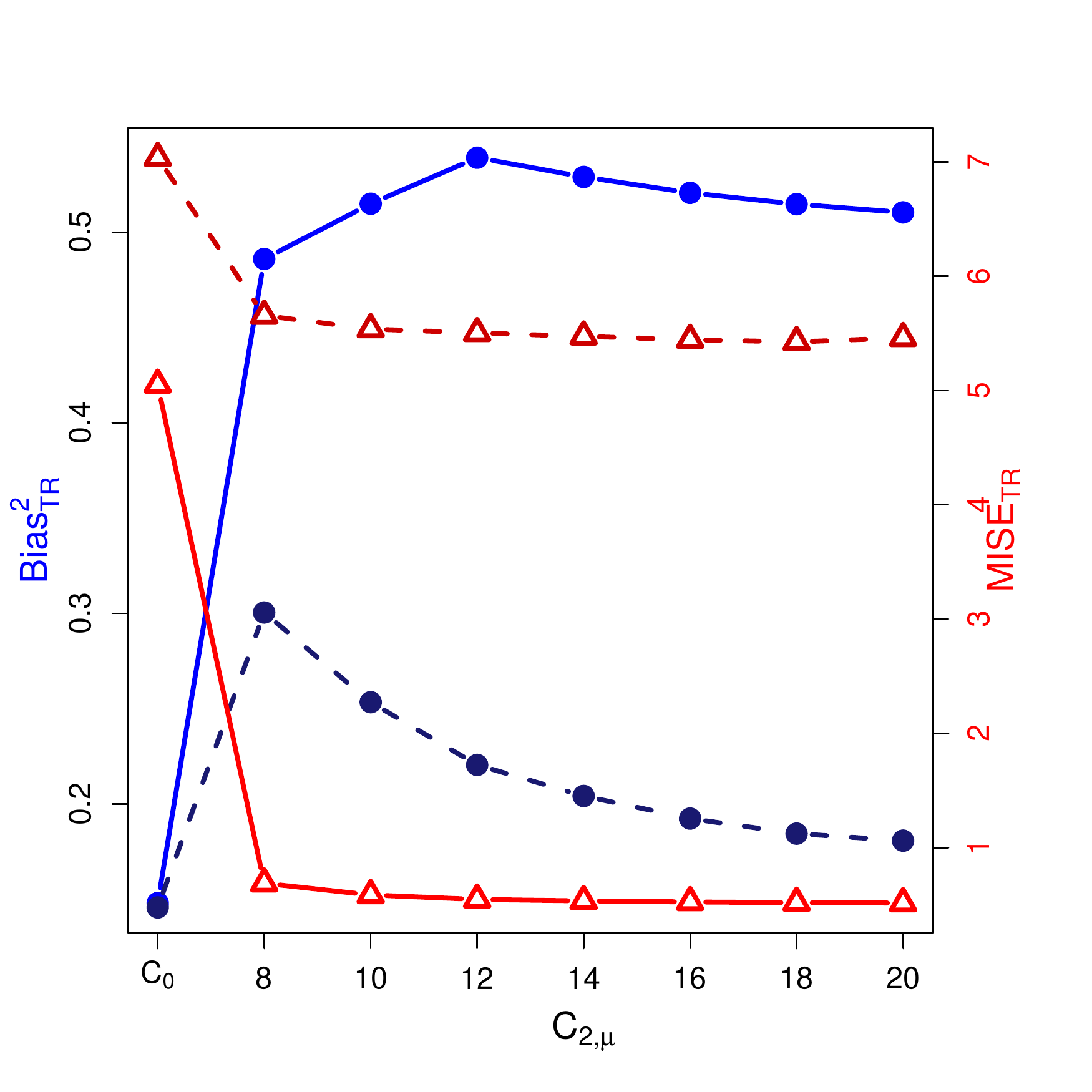}\\[-5ex]
{\small   $C_{3,2,\delta}$} &  
\includegraphics[scale=0.38]{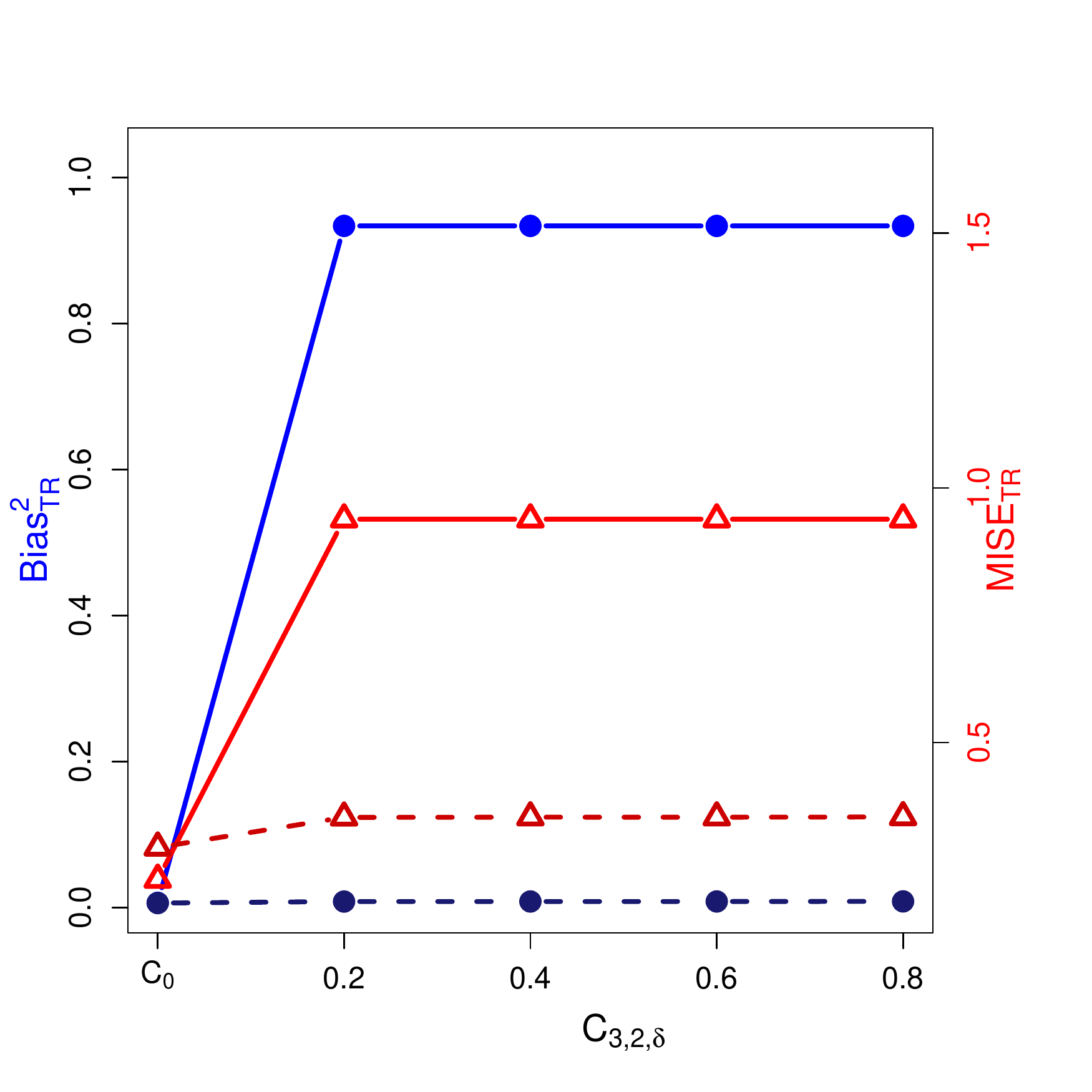} 
&  \includegraphics[scale=0.38]{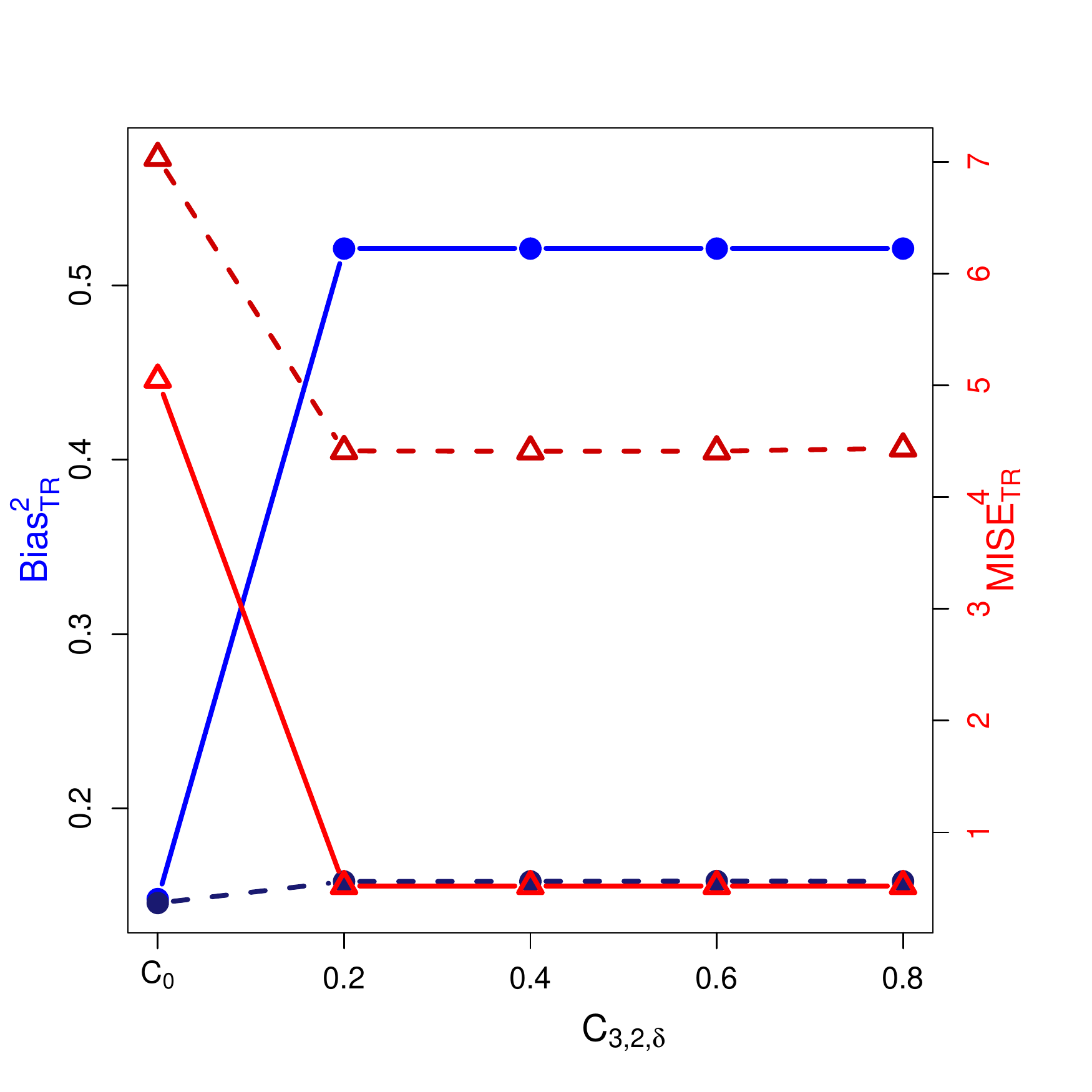}
 \end{tabular}
\caption{\small \label{fig:BIAS-MISE-Upsilon2}  Plots of the trimmed squared bias and MISE of the estimators of $\beta_0$ and $\upsilon_0$ 
as a function of  $\mu$ or $\delta$  for each contamination scenario, under \textbf{Model 1} with  $\Upsilon_0=\Upsilon_{0,2}$.  The solid and dashed   lines correspond to the least squares and $MM-$estimators, respectively. The squared bias is indicated with circles, and the MISE with triangles. }
\end{center} 
\end{figure}

The plots in Figures \ref{fig:BIAS-MISE-Upsilon0} to Figures \ref{fig:BIAS-MISE-Upsilon2} summarize the effect of the contamination scenarios for the different choices of $\Upsilon_0$ and  different values of $\mu$ when considering $C_{1,\mu}$ and $C_{2,\mu}$. For $C_{3,\mu,\delta}$, we choose $\mu=2$ and vary $\delta$. Each plot corresponds to one contamination scenario  and one parameter estimator. Within each panel, the solid and dashed lines correspond to the measures for the least squares  and $MM-$estimators, respectively. There are two lines per estimation method: the one with  triangles shows the trimmed MISE, and the one with solid circles indicates the corresponding trimmed squared  bias. We also include the values under $C_0$ for comparison, it is indicated in the horizontal axis as $C_0$.

As expected, when the data do not contain outliers, all estimators behave similarly to each other (see Table \ref{tab:tabla-1}). When estimating the quadratic kernel  $\upsilon_0$, the less efficient robust $MM-$estimator naturally results in higher MISE's. However, this efficiency loss is  smaller for the estimators of $\beta_0$. Note that the size of the squared bias and trimmed one are larger when using the quadratic operator $\Upsilon_{0,1}$ due to its shape near $(1,1)$.

The serious damage caused to the least squares estimators by  outliers can be seen in Figures \ref{fig:BIAS-MISE-Upsilon0} to  \ref{fig:BIAS-MISE-Upsilon2}. The behaviour clearly depends on the quadratic operator selected. For instance,   when considering the quadratic operator $\Upsilon_0=\Upsilon_{0,1}$, the trimmed bias and MISE of the least squares estimators of $\beta_0$ and $\upsilon_0$  are consistently much higher than those of the robust $MM-$estimators. The only exception is $C_{2,\mu}$ with $\mu=8$, which corresponds to a mild contamination and leads to larger  values of the MISE for the robust proposal, even when these values are smaller than those obtained for clean samples. The same behaviour is observed when considering  $\Upsilon_0=0$ or $\Upsilon_0=\Upsilon_{0,2}$ for the estimators of $\beta_0$, where smaller values of the MISE are obtained for the classical procedure only under $C_{2,\mu}$ with $\mu=8, 10$, even when its squared bias is consistently larger.

For the two quadratic kernels, different behaviours are obtained according to the contamination schemes. Contamination  $C_{1,\mu}$ affects both the bias and MISE of the least squares procedure  and the $MM-$estimator outperforms the classical one. When using $\Upsilon_0=\Upsilon_{0,2}$, both under  $C_{2,\mu}$  and  $C_{3,2,\delta}$ the bias of the classical procedure is enlarged, but it attains smaller values of the MISE than the robust method. Quite surprisingly, the obtained MISE are smaller than those obtained for clean samples. In contrast, when there is no quadratic term, i.e., $\Upsilon_0=0$ both the bias and the MISE of the robust procedure are larger than those obtained for the classical one, but under $C_{2,\mu}$ the contamination effect on the bias reduces as $\mu$ increases.

\begin{figure}[ht!]
 \begin{center}
 \footnotesize
 \renewcommand{\arraystretch}{0.2}
 \newcolumntype{M}{>{\centering\arraybackslash}m{\dimexpr.05\linewidth-1\tabcolsep}}
   \newcolumntype{G}{>{\centering\arraybackslash}m{\dimexpr.33\linewidth-1\tabcolsep}}
%\begin{tabular}{MGG}
\begin{tabular}{M GG}
 & $\wbeta_{\ls}$ &   $\wbeta_{\eme\eme}$ \\[-7ex]
$C_{0}$ 
&  \includegraphics[scale=0.35]{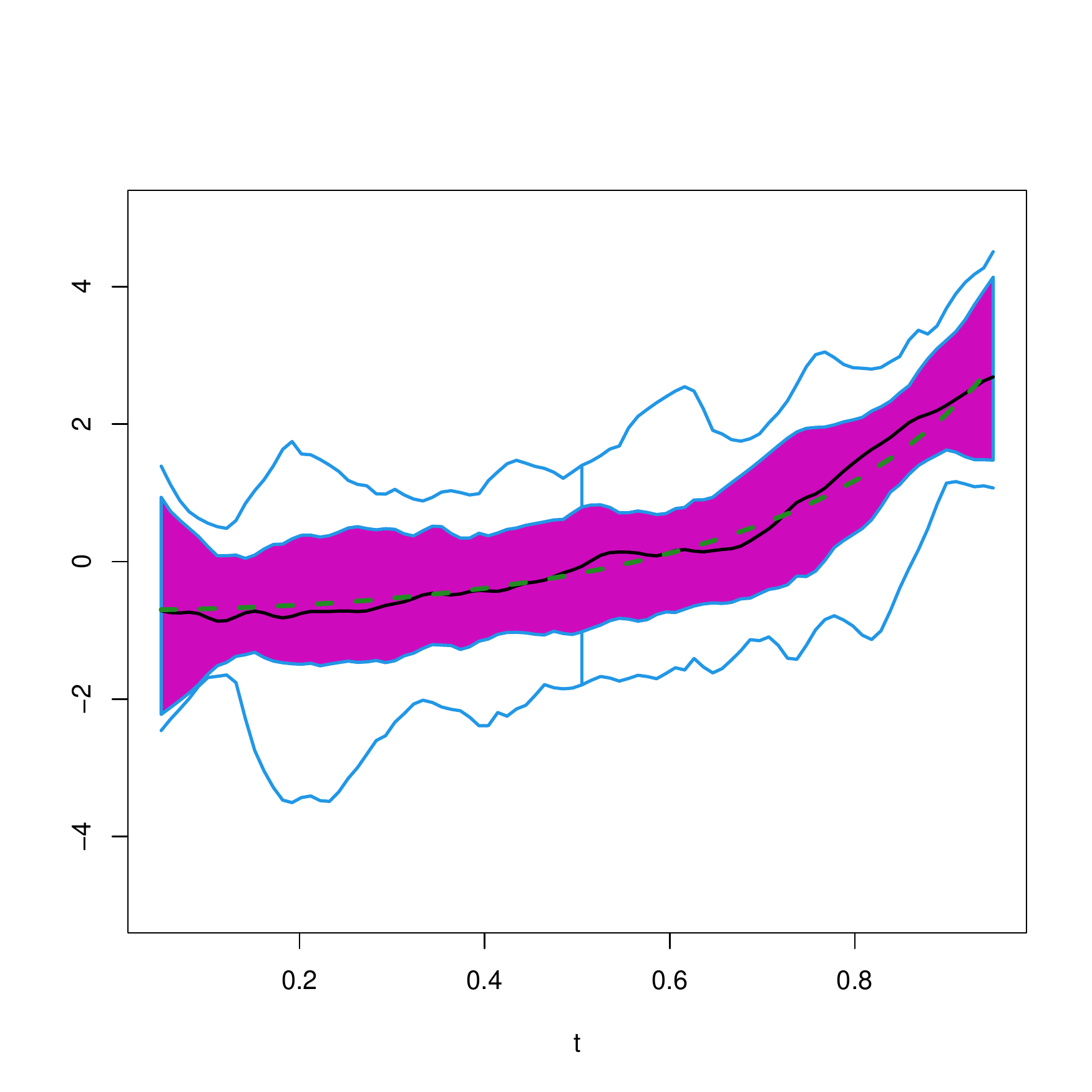} 
&  \includegraphics[scale=0.35]{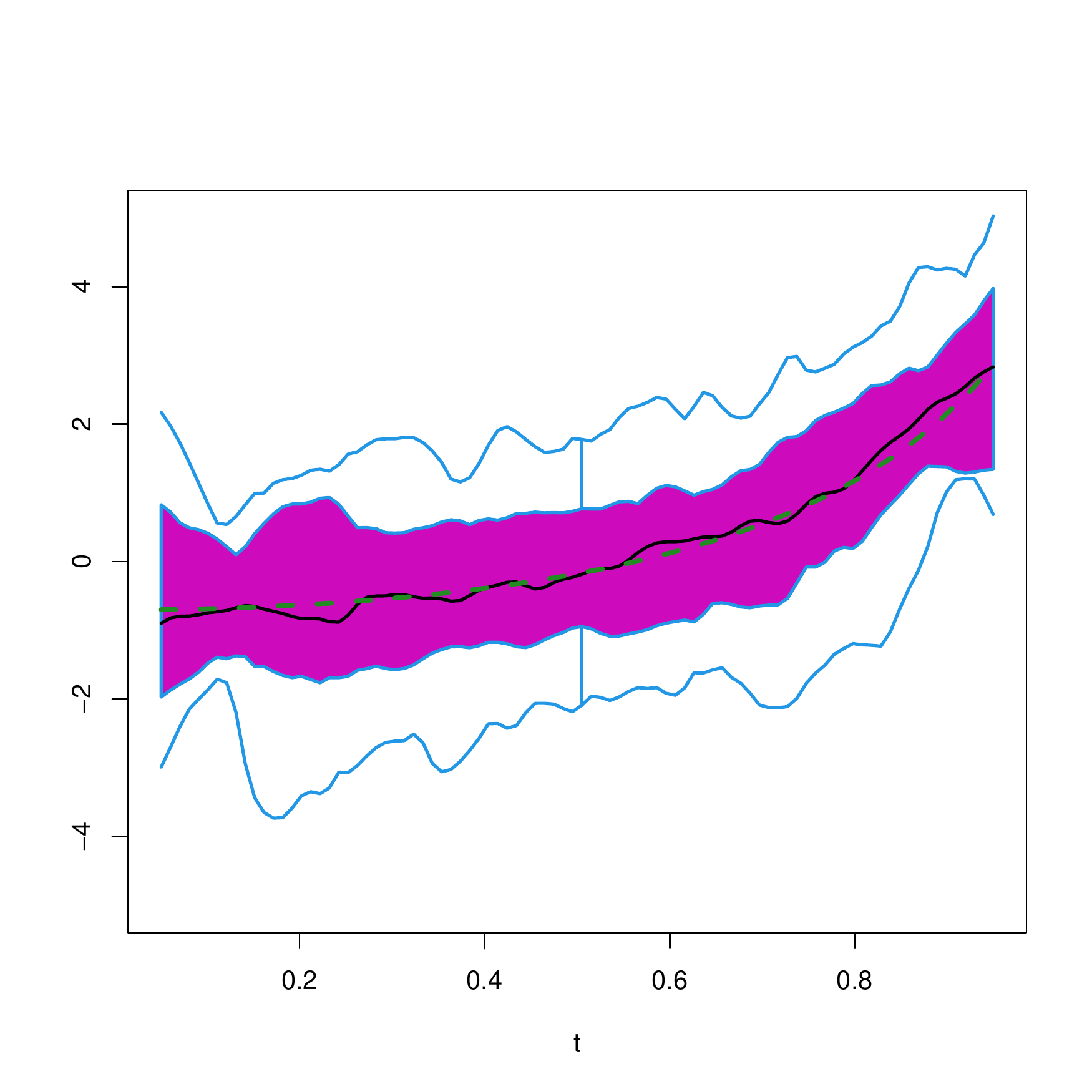} 
\\[-6ex]
$C_{1, 12}$ &  
 \includegraphics[scale=0.35]{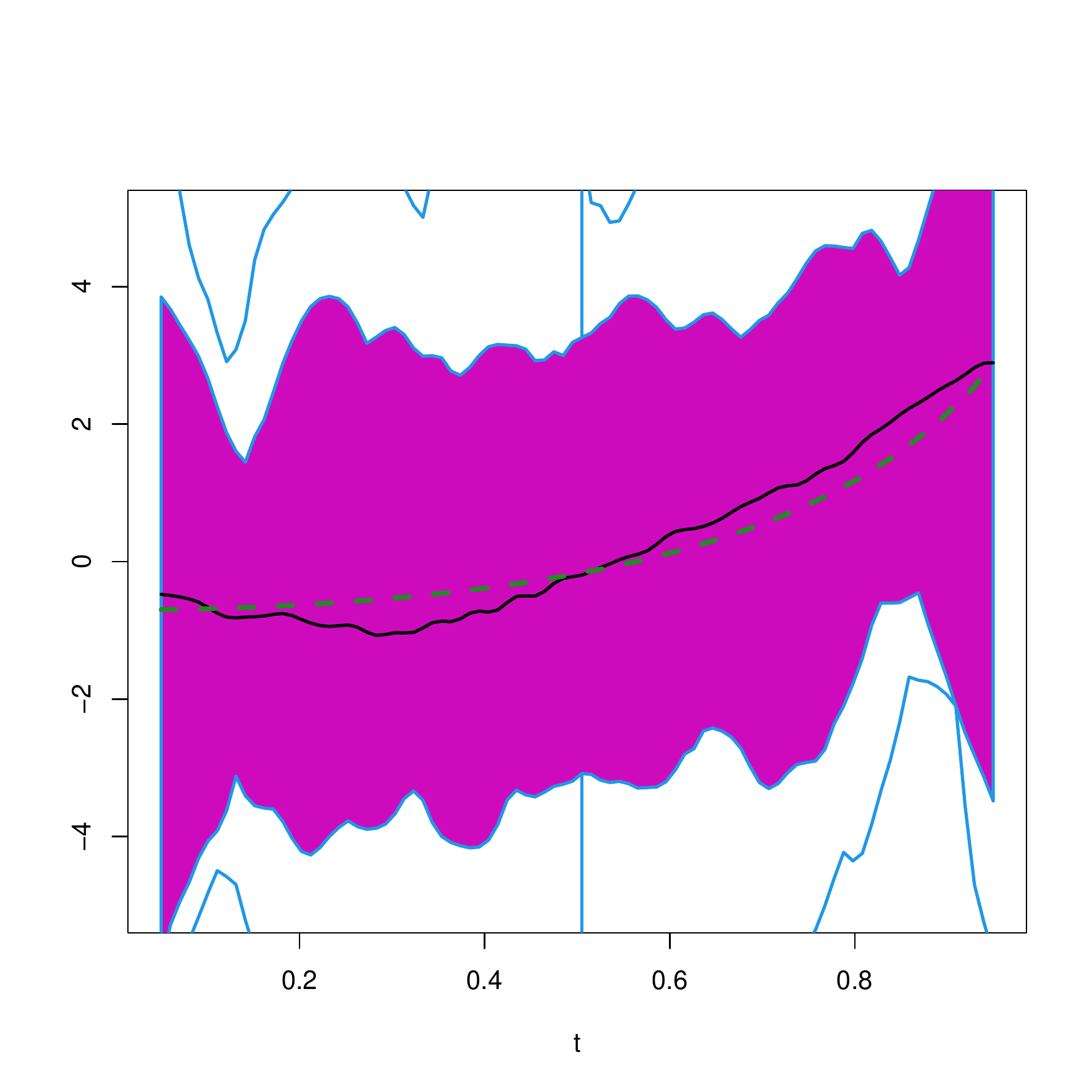}  & 
 \includegraphics[scale=0.35]{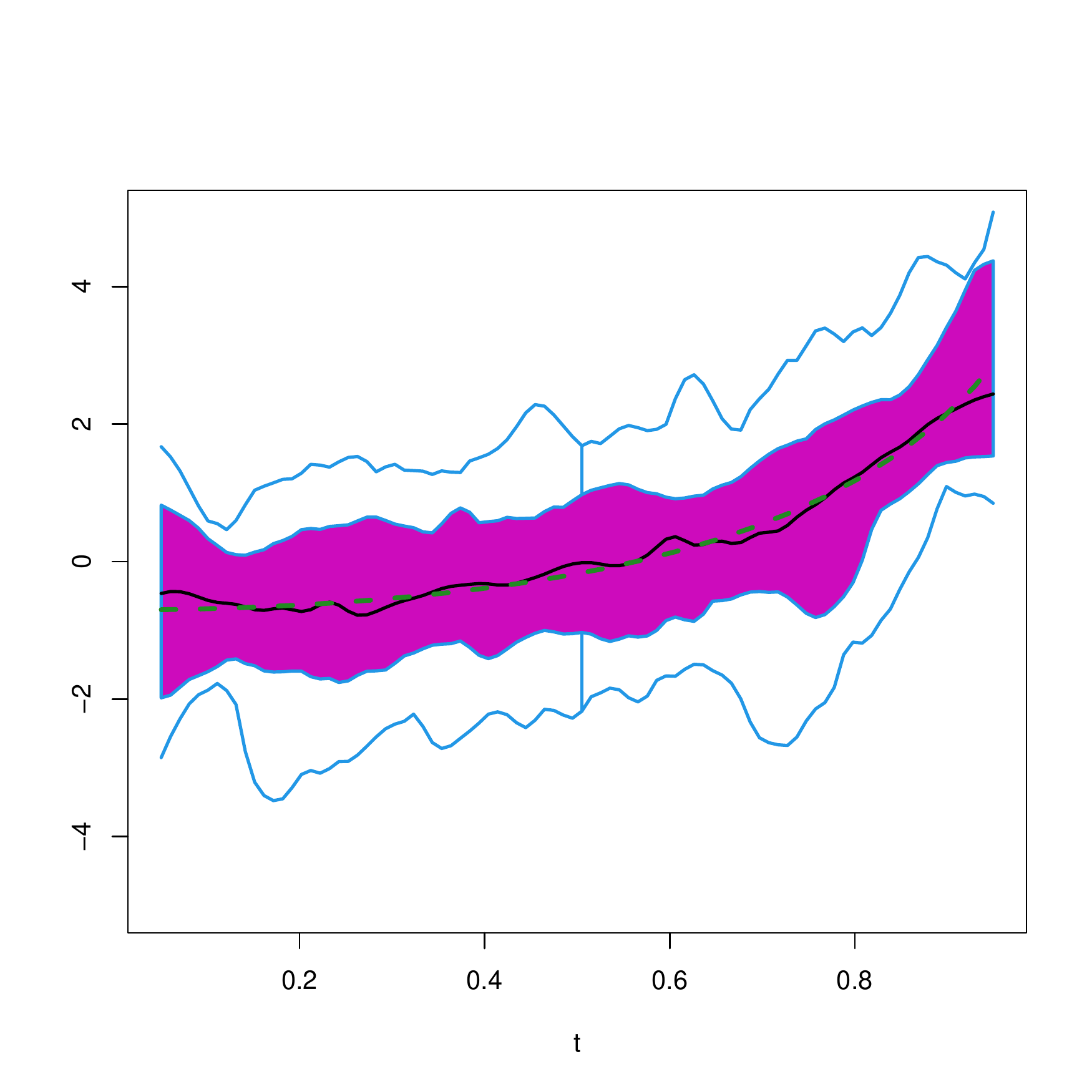} 
  \\[-6ex]
   
$C_{2, 12}$ &
 \includegraphics[scale=0.35]{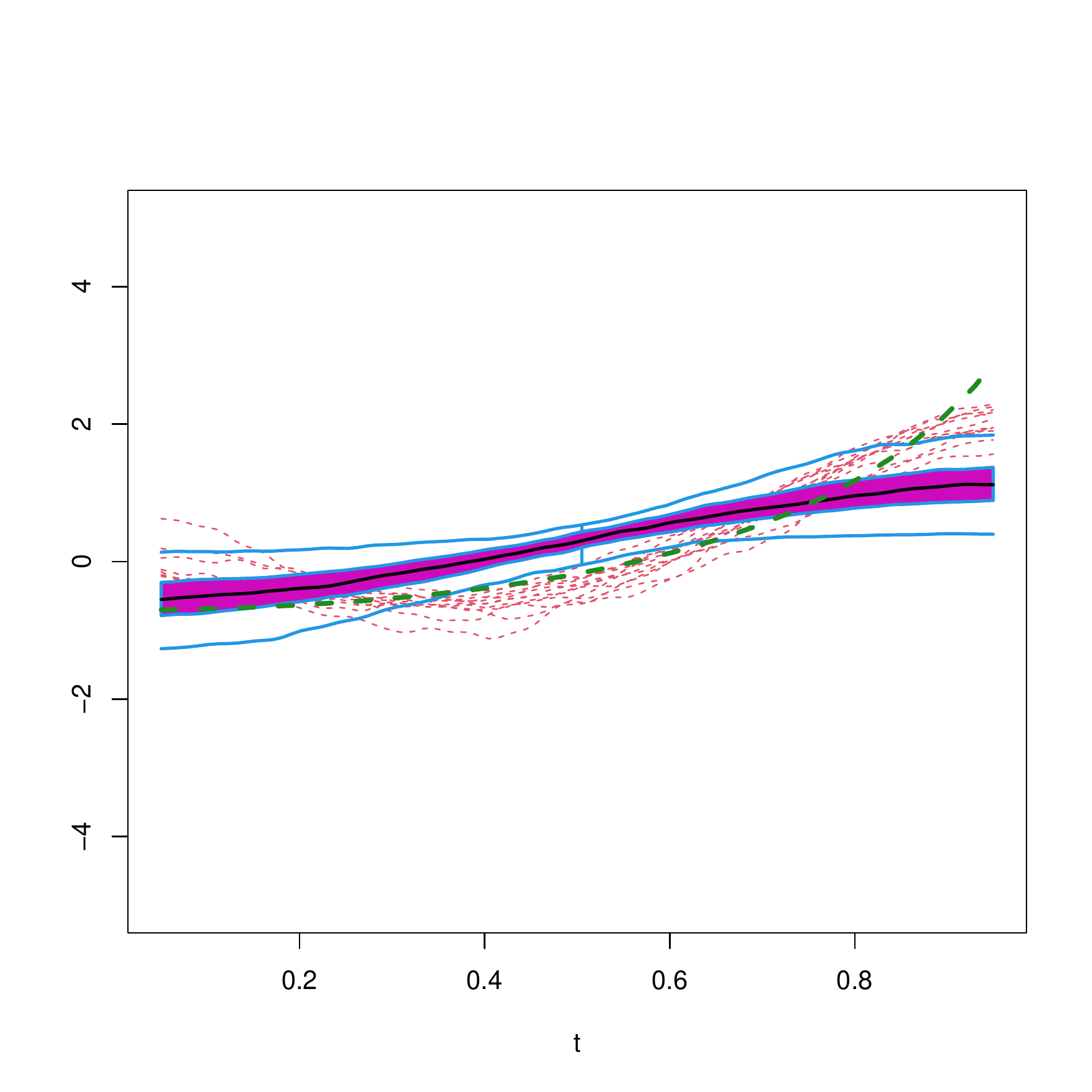}  & 
 \includegraphics[scale=0.35]{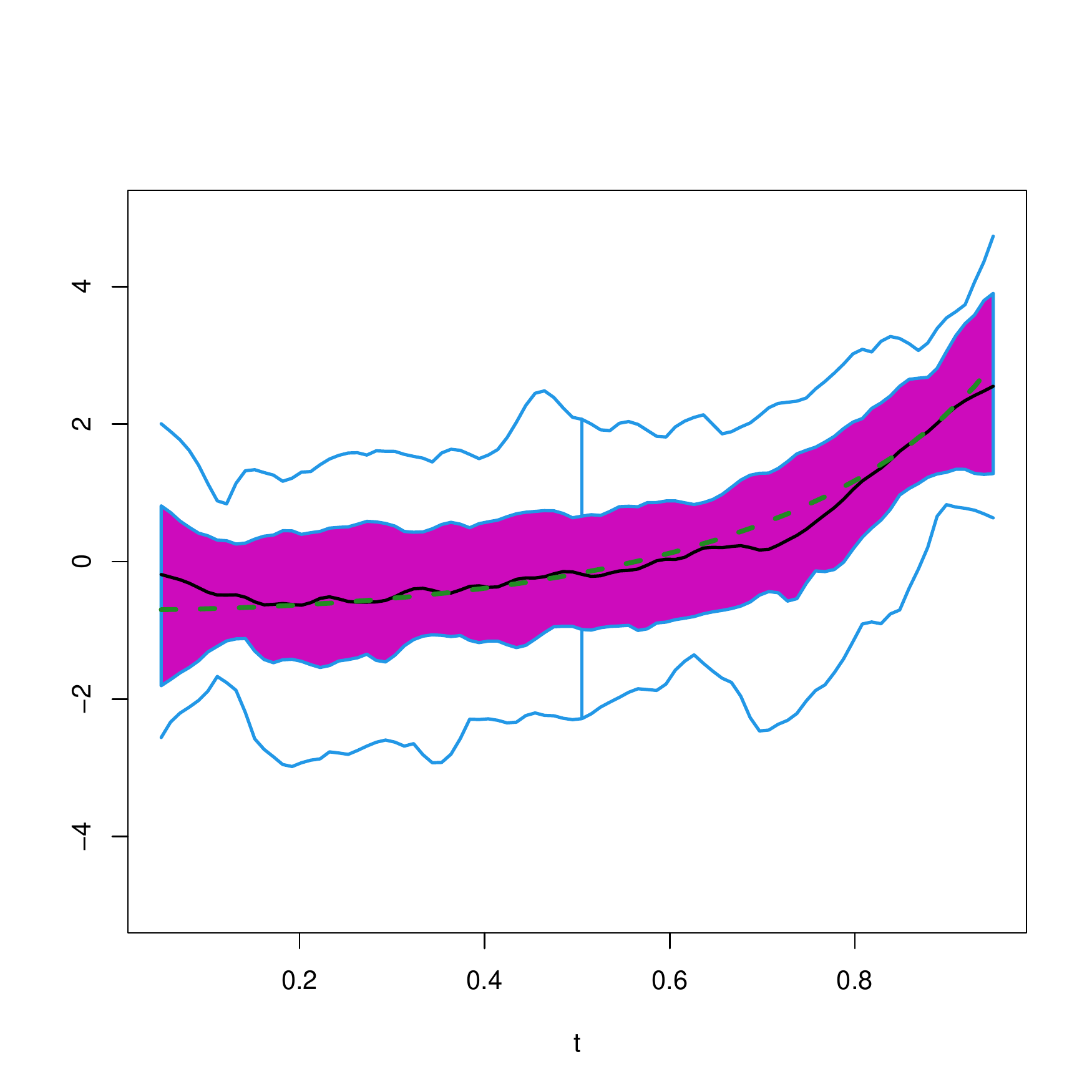} 
 \\[-6ex]
    
$C_{3, 4, 0.4}$ &
\includegraphics[scale=0.35]{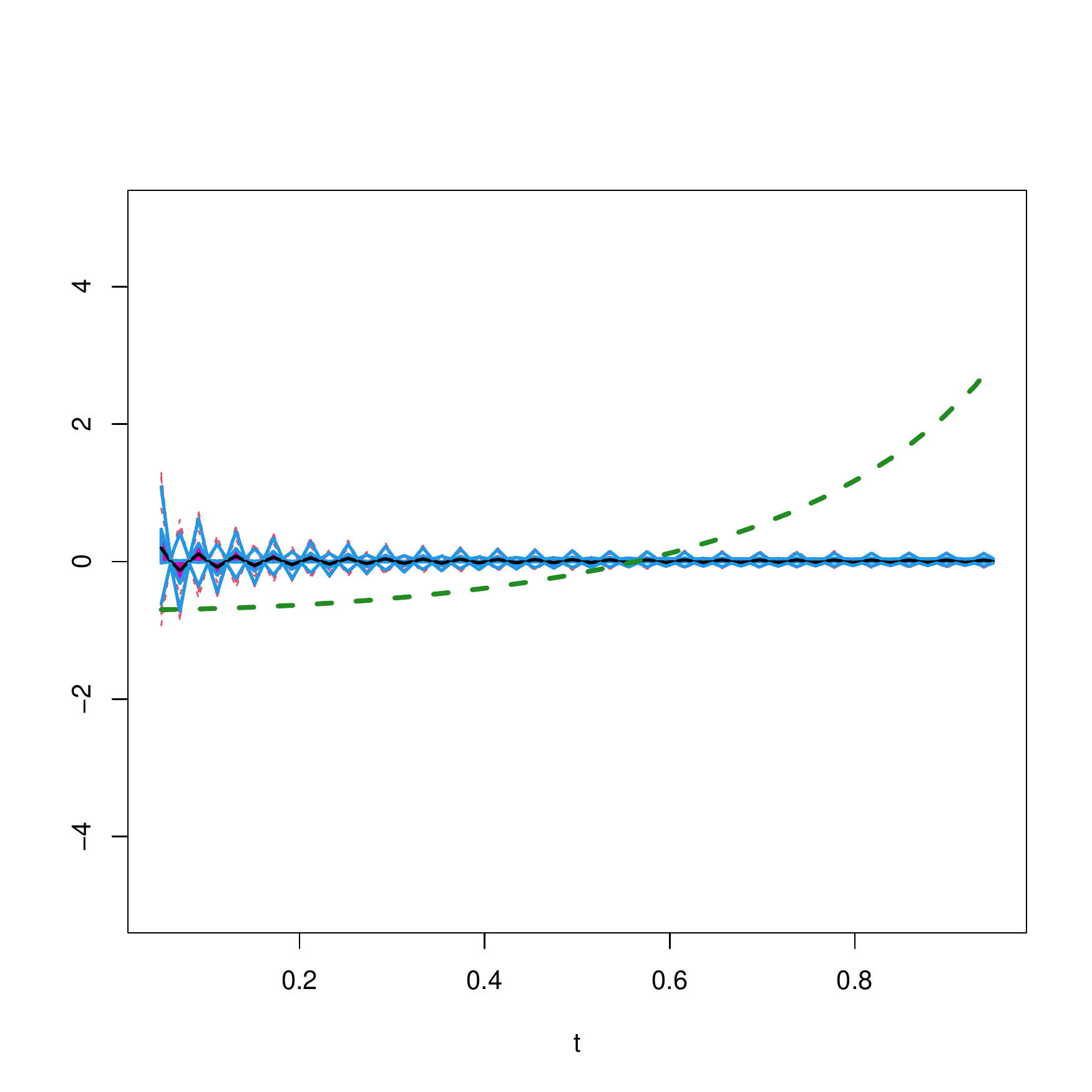}  & 
 \includegraphics[scale=0.35]{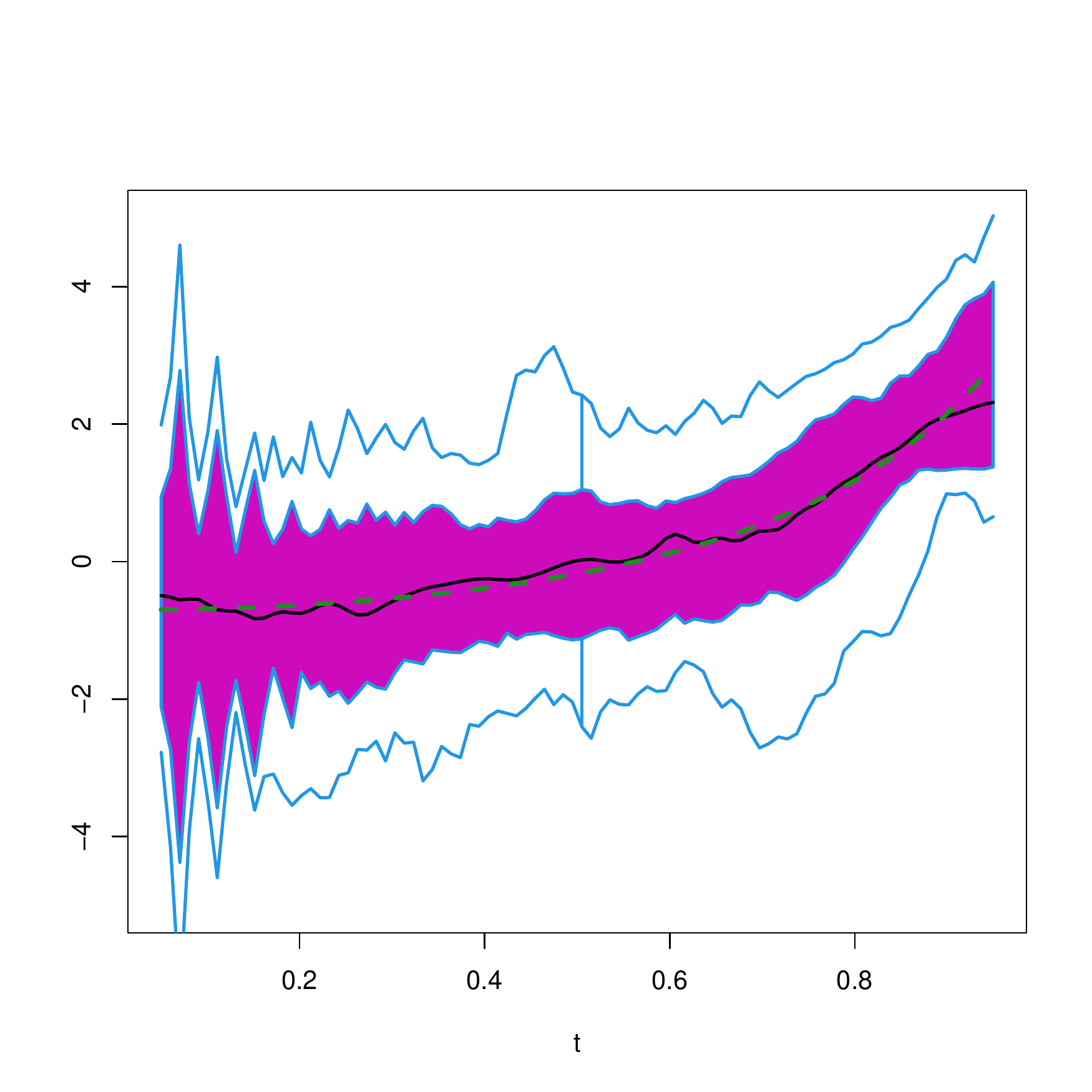}
\end{tabular}
\caption{\small \label{fig:wbeta-Upsilon0}  Functional boxplot of the estimators for $\beta_0$ under \textbf{Model 1} with  $\Upsilon_0=0$. 
The true function is shown with a green dashed line, while the black solid one is the central 
curve of the $n_R = 1000$ estimates $\wbeta$. Columns correspond to estimation 
methods, while rows to $C_0$ and   to some of the three contamination settings.}
\end{center} 
\end{figure}

\begin{figure}[ht!]
 \begin{center}
 \footnotesize
 \renewcommand{\arraystretch}{0.2}
 \newcolumntype{M}{>{\centering\arraybackslash}m{\dimexpr.05\linewidth-1\tabcolsep}}
   \newcolumntype{G}{>{\centering\arraybackslash}m{\dimexpr.33\linewidth-1\tabcolsep}}
%\begin{tabular}{MGG}
\begin{tabular}{M GG}
 & $\wbeta_{\ls}$ &   $\wbeta_{\eme\eme}$ \\[-7ex]
$C_{0}$ 
&  \includegraphics[scale=0.35]{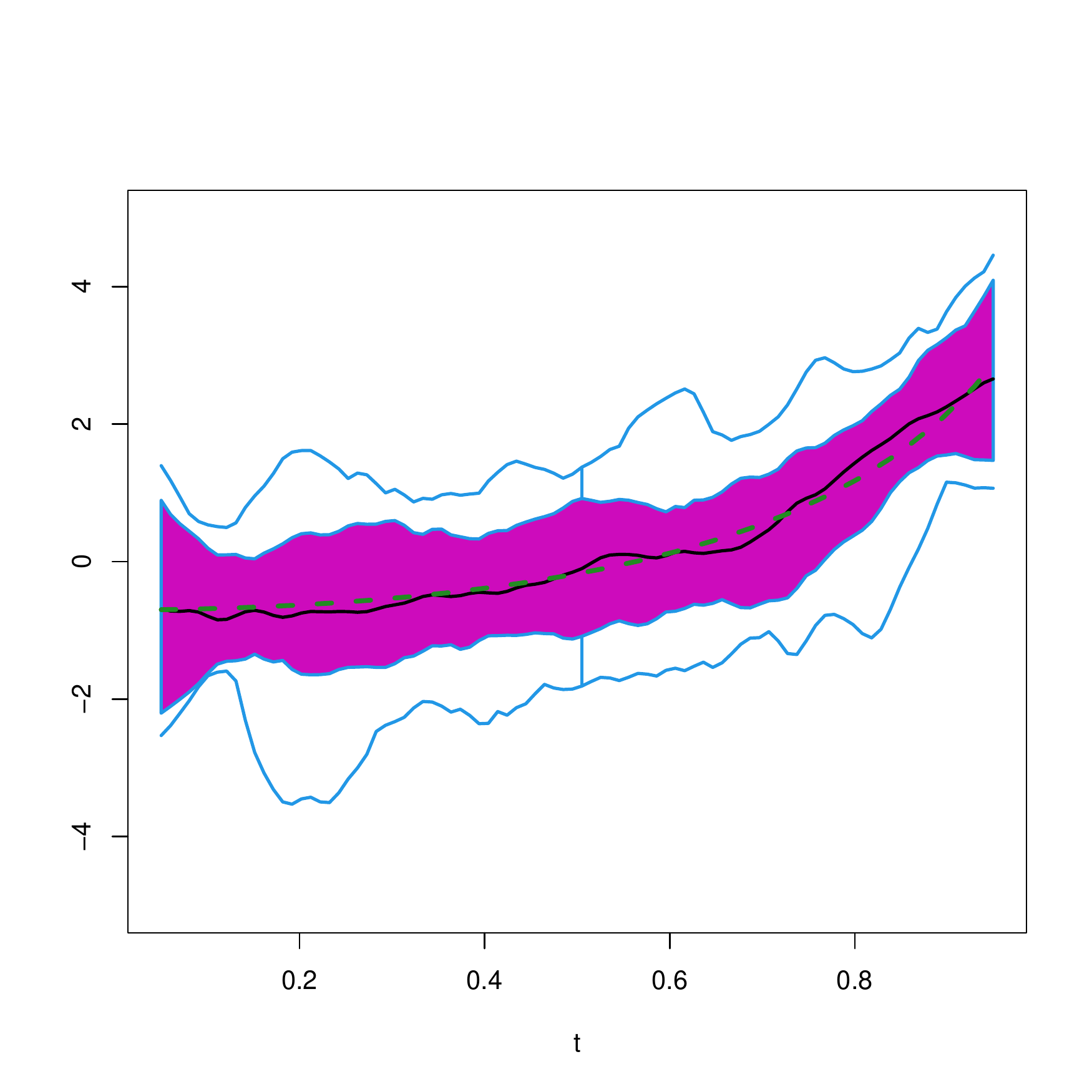} 
&  \includegraphics[scale=0.35]{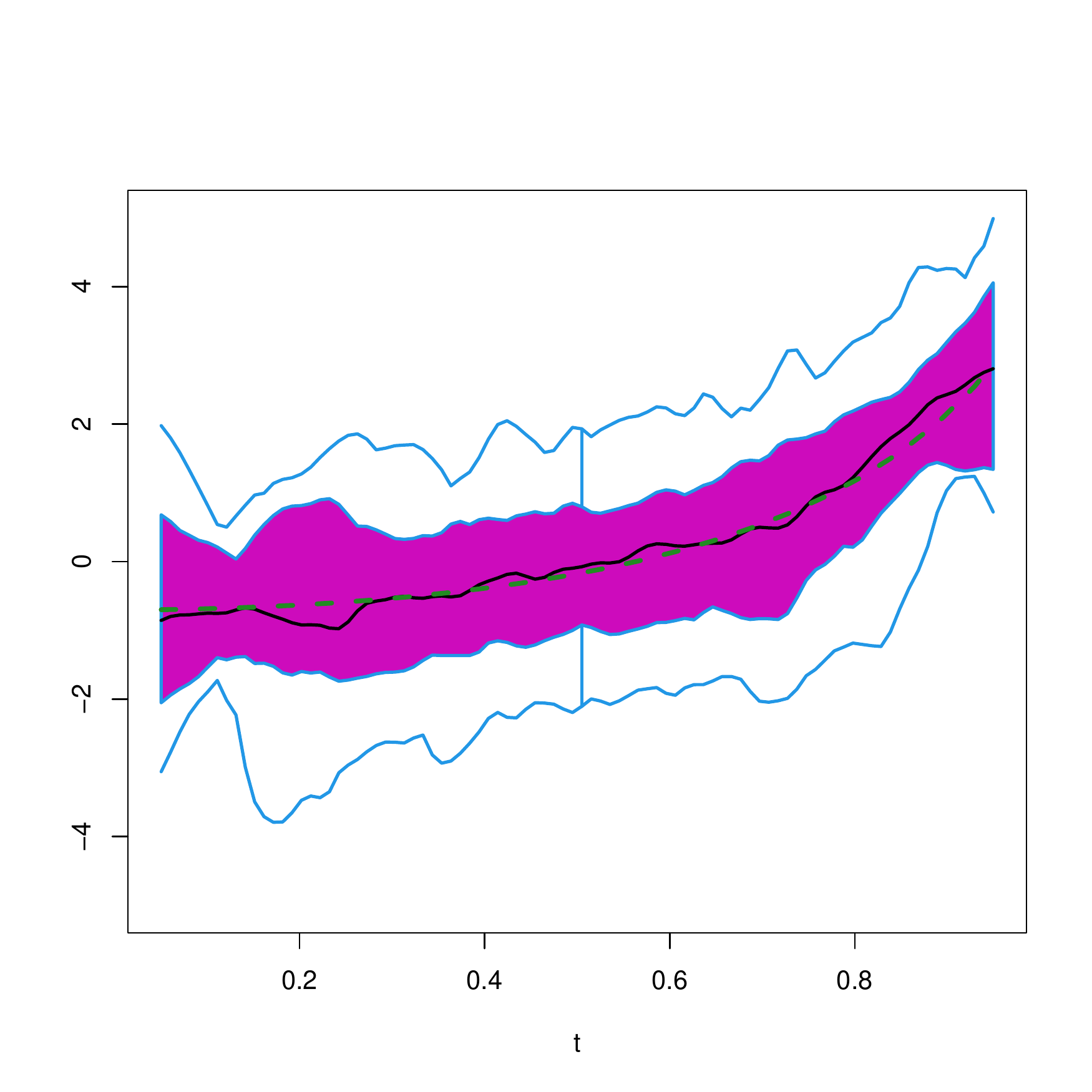} 
\\[-6ex]
$C_{1, 12}$ &  
 \includegraphics[scale=0.35]{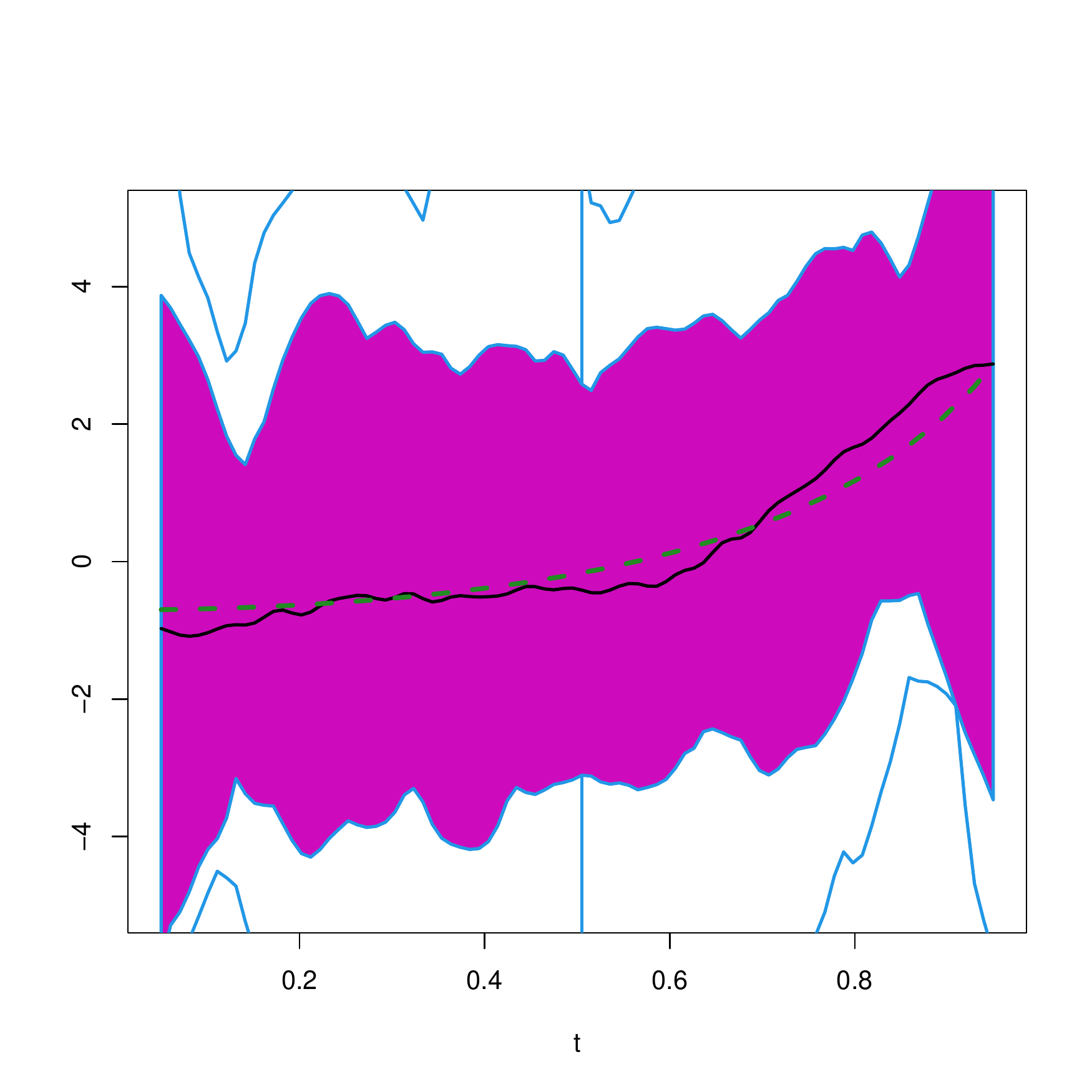}  & 
 \includegraphics[scale=0.35]{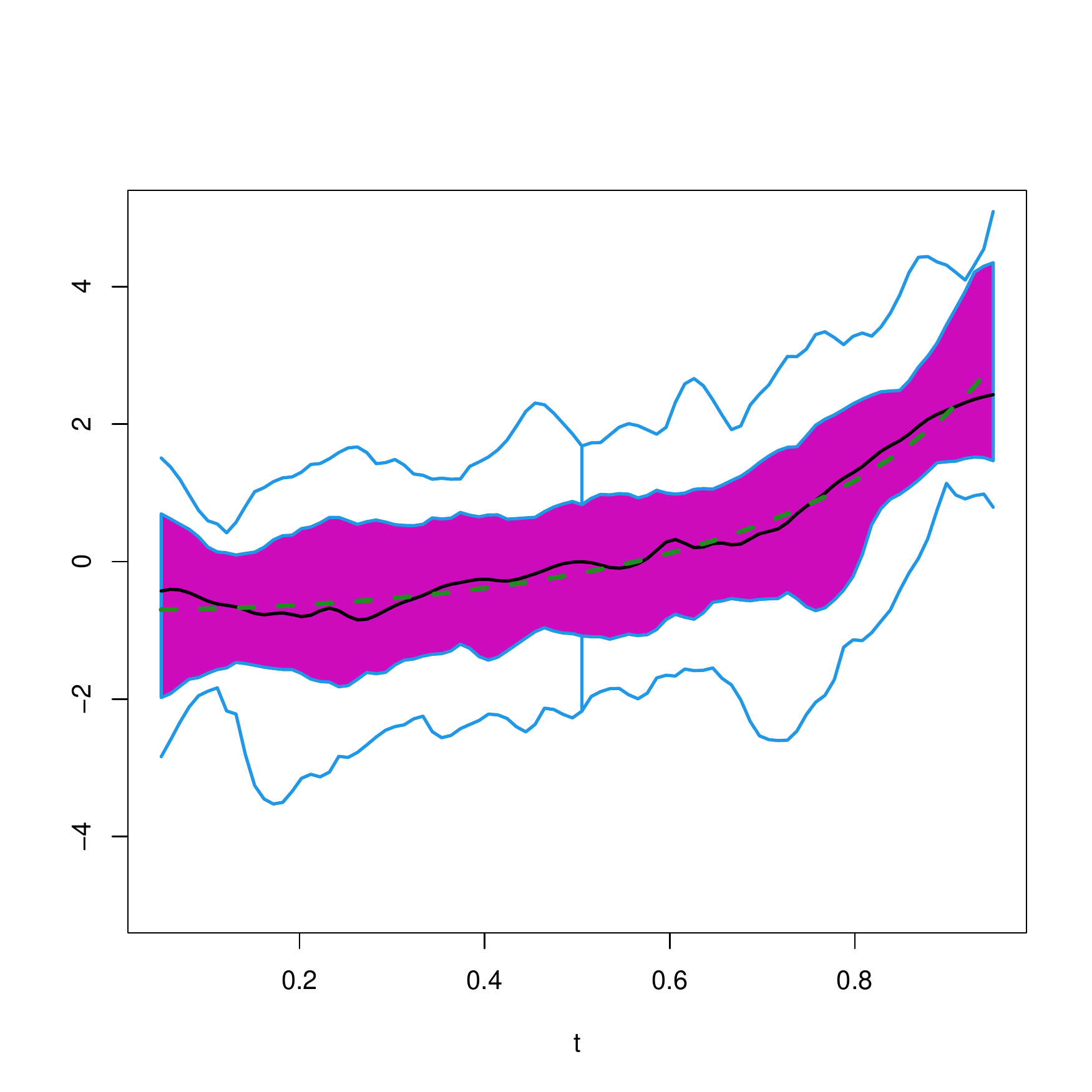} 
  \\[-6ex]
   
$C_{2, 12}$ &
 \includegraphics[scale=0.35]{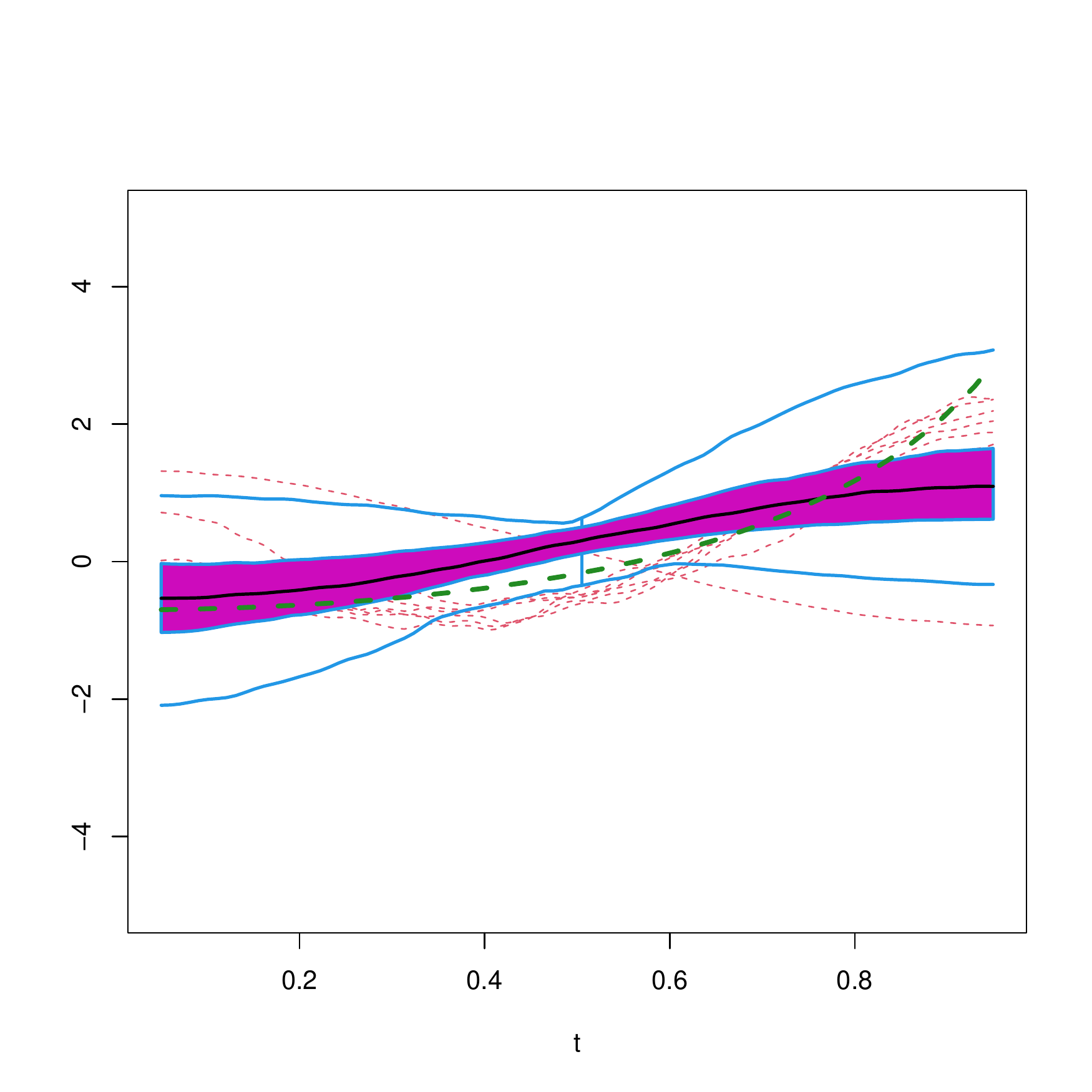}  & 
 \includegraphics[scale=0.35]{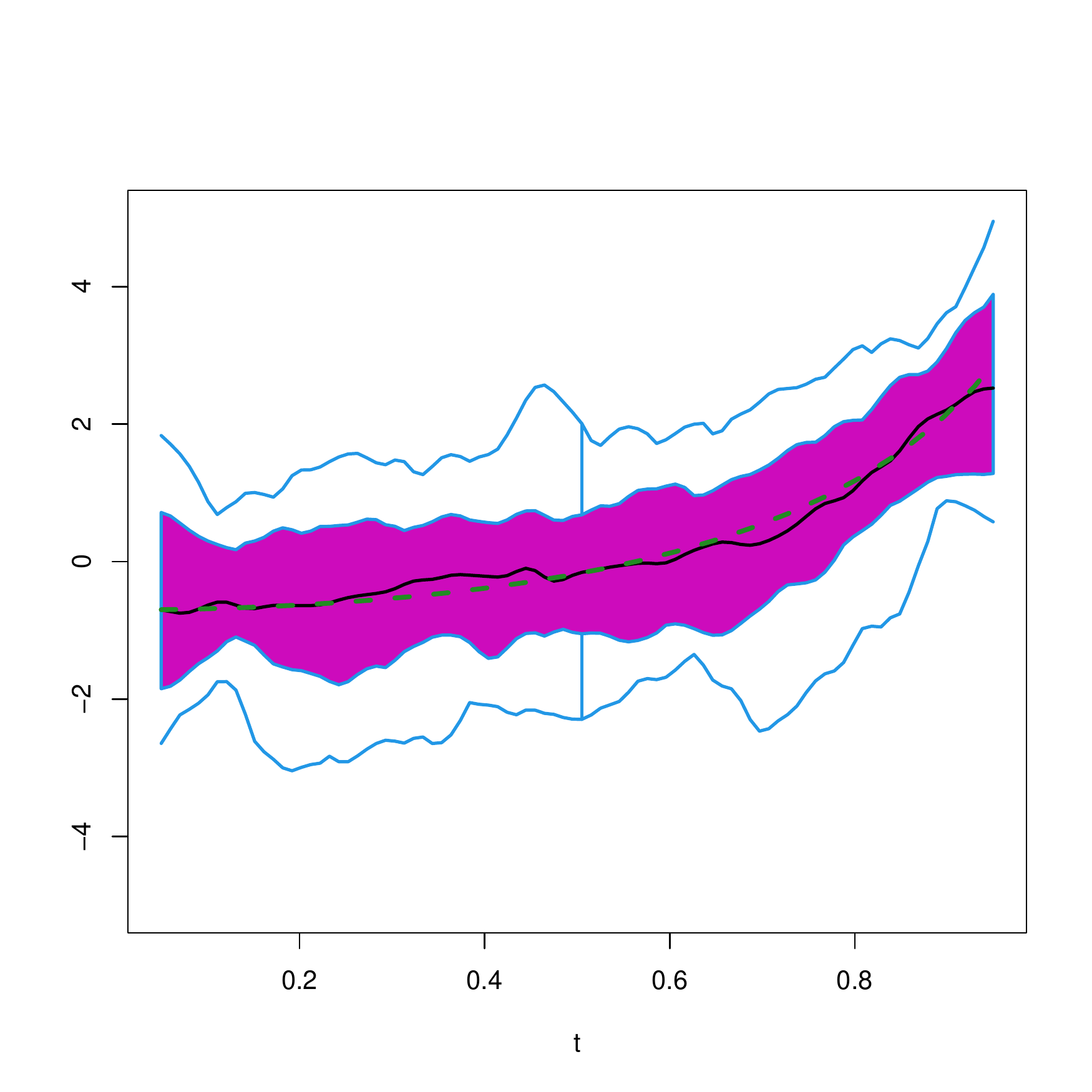} 
 \\[-6ex]
    
$C_{3, 4, 0.4}$ &
\includegraphics[scale=0.35]{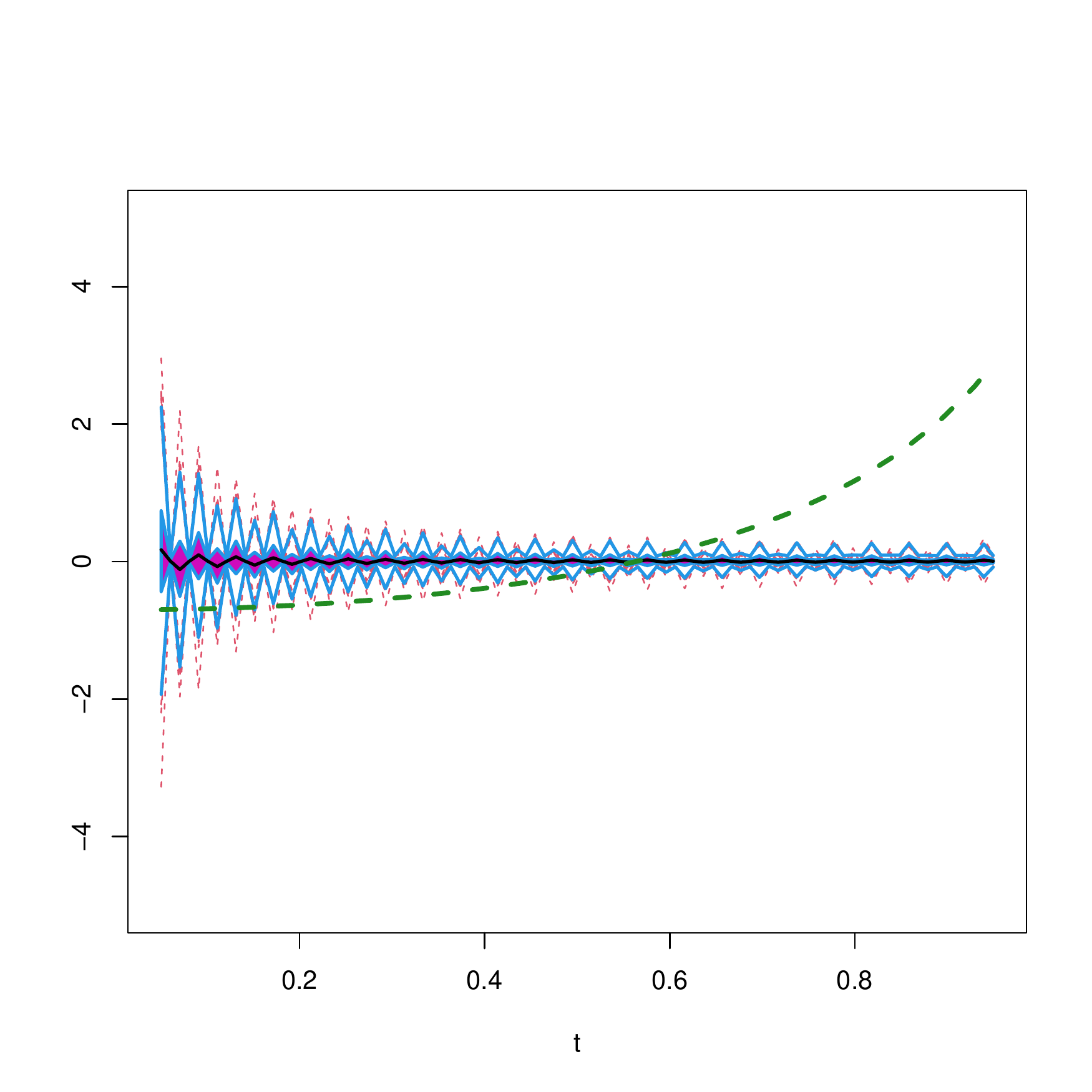}  & 
 \includegraphics[scale=0.35]{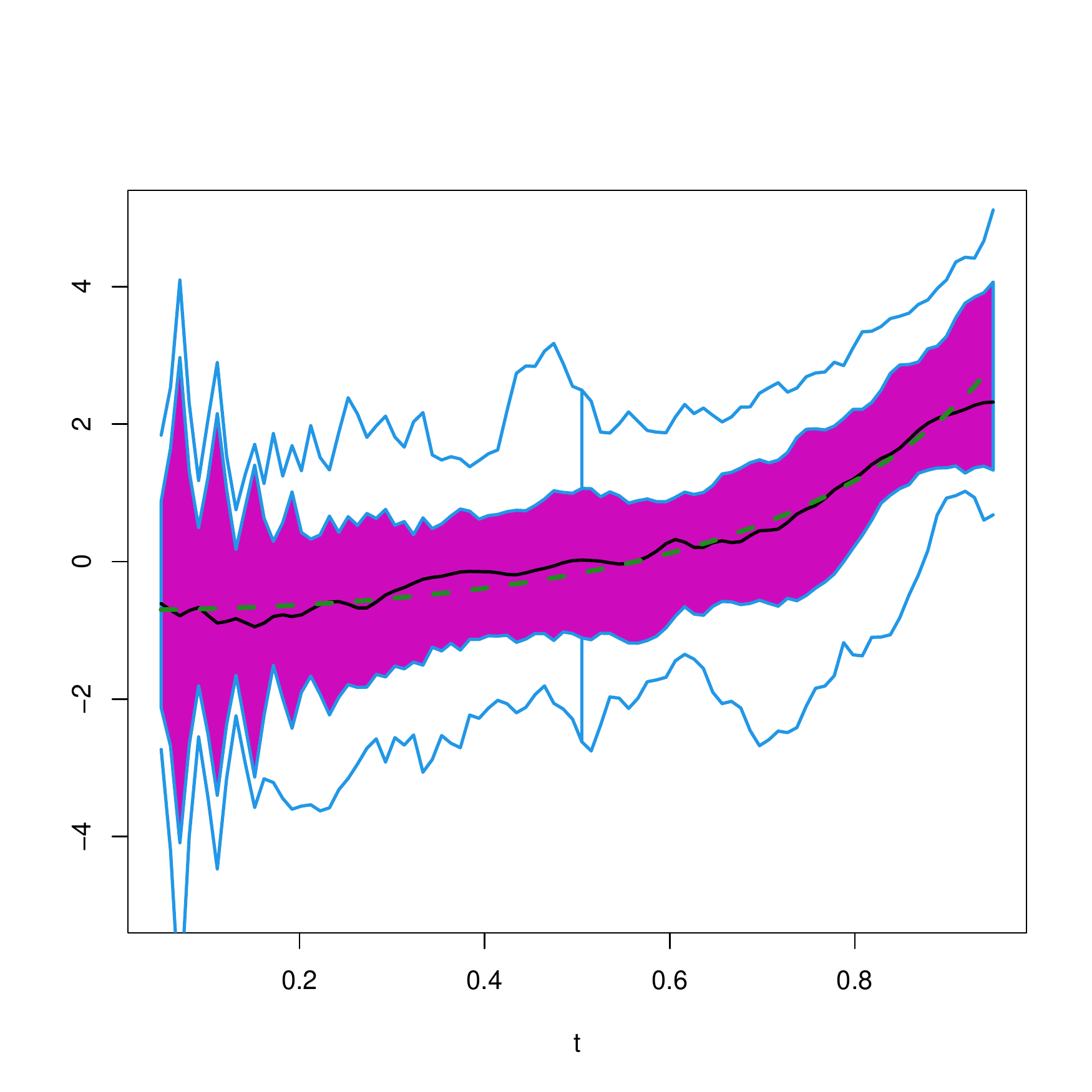}
\end{tabular}
\caption{\small \label{fig:wbeta-Upsilon1}  Functional boxplot of the estimators for $\beta_0$ under \textbf{Model 1} with  $\Upsilon_0=\Upsilon_{0,1}$. 
The true function is shown with a green dashed line, while the black solid one is the central 
curve of the $n_R = 1000$ estimates $\wbeta$. Columns correspond to estimation 
methods, while rows to $C_0$ and   to some of the three contamination settings.}
\end{center} 
\end{figure}

\begin{figure}[ht!]
 \begin{center}
 \footnotesize
 \renewcommand{\arraystretch}{0.2}
 \newcolumntype{M}{>{\centering\arraybackslash}m{\dimexpr.05\linewidth-1\tabcolsep}}
   \newcolumntype{G}{>{\centering\arraybackslash}m{\dimexpr.33\linewidth-1\tabcolsep}}
%\begin{tabular}{MGG}
\begin{tabular}{M GG}
 & $\wbeta_{\ls}$ &   $\wbeta_{\eme\eme}$ \\[-7ex]
$C_{0}$ 
&  \includegraphics[scale=0.35]{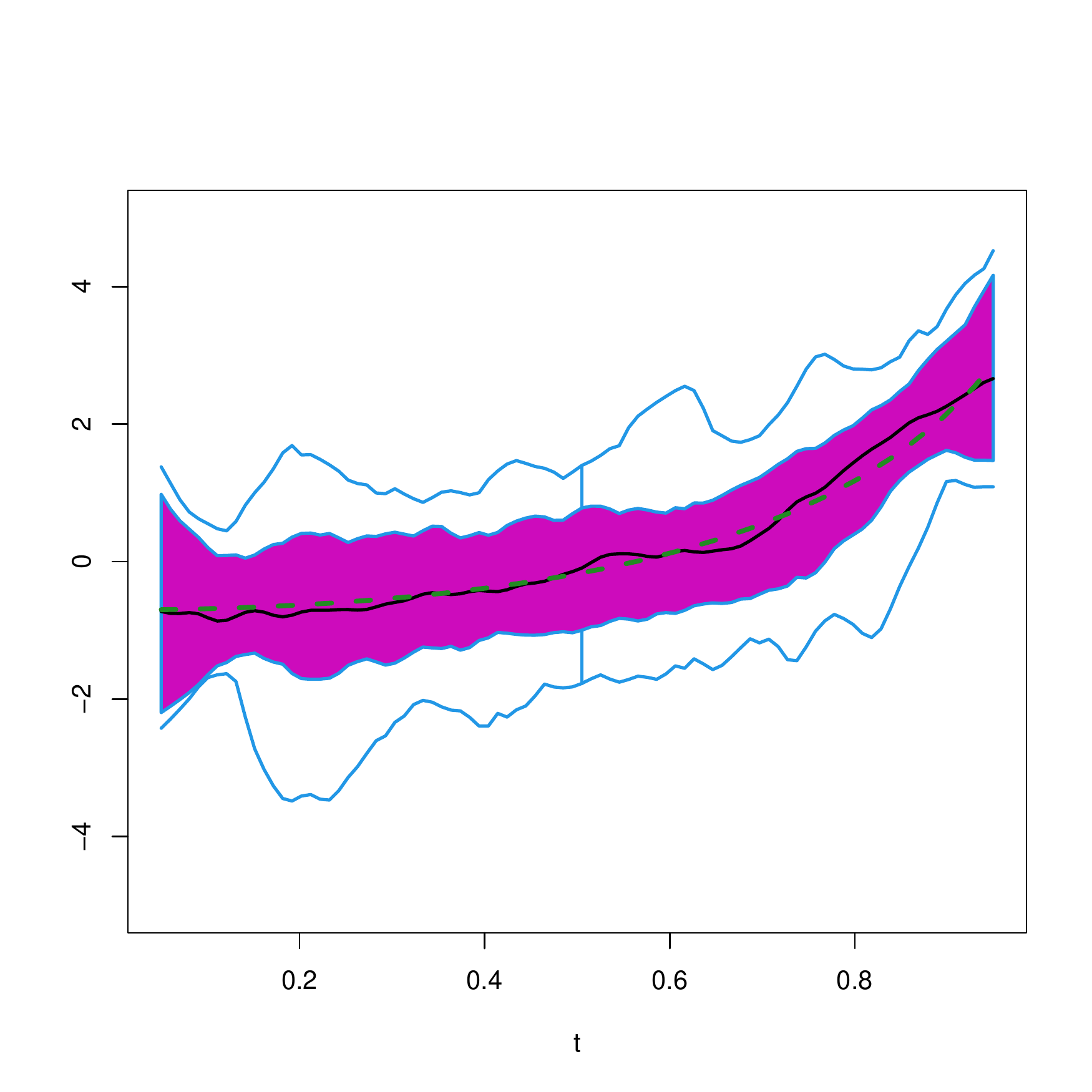} 
&  \includegraphics[scale=0.35]{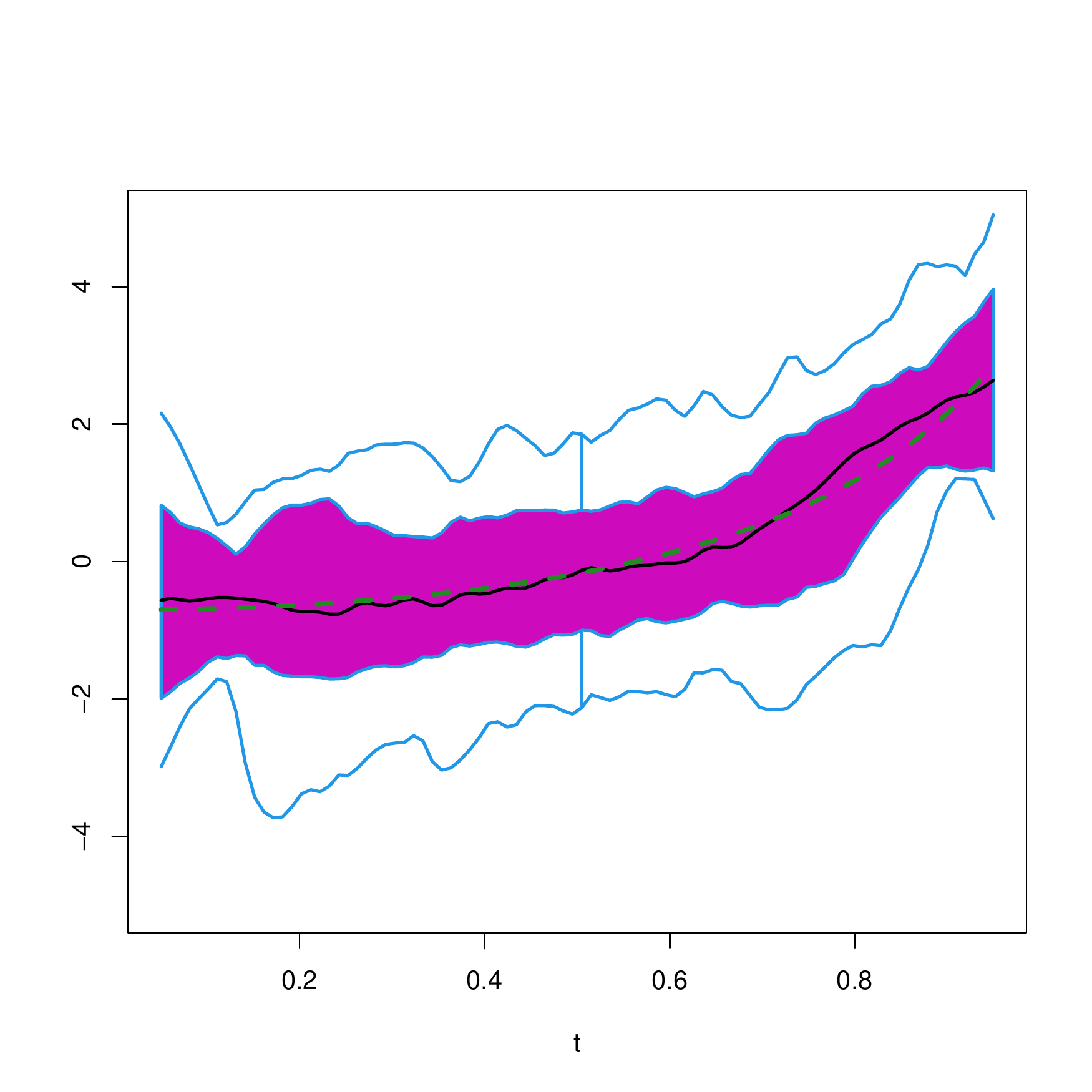} 
\\[-6ex]
$C_{1, 12}$ &  
 \includegraphics[scale=0.35]{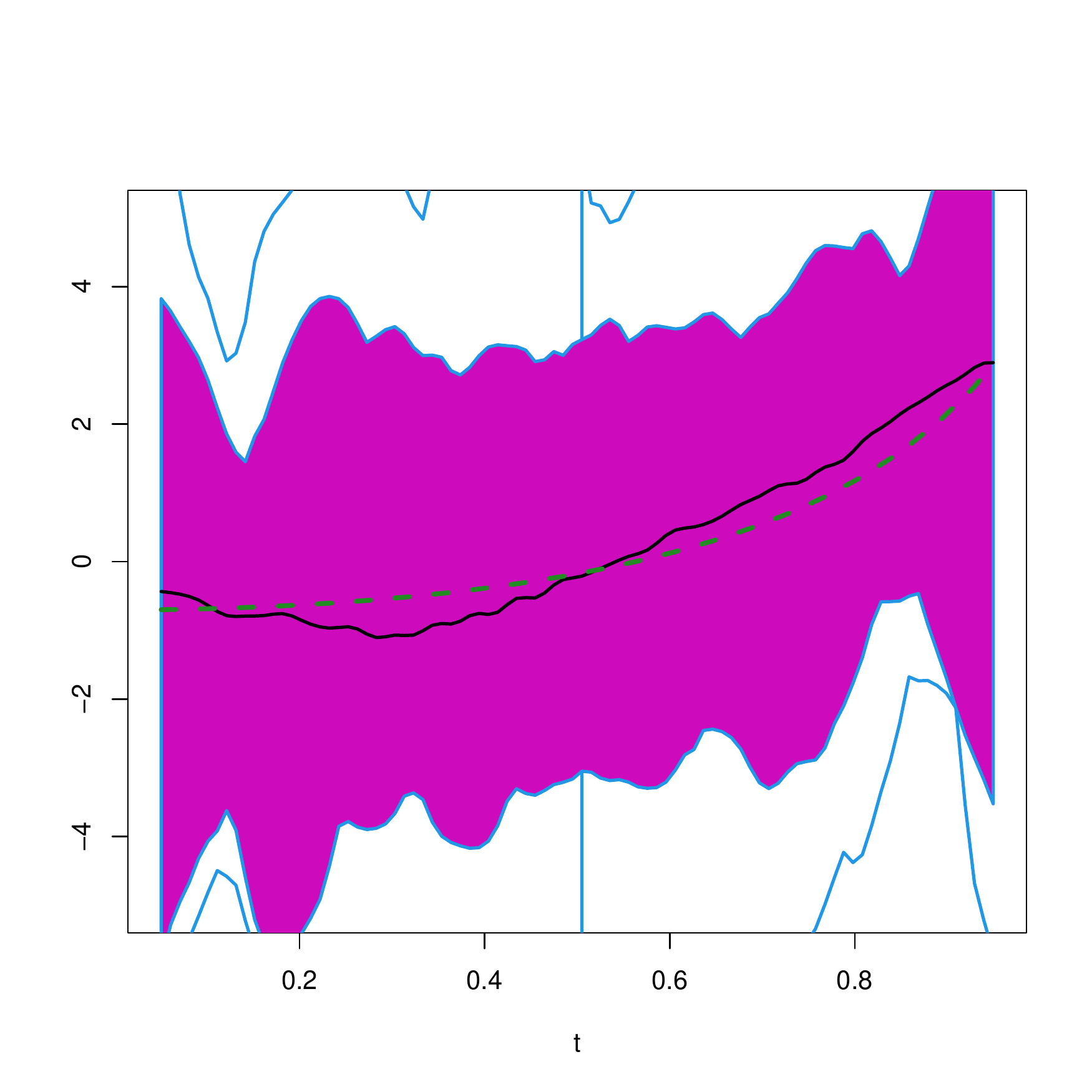}  & 
 \includegraphics[scale=0.35]{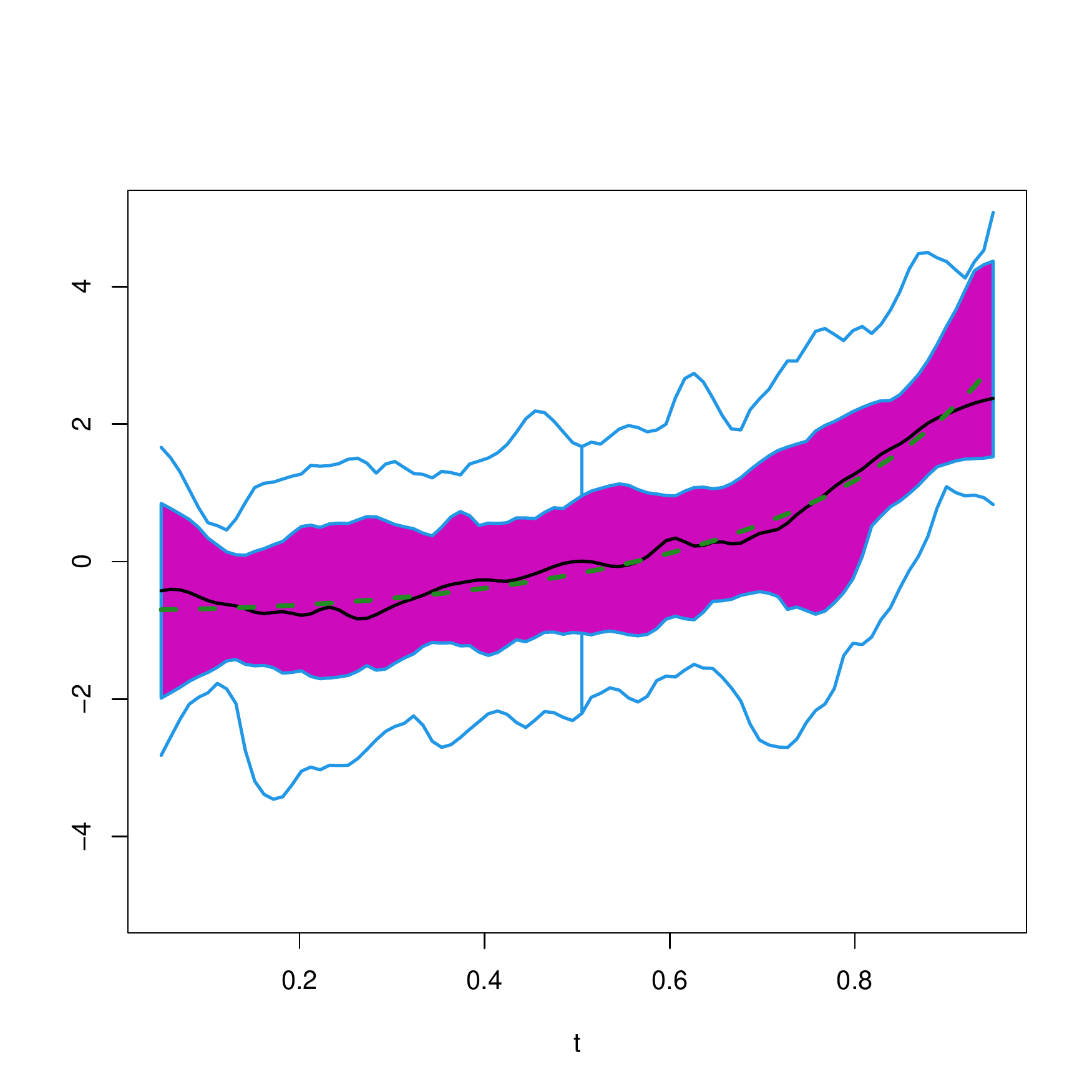} 
  \\[-6ex]
   
$C_{2, 12}$ &
 \includegraphics[scale=0.35]{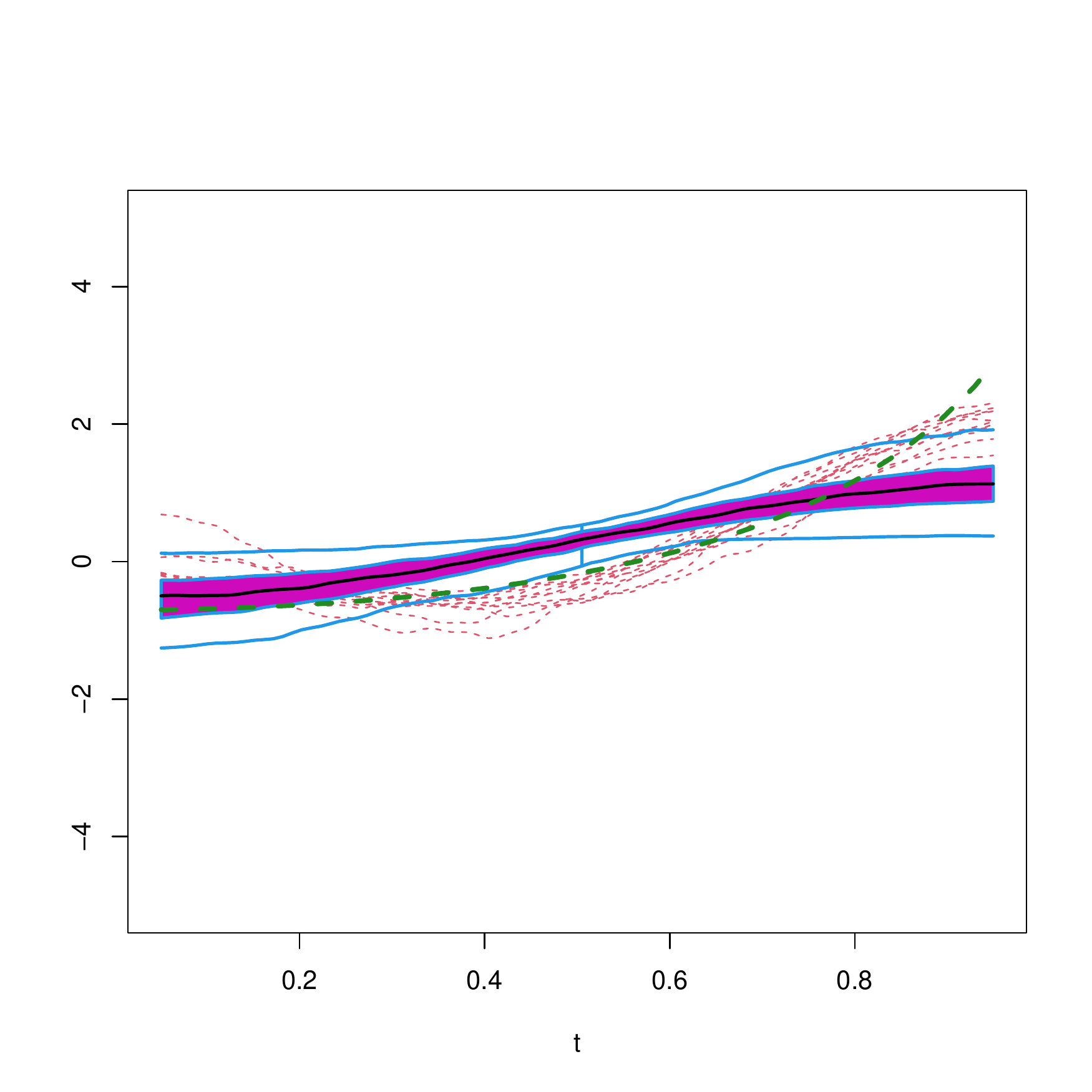}  & 
 \includegraphics[scale=0.35]{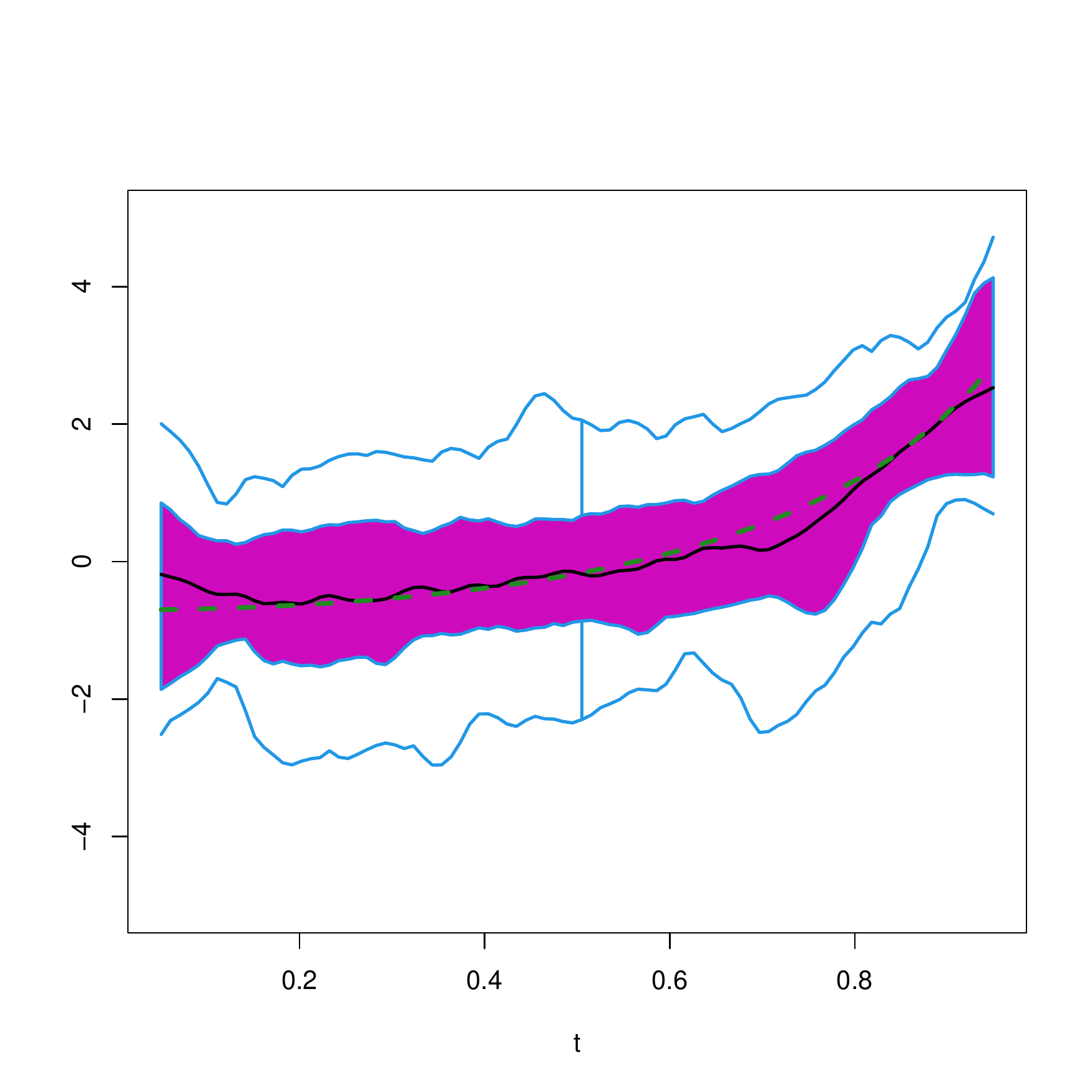} 
 \\[-6ex]
    
$C_{3, 4, 0.4}$ &
\includegraphics[scale=0.35]{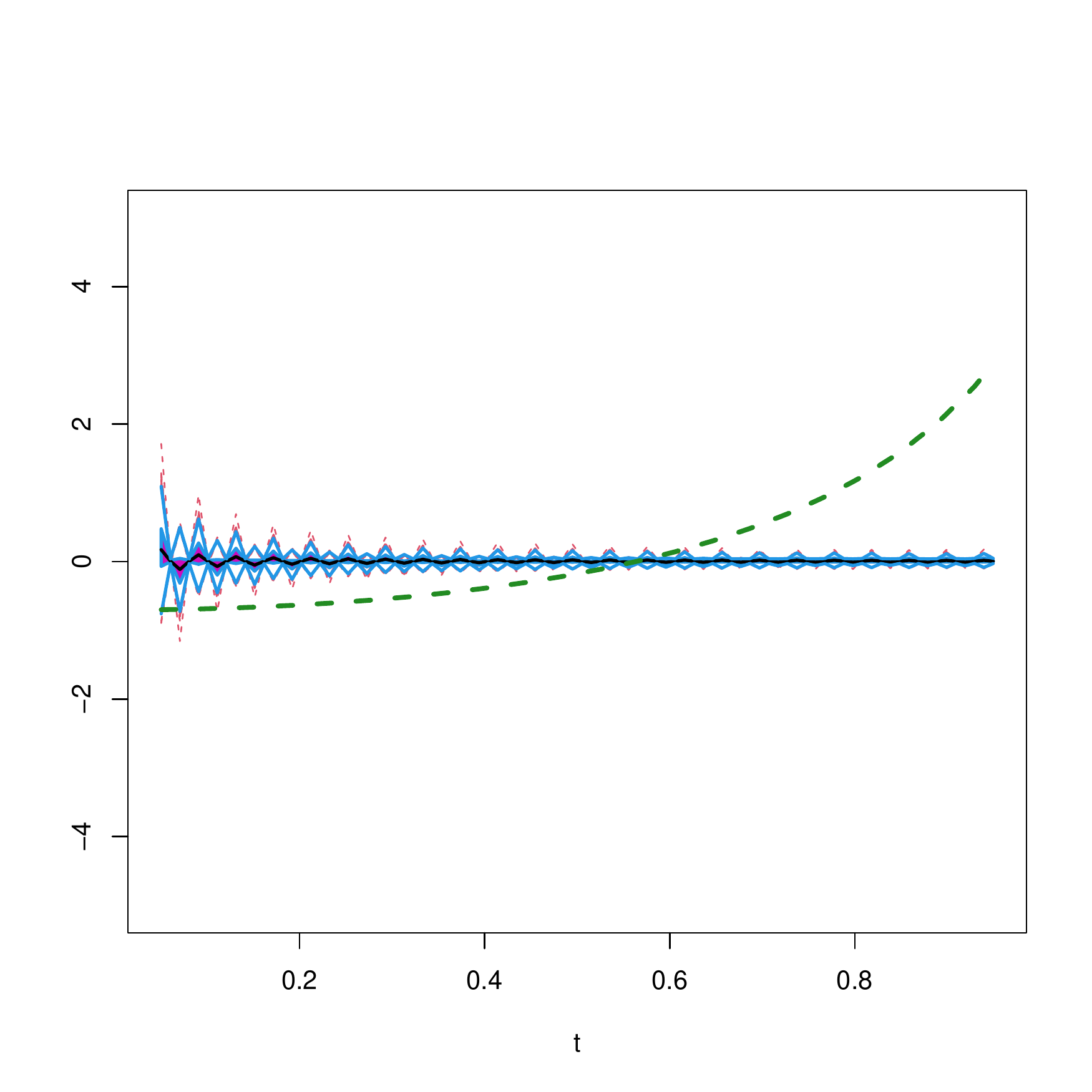}  & 
 \includegraphics[scale=0.35]{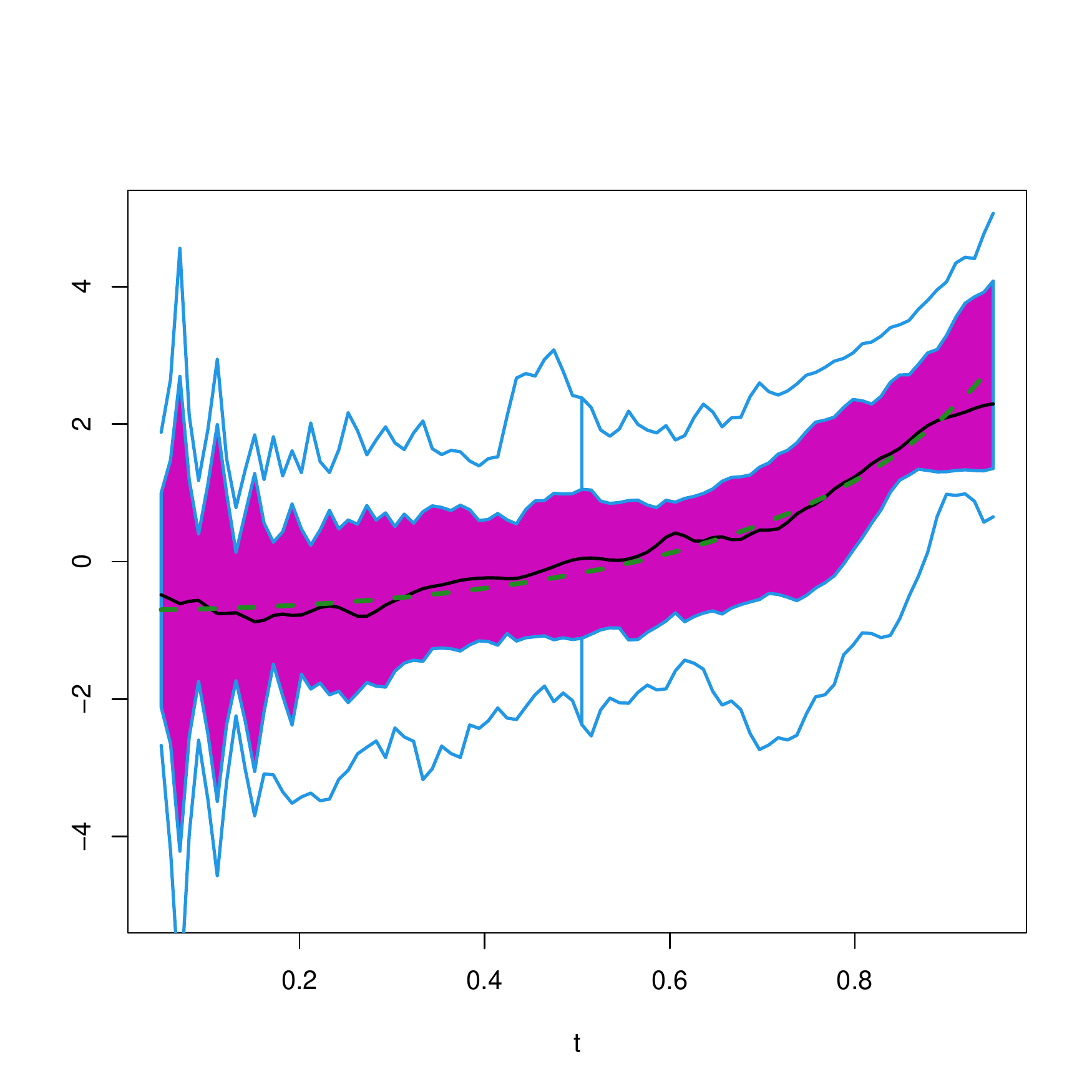}
\end{tabular}
\caption{\small \label{fig:wbeta-Upsilon2}  Functional boxplot of the estimators for $\beta_0$ under \textbf{Model 1} with $\Upsilon_0=\Upsilon_{0,2}$. 
The true function is shown with a green dashed line, while the black solid one is the central 
curve of the $n_R = 1000$ estimates $\wbeta$. Columns correspond to estimation 
methods, while rows to $C_0$ and   to some of the three contamination settings.}
\end{center} 
\end{figure}

In order to also explore visually the performance of these estimators, Figures \ref{fig:wbeta-Upsilon0} to \ref{fig:wbeta-Upsilon2} contain functional boxplots, as defined in \citet{sun2011functional}, for the $n_R = 1000$ realizations of the different estimators for $\beta_0$  under $C_0$ and some of the three contamination settings. As in standard boxplots, the magenta central box of these functional boxplots represents the 50\% inner band of curves, the solid black line indicates the central (deepest) function and the dotted red lines indicate outlying curves (in this case: outlying estimates $\wbeta_j$  for some $1 \le j \le n_R$). We also indicate the target (true) function  $\beta_0$ with a dark green dashed line. To avoid boundary effects, we show here the different estimates $\wbeta_j$ evaluated on the $90$ interior points of the grid of equally spaced points. In addition, to facilitate comparisons between contamination cases and estimation methods, the scales of the vertical axes are the same for all panels within each Figure.  
The ways in which the different outliers affect the classical estimators for $\beta_0$ can be seen in these figures. Note that under $C_{1, 12}$ the classical $\wbeta$ becomes highly variable, but mostly retains the same shape of the true $\beta_0$, which lies within the central box. However, with high--leverage outliers (as in $C_{2, 12}$ and particularly under $C_{3,4,0.4}$) the estimator becomes completely uninformative  and does not reflect the shape of the true regression coefficient $\beta_0$, for all the quadratic models considered.  
 
\begin{figure}[ht!]
 \begin{center}
 \footnotesize
 \renewcommand{\arraystretch}{0.2}
 \newcolumntype{M}{>{\centering\arraybackslash}m{\dimexpr.05\linewidth-1\tabcolsep}}
   \newcolumntype{G}{>{\centering\arraybackslash}m{\dimexpr.33\linewidth-1\tabcolsep}}
%\begin{tabular}{MGG}
\begin{tabular}{M GG}
 & $\wup_{\ls}$ &   $\wup_{\eme\eme}$ \\[-3ex]
$C_{0}$ 
&  \includegraphics[scale=0.18]{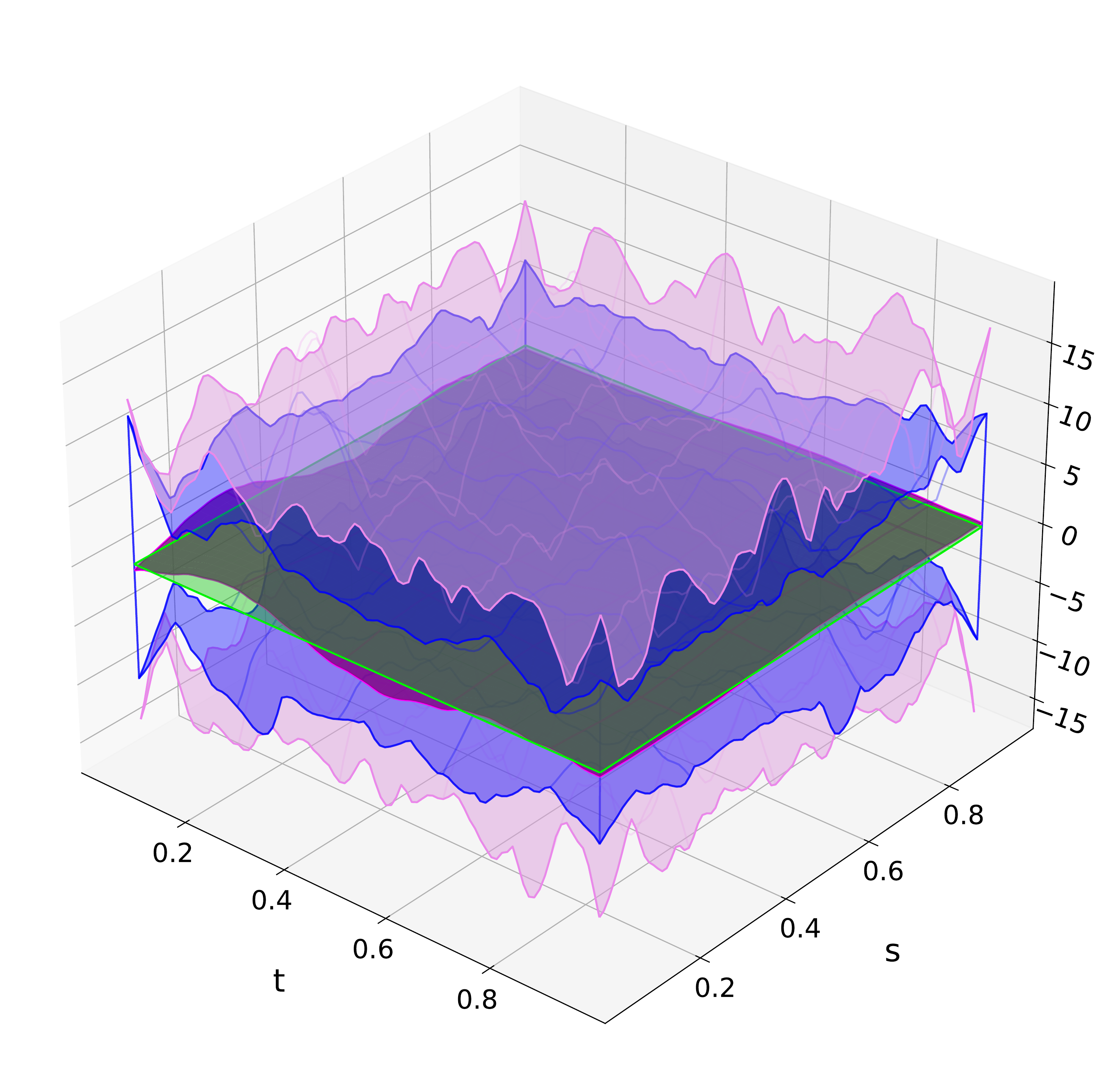} 
&  \includegraphics[scale=0.18]{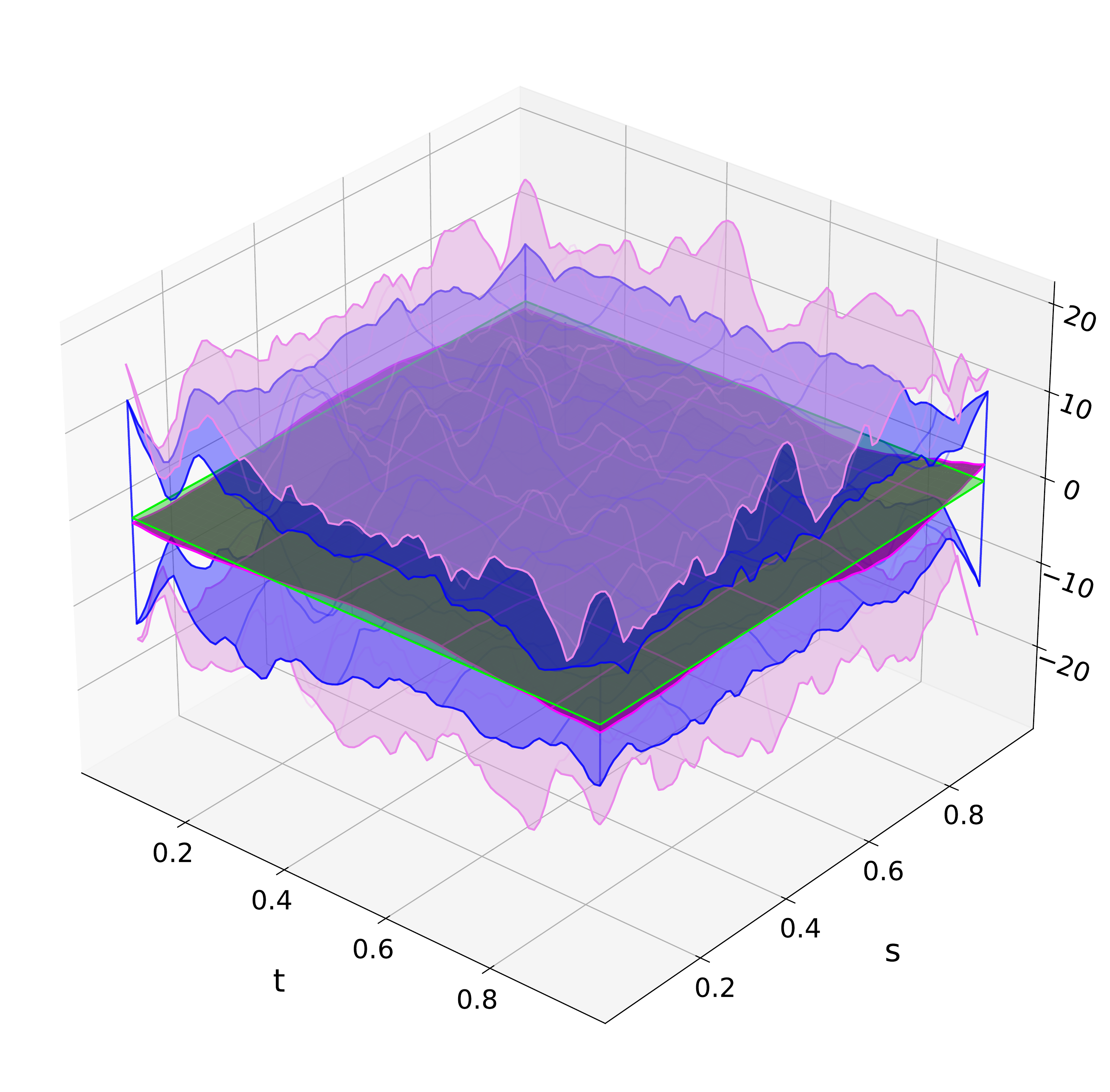} 
\\[-4ex]
$C_{1, 12}$ &  
 \includegraphics[scale=0.18]{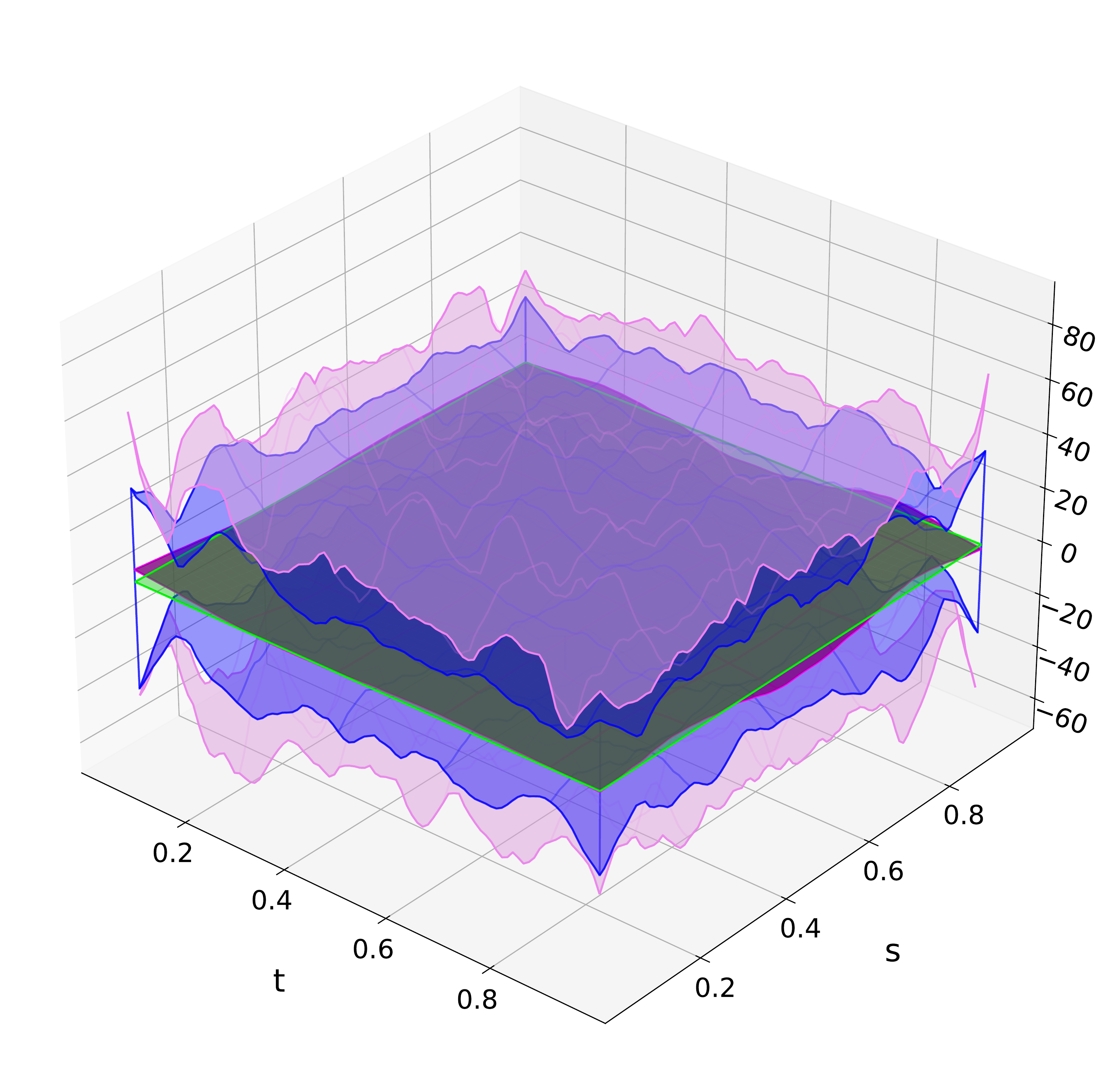}  & 
 \includegraphics[scale=0.18]{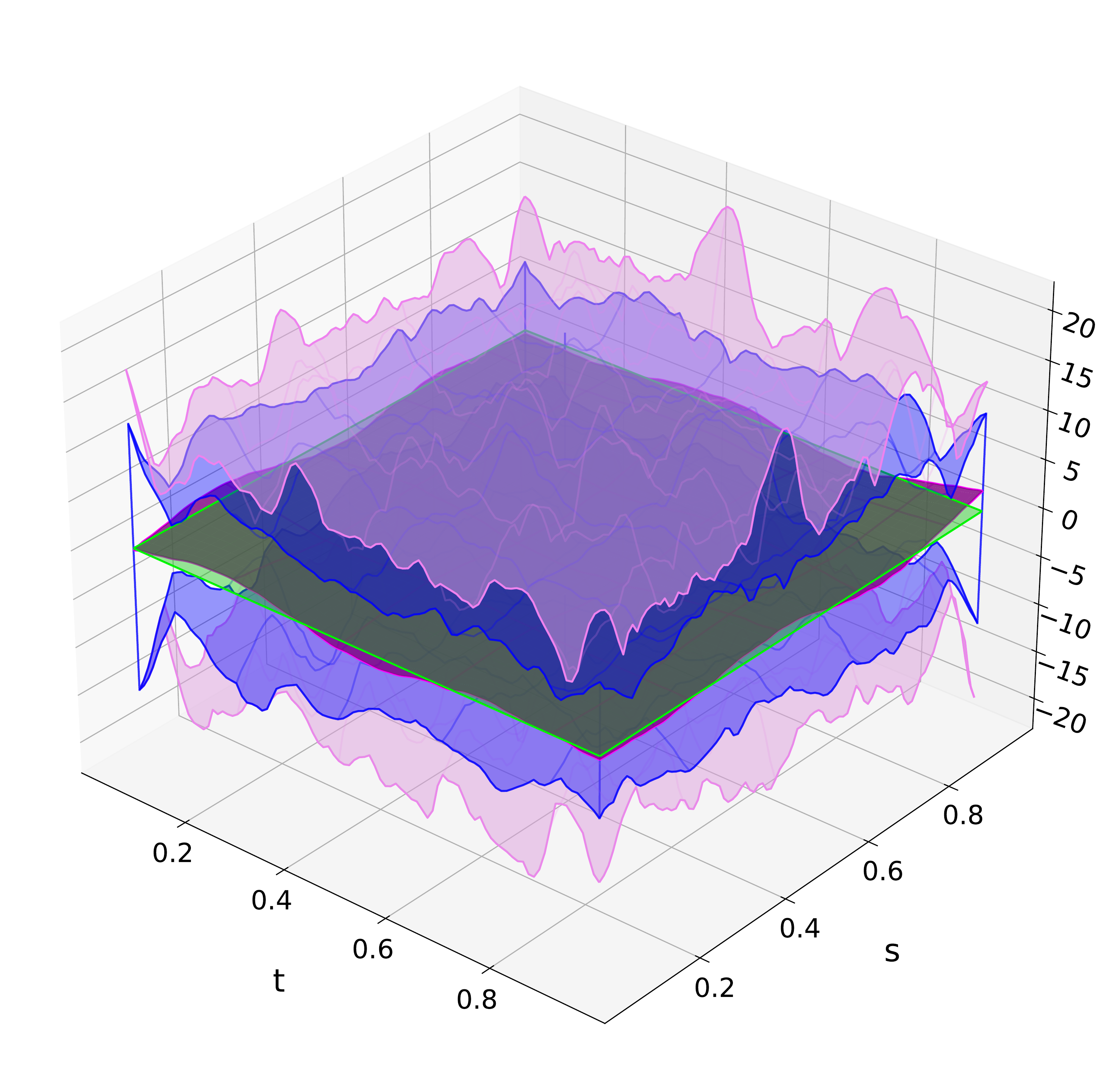} 
  \\[-4ex]
   
$C_{2, 12}$ &
 \includegraphics[scale=0.18]{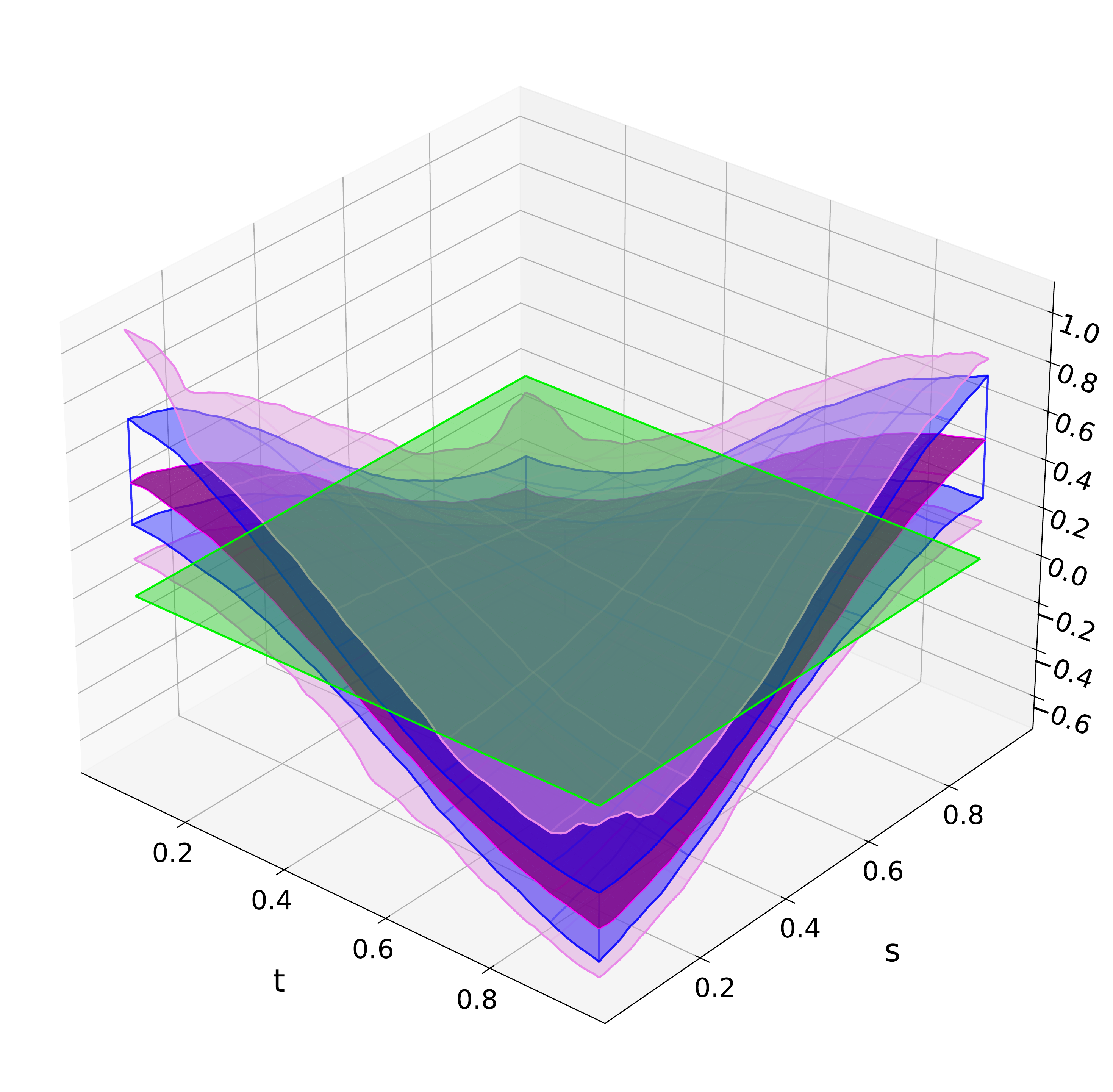}  & 
 \includegraphics[scale=0.18]{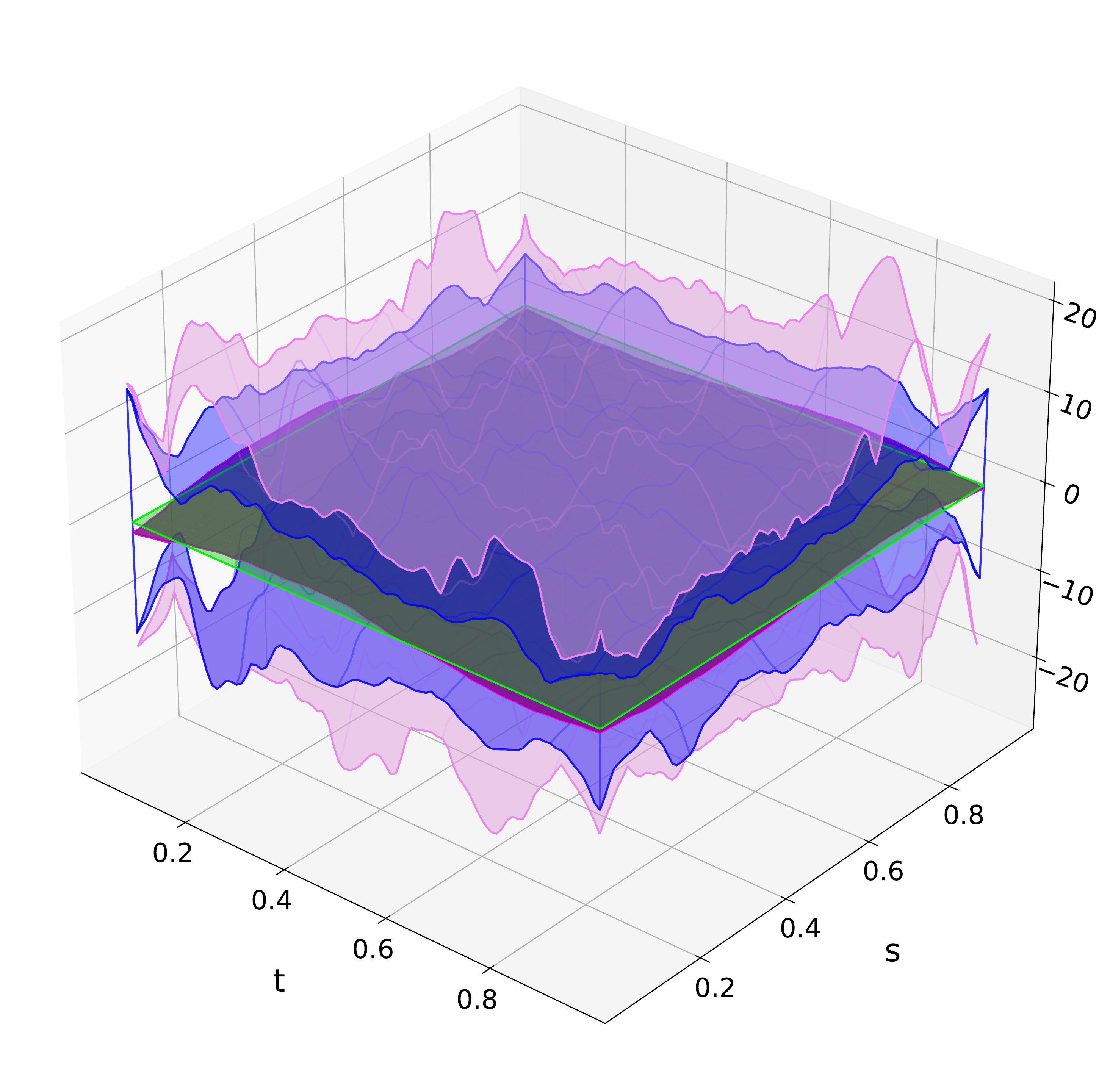} 
 \\[-4ex]
    
$C_{3, 4, 0.4}$ &
 \includegraphics[scale=0.18]{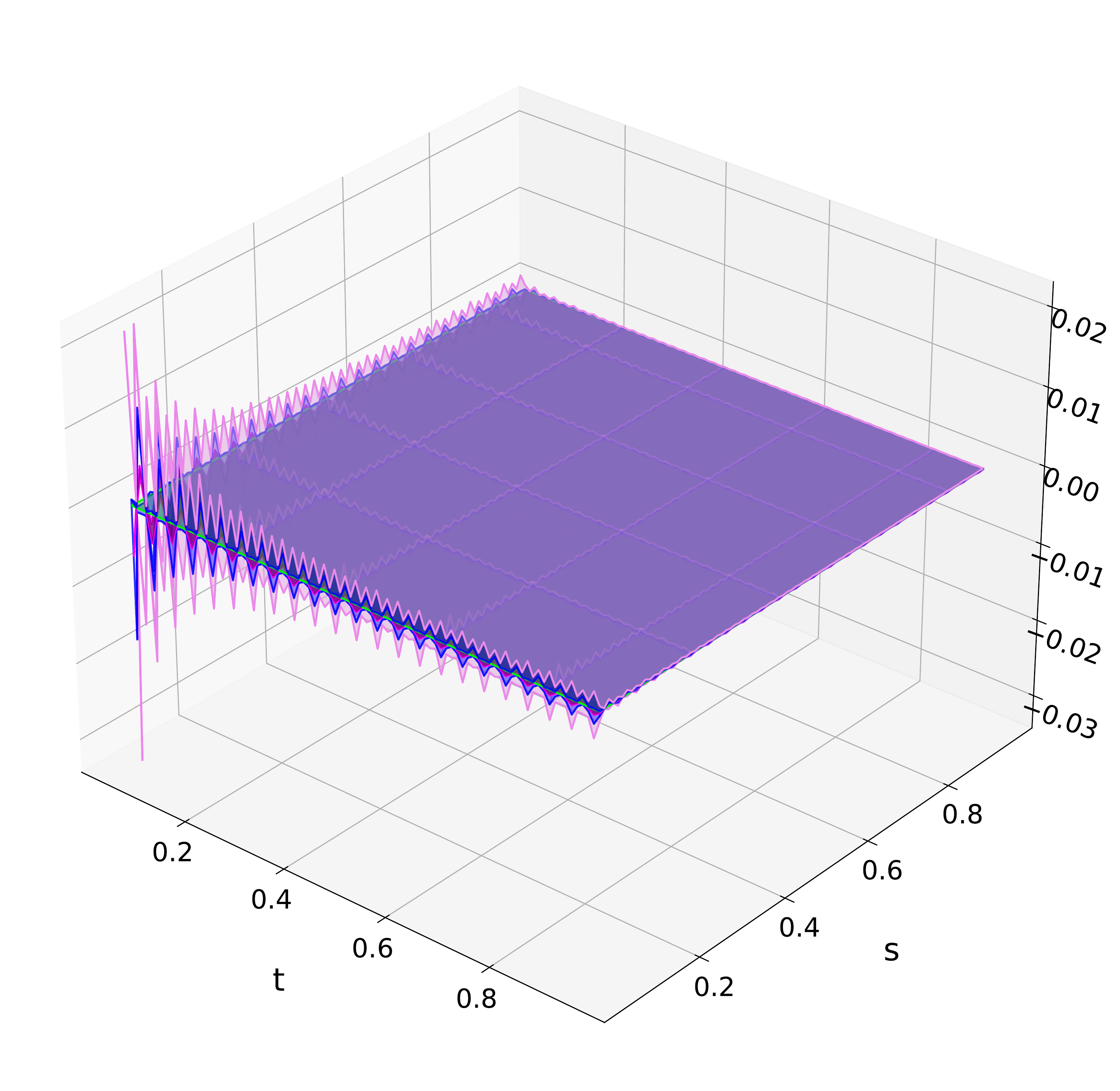}  & 
 \includegraphics[scale=0.18]{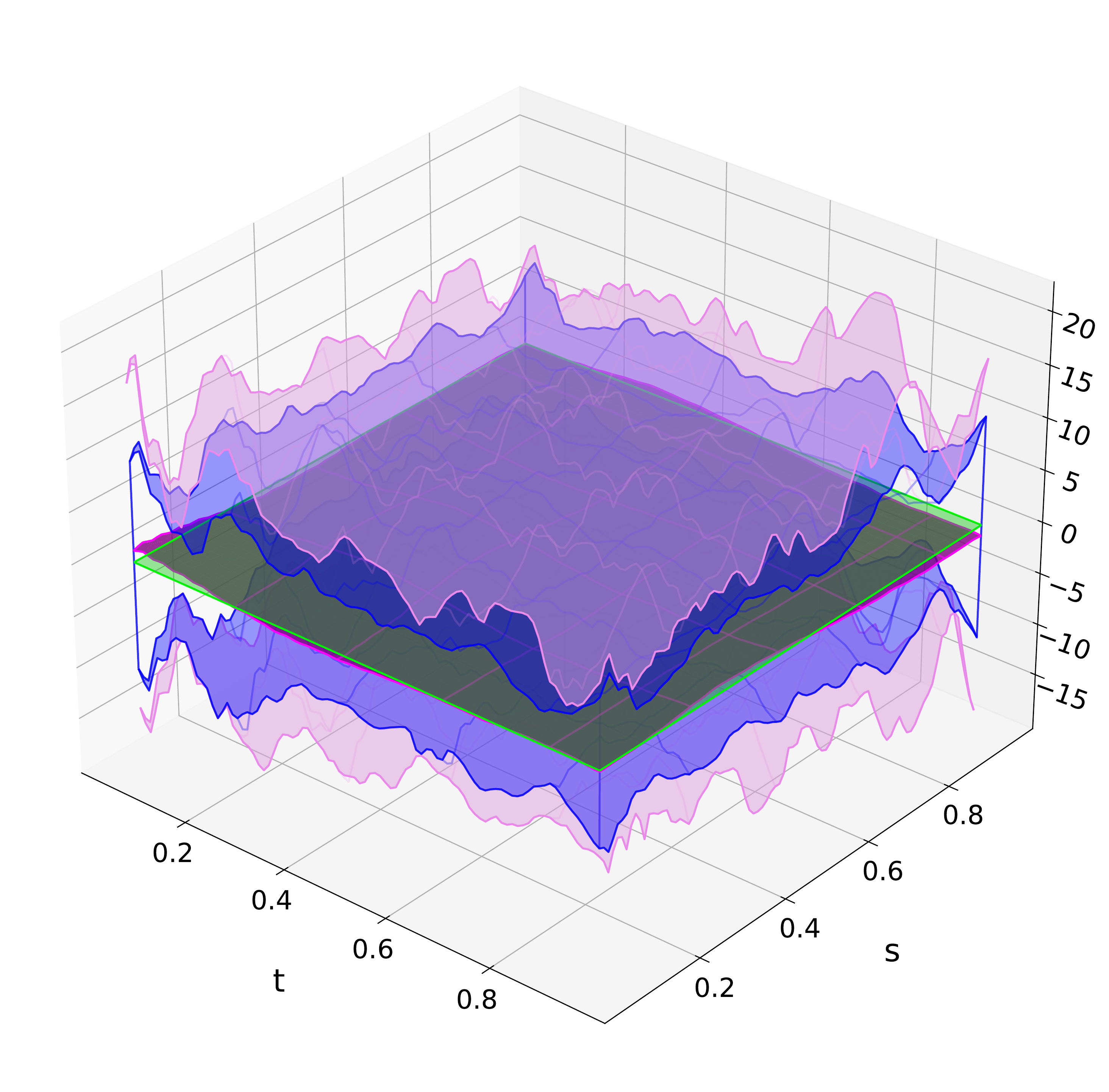}
\end{tabular}
\caption{\small \label{fig:wgamma-Upsilon0}  Surface boxplot of the estimators for $\upsilon_0$ under \textbf{Model 1} with  $\Upsilon_0=0$. 
The true function is shown in green, while the purple surface is the central 
surface of the $n_R = 1000$ estimates $\wup$. Columns correspond to estimation 
methods, while rows to $C_0$ and   to some of the three contamination settings.}
\end{center} 
\end{figure}

\begin{figure}[tp]
 \begin{center}
 \footnotesize
 \renewcommand{\arraystretch}{0.2}
 \newcolumntype{M}{>{\centering\arraybackslash}m{\dimexpr.05\linewidth-1\tabcolsep}}
   \newcolumntype{G}{>{\centering\arraybackslash}m{\dimexpr.33\linewidth-1\tabcolsep}}
%\begin{tabular}{MGG}
\begin{tabular}{M GG}
  & $\wup_{\ls}$ &   $\wup_{\eme\eme}$  \\[-3ex]
$C_{0}$ 
&  \includegraphics[scale=0.18]{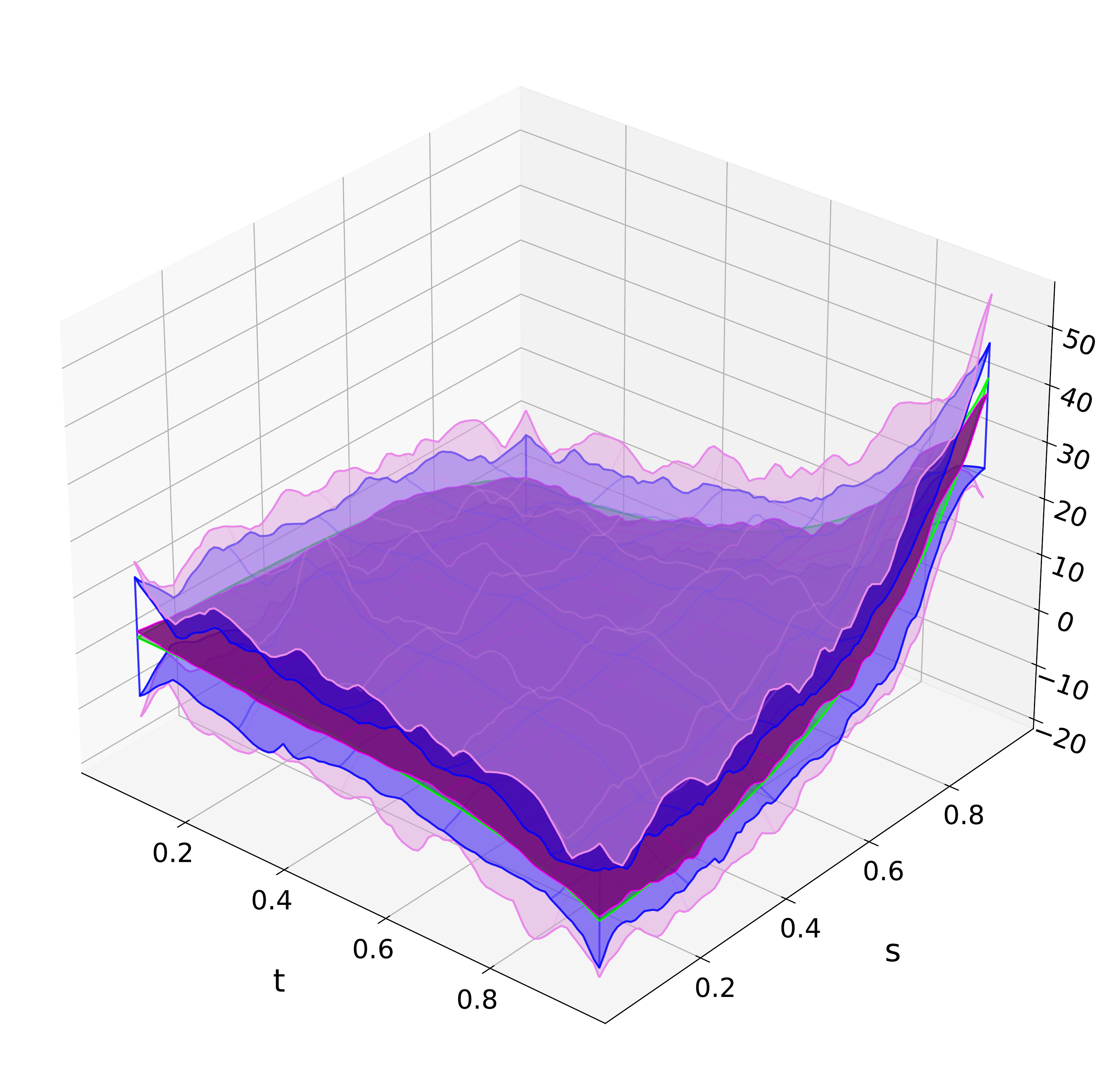} 
&  \includegraphics[scale=0.18]{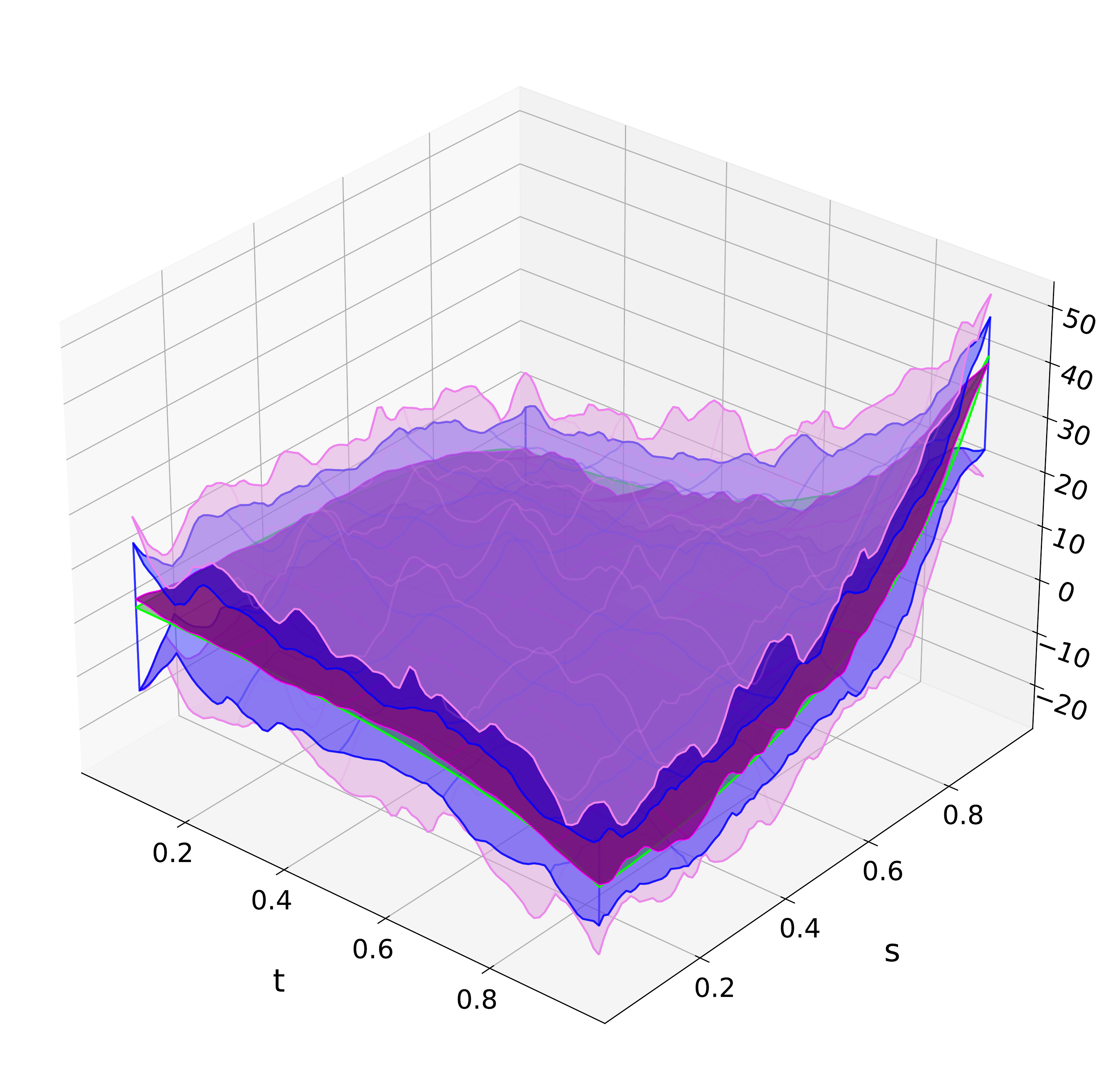} 
\\[-4ex]
$C_{1, 12}$ &  
 \includegraphics[scale=0.18]{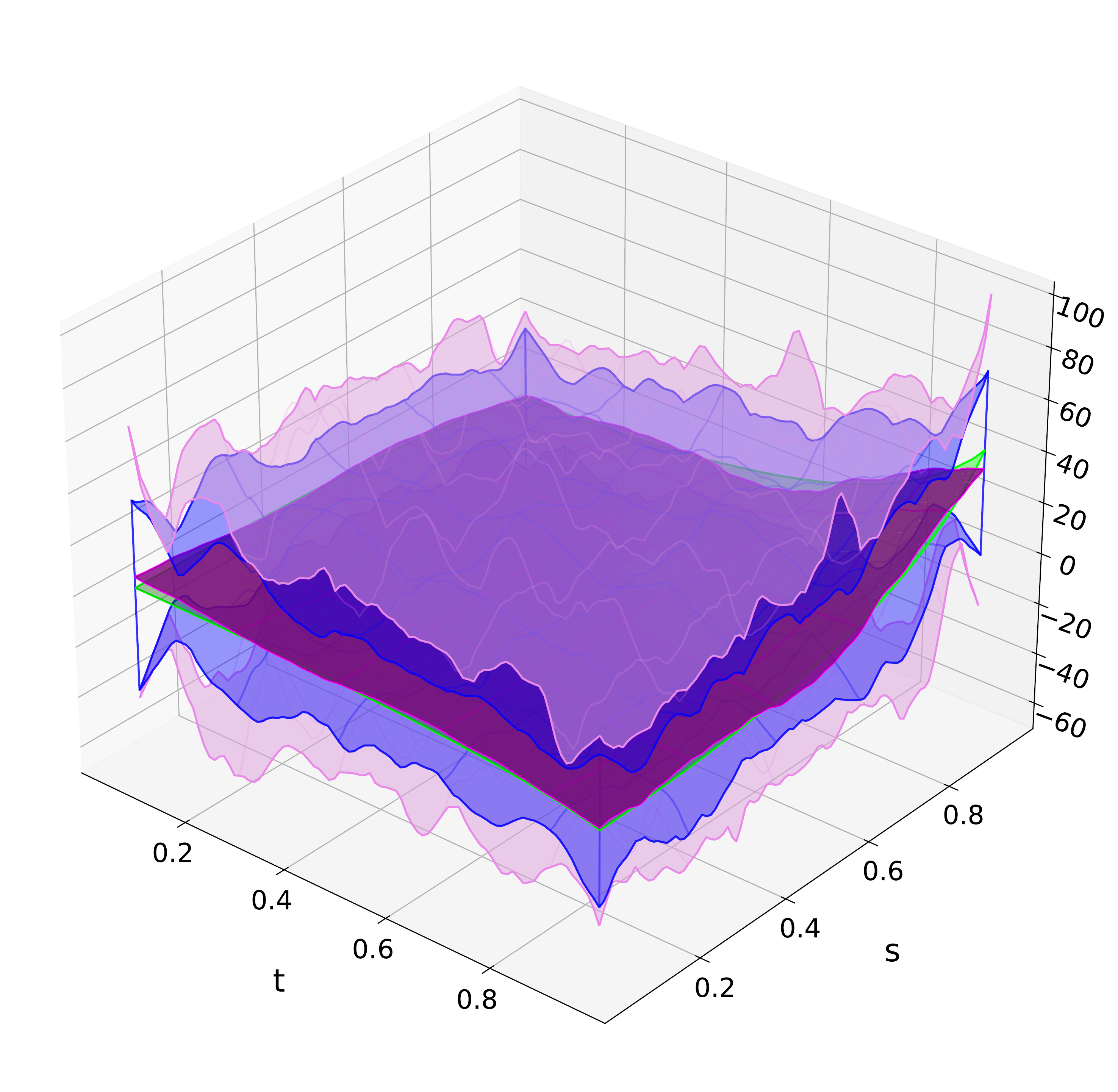}  & 
 \includegraphics[scale=0.18]{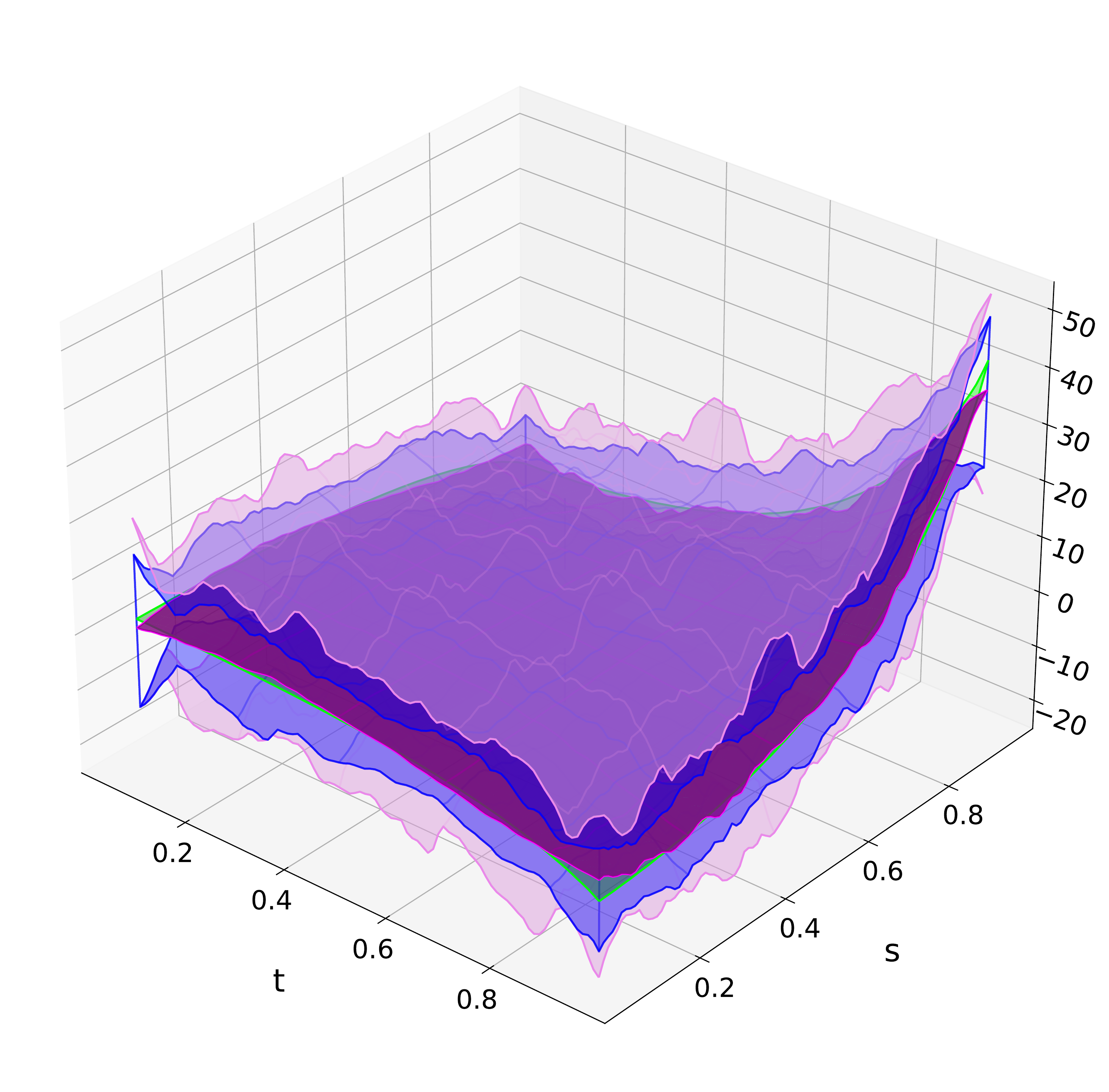} 
  \\[-4ex]
   
$C_{2, 12}$ &
 \includegraphics[scale=0.18]{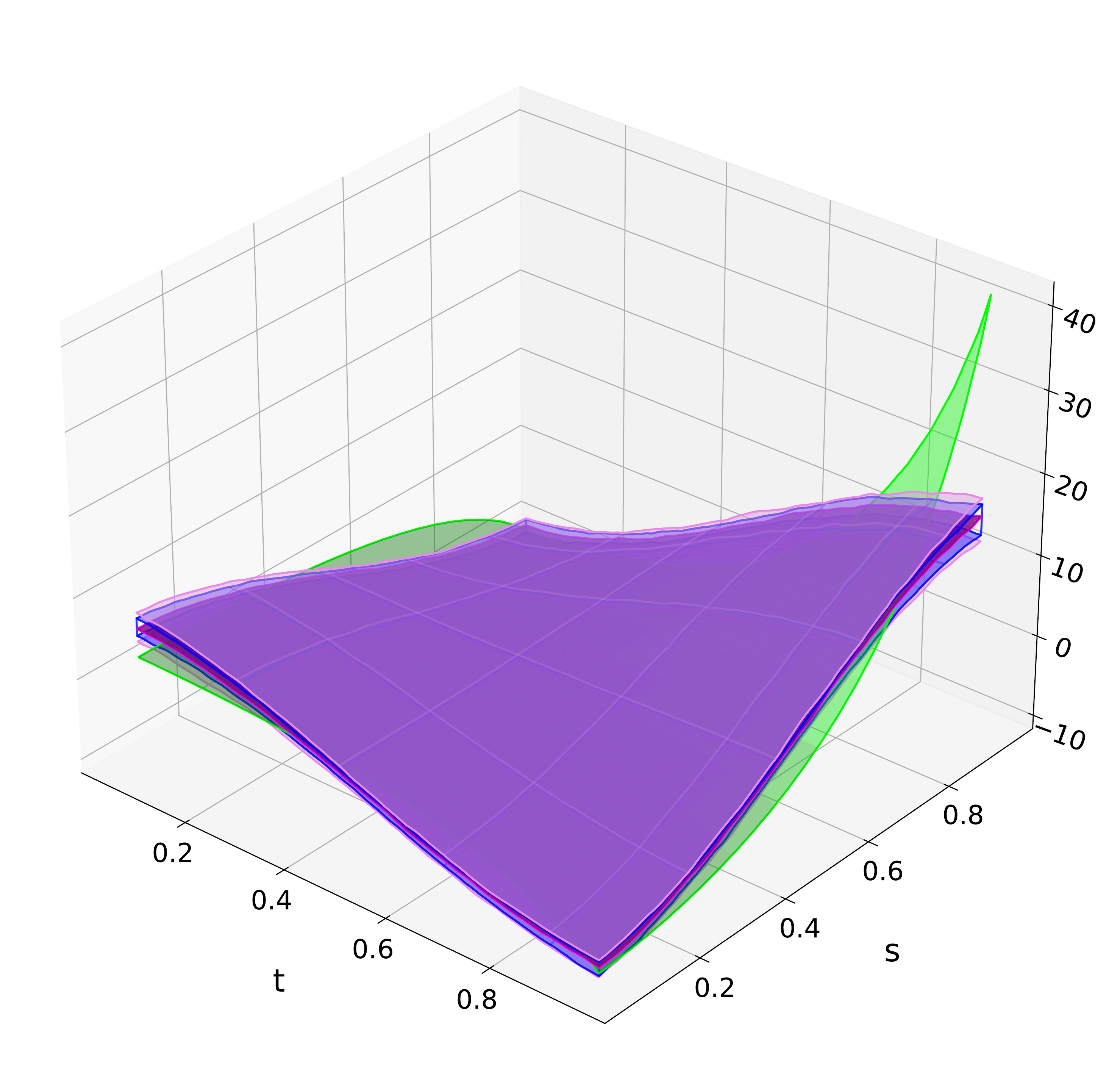}  & 
 \includegraphics[scale=0.18]{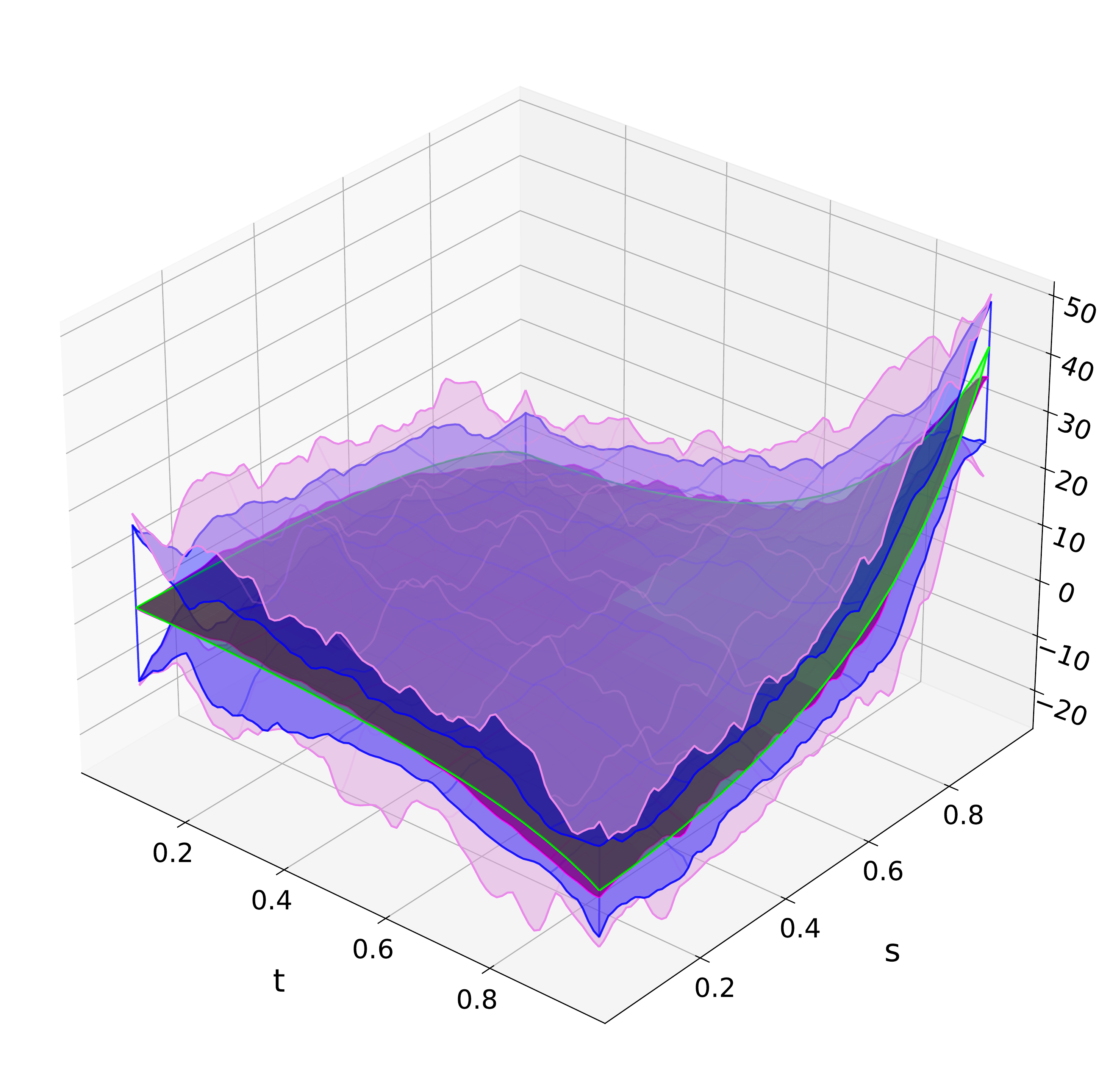} 
 \\[-4ex]
    
$C_{3, 4,0.4}$ &
 \includegraphics[scale=0.18]{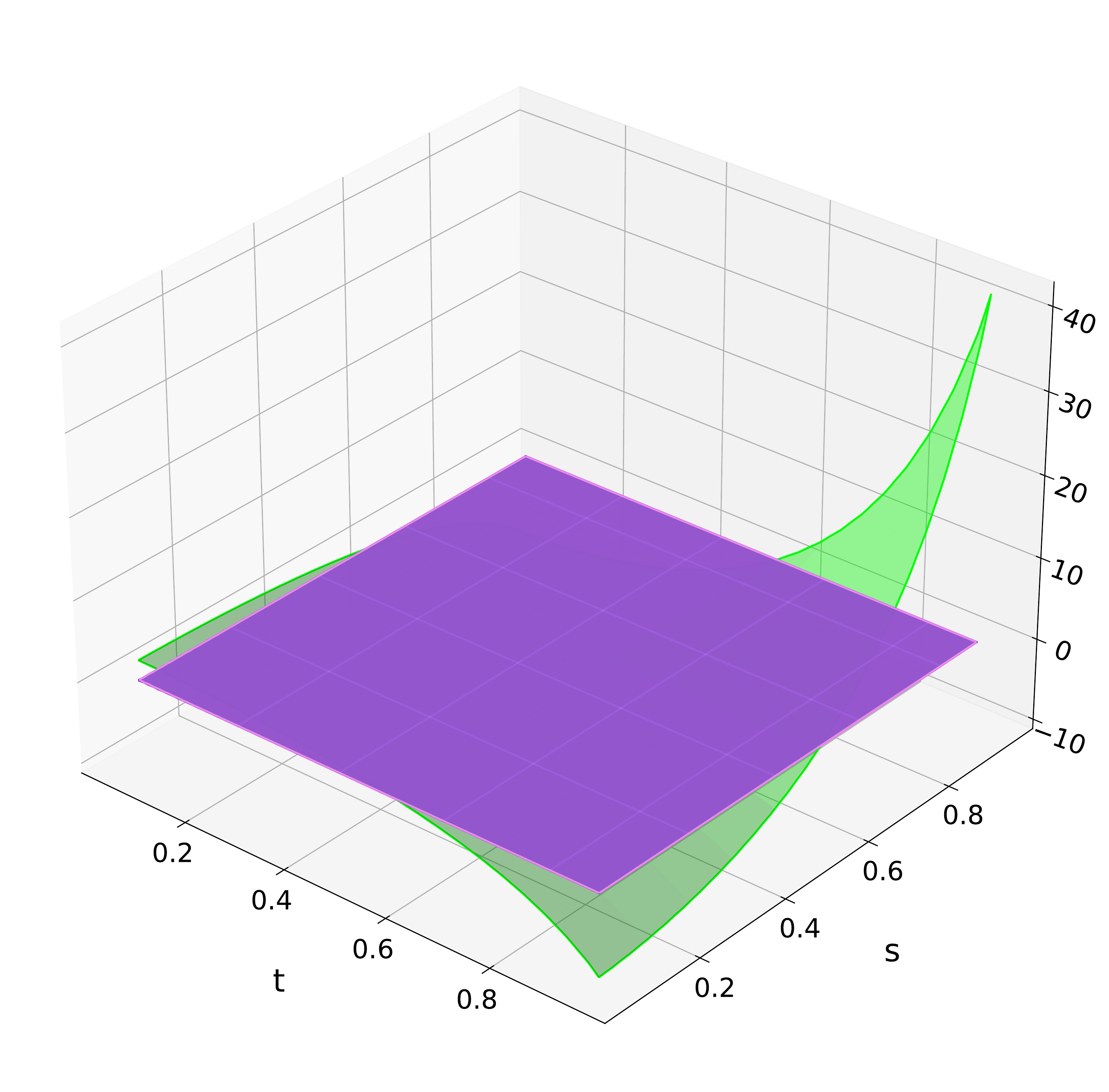}  & 
 \includegraphics[scale=0.18]{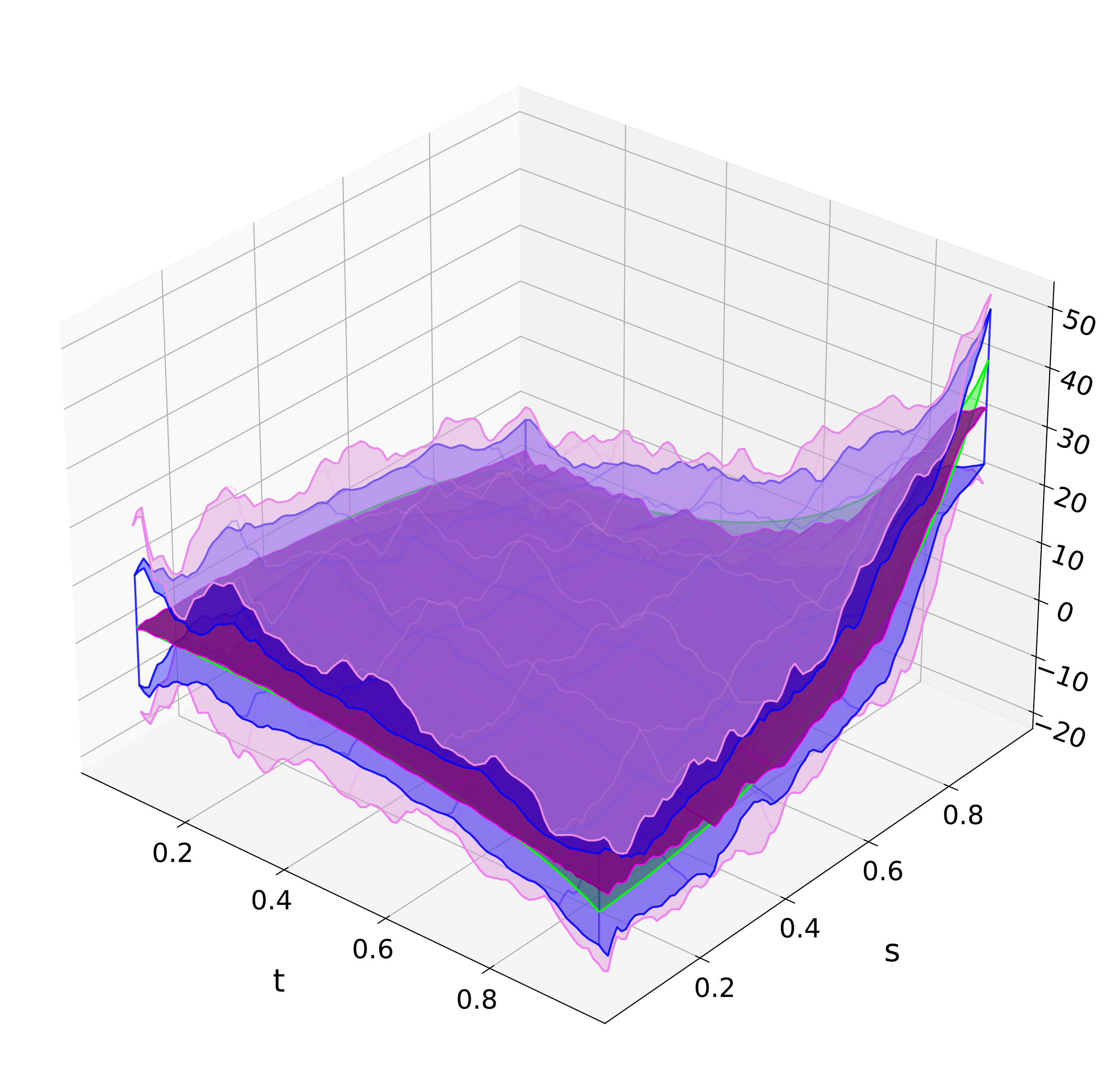}
\end{tabular}
\caption{\small \label{fig:wgamma-Upsilon1}  Surface boxplot of the estimators for $\upsilon_0$ under \textbf{Model 1} with  $\Upsilon_0=\Upsilon_{0,1}$. 
The true function is shown in green, while the purple surface is the central 
surface of the $n_R = 1000$ estimates $\wup$. Columns correspond to estimation 
methods, while rows to $C_0$ and   to some of the three contamination settings.}
\end{center} 
\end{figure}

\begin{figure}[tp]
 \begin{center}
 \footnotesize
 \renewcommand{\arraystretch}{0.2}
 \newcolumntype{M}{>{\centering\arraybackslash}m{\dimexpr.05\linewidth-1\tabcolsep}}
   \newcolumntype{G}{>{\centering\arraybackslash}m{\dimexpr.33\linewidth-1\tabcolsep}}
%\begin{tabular}{MGG}
\begin{tabular}{M GG}
 & $\wup_{\ls}$ &   $\wup_{\eme\eme}$ \\[-3ex]
$C_{0}$ 
&  \includegraphics[scale=0.18]{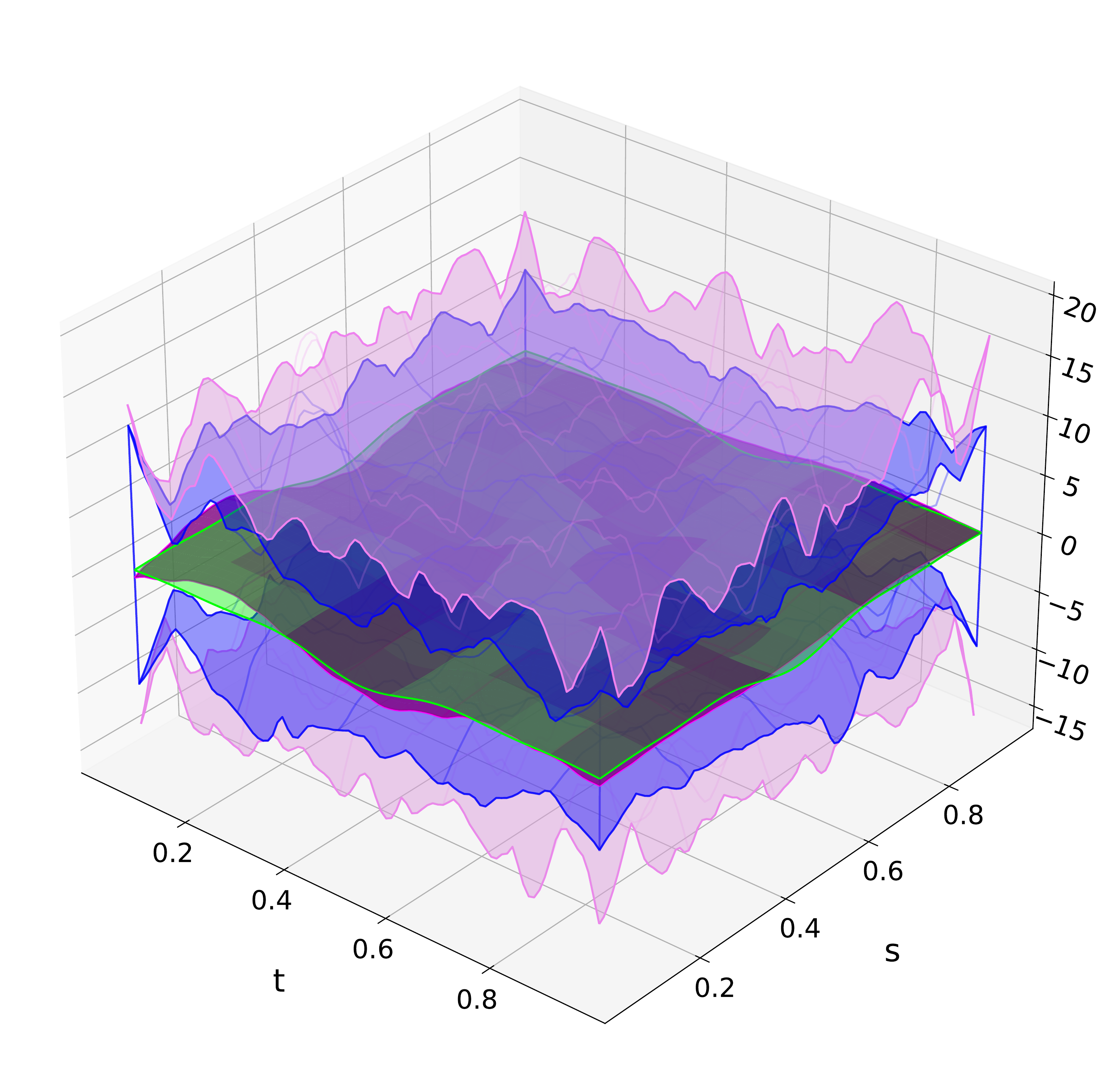} 
&  \includegraphics[scale=0.18]{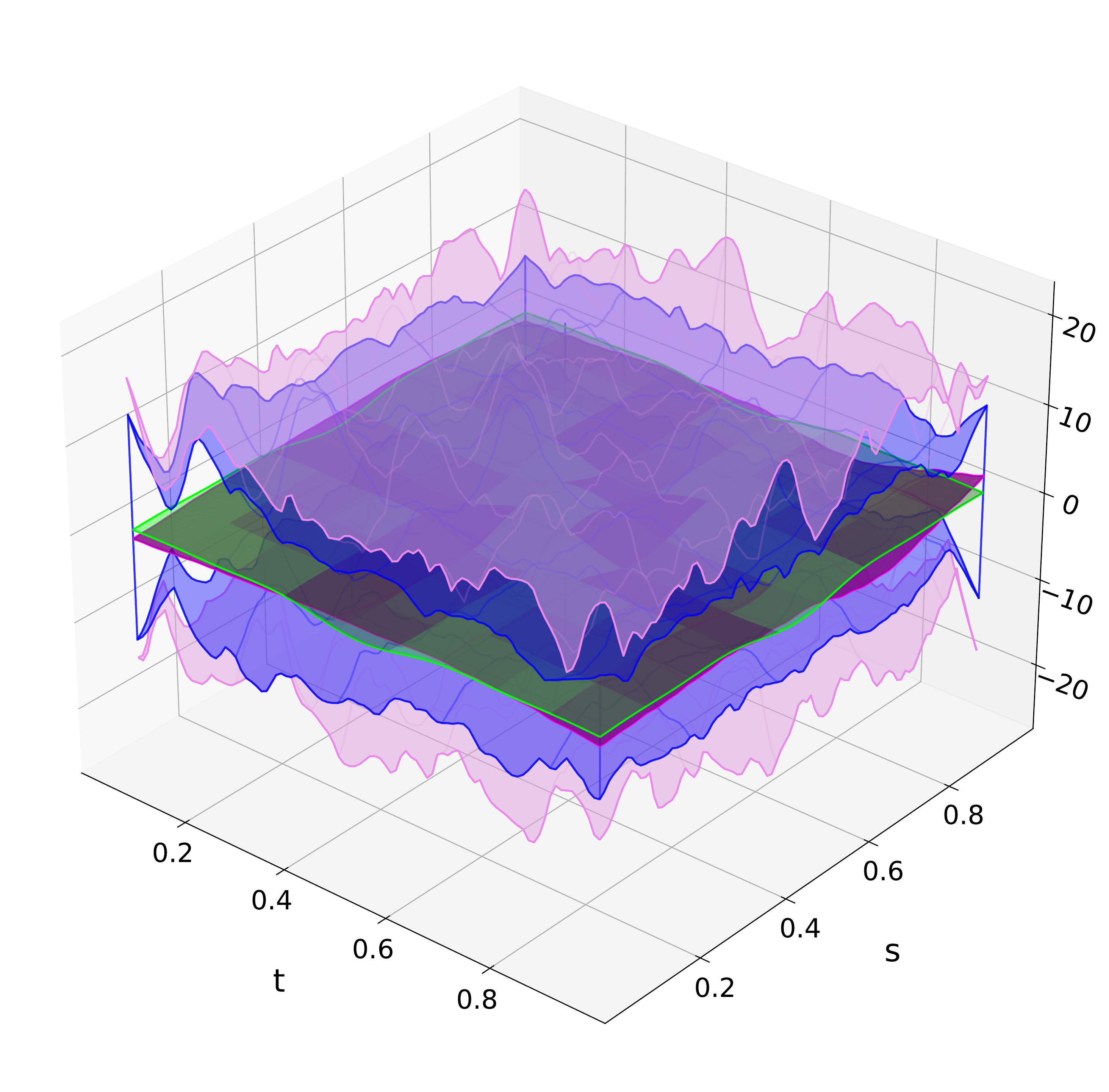} 
\\[-4ex]
$C_{1, 12}$ &  
 \includegraphics[scale=0.18]{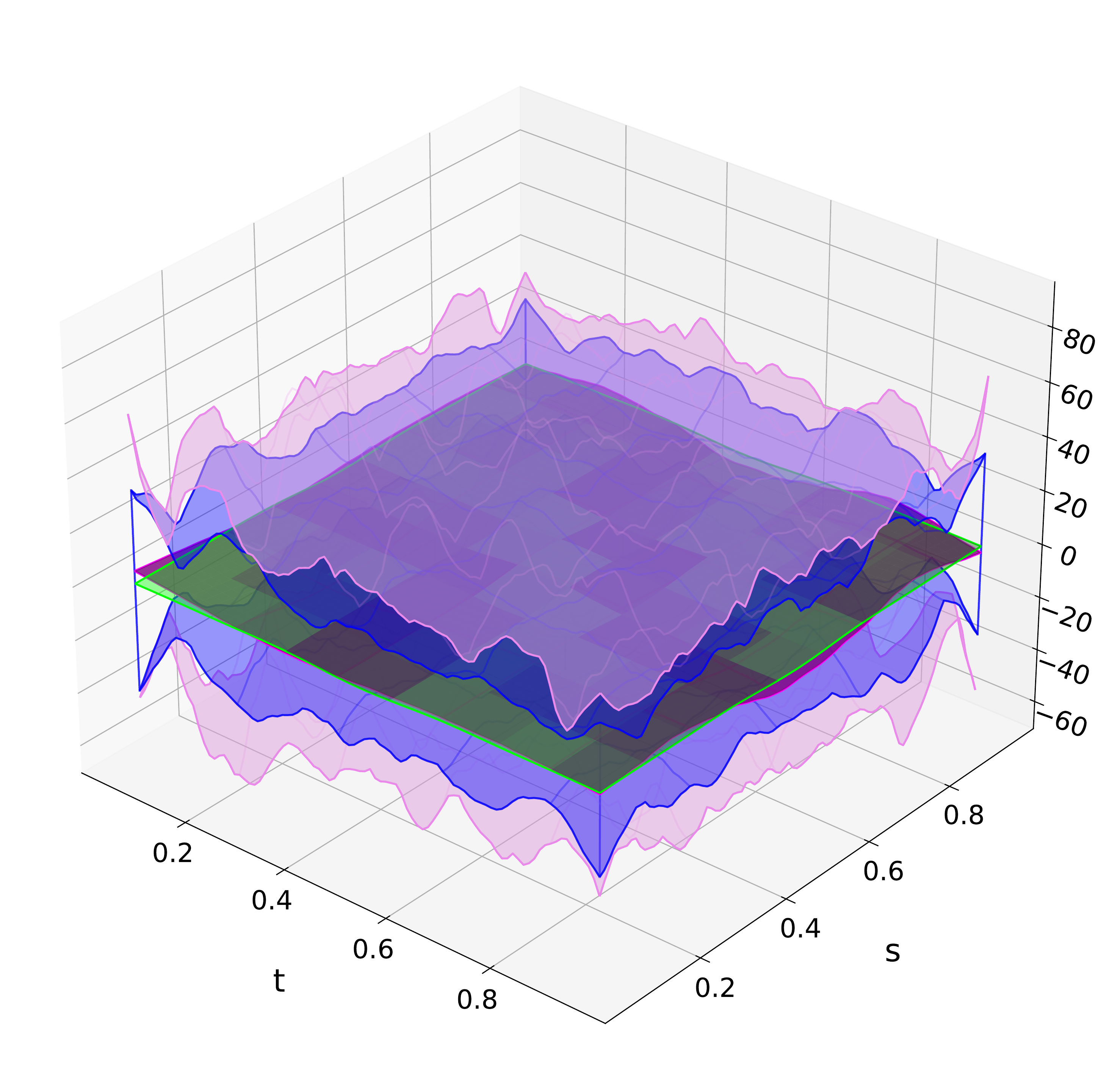}  & 
 \includegraphics[scale=0.18]{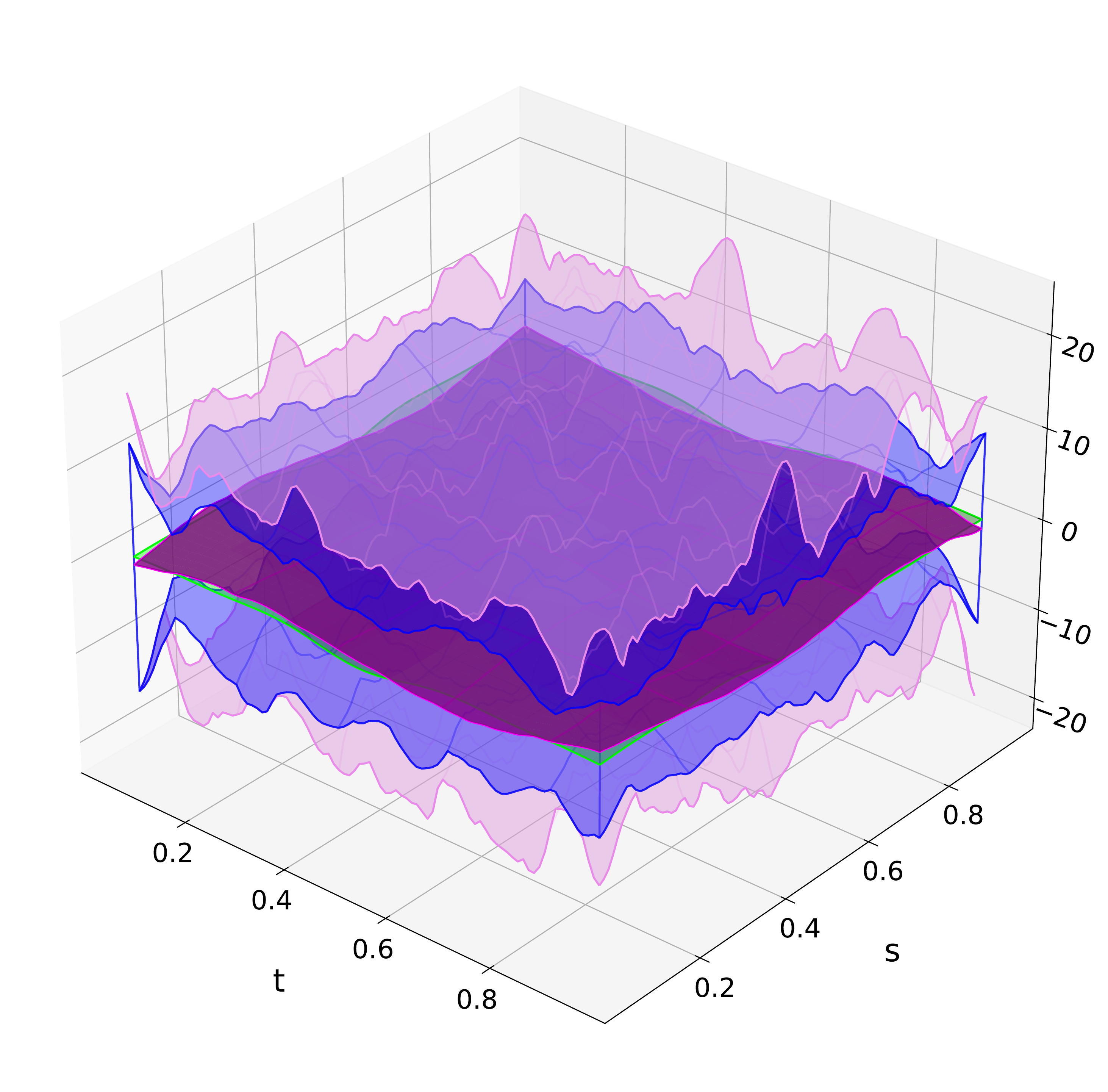} 
  \\[-4ex]
   
$C_{2, 12}$ &
 \includegraphics[scale=0.18]{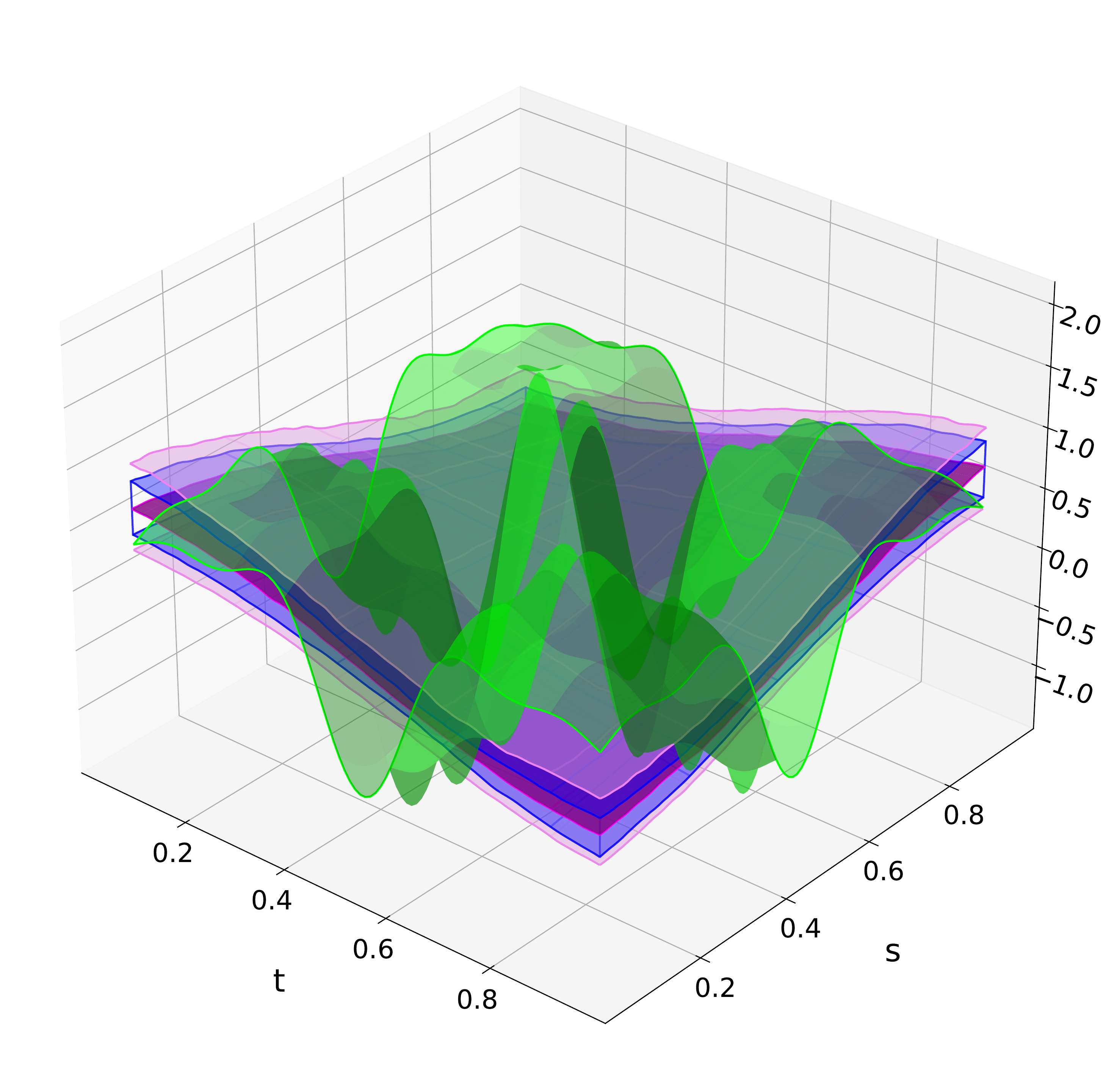}  & 
 \includegraphics[scale=0.18]{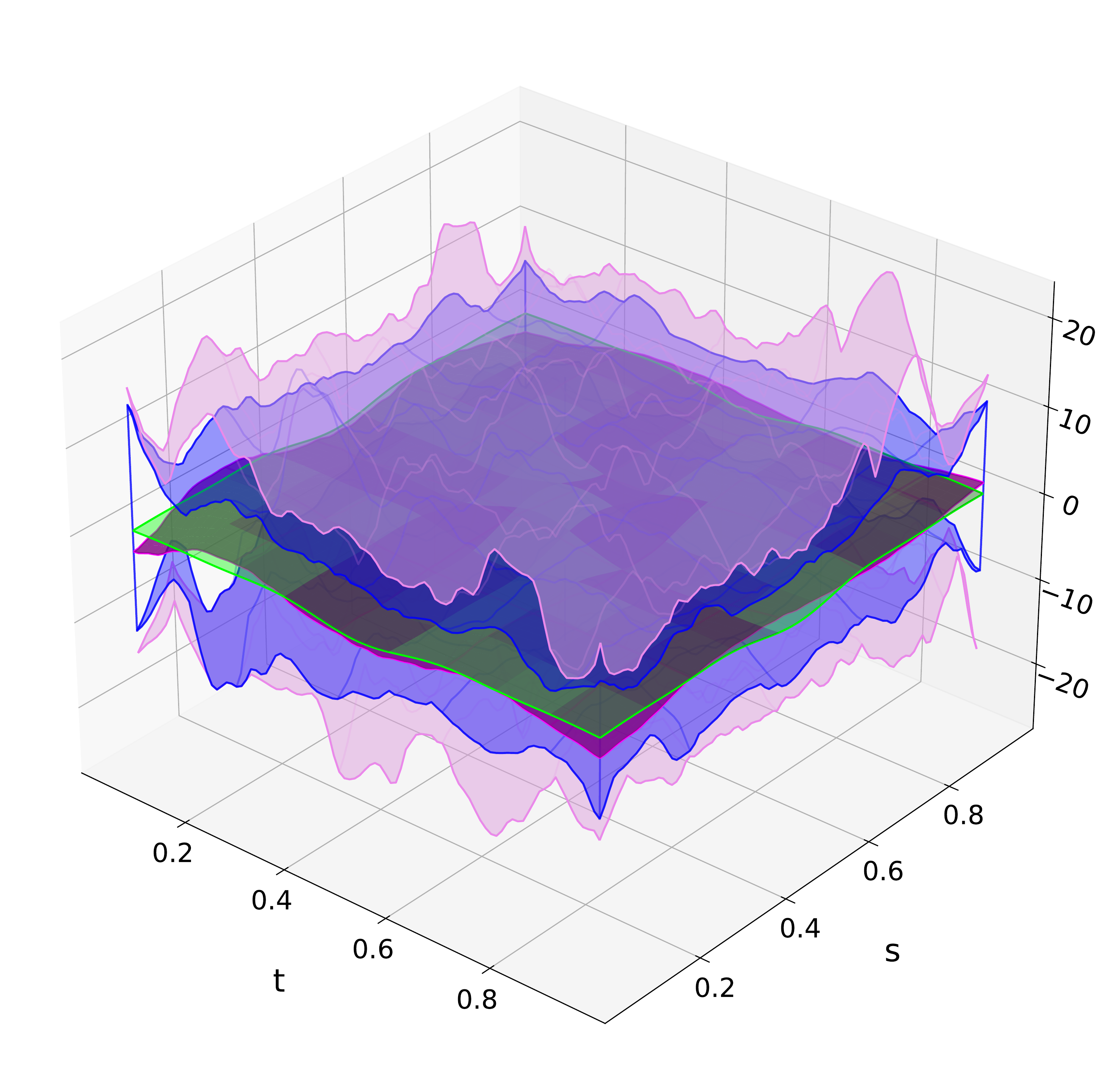} 
 \\[-4ex]
    
$C_{3,    4, 0.4}$ &
 \includegraphics[scale=0.18]{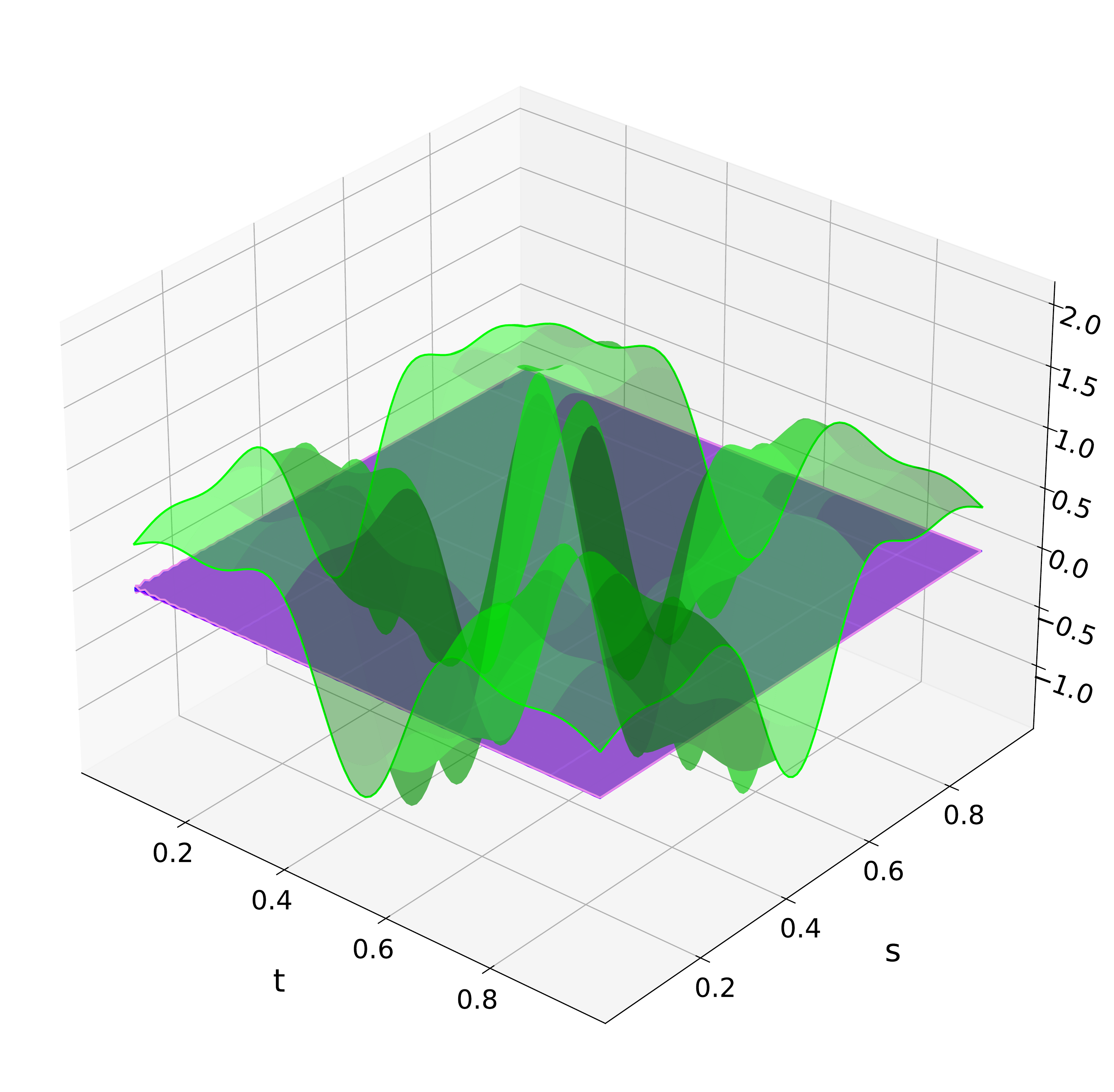}  & 
 \includegraphics[scale=0.18]{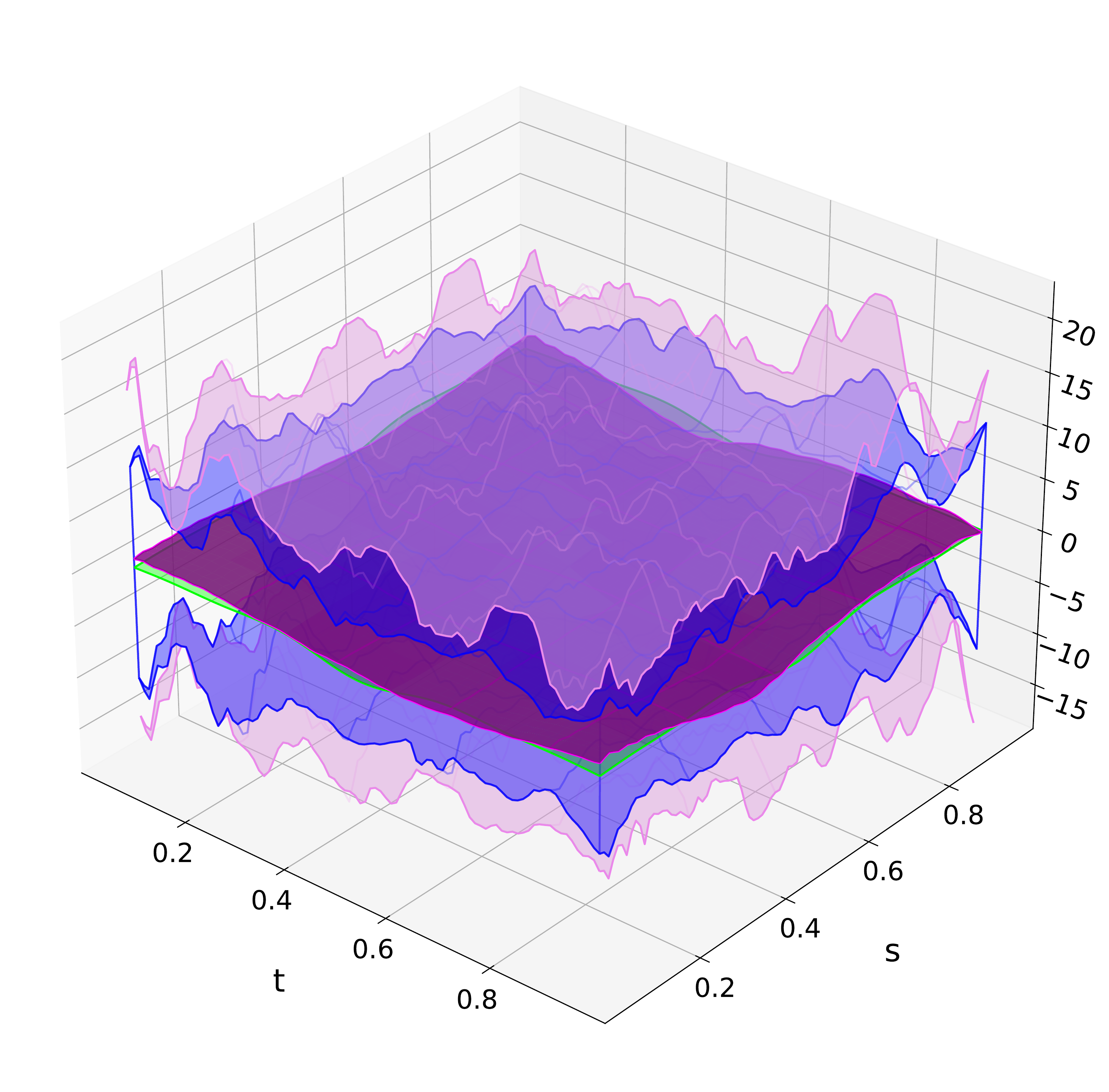}
\end{tabular}
\caption{\small \label{fig:wgamma-Upsilon2}  Surface boxplot of the estimators for $\upsilon_0$ under \textbf{Model 1} with  $\Upsilon_0=\Upsilon_{0,2}$. 
The true function is shown in green, while the purple surface is the central 
surface of the $n_R = 1000$ estimates $\wup$. Columns correspond to estimation 
methods, while rows to $C_0$ and   to some of the three contamination settings.}
\end{center} 
\end{figure}

To have a deeper insight on the effect of contamination on the quadratic component estimators, Figures \ref{fig:wgamma-Upsilon0} to \ref{fig:wgamma-Upsilon2} contain surface boxplots as defined in \citet{genton2014surface} for the $n_R = 1000$ realizations of the different estimators for $\upsilon_0$  under $C_0$ and some of the three contamination settings. For these plots the notion of volume depth is used to order the   surfaces. The median surface is represented in dark violet, the central region containing the 50\% deepest surfaces is represented in blue, while the surfaces in pink indicate the whiskers, beyond whose limits a surface is declared as outlier. The green surface represents the true function $\upsilon_0$.

The effect on the quadratic operator least squares estimators is observed in Figures \ref{fig:wgamma-Upsilon0} to \ref{fig:wgamma-Upsilon2}, where the enlargement of the central 50\% region  under $C_{1, 12}$ and the damaging effect of contamination  $C_{2, 12}$  become  evident. This effect is also observed under $C_{3,4,0.4}$ when considering quadratic terms in the model, that is, when choosing $\Upsilon_0=\Upsilon_{0,1}$ or $\Upsilon_0=\Upsilon_{0,2}$. In the first case, the true surface lies beneath the surface plot limits while in the former one it crosses the limiting surfaces.

In contrast, the $MM-$estimators display a remarkably stable behaviour across contamination settings. Their bias and MISE curves show that the $MM-$estimators for $\beta_0$  are highly robust against the considered contamination scenarios. If we look at the behaviour of these estimators in Figures \ref{fig:wbeta-Upsilon0} to \ref{fig:wbeta-Upsilon2} we note that the central box and the ``whiskers'' for the $MM-$estimators remain almost constant for all the contamination schemes in considered simulation scenarios ($\Upsilon_0=0$, $\Upsilon_0=\Upsilon_{0,1}$ or $\Upsilon_0=\Upsilon_{0,2}$), in sharp contrast to what happens to the classical method. Contamination $C_{3,4,0.4}$ affects the boxplot for values smaller than 2, where they mymic the distorting effect introduced in the model but even in this case,  the band preserves the shape.  The results in Figures \ref{fig:wgamma-Upsilon0} to \ref{fig:wgamma-Upsilon2} show that the $MM-$estimators for $\upsilon_0$ are almost unaffected by the different types of outliers, and the surface boxplots remain very similar to each other.

\begin{table}[ht!]
	\centering
	\footnotesize
	\renewcommand{\arraystretch}{1.4}   
	\setlength{\tabcolsep}{3pt}
	\begin{tabular}{r  c  c:  c | c:  c | c:  c  }
		\hline    
		&& \multicolumn{2}{c|}{$\Upsilon_{0,0}$}& \multicolumn{2}{c|}{$\Upsilon_{0,1}$}& \multicolumn{2}{c }{$\Upsilon_{0,2}$}\\
		\hline

		& & $|$Mean$|$  & SD     
		& $|$Mean$|$  & SD   
		& $|$Mean$|$  & SD   \\
		\hline
		$C_0$  & \textsc{ls} & 1.74 &  114.45  
		& 0.07   & 115.27  
		&  0.01  & 114.64  
		\\  
		
		& \textsc{mm} & 3.26  & 130.02 
		& 7.86  & 131.10 
		& 1.49   & 130.33 
		\\ 
		\hline
		$C_{1}$  & \textsc{ls} & 2019.87 &  723.84  
		& 2021.67  & 723.88  
		& 2021.59   & 723.88 
		\\ 
		& \textsc{mm} & 2.58 &  130.92  
		& 6.85 & 131.26 
		& 0.85 &  131.28  
		\\ 
		\hline
		$C_{2}$  & \textsc{ls} & 45.15 &   98.32 
		& 129.87  & 233.93  
		& 33.42   & 98.47 
		\\ 
		& \textsc{mm} & 90.01 &   128.36 
		& 95.61 &  129.93  
		& 87.82 &   129.45  
		\\ 
		\hline
		$C_{3}$ & \textsc{ls} & 1.52 &   74.54 
		& 1825.23   & 172.97  
		& 208.23 &   78.36  
		\\   %ES LA C4 
		& \textsc{mm} & 2.68 &   132.44  
		& 9.82 &   132.77  
		& 0.68 &   132.51  
		\\ 
		\hline
	\end{tabular}
	\caption{ \small \label{tab:res-alphas-M1} Summary measures (multiplied by 1000) for $\alpha_0$ estimates over  clean and contaminated samples, \textbf{Model 1}. The reported values under contamination correspond to the worst situation.}
\end{table}
Regarding the estimation of $\alpha_0$, Table \ref{tab:res-alphas-M1} reports as summary measures: the absolute value of the mean which is a measure of the bias  since $\alpha_0=0$; and the standard deviation   as variability measure. For the contamination settings, the maximum over the different values of $\mu$ and/or $\delta$ are reported. In all cases, the reported values correspond to the summary measures multiplied by 1000. Figure \ref{fig:alphas-boxplots-M1}  presents the boxplots of the estimators for $\alpha_0$ for clean  samples and for some contamination scenarios. The true value is plotted with a green dashed line, for reference. Each row  corresponds to different $\Upsilon_0$ scenarios, while each column to a contamination setting. The boxplots of the classical estimators are given in magenta, while those of the robust ones are presented in blue. The reported results show that scheme $C_{1,\mu}$ affects the classical estimator of the intercept for any choice of the quadratic operator with maximum biases increased more than 1000 times and standard deviations enlarged more than 7 times. In contrast, under $C_{3,4,0.4}$, the largest effect is observed when considering $\Upsilon_{0,1}$. For this choice of the quadratic kernel, the classical procedure is also affected under $C_{2,12}$. The robust procedure is stable over the contaminations considered, even though some effect in the bias is observed under $C_2$ (see Table  \ref{tab:res-alphas-M1}) it is much smaller than that of its classical counterpart when the quadratic operator is $\Upsilon_{0,1}$.

\begin{figure}[ht!]
		\centering
		\footnotesize
		\renewcommand{\arraystretch}{1.3}   
		\setlength{\tabcolsep}{2pt}
		\newcolumntype{M}{>{\centering\arraybackslash}m{\dimexpr.05\linewidth-1\tabcolsep}}
		\newcolumntype{G}{>{\centering\arraybackslash}m{\dimexpr.16\linewidth}}
		\begin{tabular}{M GGGG} 
			& {\hspace{2cm} \small   $C_0$} &{\hspace{1.5cm} \small   $C_{1,12}$} &{\hspace{0.8cm} \small $C_{2,12}$} &{\hspace{0.1cm} \small $C_{3,4,0.4}$ } \\[-5ex]
			% La chanchada de los hspace es porque NO logro centrarlos...
			% Probé con el multicolumn 1 c pero pero queda peor...
			%& \multicolumn{1}{c} {\small   $C_0$} & \multicolumn{1}{c} {\small   $C_{1,12}$} & \multicolumn{1}{c} {\small $C_{2,12}$} & \multicolumn{1}{c}{\small $C_{3,4,0.4}$ } \\[-5ex]
			{\small   $\Upsilon_{0,0}$} & \multicolumn{4}{G}
			{\includegraphics[scale=0.57]{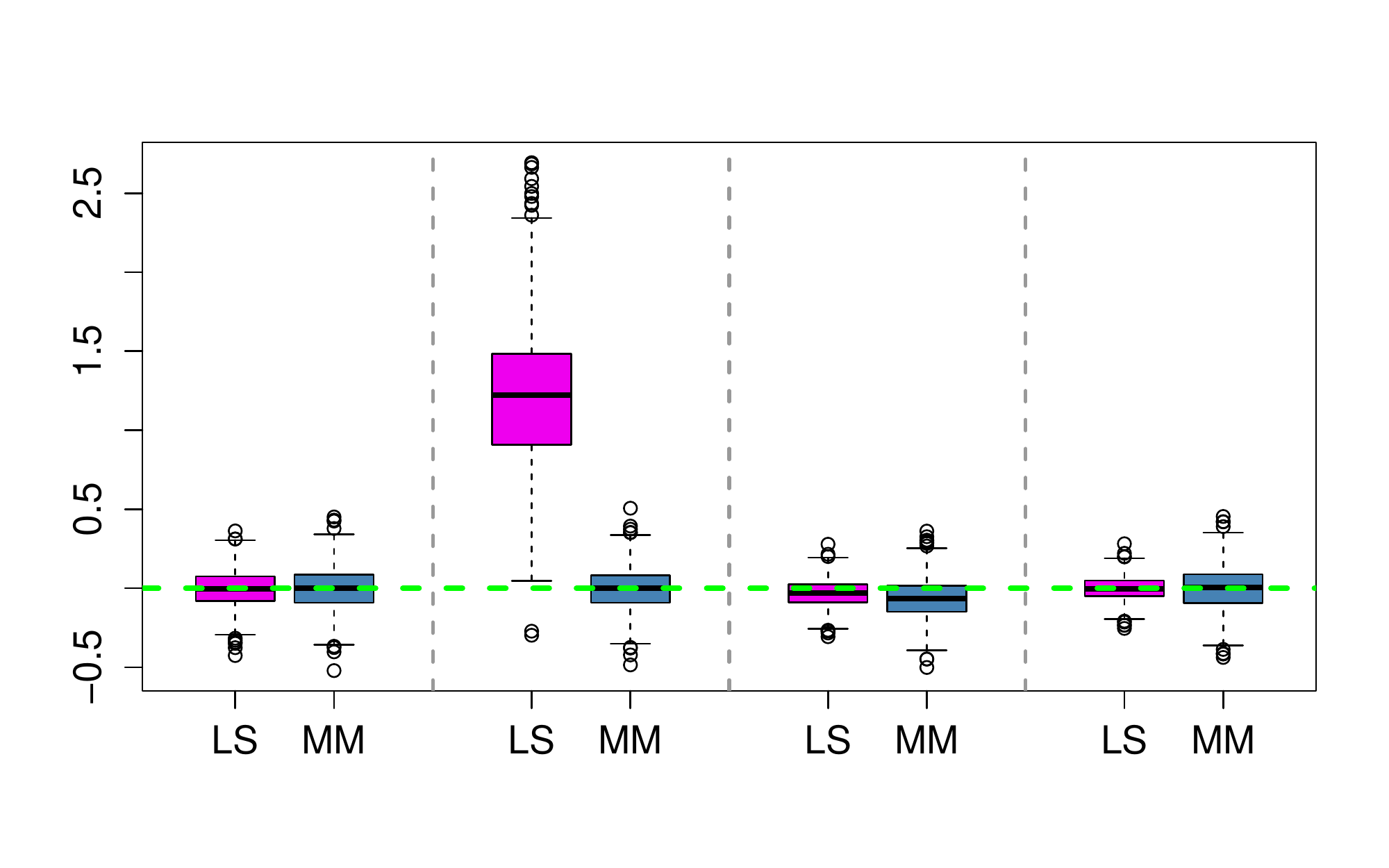}}\\[-8ex]
			{\small   $\Upsilon_{0,1}$} & \multicolumn{4}{G}
			{\includegraphics[scale=0.57]{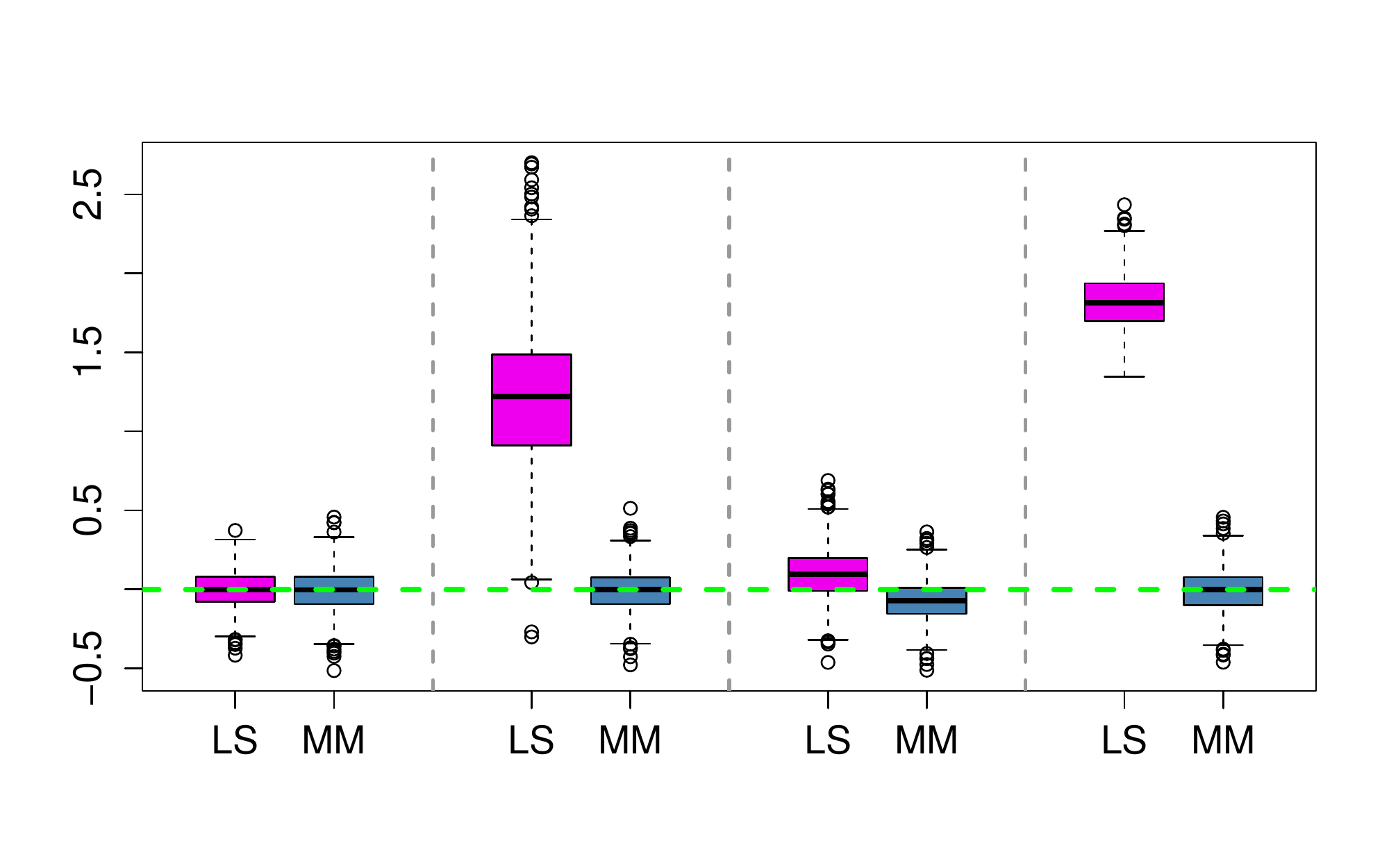}}\\[-8ex]
			{\small   $\Upsilon_{0,2}$} & \multicolumn{4}{G}
			{\includegraphics[scale=0.57]{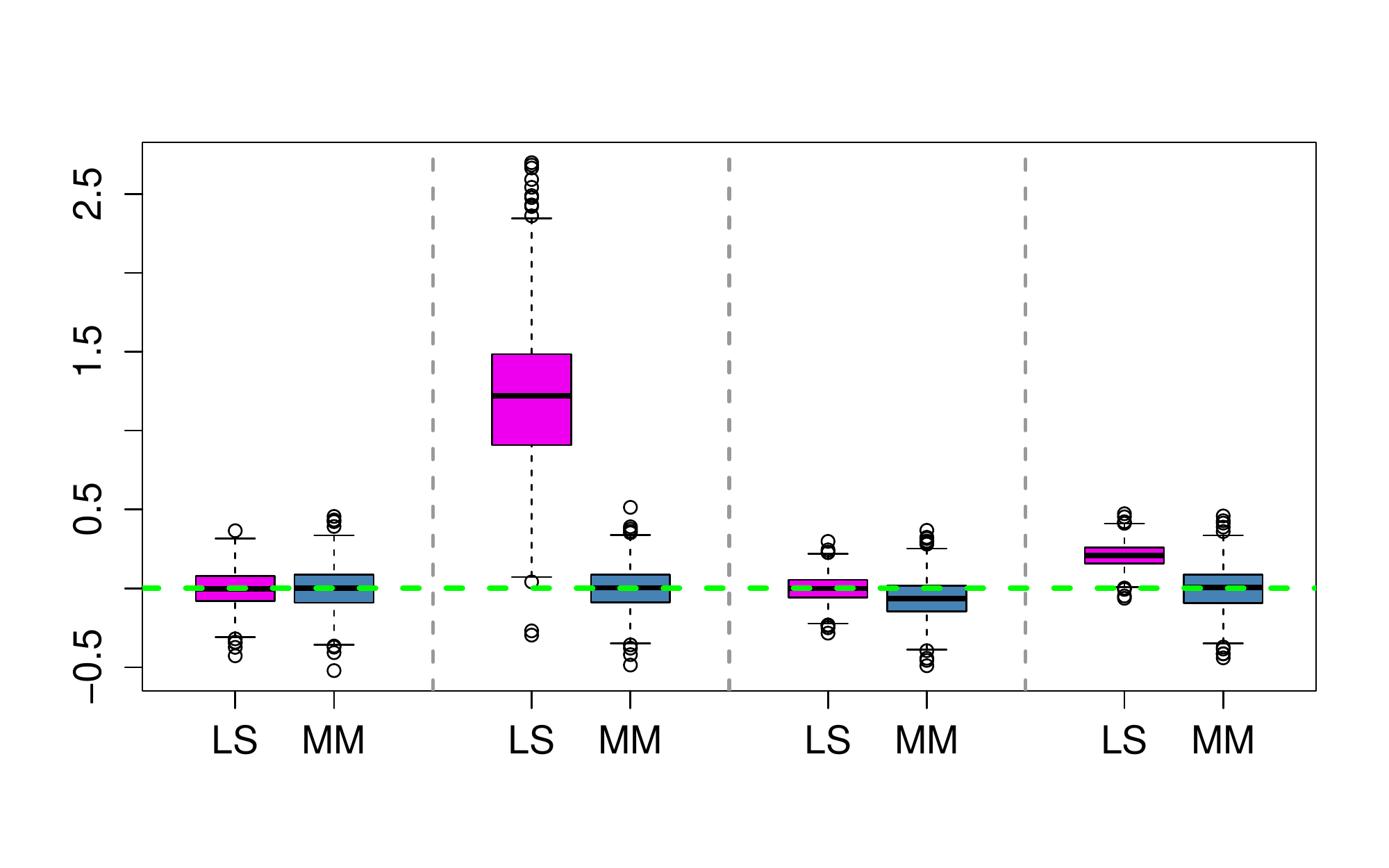}}\\[-4ex]
		\end{tabular}
		\caption{\small \label{fig:alphas-boxplots-M1}  Boxplots of the estimators for $\alpha_0$ for clean and contaminated samples, under \textbf{Model 1}. The true value is shown with a green dashed line. Rows correspond to different $\Upsilon_0$ scenarios. Columns correspond to $C_0$ and to some of the three contamination  settings. Magenta and blue boxplots correspond to classical and robust methods, respectively.}
\end{figure}
 \clearpage
 
\subsection{Model 2}
\begin{figure}[ht!] %[tp]
  \begin{center} 
  \begin{tabular}{cc  }
  Linear Model & \multicolumn{1}{c}{Quadratic Model} \\
   $\beta_0$ & $\beta_0$     \\
   \includegraphics[width=0.3\textwidth]{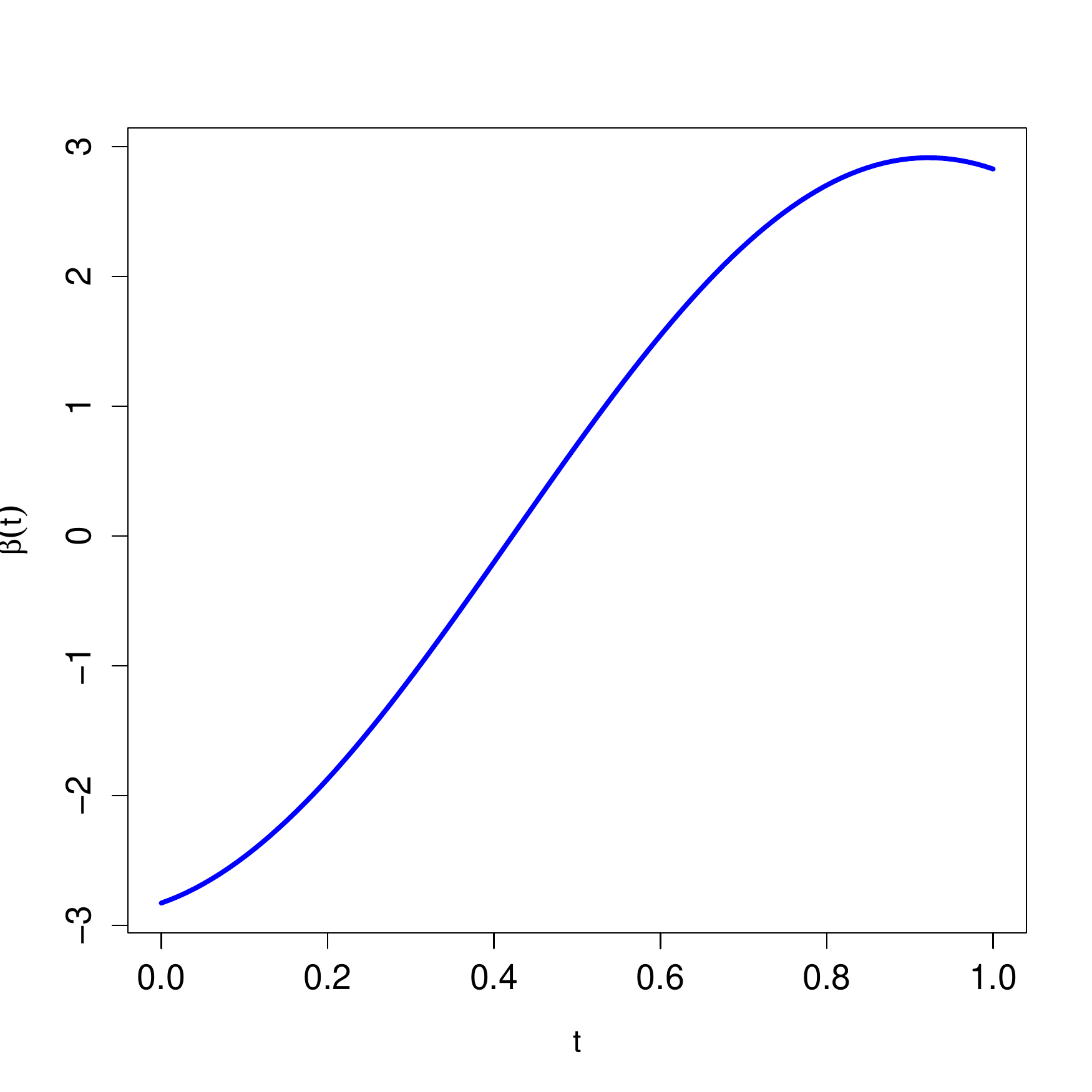} &
   \includegraphics[width=0.3\textwidth]{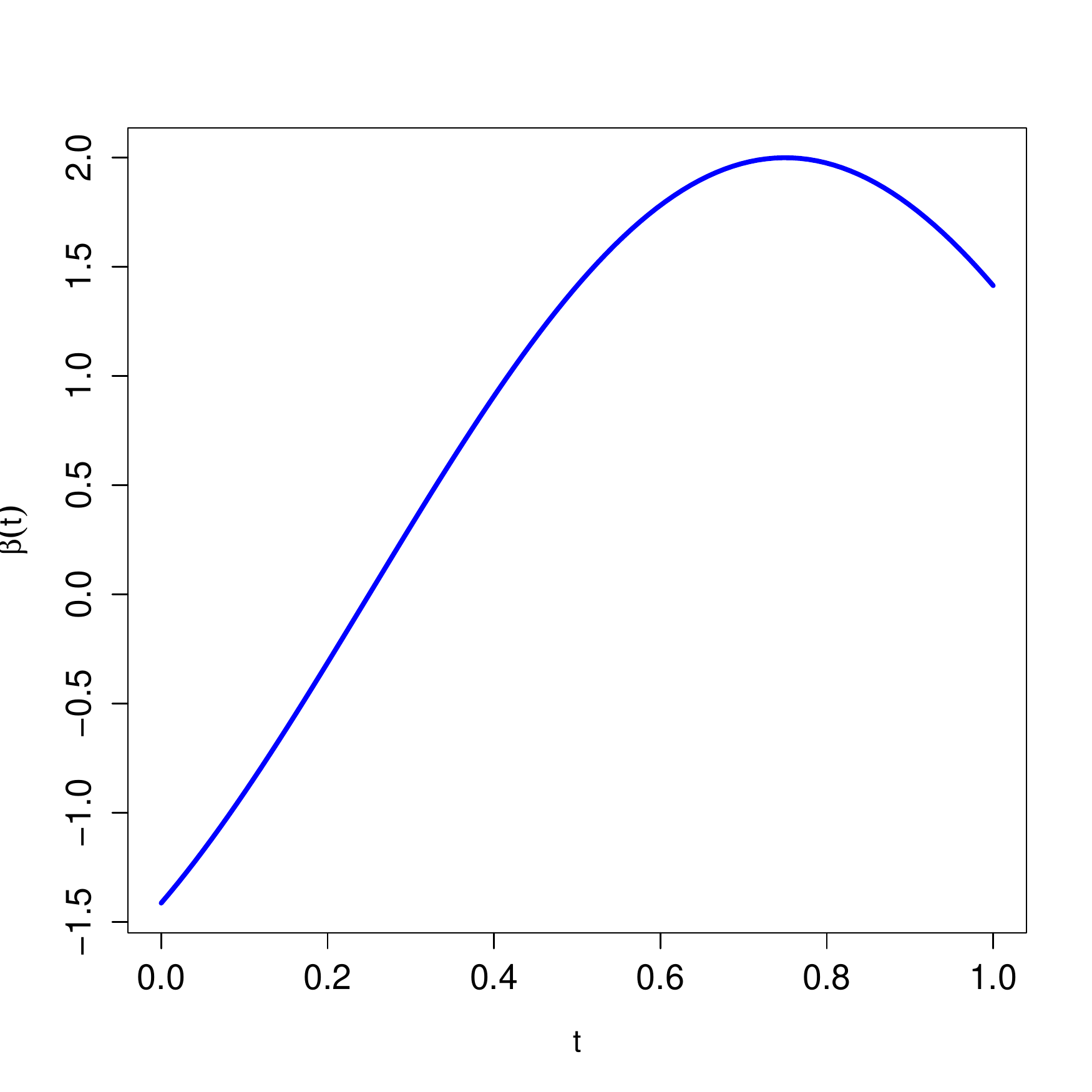} \\
   \multicolumn{2}{c}{Quadratic Model: $\upsilon_{0,1}$ } \\
    \includegraphics[width=0.29\textwidth]{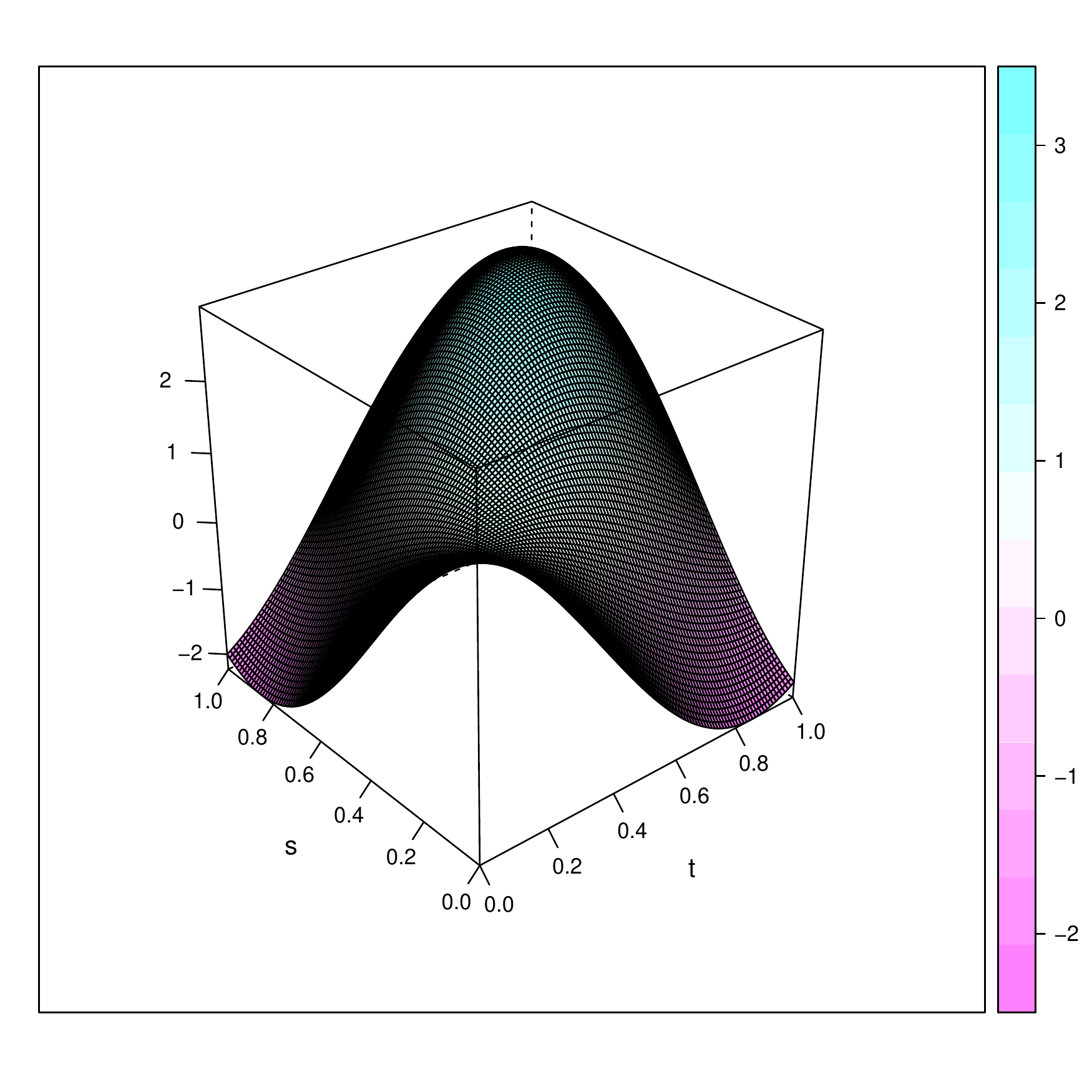} &  
    \includegraphics[width=0.29\textwidth]{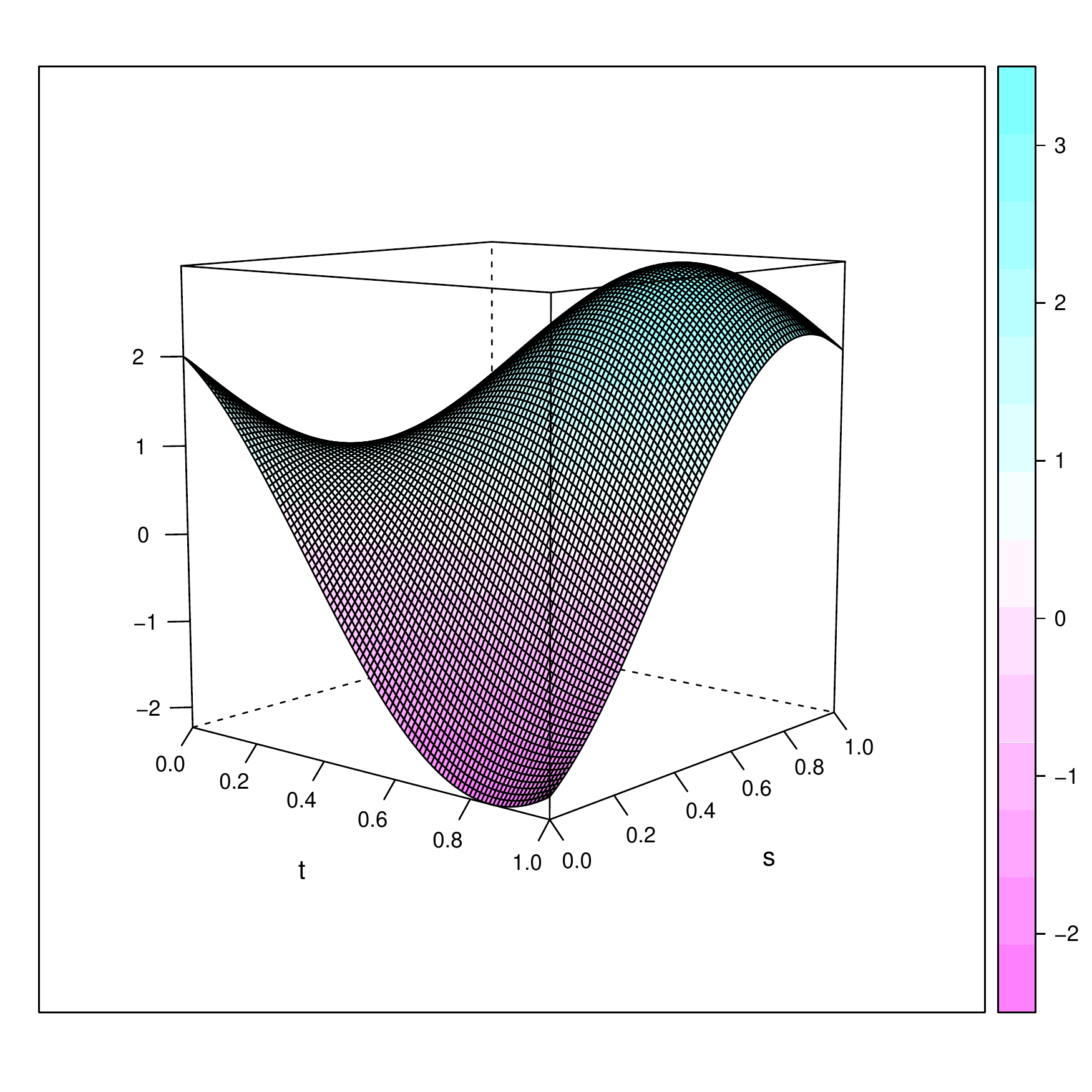} 
\end{tabular}
  \caption{\small \label{fig:parametros-verdaderos-YM} True parameters under Model 2. For the quadratic kernel we present two view points.}
\end{center}
\end{figure}

{\label{sec:model2}}
In this section, we considered the functional quadratic regression model \eqref{eq:monte-carlo-modelo}, where  $\sigma_0=0.5$ and the regression parameter equals the one used in \citet{yao2010functional}, that is, $\beta_0  = a_1 \phi_1 +a_2 \phi_2$ with 
$\phi_1(t) = -\,\sqrt{2} \cos (\pi t)$  and  $\phi_2(t) =  \sqrt{2} \sin (\pi t)$. We label this model as \textbf{Model 2}. The choice of the coefficients $a_1$ and $a_2$ depend on the quadratic operator selected. More precisely,   when  $\Upsilon_{0,0}=0$, that is under a functional linear model, we chose  $a_1=2$ and $a_2=0.5$. When considering a functional quadratic model with  $\Upsilon_{0,1}=   \phi_1\otimes  \phi_1+ \phi_2\otimes  \phi_2+ (1/2)\left(\phi_1\otimes  \phi_2+ \phi_2\otimes  \phi_1 \right) $, the coefficients of $\beta_0$ equal $a_1=a_2=1$.
Figure \ref{fig:parametros-verdaderos-YM} shows the functions $\beta_0$ related to the linear and quadratic model and the surface $\upsilon_{0,1}(s,t)$   associated to the non--null quadratic operator $\Upsilon_{0,1}$. For the quadratic kernel we present two viewpoints, since the surface plots are better appreciated with one of them due to the surface shape.

For clean samples, denoted $C_0$, the errors are normally distributed $\epsilon_i \sim N(0,1)$, independent from $X_i$. The   functional covariates $X_i(t)$ were generated as  Gaussian processes with mean 0 and covariance operator $\Gamma= 4 \phi_1\otimes \phi_1+ \phi_2\otimes \phi_2$, that is, $X_{ij}=  \xi_{i1} \phi_1 + \xi_{i2} \phi_2  $ where $\xi_{ij} $ are independent Gaussian random variables such that $\xi_{i1}\sim N(0,4)$ and  $\xi_{i2}\sim N(0,1)$. 
 
As in Model 1, we considered three contamination scenarios. Contaminations $C_{1,\mu}$ and $C_{2,\mu}$ are identical to the ones used   in Model 1, and we include a third contamination, labelled $C_{3,\mu}$. The selected contamination schemes are then
\begin{itemize}
	\item $C_{1,\mu}$: corresponds to ``vertical outliers''. The  distribution $G$ of the errors $\epsilon$ is given by $G(u)=0.9 \, \Phi(u )+0.1\, \Phi\left((u-\mu)/0.5 \right)$, with $\Phi$ the standard normal distribution function.   We chose $\mu$ varying between 8 and 20 with a step of 2.
	\item   $C_{2,\mu}$: we sample $v_i\sim Bi(1, 0.10)$ and then:
	 	\begin{itemize}
	  	\item if $v_i = 0$, let $\epsilon_i^{(\cont)}=\epsilon_i$ and $X_i^{(\cont)}=X_i$;
	  	\item if $v_i=1$, let  $\sigma_0 \epsilon_i^{(\cont)} \sim N( \mu, \sigma_0^2/4)$ and $X_i^{(\cont)}=   \xi_{i1}^{(\cont)} \phi_1 + \xi_{i2}^{(\cont)} \phi_2    $, with $\xi_{i1}^{(\cont)}\sim N(0,4)$   and  $\xi_{i2}^{(\cont)}\sim N(\mu/2, 0.25 ) $. 
	   	\end{itemize} 
 	The responses are generated as $y_i^{(\cont)} = \langle \beta_0, X_i^{(\cont)}\rangle + \langle X_i^{(\cont)}\Upsilon_0 , X_i^{(\cont)}\rangle  + \sigma_0  \epsilon_i^{(\cont)}$. The values of $\mu$ vary between 8 and 20 with a step of 2.
	\item $C_{3,\mu}$: we sample $v_i\sim Bi(1, 0.10)$ and then:
  	\begin{itemize}
  		\item if $v_i = 0$, let $y_i^{(\cont)}=y_i$ and $X_i^{(\cont)}= X_i$;
  		\item if $v_i=1$, let   $X_i^{(\cont)}=  \xi_{i1}^{(\cont)} \phi_1 + \xi_{i2}^{(\cont)} \phi_2    $ where $\xi_{ij}^{(\cont)}=2\, |\xi_{ij}|$, while    $y_i^{(\cont)}=2\, \mu\, |y_i|$.
 	\end{itemize} 
 	The values of $\mu$ vary in $\{ 0.2, 0.4, 0.6, 0.8, 1.2, 1.4, 1.6, 1.8, 2.0, 2.2, 2.4\}$
\end{itemize}
 
\begin{figure}[ht!]%[tp]
  \small
  \begin{center} 
   \newcolumntype{M}{>{\centering\arraybackslash}m{\dimexpr.05\linewidth-1\tabcolsep}}
   \newcolumntype{G}{>{\centering\arraybackslash}m{\dimexpr.3\linewidth-1\tabcolsep}}
\begin{tabular}{GGG}
 $C_0$ & $C_{2,8}$ & $C_{3,1.2}$\\
      \includegraphics[width=0.3\textwidth]{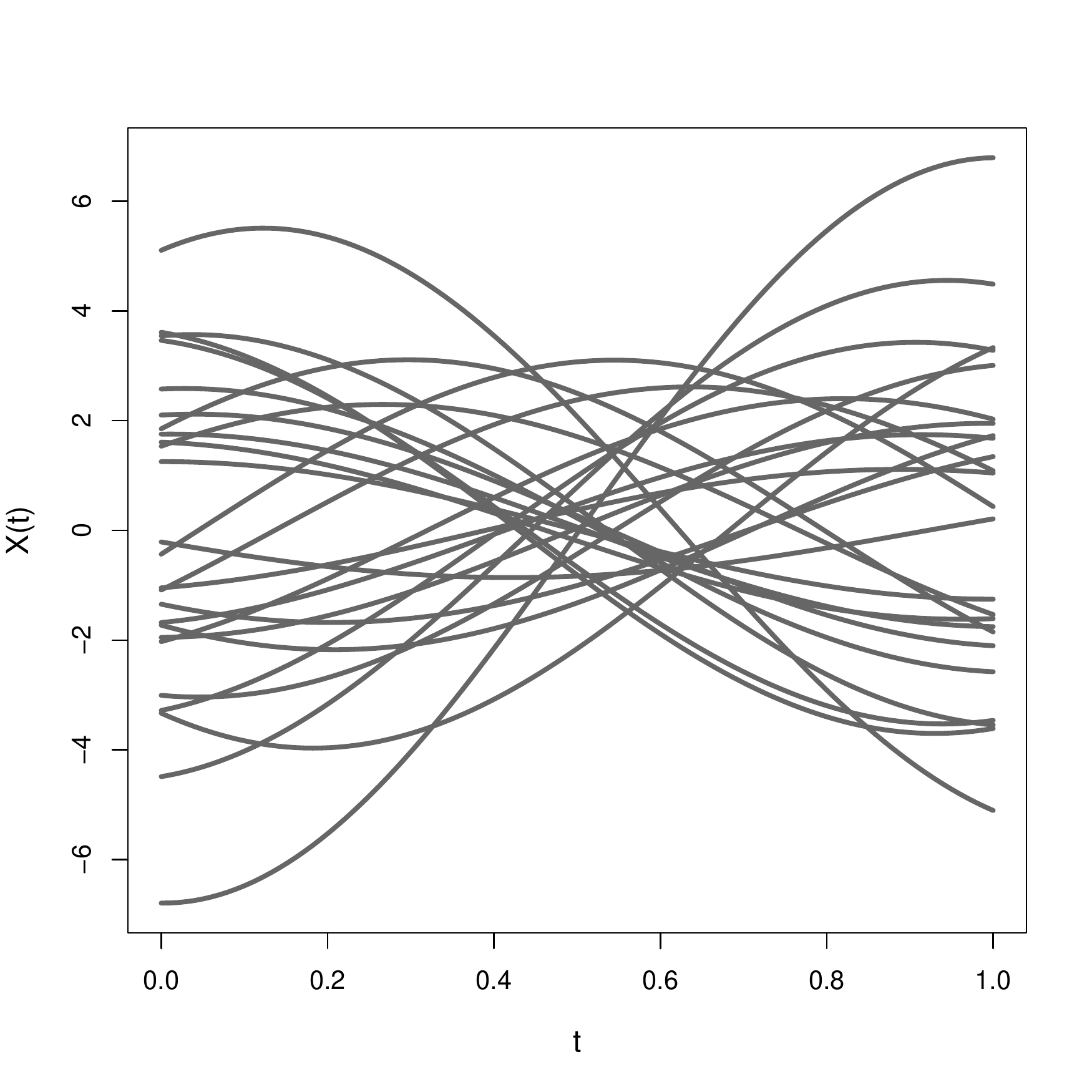} & 
      \includegraphics[width=0.3\textwidth]{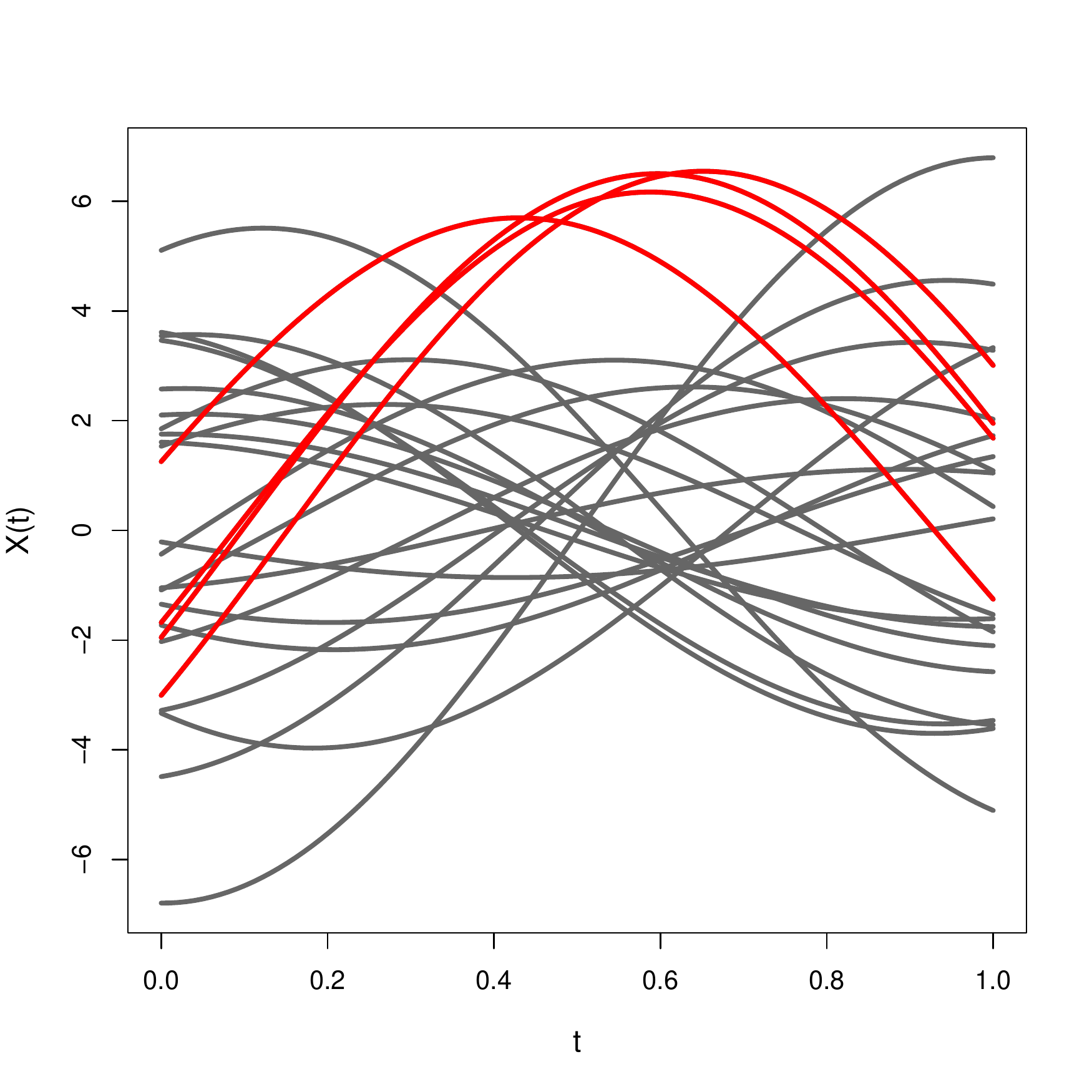} &
      \includegraphics[width=0.3\textwidth]{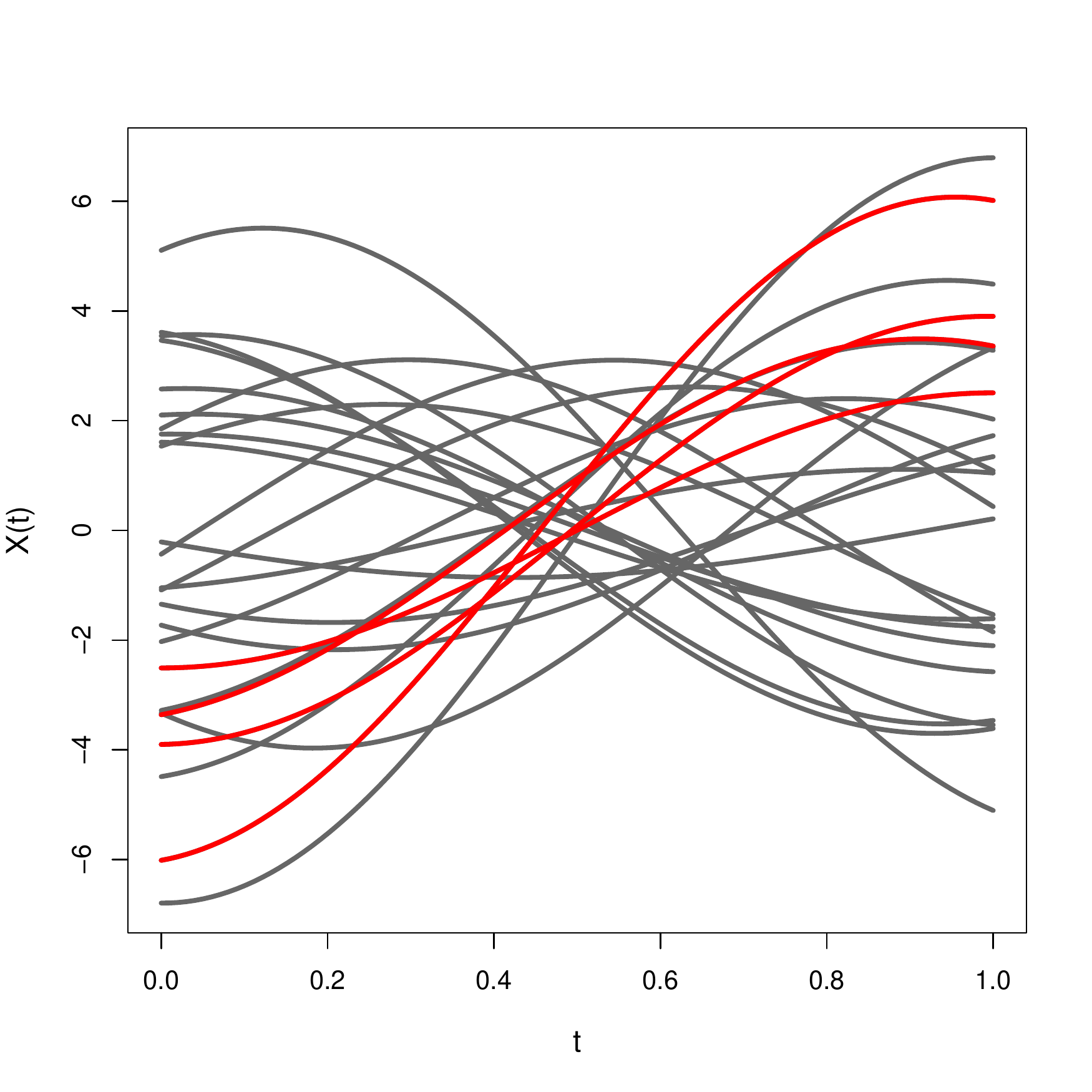}
      \end{tabular}
   \end{center}
  \vskip-0.2in
  \caption{ \label{fig:trayectorias-YM} 25 trajectories $X_i(t)$ under \textbf{Model 2} with and without contamination.}
\end{figure}

As an illustration of the type of outliers generated with the second setting above, Figure \ref{fig:trayectorias-YM} shows 25 randomly chosen functional covariates $X_i(t)$, for one sample generated under $C_0$ (with no outliers), one obtained under $C_{2,8}$ and the other one under  $C_{3,1.2}$.
 
The same summary measures as in Section \ref{sec:model1} were computed. As in \textbf{Model 1}, Table \ref{tab:tabla-Upsilon-YM} reports the maximum value of the squared bias and of the $\mbox{MISE}$ over the different values of $\mu$ and/or $\delta$ and we simply label the situation as $C_j$, for $j=1,2,3$ to avoid burden notation.

\begin{table}
\begin{center}
\footnotesize
   \renewcommand{\arraystretch}{1.4}   
  \setlength{\tabcolsep}{2pt}
\begin{tabular}{rr rr :  rr |rr  :  rr    }
\hline  
&& \multicolumn{4}{c|}{\textbf{Linear Model}}& \multicolumn{4}{c}{\textbf{Quadratic Model}} \\
\hline
&&\multicolumn{2}{c:}{$\wbeta$} &\multicolumn{2}{ c|}{$\wup$}&
\multicolumn{2}{c:}{$\wbeta$} &\multicolumn{2}{ c}{$\wup$} \\
\hline

& & Bias$^2_{\trim}$ & MISE$_{\trim}$ & Bias$^2_{\trim}$ & MISE$_{\trim}$   
& Bias$^2_{\trim}$ & MISE$_{\trim}$ & Bias$^2_{\trim}$ & MISE$_{\trim}$ 
\\
\hline
$C_0$  & \textsc{ls} & 0.0145 & 0.0256 & 0.0000 & 0.0066
					& 0.0037 & 0.0148 & 0.0142 & 0.0208
					\\  
   
 & \textsc{mm} & 0.0148 & 0.0281 & 0.0000 & 0.0082  
 			   & 0.0038 & 0.0172 & 0.0142 & 0.0224
 			\\ 
\hline
 $C_{1}$  & \textsc{ls} & 0.0136 & 1.6399 & 0.0005 & 1.0112
 						 & 0.0033 & 1.6302 & 0.0149 & 1.0257 
 						\\ 
 & \textsc{mm}  & 0.0148 & 0.0285 & 0.0000 & 0.0083
 			    & 0.0039 & 0.0176 & 0.0142 & 0.0224
 			    \\ 
\hline
 $C_{2}$  & \textsc{ls}  & 0.9683 & 1.0307 & 1.7754 & 1.7874 
 						 & 0.9578 & 1.0201 & 1.7892 & 1.8012
 						 \\ 
 & \textsc{mm} & 0.0161 & 0.0391 & 0.0139 & 0.0923
 			   & 0.0052 & 0.0282 & 0.0281 & 0.1064
 			   \\ 
\hline
$C_{3}$ & \textsc{ls}  & 1.7854 & 2.0006 & 0.5086 & 0.6485
					  & 13.8164 & 15.4307 & 15.5136 & 16.2597
					  \\   
 & \textsc{mm} & 0.0172 & 0.0329 & 0.0010 & 0.0134 
 			   & 0.0040 & 0.0178 & 0.0141 & 0.0223 
 			   \\ 
 \hline
\end{tabular}
\caption{ \small \label{tab:tabla-Upsilon-YM} Trimmed versions of the integrated squared bias  and mean integrated squared errors (multiplied by 10) for clean and contaminated samples, \textbf{Model 2}. The reported values under contamination correspond to the worst situation.}
\end{center}
\end{table}

\begin{figure}[ht!]
 \begin{center}
 \newcolumntype{M}{>{\centering\arraybackslash}m{\dimexpr.05\linewidth-1\tabcolsep}}
   \newcolumntype{G}{>{\centering\arraybackslash}m{\dimexpr.4\linewidth-1\tabcolsep}}
\begin{tabular}{M GG}
 &  $\wbeta$  & $\wup$  \\[-4ex]
{\small   $C_{1,\mu}$} &  
\includegraphics[scale=0.38]{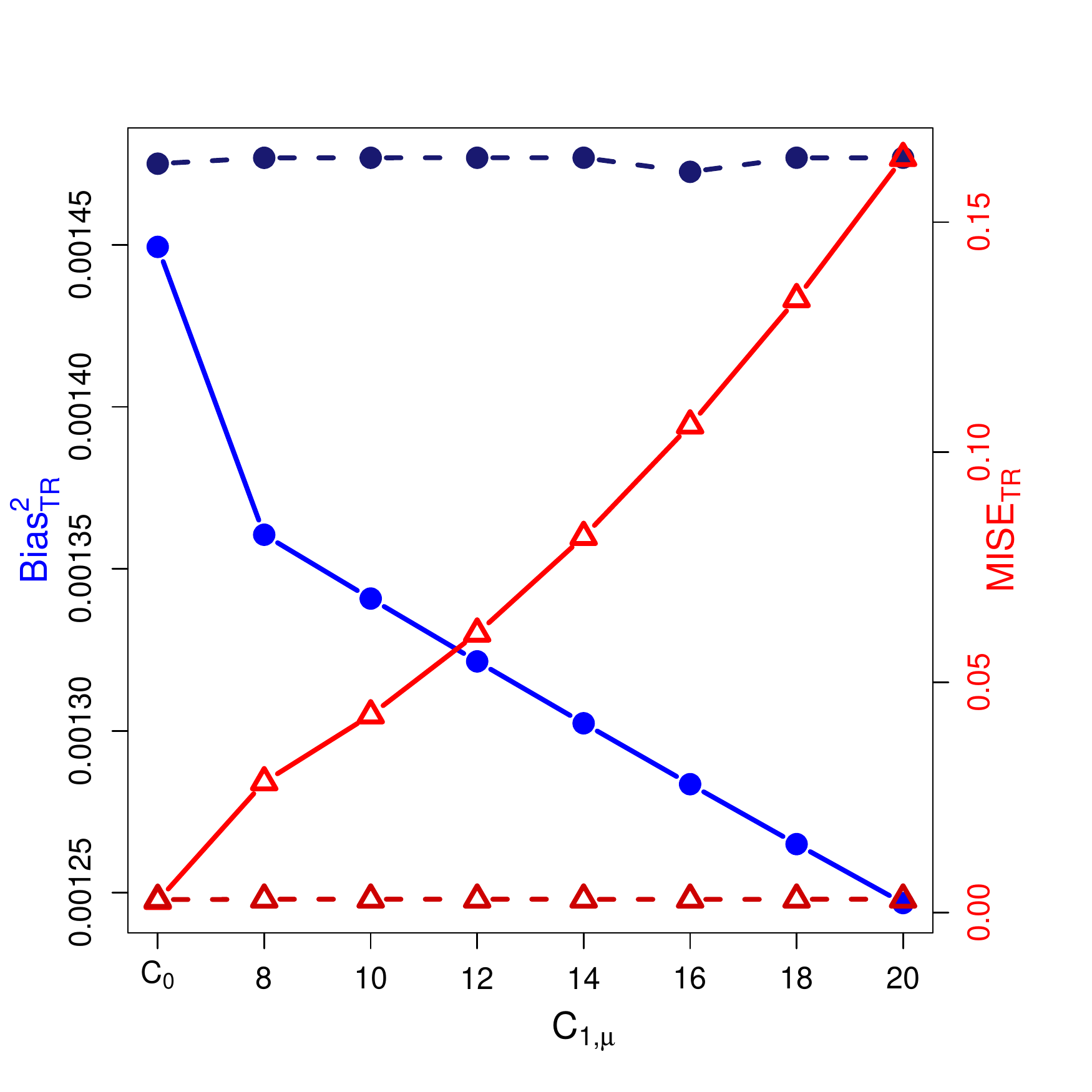} 
&  \includegraphics[scale=0.38]{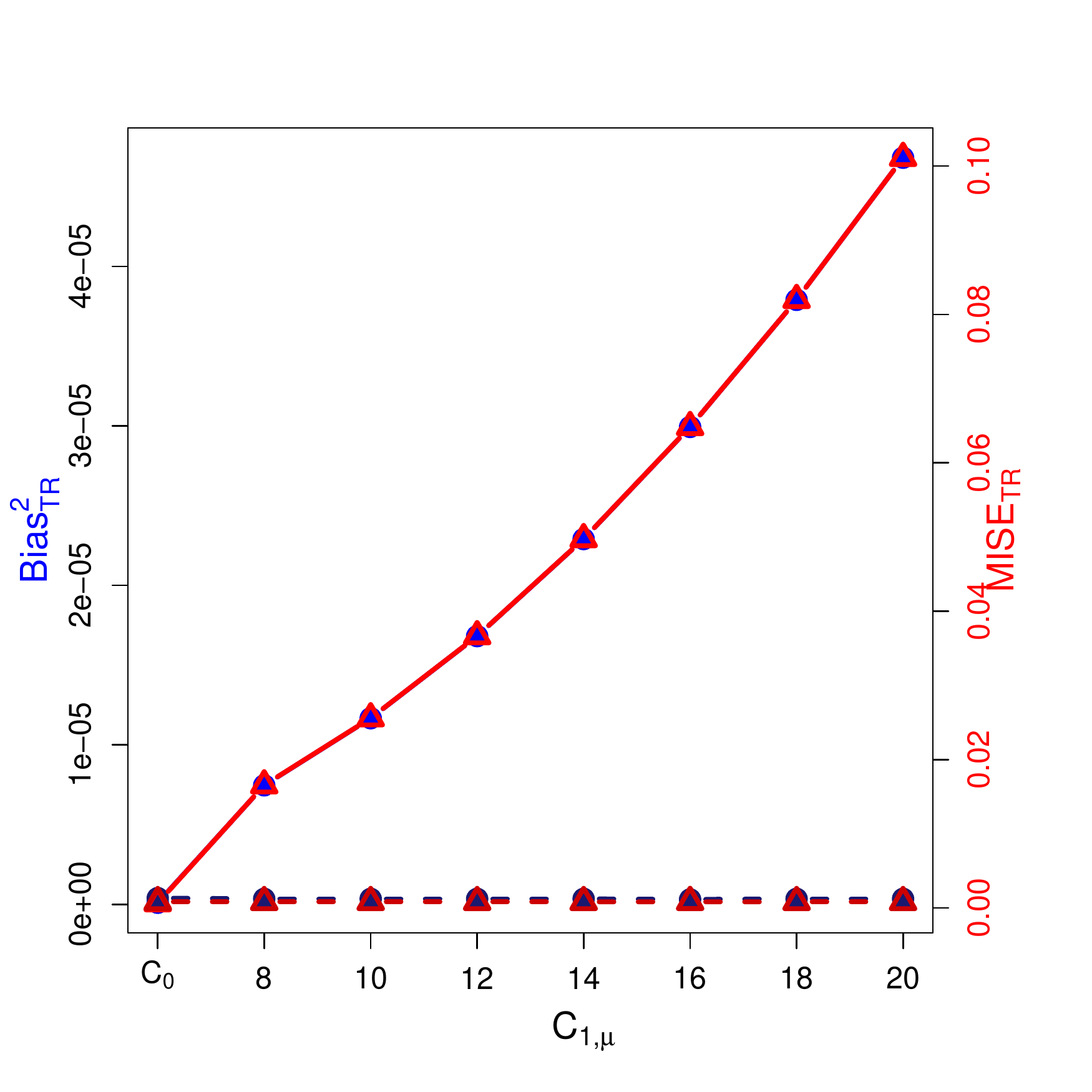}
 \\[-5ex]
 {\small   $C_{2,\mu}$} &  
 \includegraphics[scale=0.38]{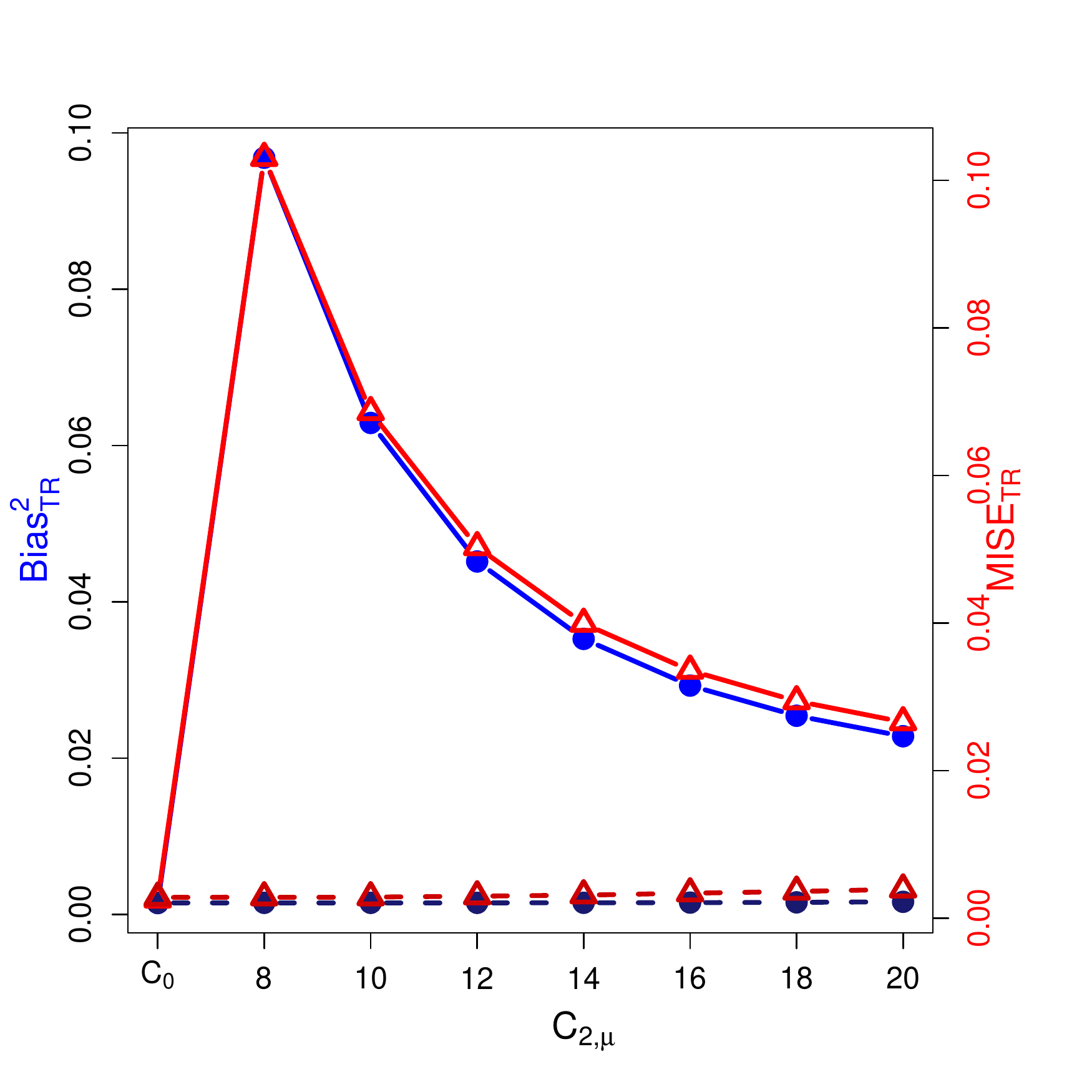} 
&  \includegraphics[scale=0.38]{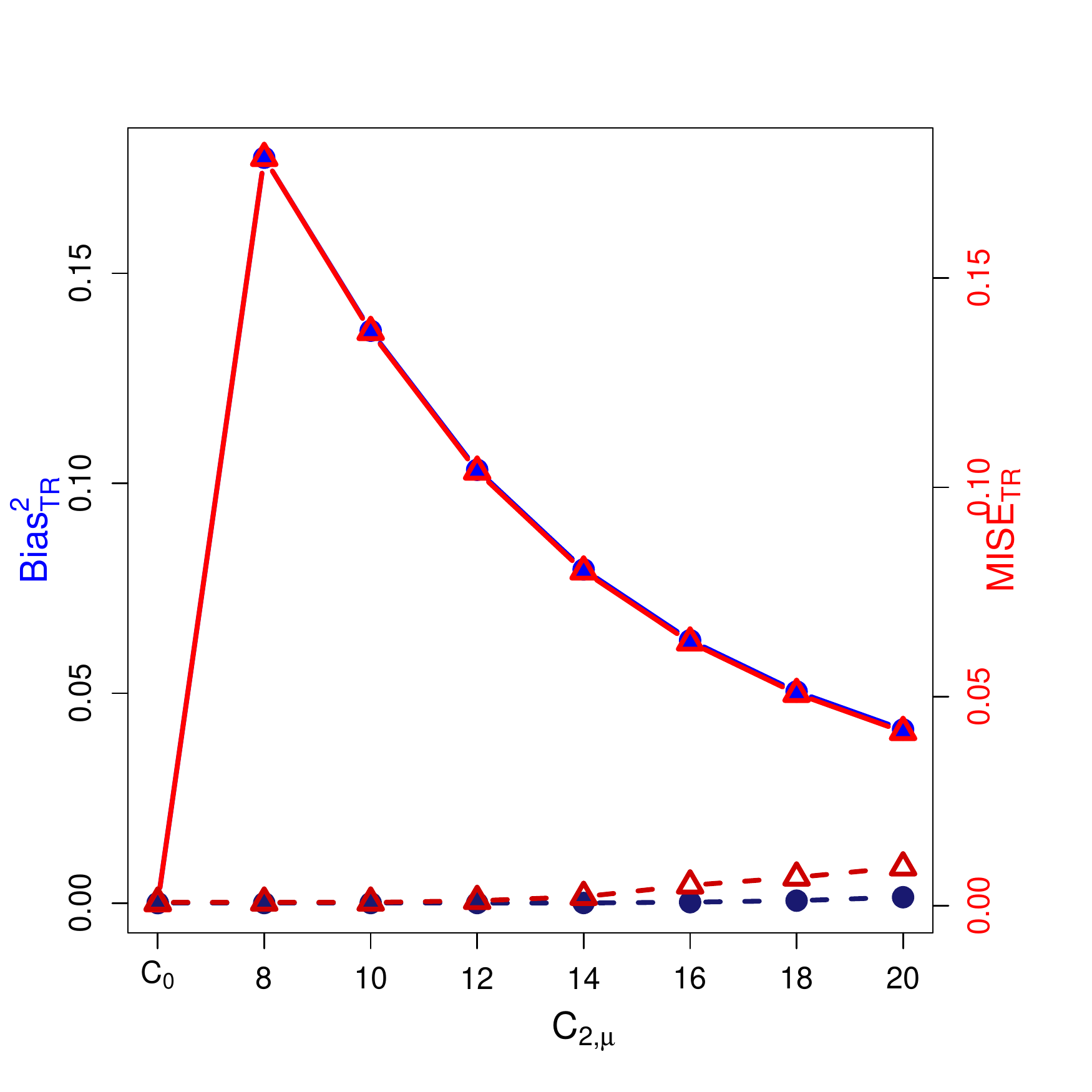}\\[-5ex]
{\small   $C_{3,\mu }$} &  
\includegraphics[scale=0.38]{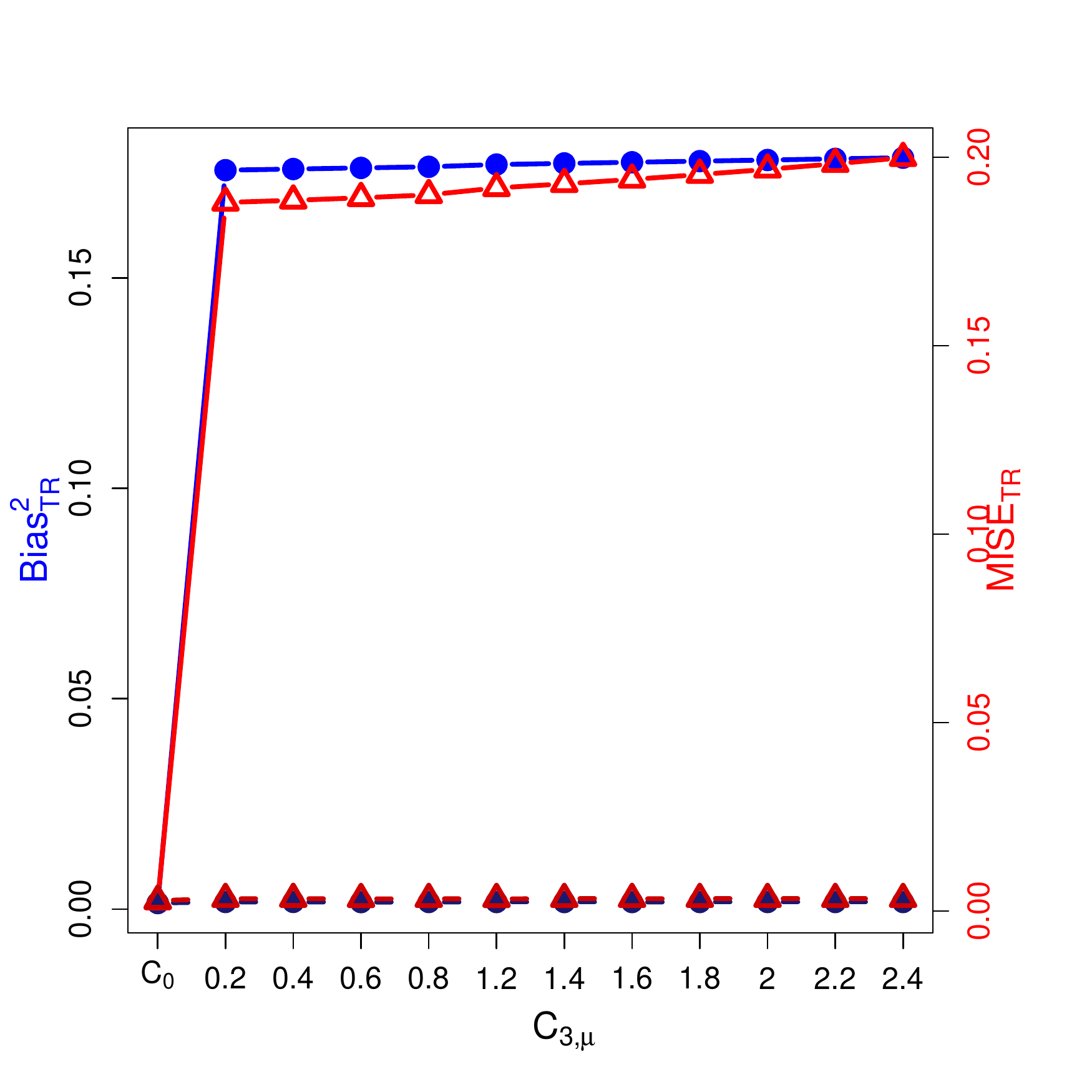} 
&  \includegraphics[scale=0.38]{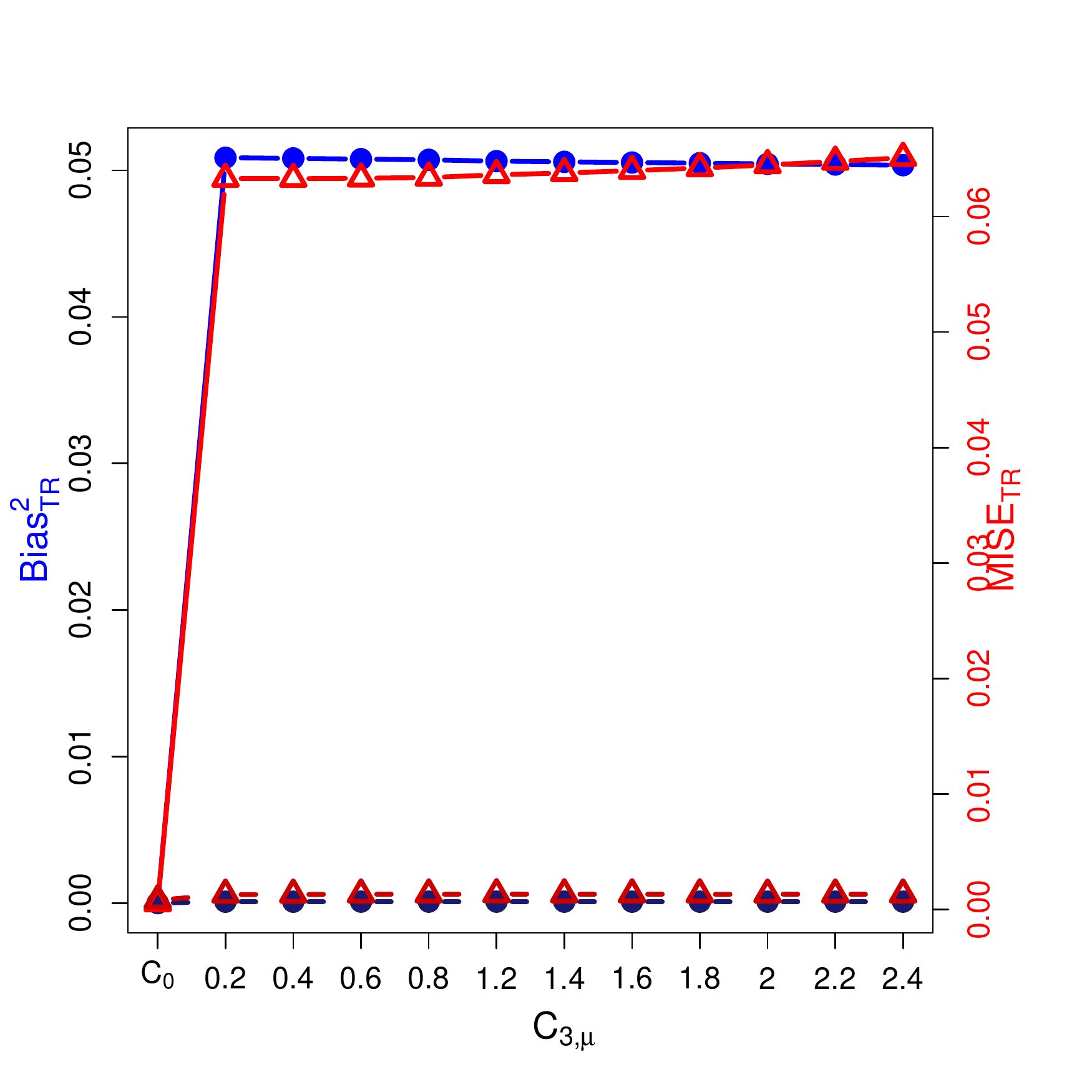}
 \end{tabular}
\caption{\small \label{fig:BIAS-MISE-Upsilon0-YM}  Plots of the trimmed squared bias and MISE of the estimators of $\beta_0$ and $\upsilon_0$ 
as a function of $\mu$ for each contamination scenario, under \textbf{Model 2} with $\Upsilon_0=0$.  The solid and dashed   lines correspond to the least squares and $MM-$estimators, respectively. The squared bias is indicated with circles, and the MISE with triangles. }
\end{center} 
\end{figure}

\begin{figure}[ht!]
 \begin{center}
 \newcolumntype{M}{>{\centering\arraybackslash}m{\dimexpr.05\linewidth-1\tabcolsep}}
   \newcolumntype{G}{>{\centering\arraybackslash}m{\dimexpr.4\linewidth-1\tabcolsep}}
\begin{tabular}{M GG}
 &  $\wbeta$  & $\wup$  \\[-4ex]
{\small   $C_{1,\mu}$} &  
\includegraphics[scale=0.38]{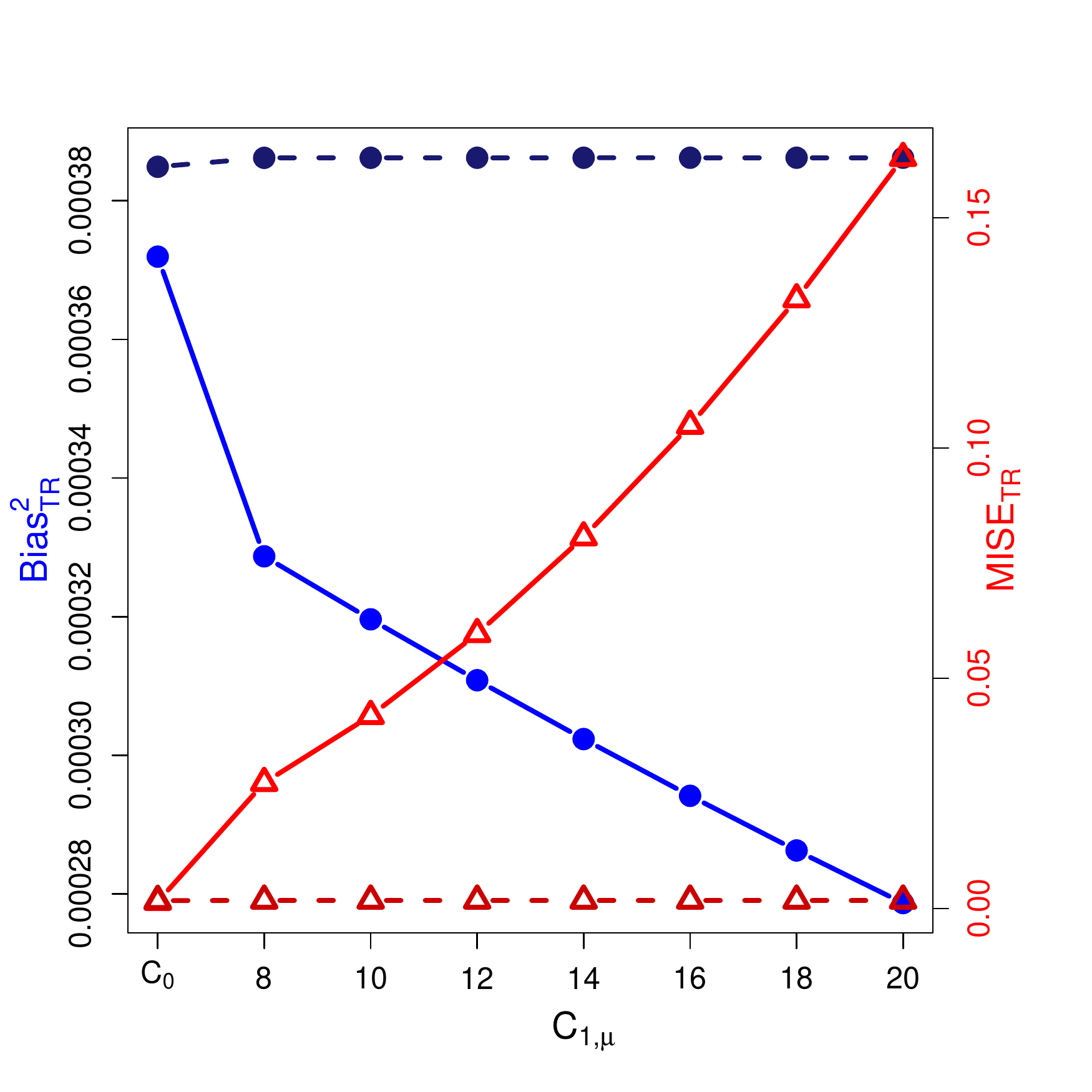} 
&  \includegraphics[scale=0.38]{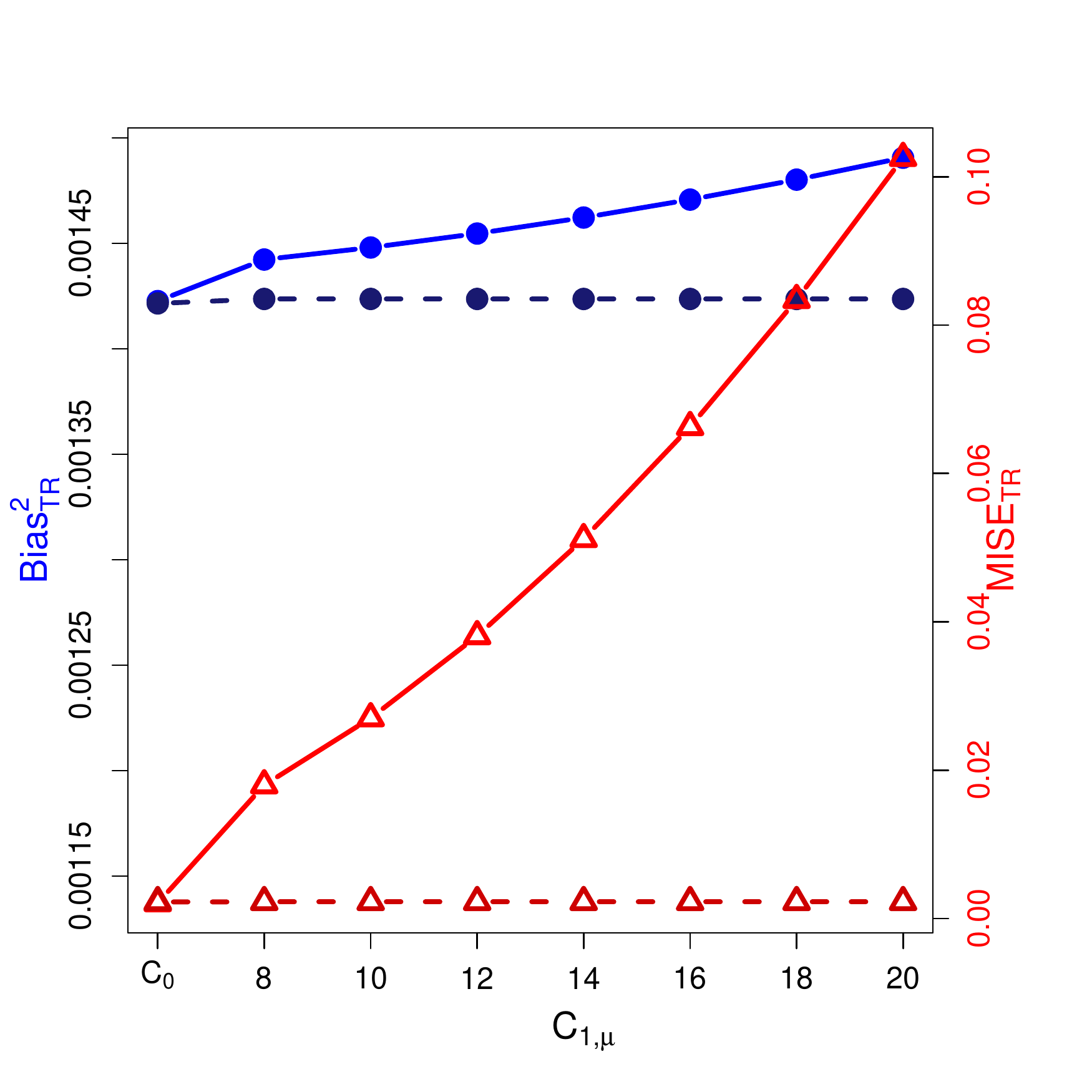}
 \\[-5ex]
 {\small   $C_{2,\mu}$} &  
 \includegraphics[scale=0.38]{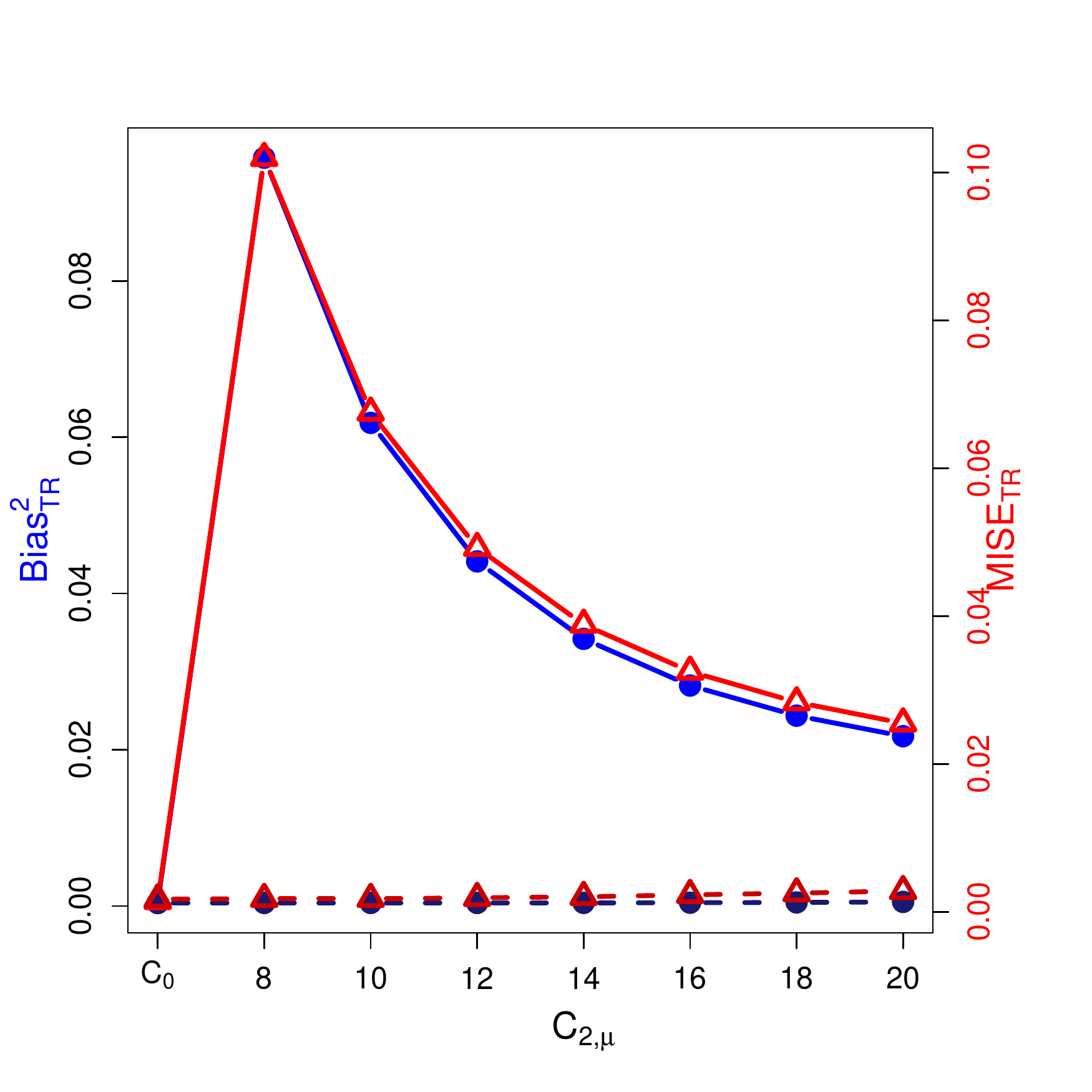} 
&  \includegraphics[scale=0.38]{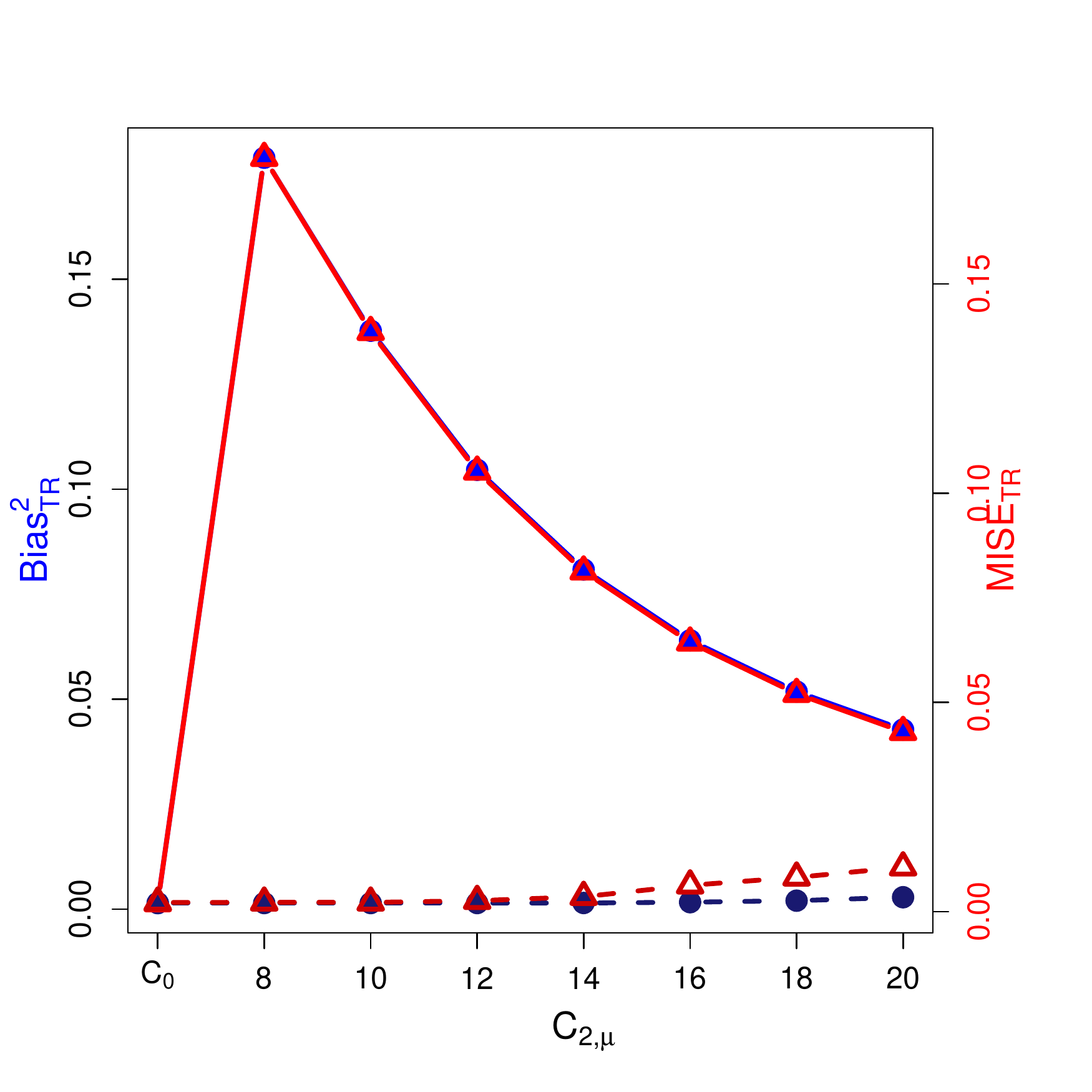}\\[-5ex]
{\small   $C_{3,\mu }$} &  
\includegraphics[scale=0.38]{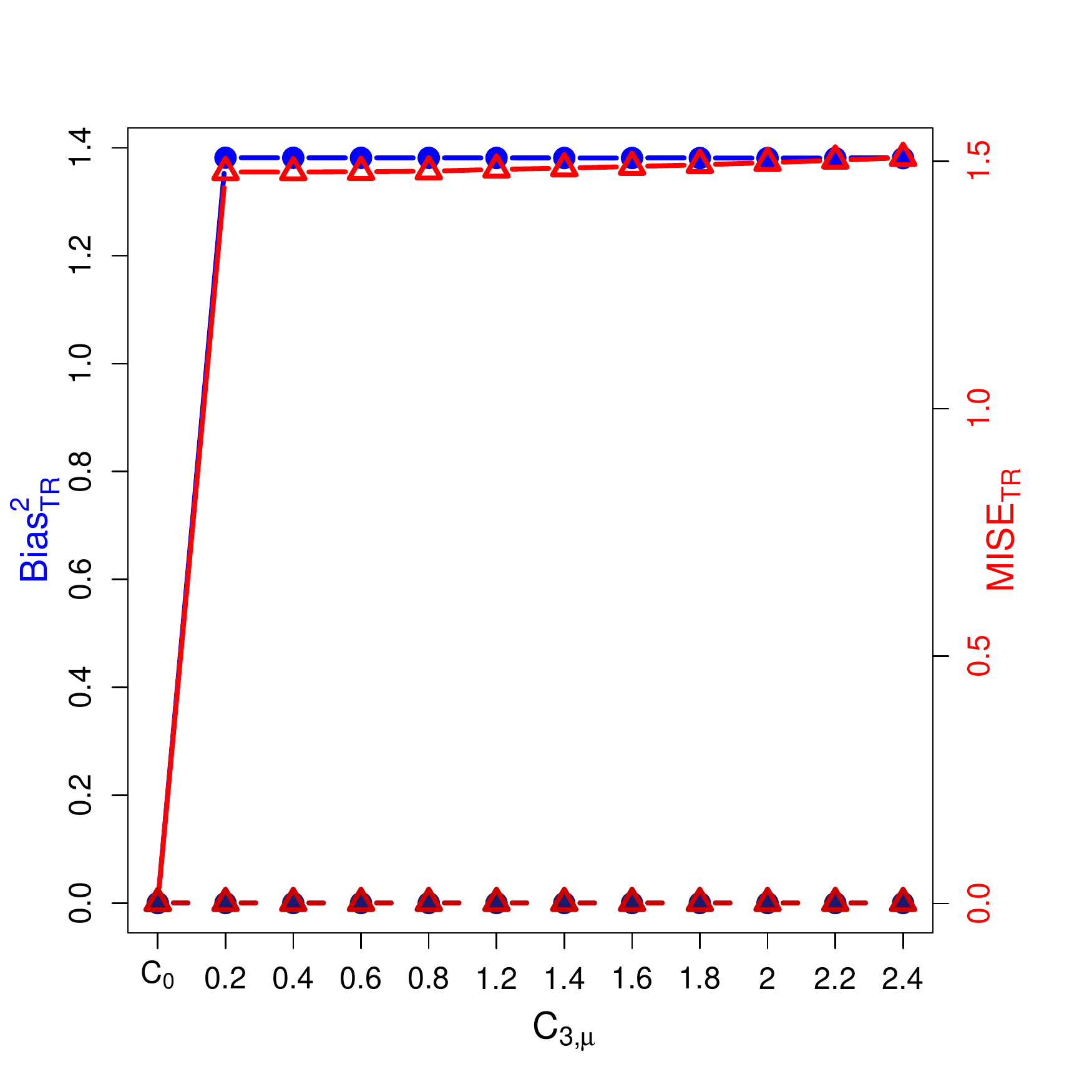} 
&  \includegraphics[scale=0.38]{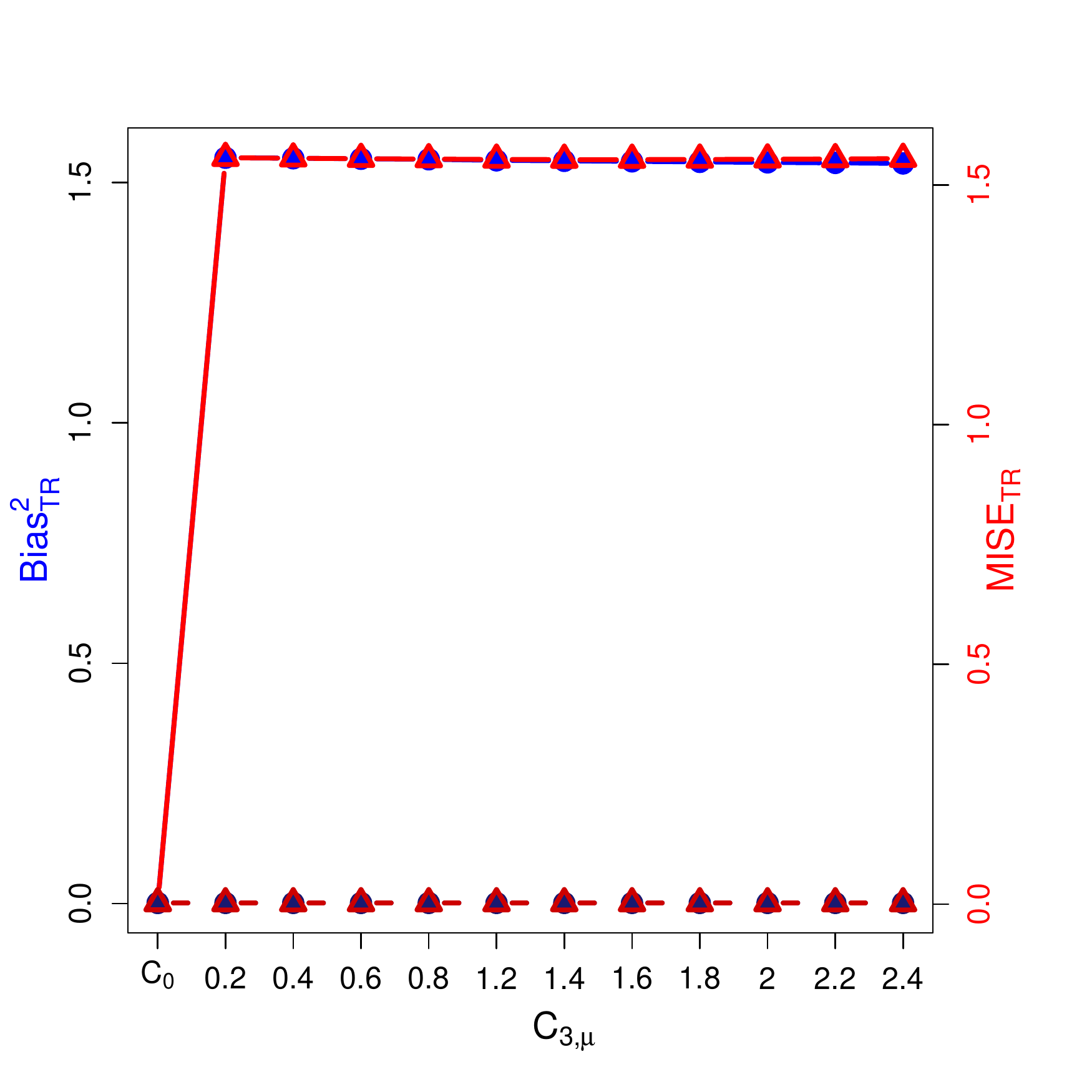}
 \end{tabular}
\caption{\small \label{fig:BIAS-MISE-Upsilon1-YM}  Plots of the trimmed squared bias and MISE of the estimators of $\beta_0$ and $\upsilon_0$ 
as a function of $\mu$ for each contamination scenario, under \textbf{Model 2} with  $\Upsilon_0=\Upsilon_{0,1}$.  The solid and dashed   lines correspond to the least squares and $MM-$estimators, respectively. The squared bias is indicated with circles, and the MISE with triangles. }
\end{center} 
\end{figure}

The plots in Figures \ref{fig:BIAS-MISE-Upsilon0-YM} and Figures \ref{fig:BIAS-MISE-Upsilon1-YM} summarize the effect of the contamination scenarios for the different choices of $\Upsilon_0$ and  different values of $\mu$ when considering $C_{1,\mu}$ and $C_{2,\mu}$. For $C_{3,\mu}$. Each plot corresponds to one contamination scenario  and one parameter estimator. Within each panel, the solid and dashed lines correspond to the measures for the least squares  and $MM-$estimators, respectively. As above, the line with  triangles shows the trimmed MISE, and the one with solid circles indicates the corresponding trimmed bias squared. For this model, under all contamination schemes, the MISE of the classical procedure is much larger than those obtained for the robust procedure, both when estimating the regression parameter or the quadratic operator. Regarding the Bias, except for $C_{1,\mu}$, the bias of the method based on least squares is highly affected by the contaminations considered. In particular, under $C_{3,\mu}$ the bias is increased more than 1000 times with respect to those obtained for clean samples. The robust proposal given in this paper is quite stable across all contaminations.

As when considering \textbf{Model 1}, to visualize  the performance of the  estimators of $\beta_0$, Figures \ref{fig:wbeta-Upsilon0-YM} and \ref{fig:wbeta-Upsilon1-YM} contain functional boxplots  of the estimators, under $C_0$ and some of the three contamination settings. To avoid boundary effects, we show here the different estimates $\wbeta_j$ evaluated on central 90\% interior points of the grid. 
In addition, to facilitate comparisons between contamination cases and estimation methods, the scales of the vertical axes are the same for all panels within each Figure.  
 
The effect of contamination in this model is less striking than under \textbf{Model 1}. However, as in that model, under  $C_{1, 12}$ the classical estimator $\wbeta$ becomes highly variable, but retains the   shape of  $\beta_0$. In contrast, under   $C_{3,1.2}$ when the data are generated according to a functional quadratic model  the estimator becomes completely uninformative due to the distorted region containing the non--outlying curve estimators (see Figure \ref{fig:wbeta-Upsilon1-YM}). It is worth mentioning that  when $\Upsilon_0=0$ and under  the scheme $C_{3,1.2}$, the true curve is beyond the limits of the functional boxplot for values larger than 0.6, while under $C_{2,8}$ it lies near the boundary that limits the non-outlying curves but outside this region. It is worth mentioning that, for clean samples, the obtained estimators of $\beta_0$ have smaller variability than under \textbf{Model 1}, which is reflected on narrower bands. This fact may be explained by the ratio noise signal, which is much smaller under the model considered in \citet{yao2010functional} than under \textbf{Model 1}.
 
As under \textbf{Model 1}, the classical estimators of $\alpha_0$ are sensitive to the contaminations considered. Table \ref{tab:res-alphas-M2} reports, multiplied by 1000,  the absolute  bias, which in this case equals the absolute value of the mean,  and the standard deviations over replications. For the contamination settings, the maximum over the different values of $\mu$ is reported. Figure \ref{fig:alphas-boxplots-M2}  presents the boxplots of the estimators for $\alpha_0$ for clean  samples and for some contamination scenarios. The true value is plotted with a green dashed line, for reference. The first row  corresponds to the case where the observations are generated according to a  linear model ($\Upsilon_0=0$), while the second one to the quadratic model considered ($\Upsilon_0=\Upsilon_{0,1}$). Each column corresponds to a contamination setting. The boxplots of the classical and robust estimators are given in magenta and  blue, respectively. The reported results show that schemes $C_{1,\mu}$ and $C_{3,\mu}$ affect  the classical estimator of the intercept for any choice of the quadratic operator with maximum biases increased more than ten thousand times and standard deviations enlarged more than one thousand times. In contrast,  a smaller effect is observed under $C_{2,\mu}$ (see Figure \ref{fig:alphas-boxplots-M2}). The robust procedure is stable over the contaminations considered, even though some effect in the bias is observed again under $C_2$ (see Table  \ref{tab:res-alphas-M2}).

\begin{table}[ht!]
	\centering
	\footnotesize
	\renewcommand{\arraystretch}{1.4}   
	\setlength{\tabcolsep}{3pt}
	\begin{tabular}{r  c  c:  c | c:  c    }
		\hline    
		&& \multicolumn{2}{c|}{$\Upsilon_{0,0}$}& \multicolumn{2}{c}{$\Upsilon_{0,1}$} \\
		\hline

		& & $|$Mean$|$  & SD     
		& $|$Mean$|$  & SD   \\
		\hline
		$C_0$  & \textsc{ls} & 0.20 &  40.86   
		& 0.20 &   40.86  
		\\  
		
		& \textsc{mm} & 0.34 &   44.36 
		& 0.34 & 44.36  
		\\ 
		\hline
		$C_{1}$  & \textsc{ls} & 1989.93  & 496.22  
		& 1989.93  & 496.22  
		\\ 
		& \textsc{mm} & 0.65   & 44.85  
		& 0.13   & 44.81 
		\\ 
		\hline
		$C_{2}$  & \textsc{ls} & 286.71   & 67.61  
		& 286.71  & 67.61 
		\\ 
		& \textsc{mm} & 28.63   & 78.87  
		& 28.63   & 78.87  
		\\ 
		\hline
		$C_{3}$ & \textsc{ls} & 340.05  & 200.35 
		& 2281.67  & 528.91  
		\\  
		& \textsc{mm} & 4.86   & 47.67  
		& 5.70   & 45.32 
		\\ 
		\hline
	\end{tabular}
	\caption{ \small \label{tab:res-alphas-M2} Summary measures (multiplied by 1000) for $\alpha_0$ estimates over  clean and contaminated samples, \textbf{Model 1}. The reported values under contamination correspond to the worst situation.}
\end{table}

\begin{figure}[ht!]
		\centering
		\footnotesize
		\renewcommand{\arraystretch}{1.3}   
		\setlength{\tabcolsep}{2pt}
		\newcolumntype{M}{>{\centering\arraybackslash}m{\dimexpr.05\linewidth-1\tabcolsep}}
		\newcolumntype{G}{>{\centering\arraybackslash}m{\dimexpr.16\linewidth}}
		\begin{tabular}{M GGGG} 
			& {\hspace{2cm} \small   $C_0$} &{\hspace{1.5cm} \small   $C_{1,12}$} &{\hspace{0.8cm} \small $C_{2,8}$} &{\hspace{0.1cm} \small  $C_{3,1.2}$  } \\[-5ex]
			% La chanchada de los hspace es porque NO logro centrarlos...
			% Probé con el multicolumn 1 c pero pero queda peor...
			%& \multicolumn{1}{c} {\small   $C_0$} & \multicolumn{1}{c} {\small   $C_{1,12}$} & \multicolumn{1}{c} {\small $C_{2,12}$} & \multicolumn{1}{c}{\small $C_{3,4,0.4}$ } \\[-5ex]
				{\small   $\Upsilon_{0,0}$} & \multicolumn{4}{G}
		{\includegraphics[scale=0.57]{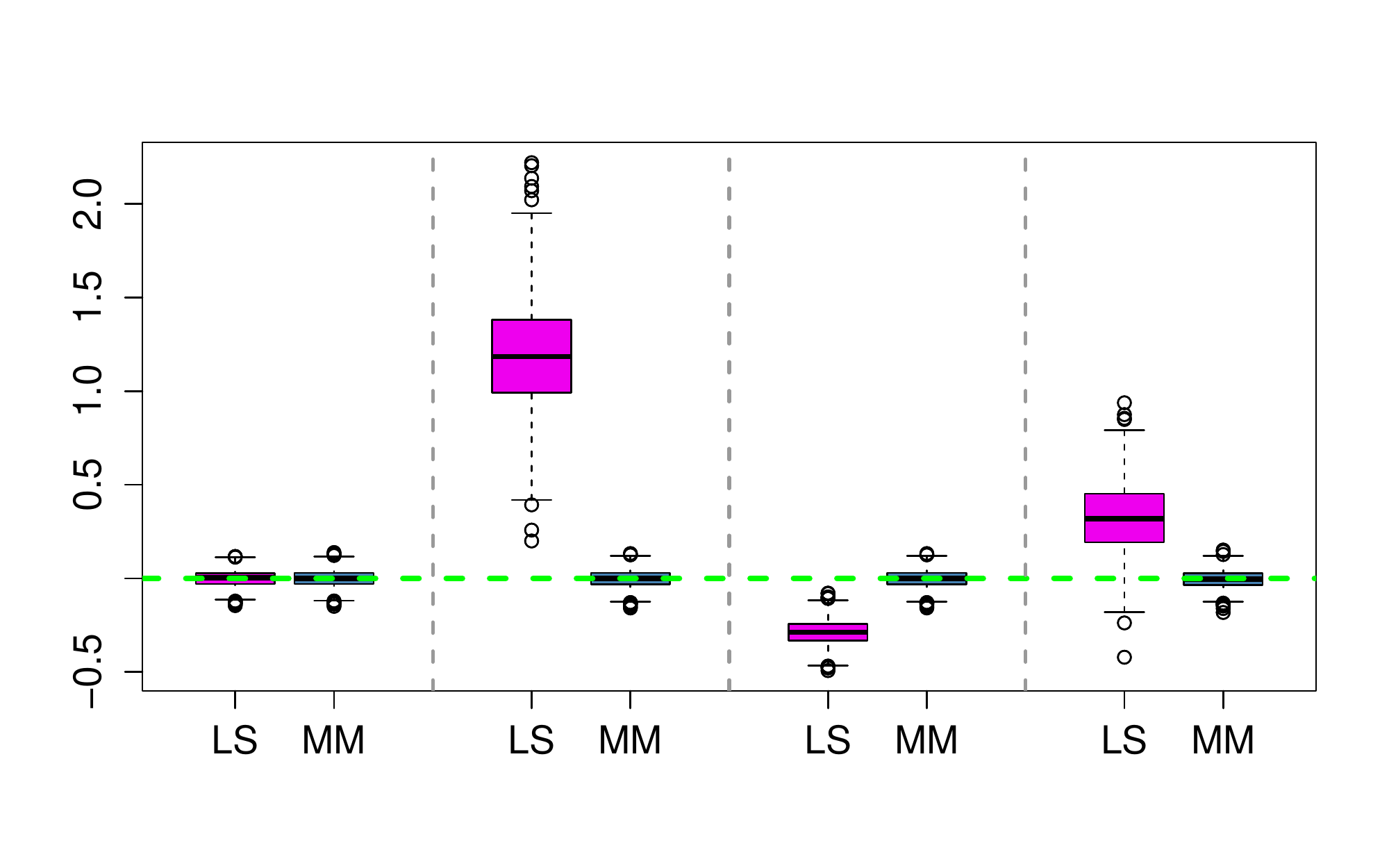}}\\[-8ex]
		{\small   $\Upsilon_{0,1}$} & \multicolumn{4}{G}
		{\includegraphics[scale=0.57]{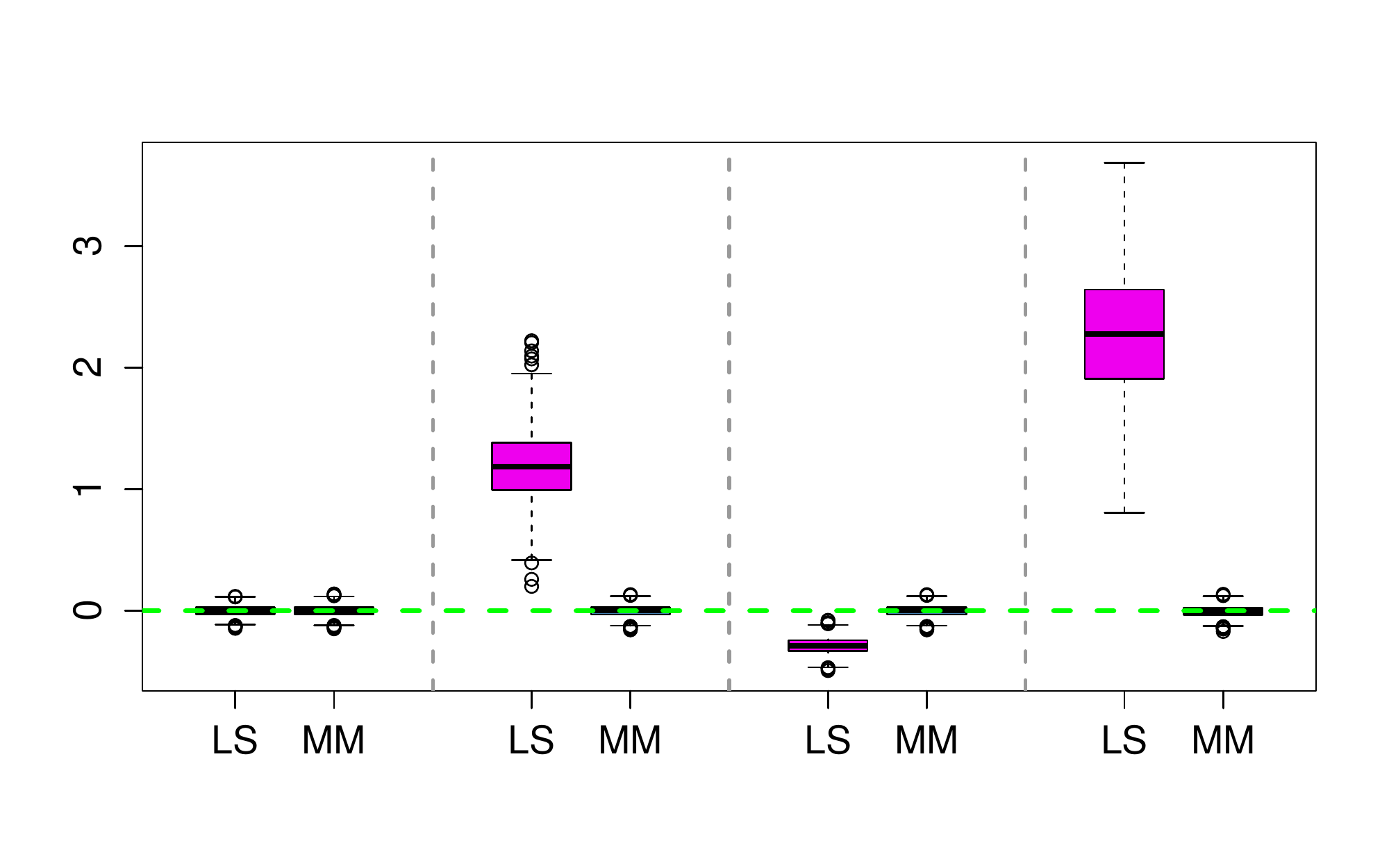}}\\[-4ex]
	\end{tabular}
		\caption{\small \label{fig:alphas-boxplots-M2}  Boxplots of the estimators for $\alpha_0$ for clean and contaminated samples, under \textbf{Model 2}. The true value is shown with a green dashed line. Rows correspond to different $\Upsilon$ scenarios. Columns correspond to $C_0$ and to some of the three contamination  settings. Magenta and blue boxplots correspond to classical and robust methods, respectively.}
\end{figure}
 
\begin{figure}[ht!]
 \begin{center}
 \footnotesize
 \renewcommand{\arraystretch}{0.2}
 \newcolumntype{M}{>{\centering\arraybackslash}m{\dimexpr.05\linewidth-1\tabcolsep}}
   \newcolumntype{G}{>{\centering\arraybackslash}m{\dimexpr.33\linewidth-1\tabcolsep}}
%\begin{tabular}{MGG}
\begin{tabular}{M GG}
 & $\wbeta_{\ls}$ &   $\wbeta_{\eme\eme}$ \\[-7ex]
$C_{0}$ 
&  \includegraphics[scale=0.35]{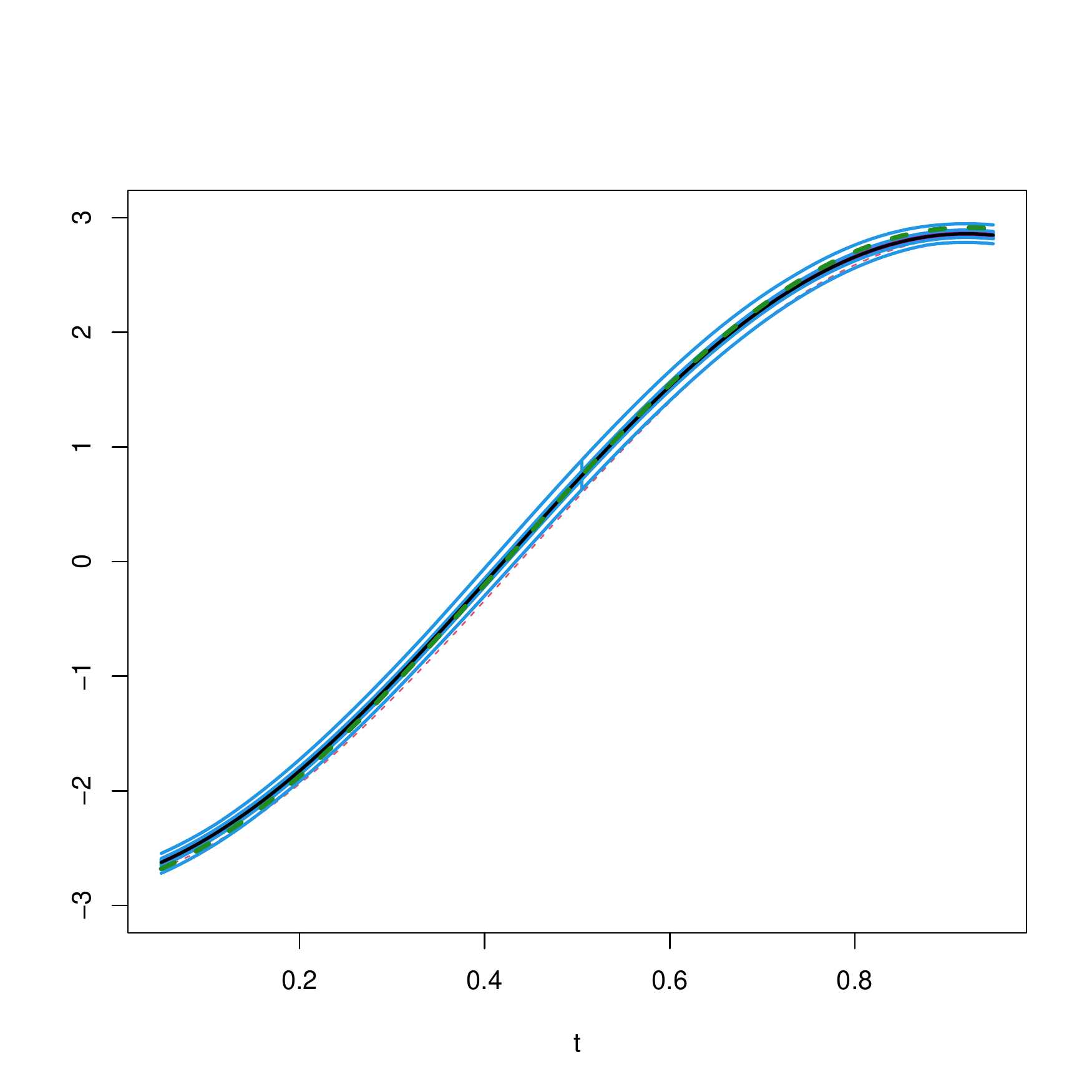} 
&  \includegraphics[scale=0.35]{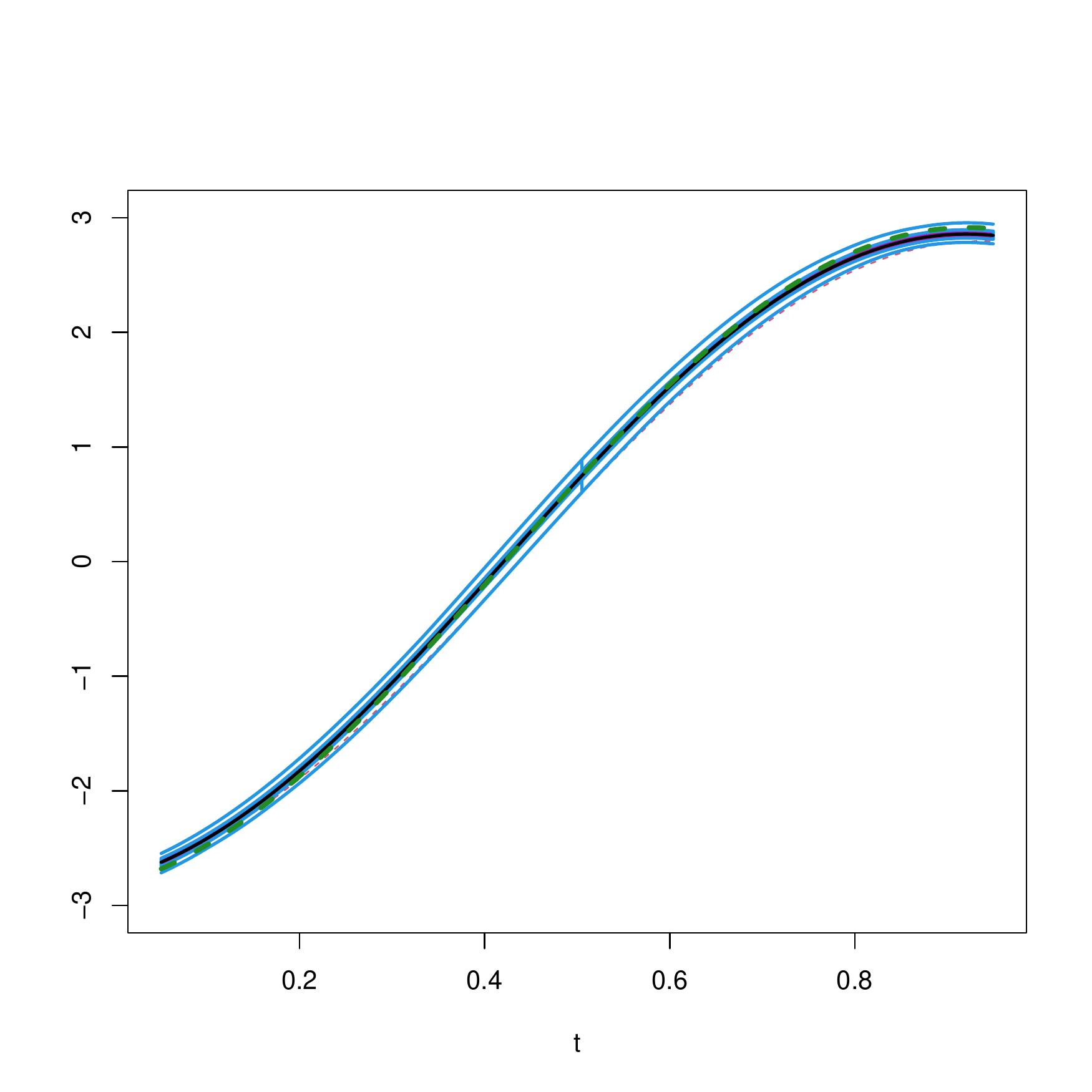} 
\\[-6ex]
$C_{1, 12}$ &  
 \includegraphics[scale=0.35]{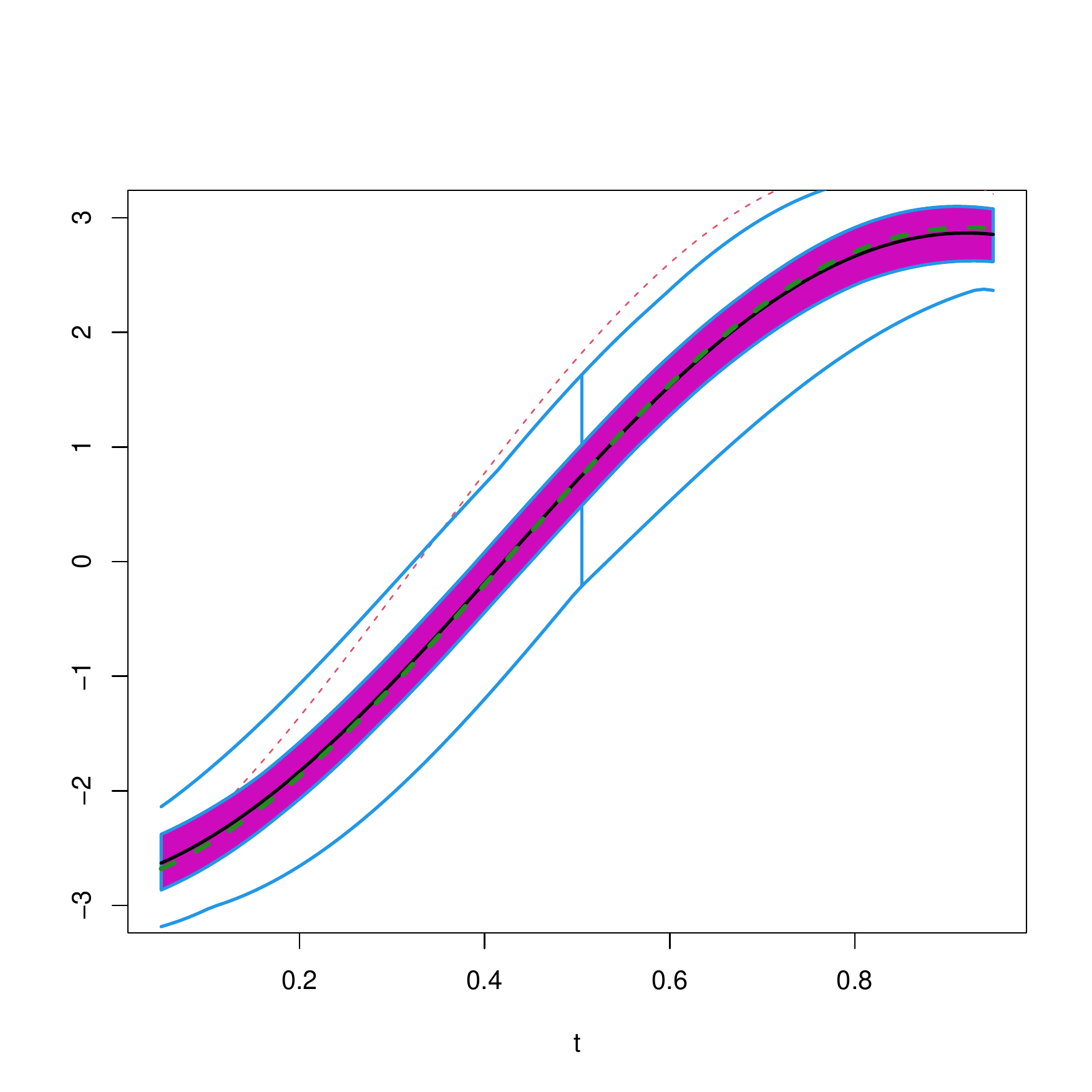}  & 
 \includegraphics[scale=0.35]{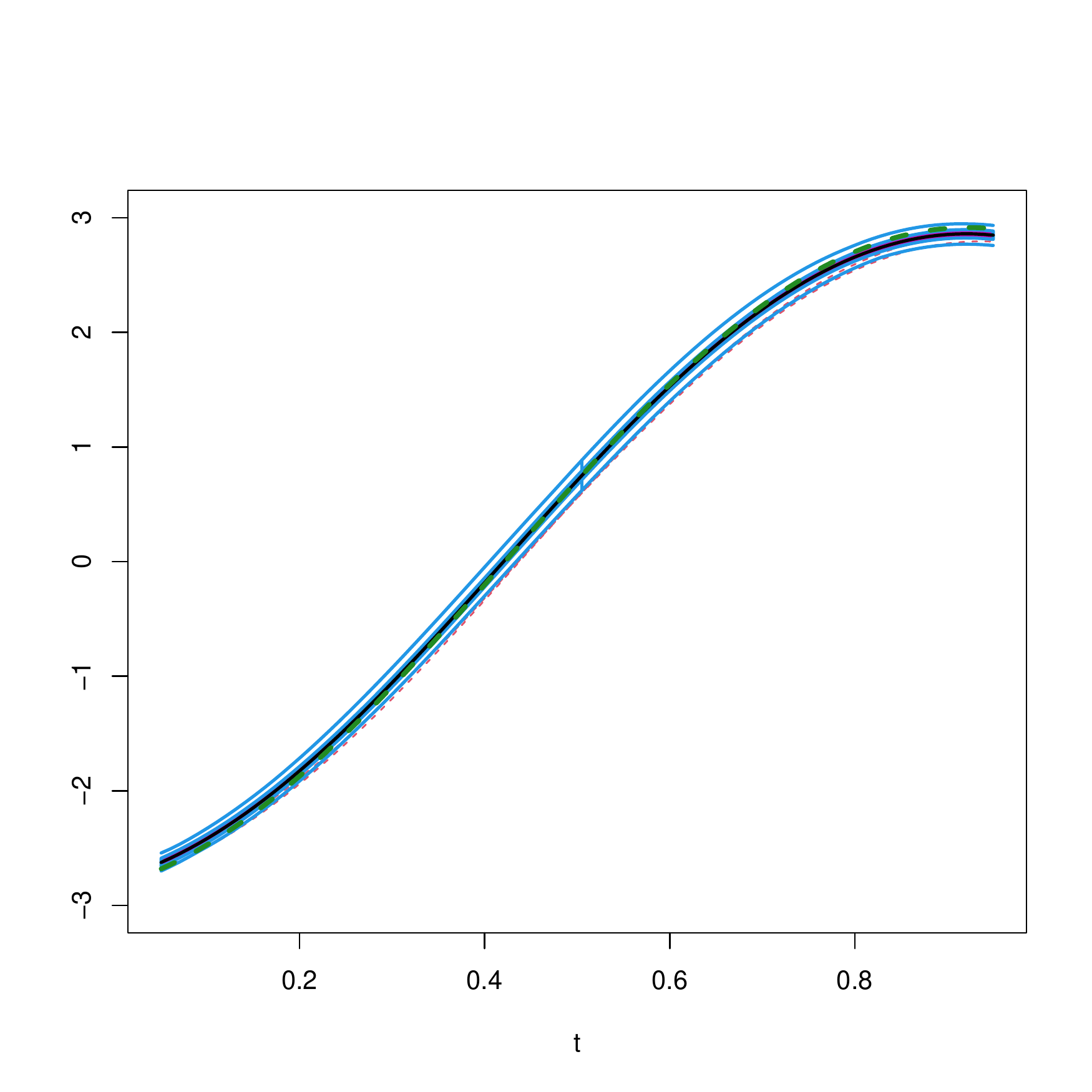} 
  \\[-6ex]
   
$C_{2, 8}$ &
 \includegraphics[scale=0.35]{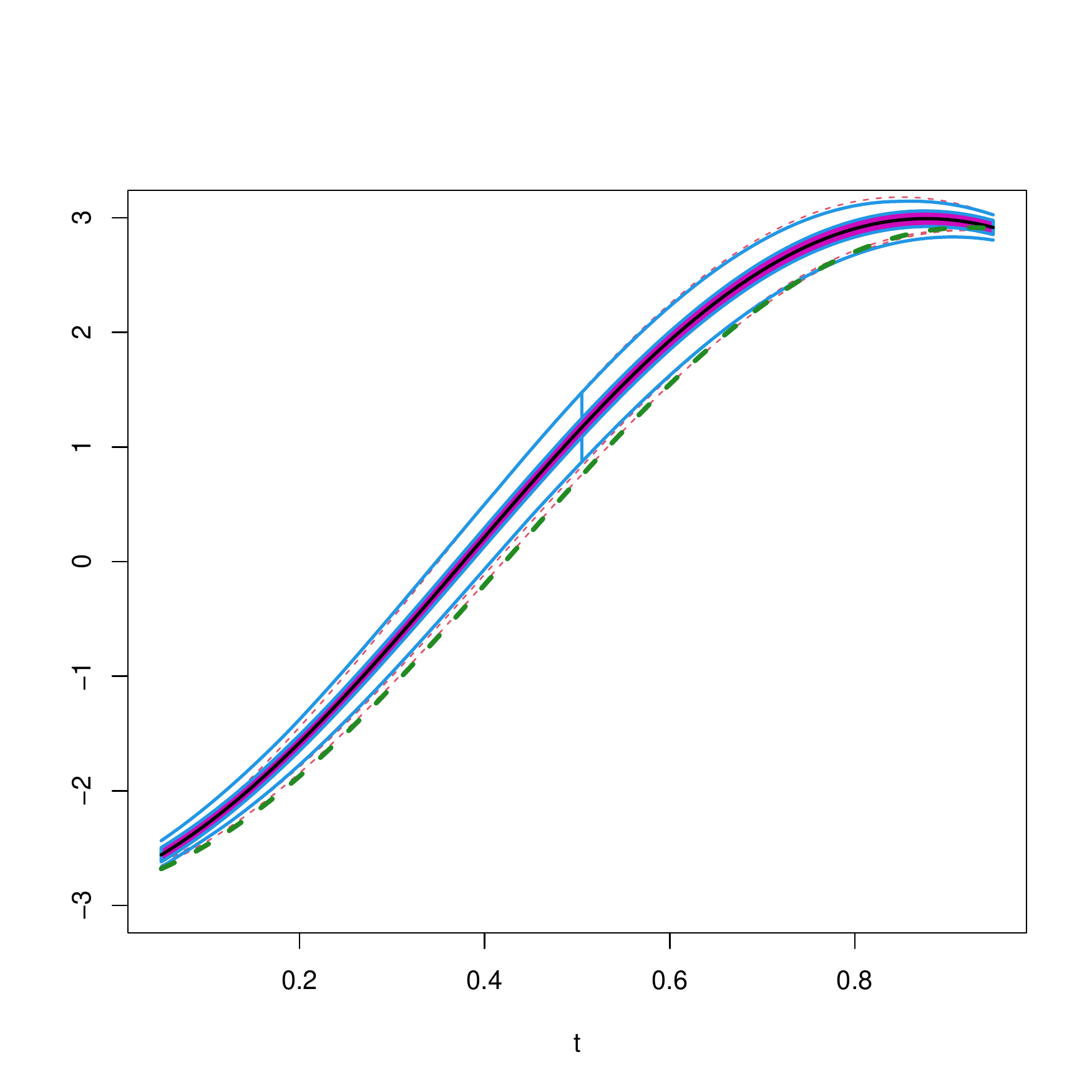}  & 
 \includegraphics[scale=0.35]{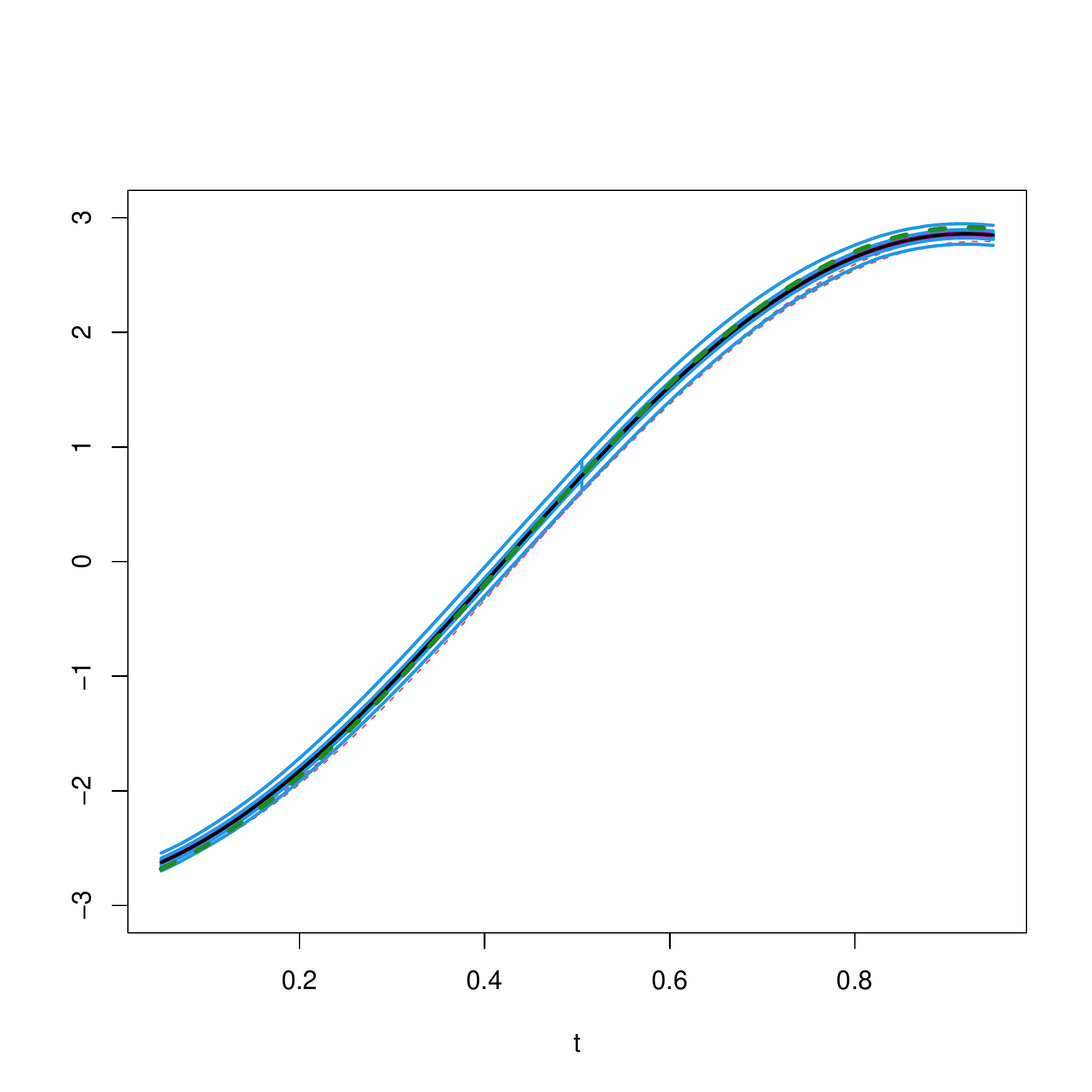} 
 \\[-6ex]
    
$C_{3, 1.2}$ &
\includegraphics[scale=0.35]{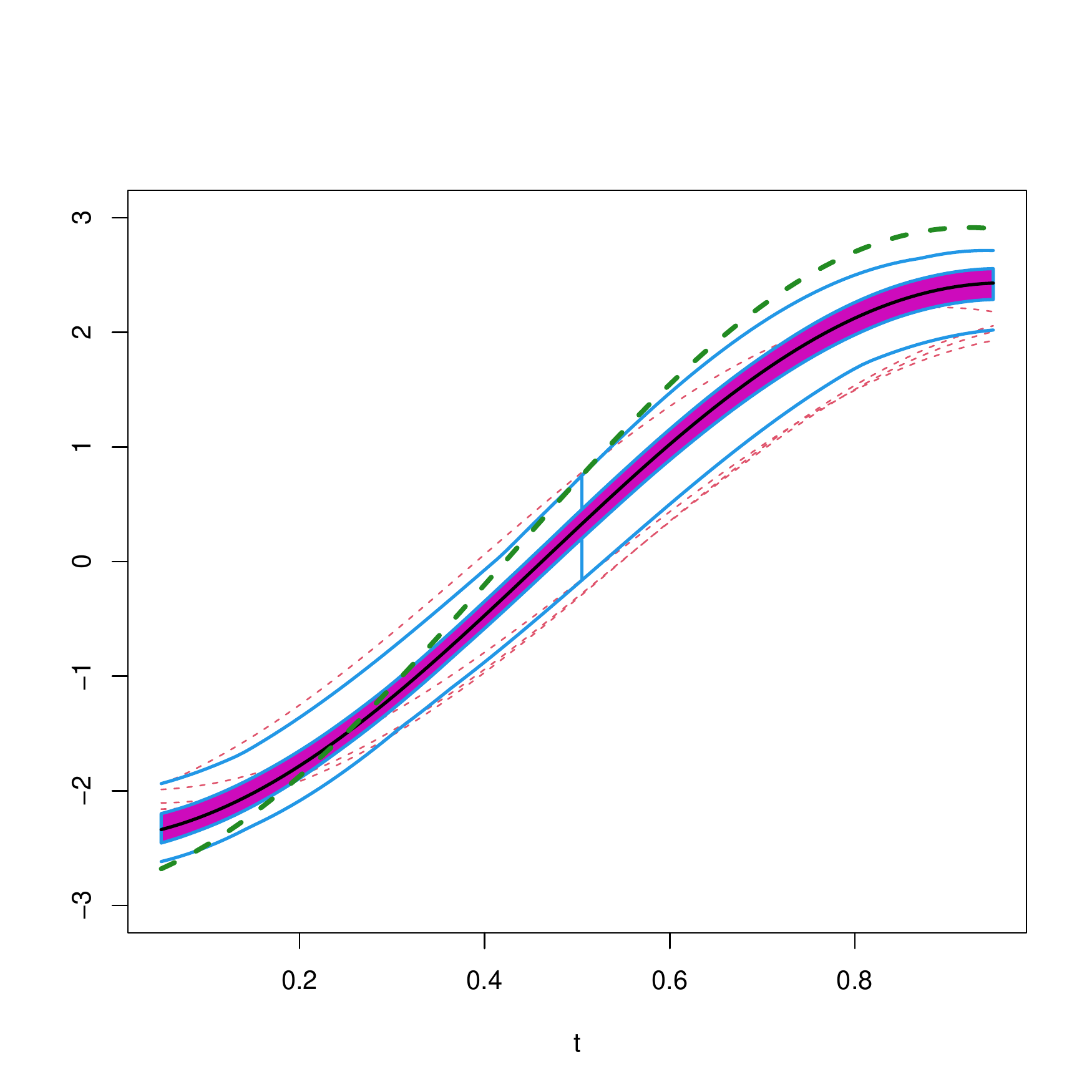}  & 
 \includegraphics[scale=0.35]{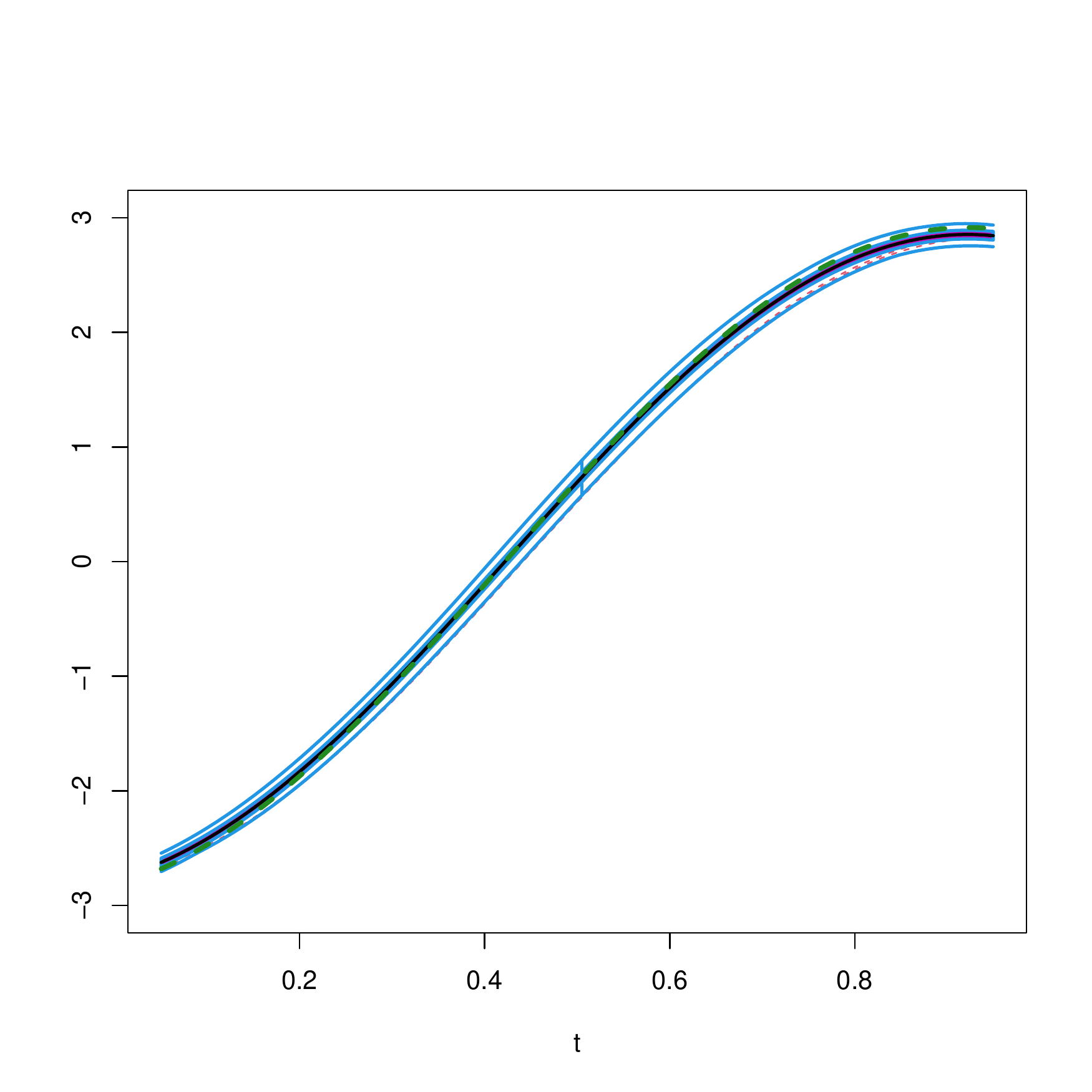}
\end{tabular}
\caption{\small \label{fig:wbeta-Upsilon0-YM}  Functional boxplot of the estimators for $\beta_0$ under \textbf{Model 2} with  $\Upsilon_0=0$. 
The true function is shown with a green dashed line, while the black solid one is the central 
curve of the $n_R = 1000$ estimates $\wbeta$. Columns correspond to estimation 
methods, while rows to $C_0$ and   to some of the three contamination settings.}
\end{center} 
\end{figure}

\begin{figure}[ht!]
 \begin{center}
 \footnotesize
 \renewcommand{\arraystretch}{0.2}
 \newcolumntype{M}{>{\centering\arraybackslash}m{\dimexpr.05\linewidth-1\tabcolsep}}
   \newcolumntype{G}{>{\centering\arraybackslash}m{\dimexpr.33\linewidth-1\tabcolsep}}
%\begin{tabular}{MGG}
\begin{tabular}{M GG}
 & $\wbeta_{\ls}$ &   $\wbeta_{\eme\eme}$ \\[-7ex]
$C_{0}$  
&  \includegraphics[scale=0.35]{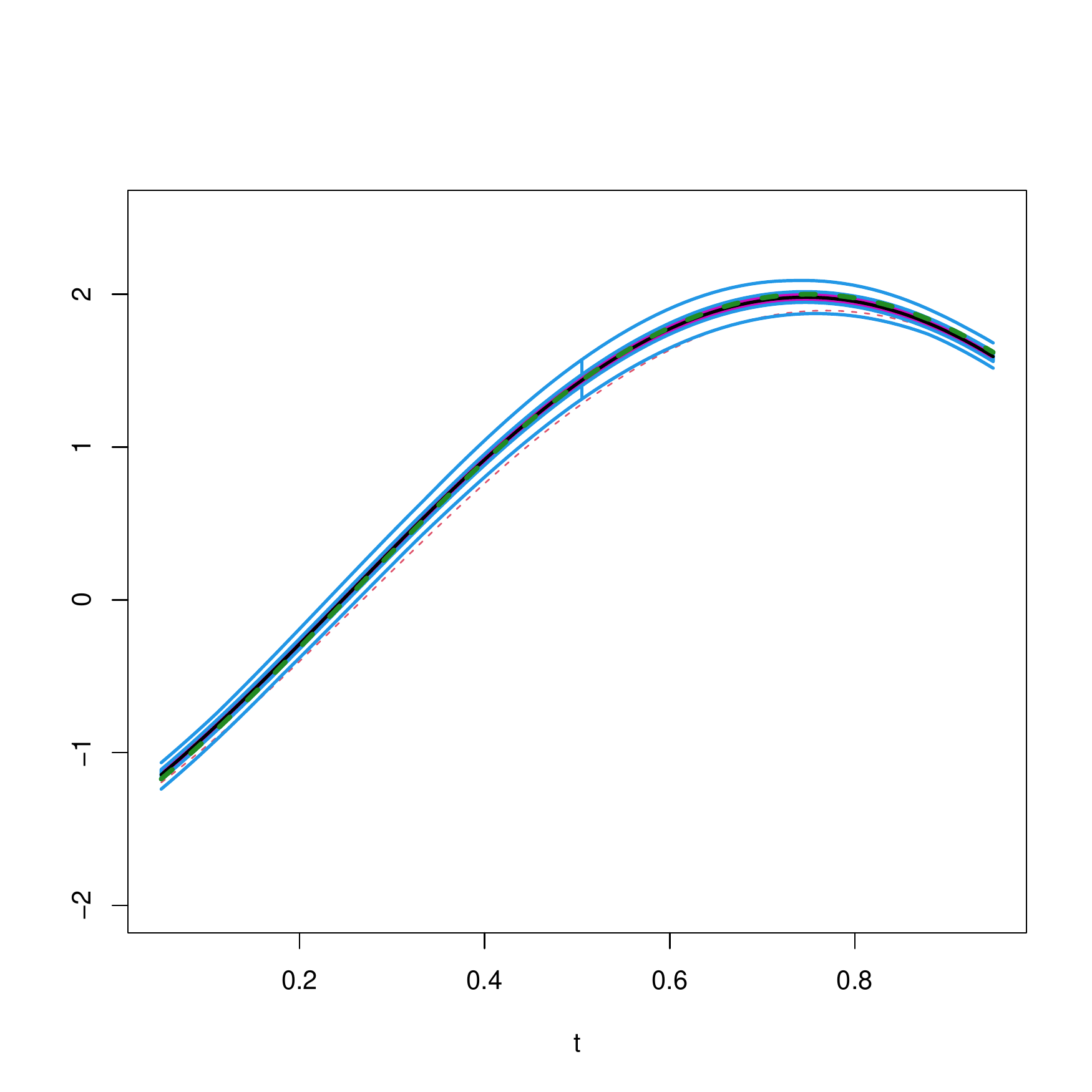} 
&  \includegraphics[scale=0.35]{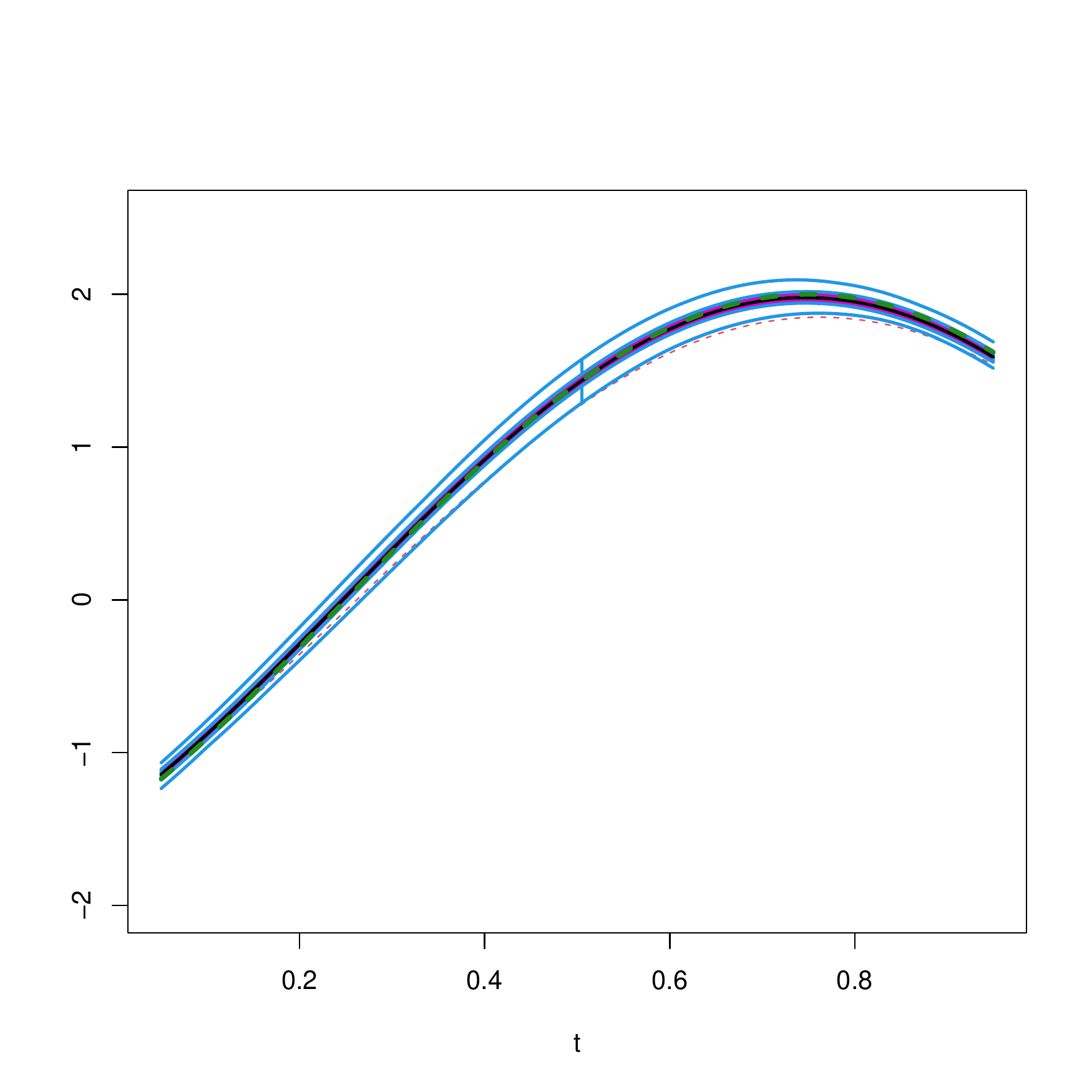} 
\\[-6ex]
$C_{1, 12}$ &  
 \includegraphics[scale=0.35]{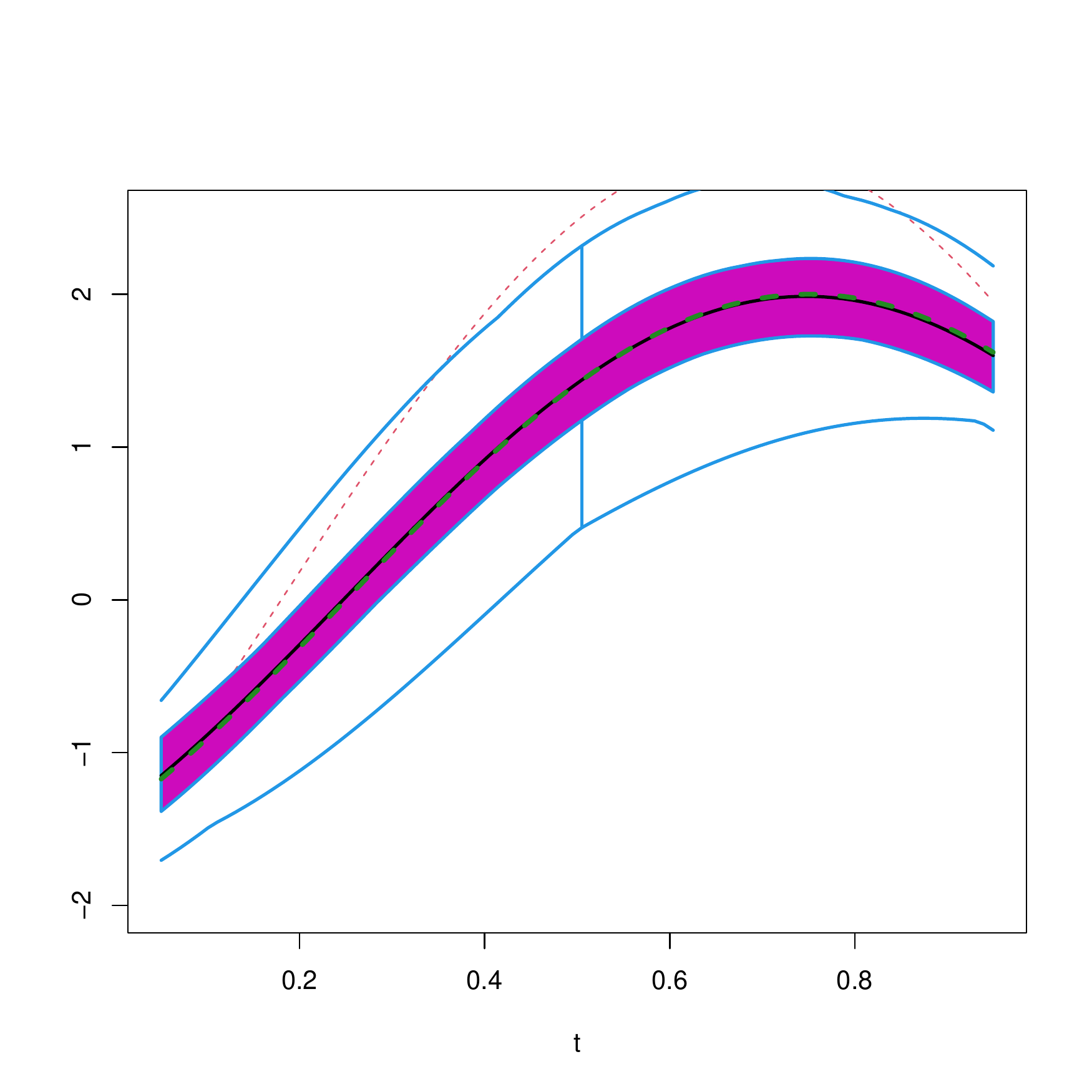}  & 
 \includegraphics[scale=0.35]{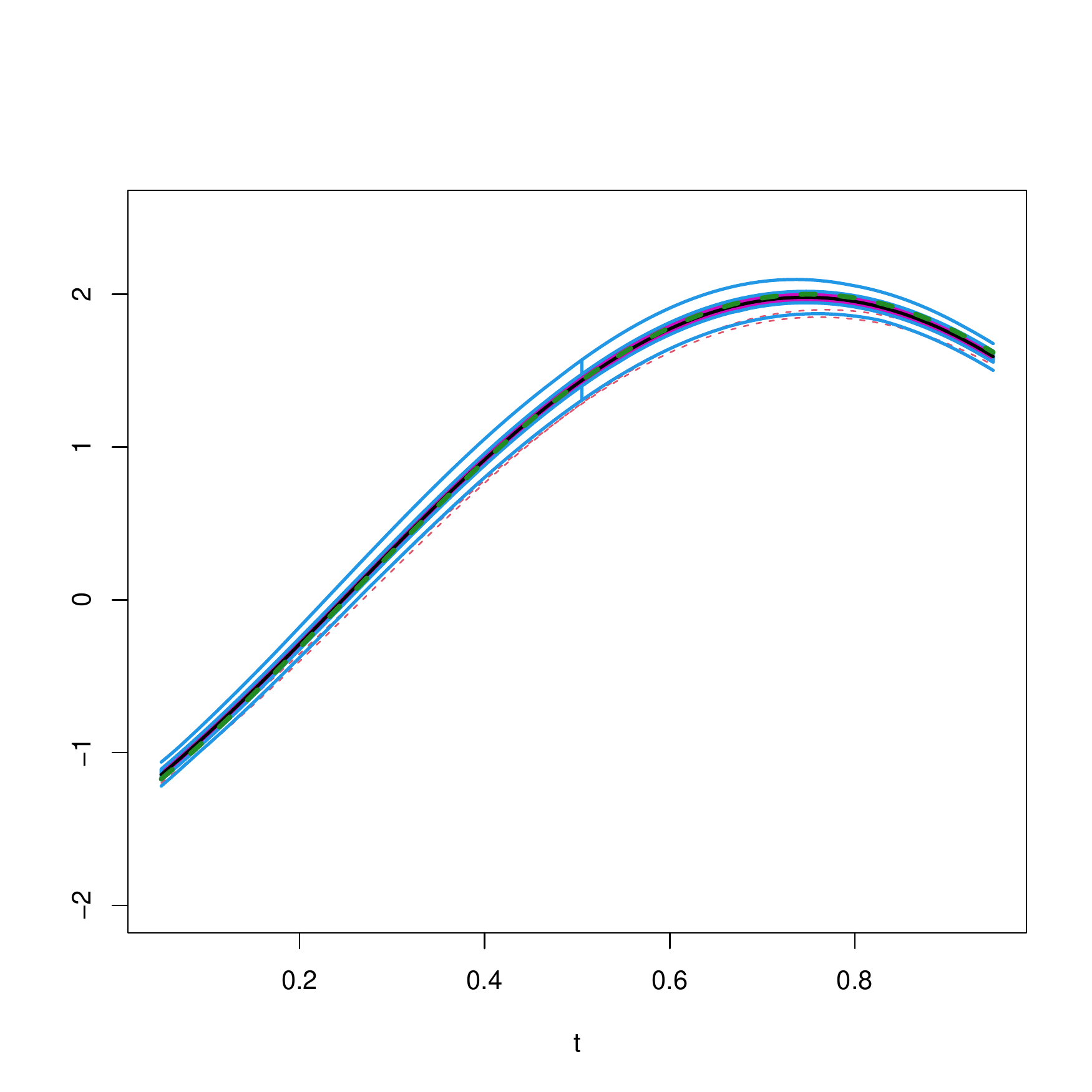} 
  \\[-6ex]
   
$C_{2, 8}$ &
 \includegraphics[scale=0.35]{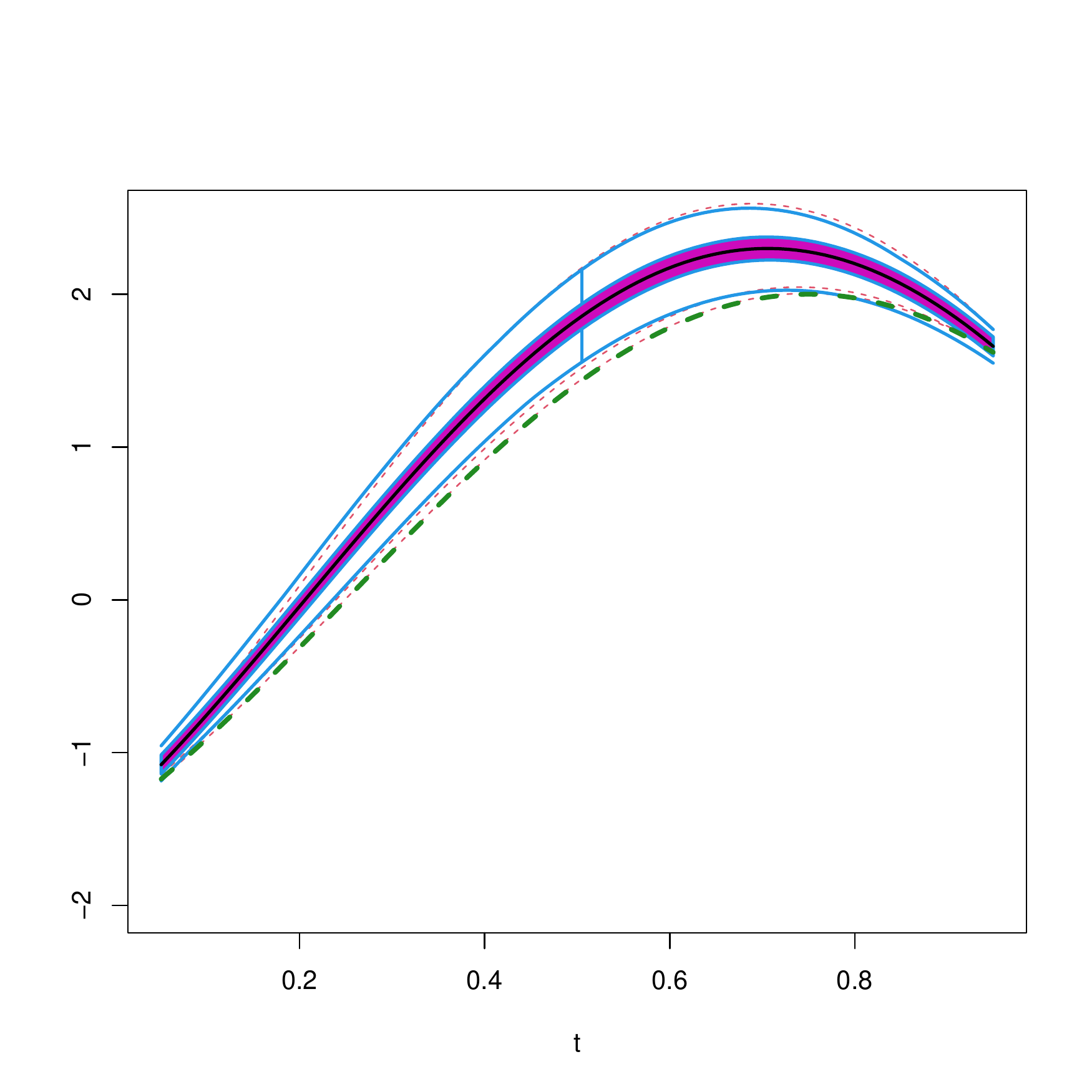}  & 
 \includegraphics[scale=0.35]{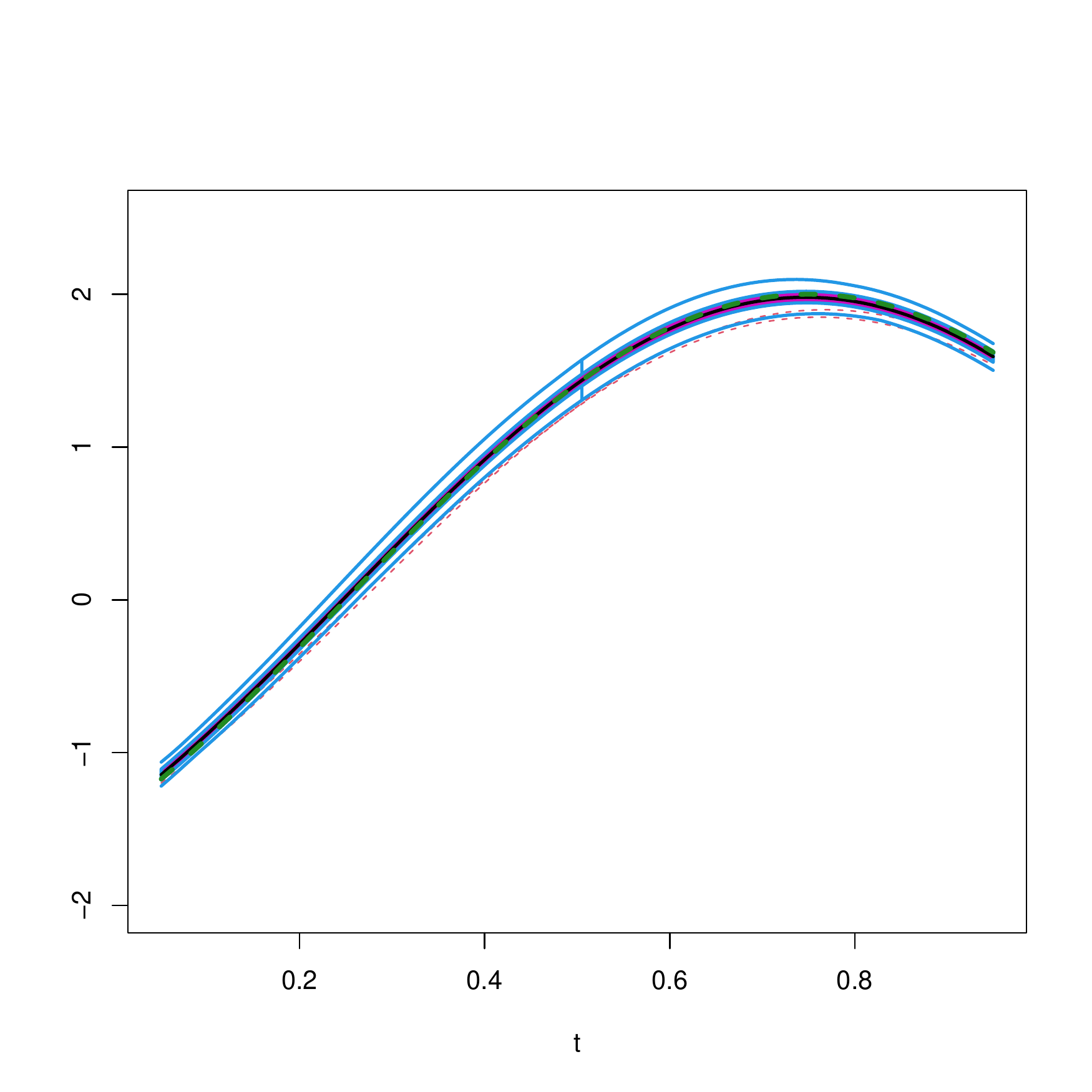} 
 \\[-6ex]
    
$C_{3, 1.2}$ &
\includegraphics[scale=0.35]{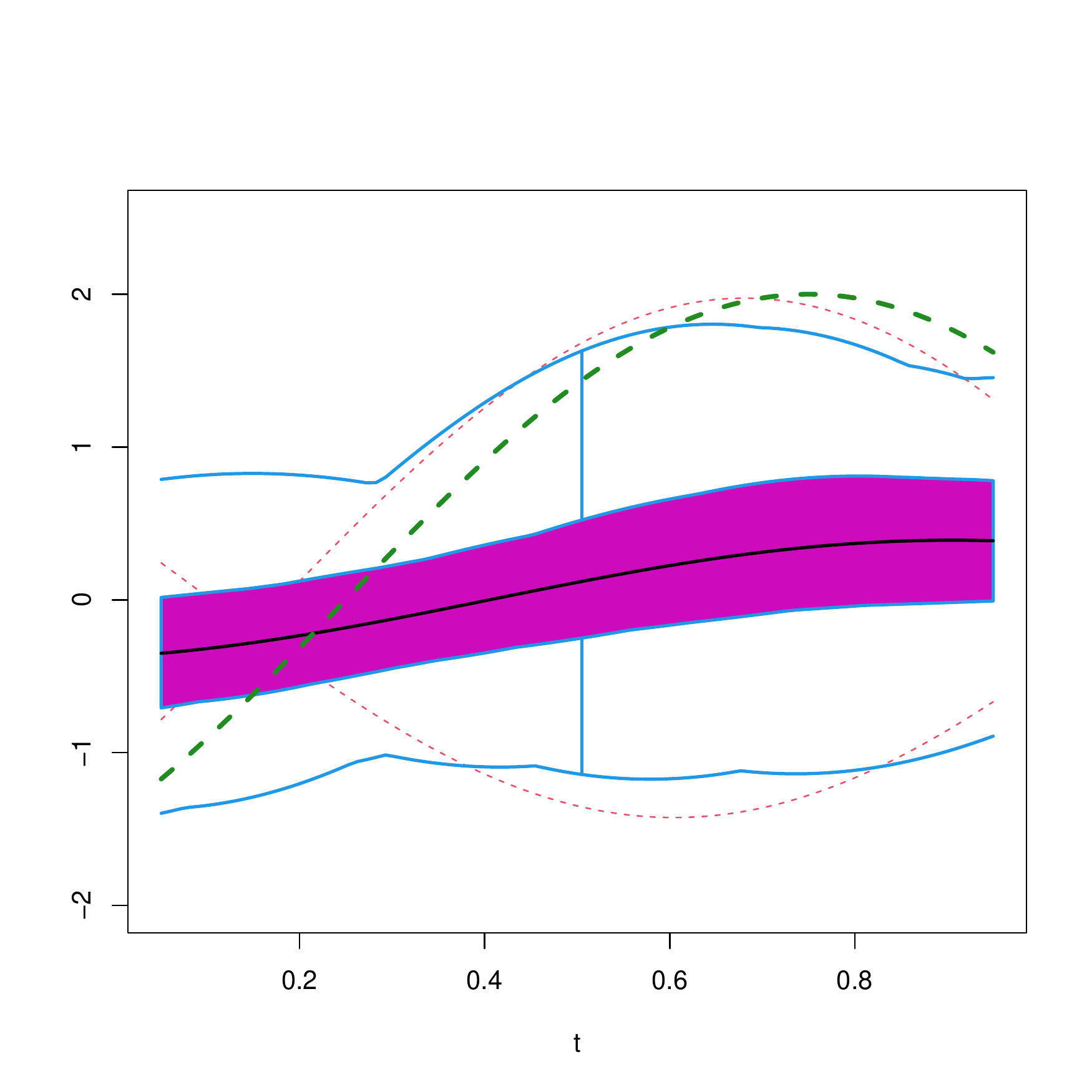}  & 
 \includegraphics[scale=0.35]{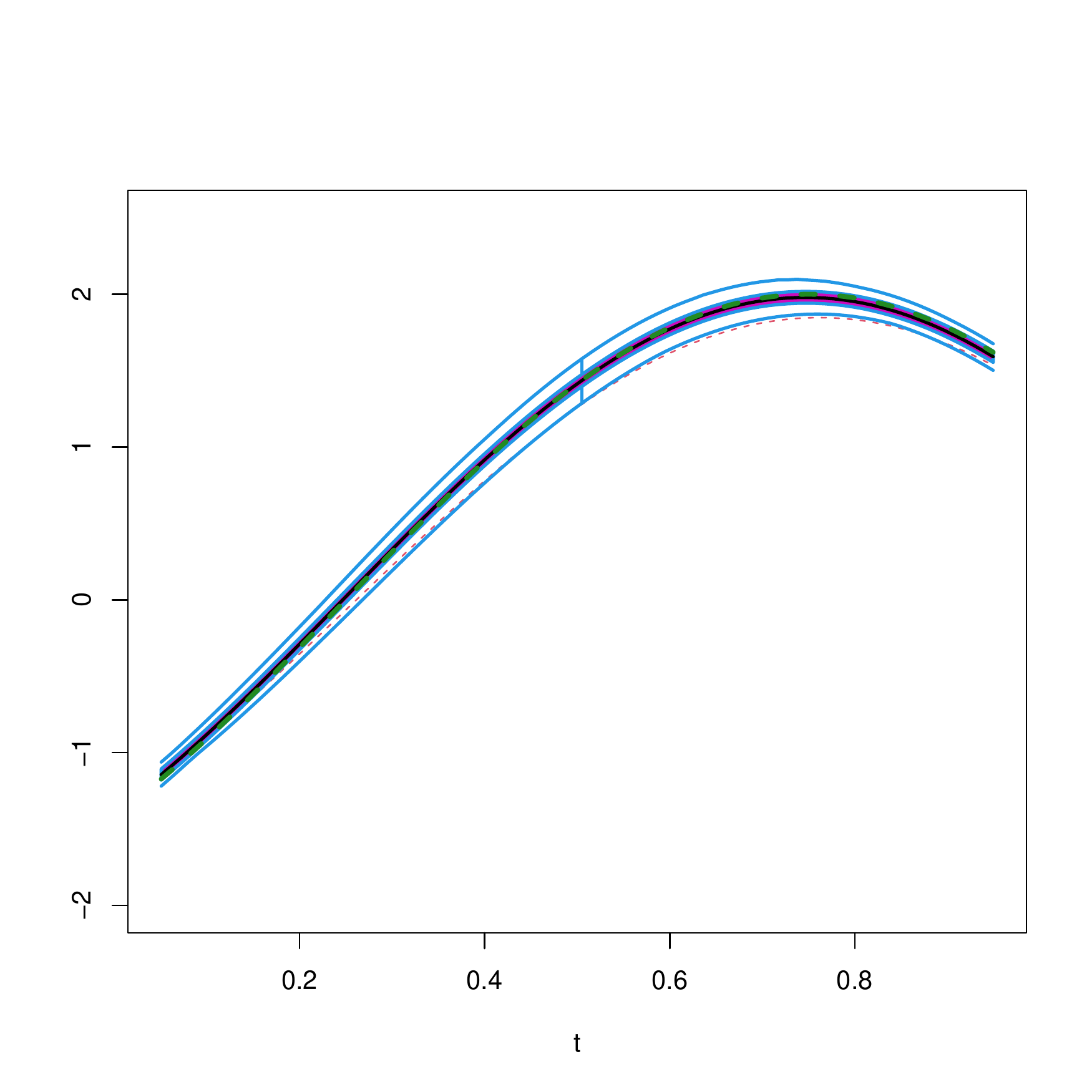}
\end{tabular}
\caption{\small \label{fig:wbeta-Upsilon1-YM}  Functional boxplot of the estimators for $\beta_0$ under \textbf{Model 2} with  $\Upsilon_0=\Upsilon_{0,1}$. 
The true function is shown with a green dashed line, while the black solid one is the central 
curve of the $n_R = 1000$ estimates $\wbeta$. Columns correspond to estimation 
methods, while rows to $C_0$ and   to some of the three contamination  settings.}
\end{center} 
\end{figure}

\begin{figure}[tp]
 \begin{center}
 \footnotesize
 \renewcommand{\arraystretch}{0.2}
 \newcolumntype{M}{>{\centering\arraybackslash}m{\dimexpr.05\linewidth-1\tabcolsep}}
   \newcolumntype{G}{>{\centering\arraybackslash}m{\dimexpr.33\linewidth-1\tabcolsep}}
%\begin{tabular}{MGG}
\begin{tabular}{M GG}
 & $\wup_{\ls}$ &   $\wup_{\eme\eme}$  \\[-3ex]
$C_{0}$ 
&  \includegraphics[scale=0.18]{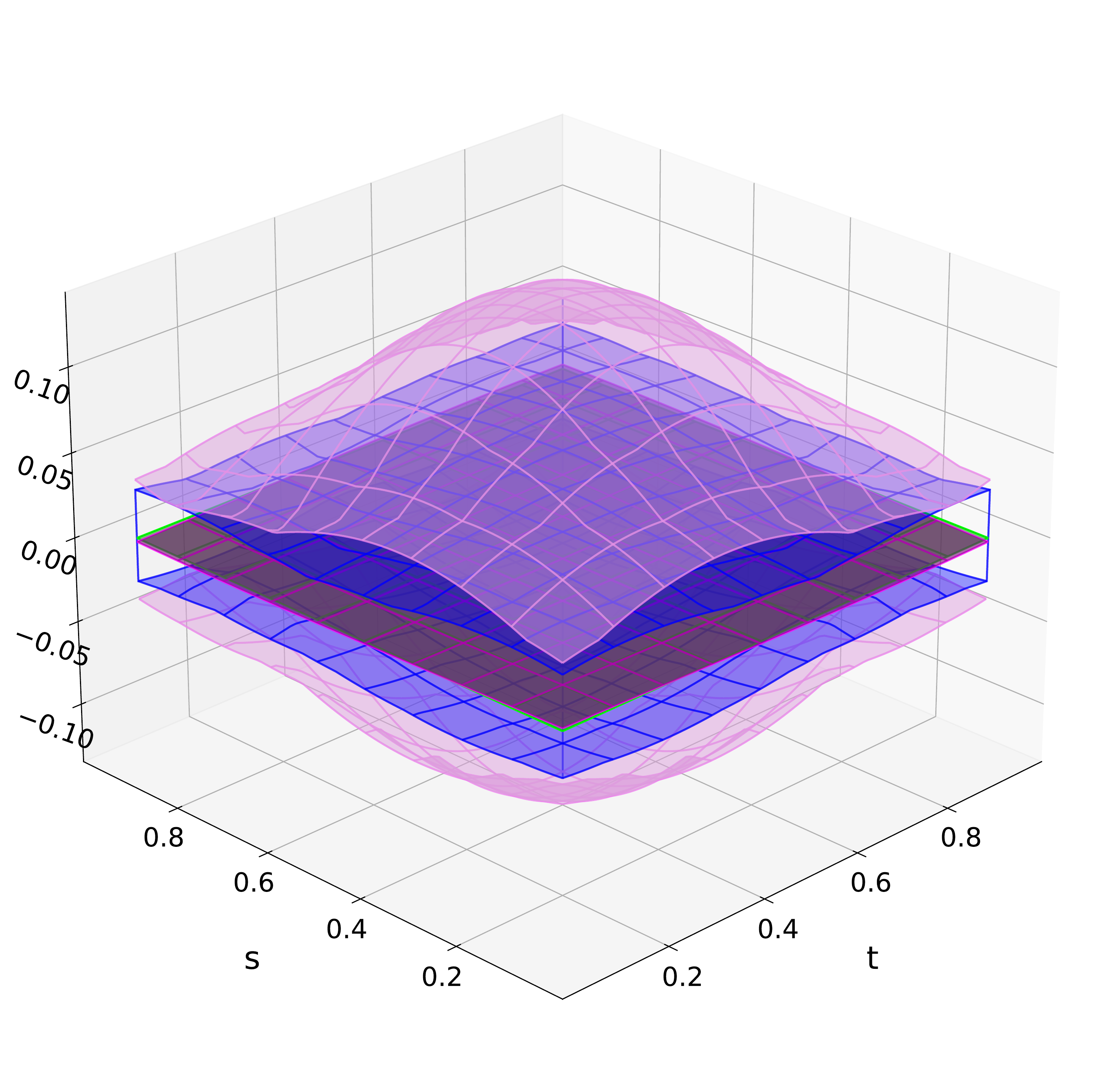} 
&  \includegraphics[scale=0.18]{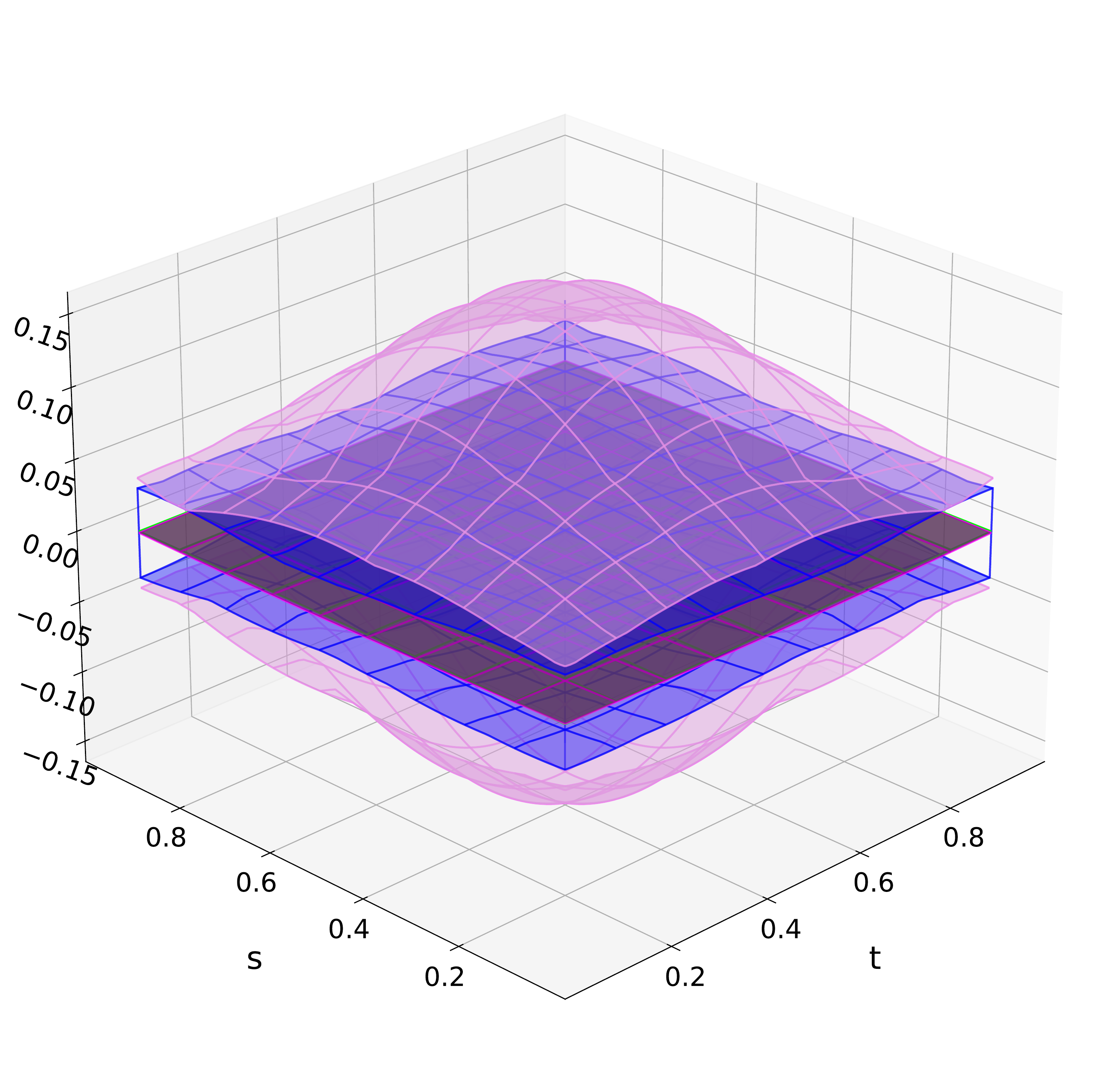} 
\\[-4ex]
$C_{1, 12}$ &  
 \includegraphics[scale=0.18]{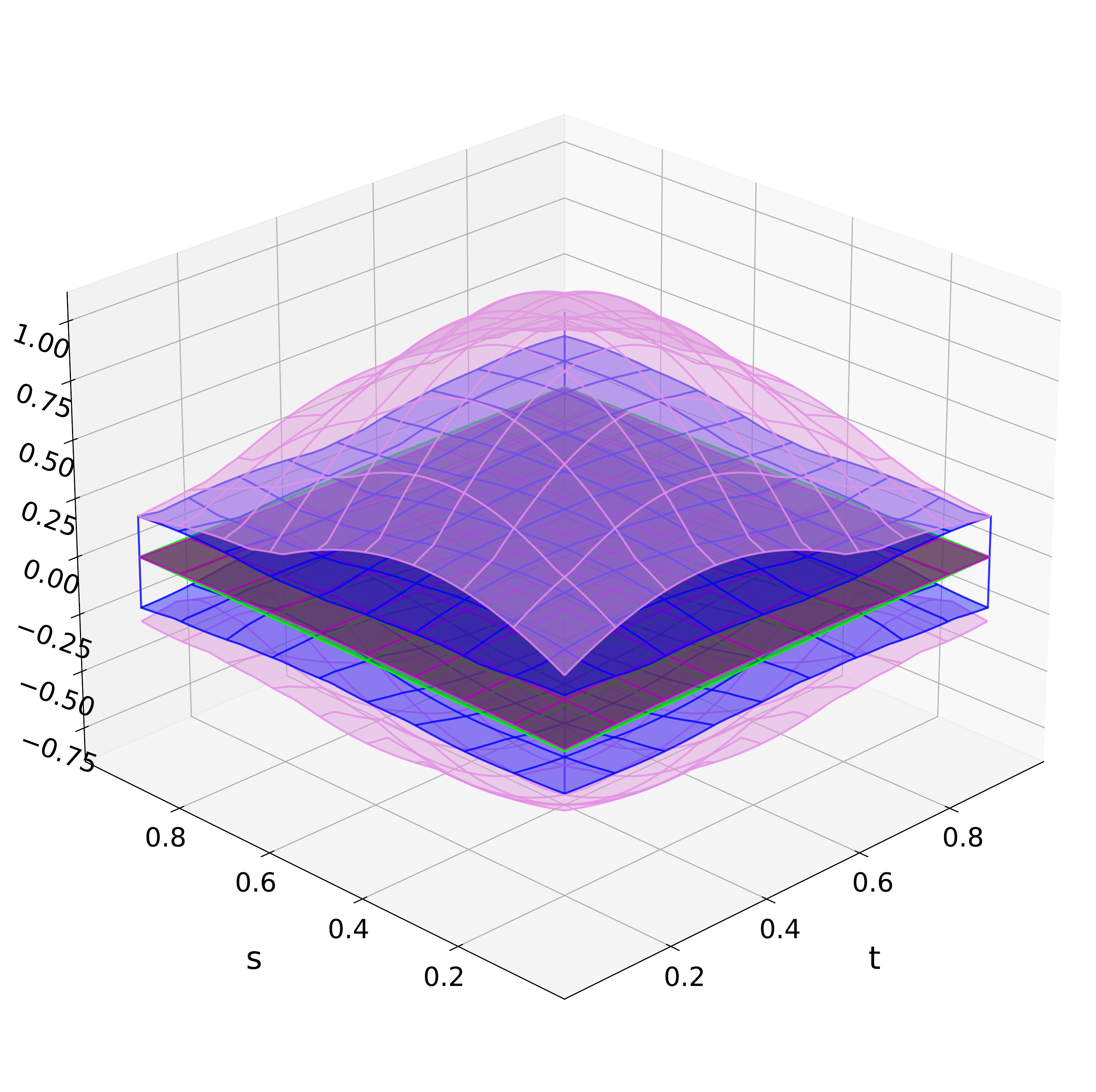}  & 
 \includegraphics[scale=0.18]{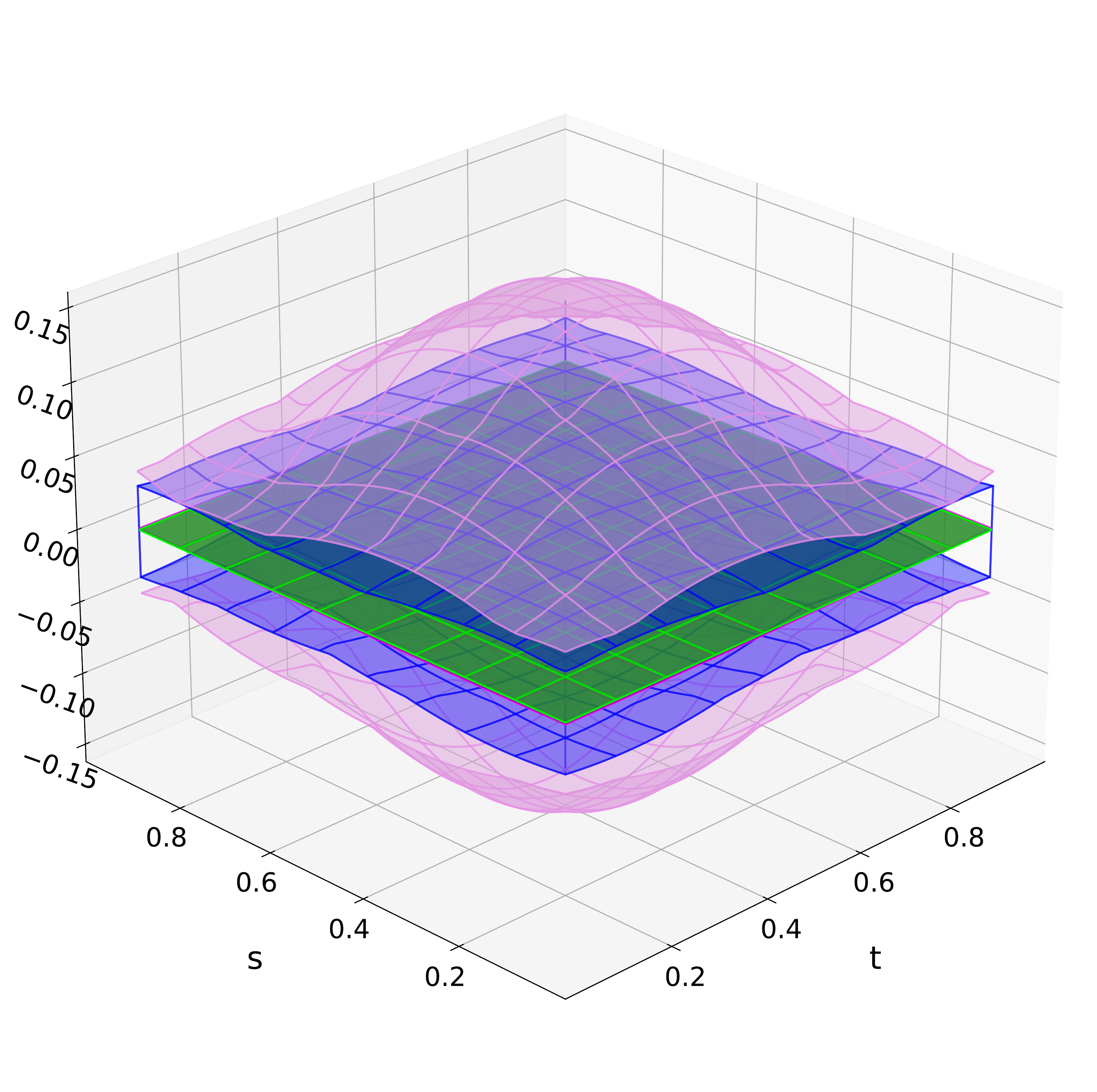} 
  \\[-4ex]
   
$C_{2, 8}$ &  
 \includegraphics[scale=0.18]{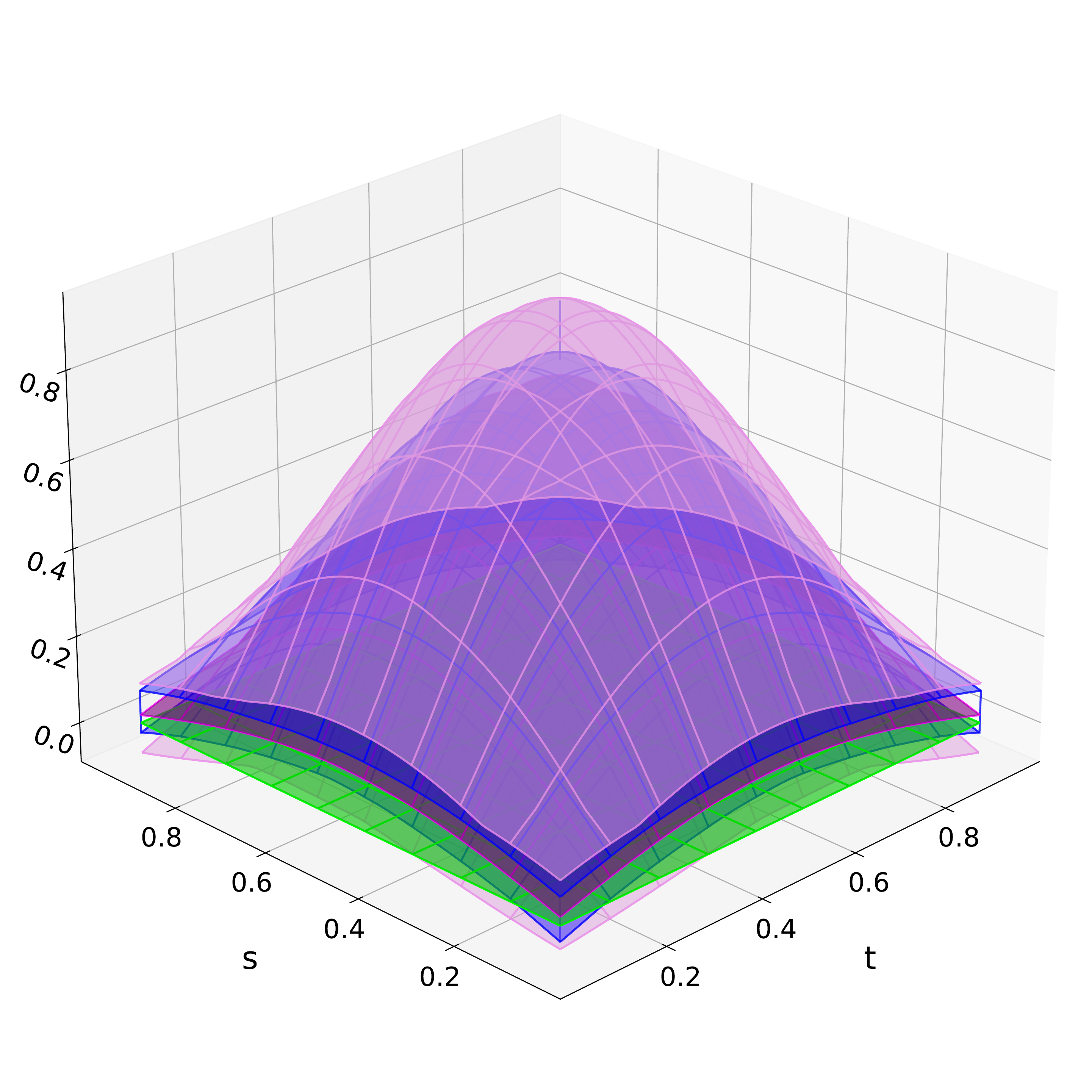}  & 
 \includegraphics[scale=0.18]{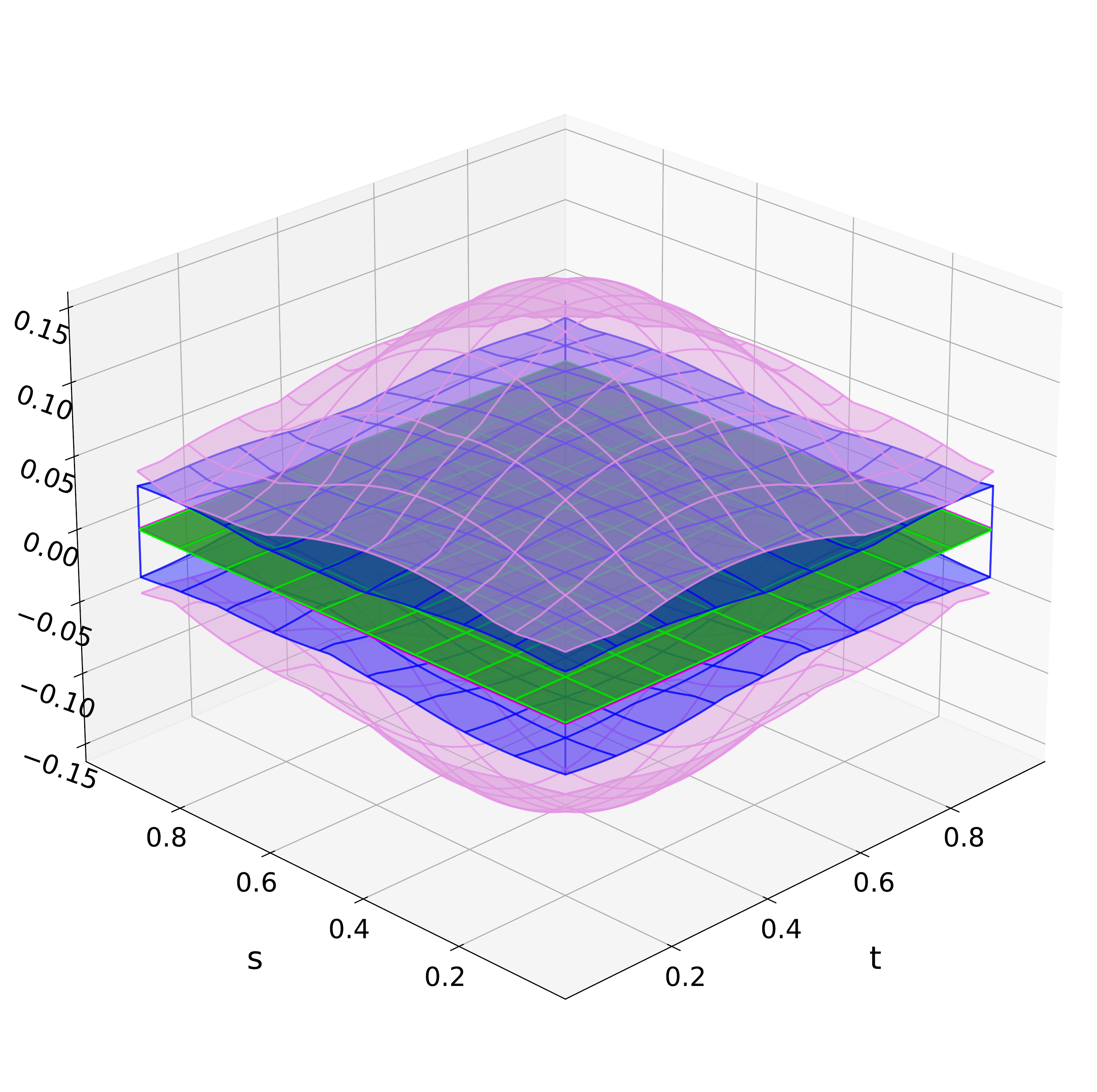} 
 \\[-4ex]
    
$C_{3, 1.2}$ &  
\includegraphics[scale=0.18]{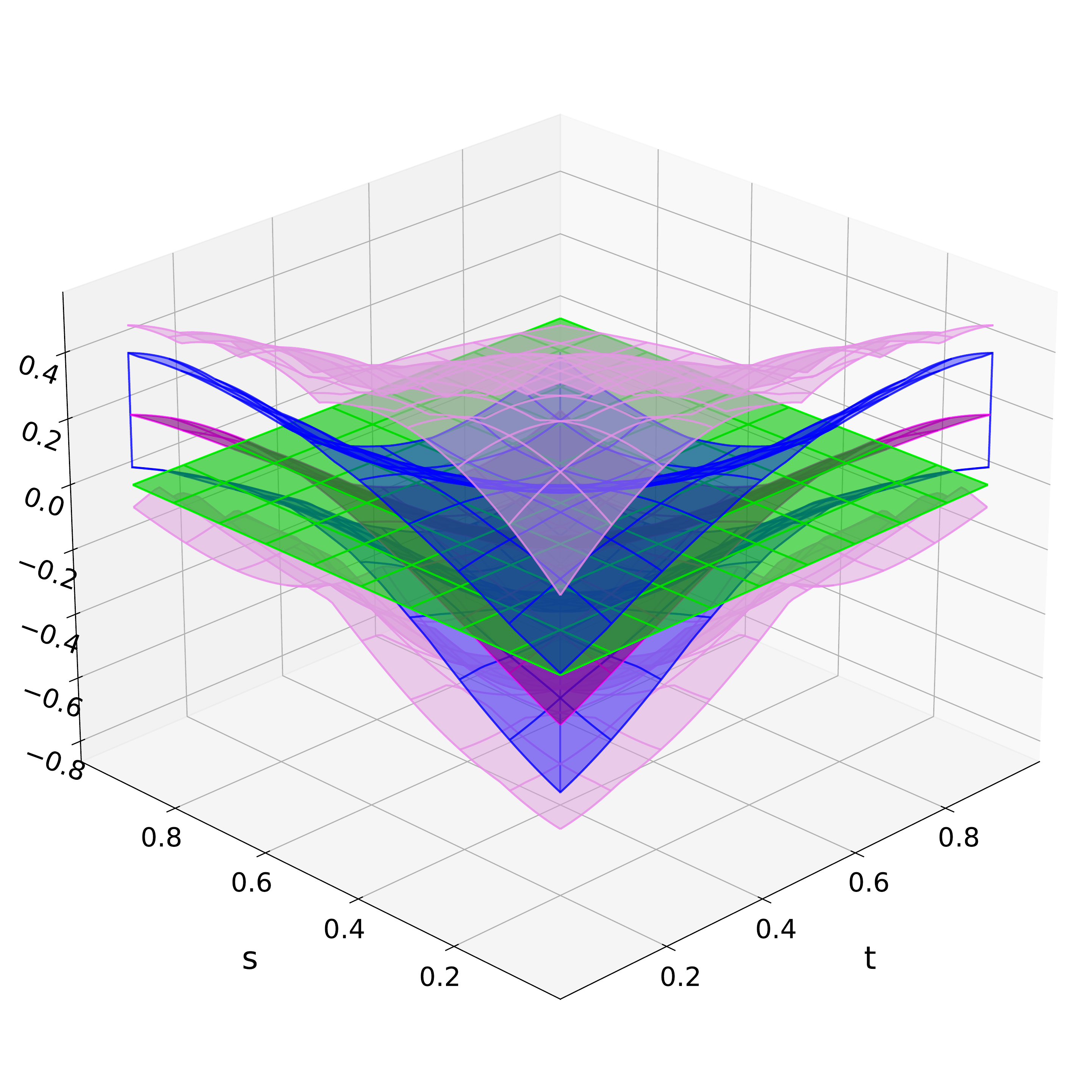}  & 
\includegraphics[scale=0.18]{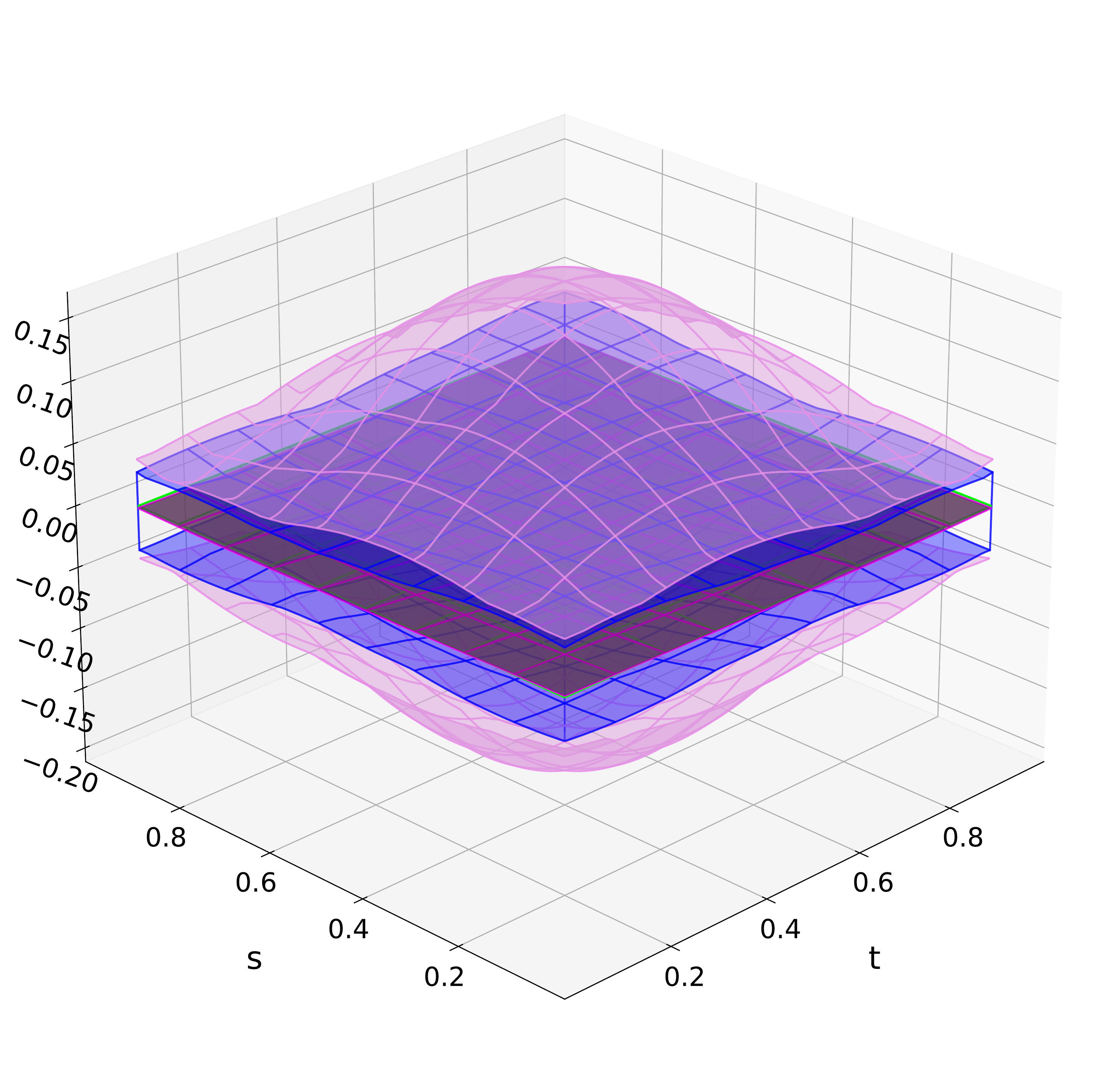}
\end{tabular}
\caption{\small \label{fig:wgamma-Upsilon0-YM}  Surface boxplot of the estimators for $\upsilon_0$ under \textbf{Model 2} with  $\Upsilon_0=0$. 
The true function is shown in green, while the purple surface is the central 
surface of the $n_R = 1000$ estimates $\wup$. Columns correspond to estimation 
methods, while rows to $C_0$ and   to some of the three contamination  settings.}
\end{center} 
\end{figure}

\begin{figure}[tp]
 \begin{center}
 \footnotesize
 \renewcommand{\arraystretch}{0.2}
 \newcolumntype{M}{>{\centering\arraybackslash}m{\dimexpr.05\linewidth-1\tabcolsep}}
   \newcolumntype{G}{>{\centering\arraybackslash}m{\dimexpr.33\linewidth-1\tabcolsep}}
%\begin{tabular}{MGG}
\begin{tabular}{M GG}
& $\wup_{\ls}$ &   $\wup_{\eme\eme}$  \\[-3ex]
$C_{0}$ 
&  \includegraphics[scale=0.18]{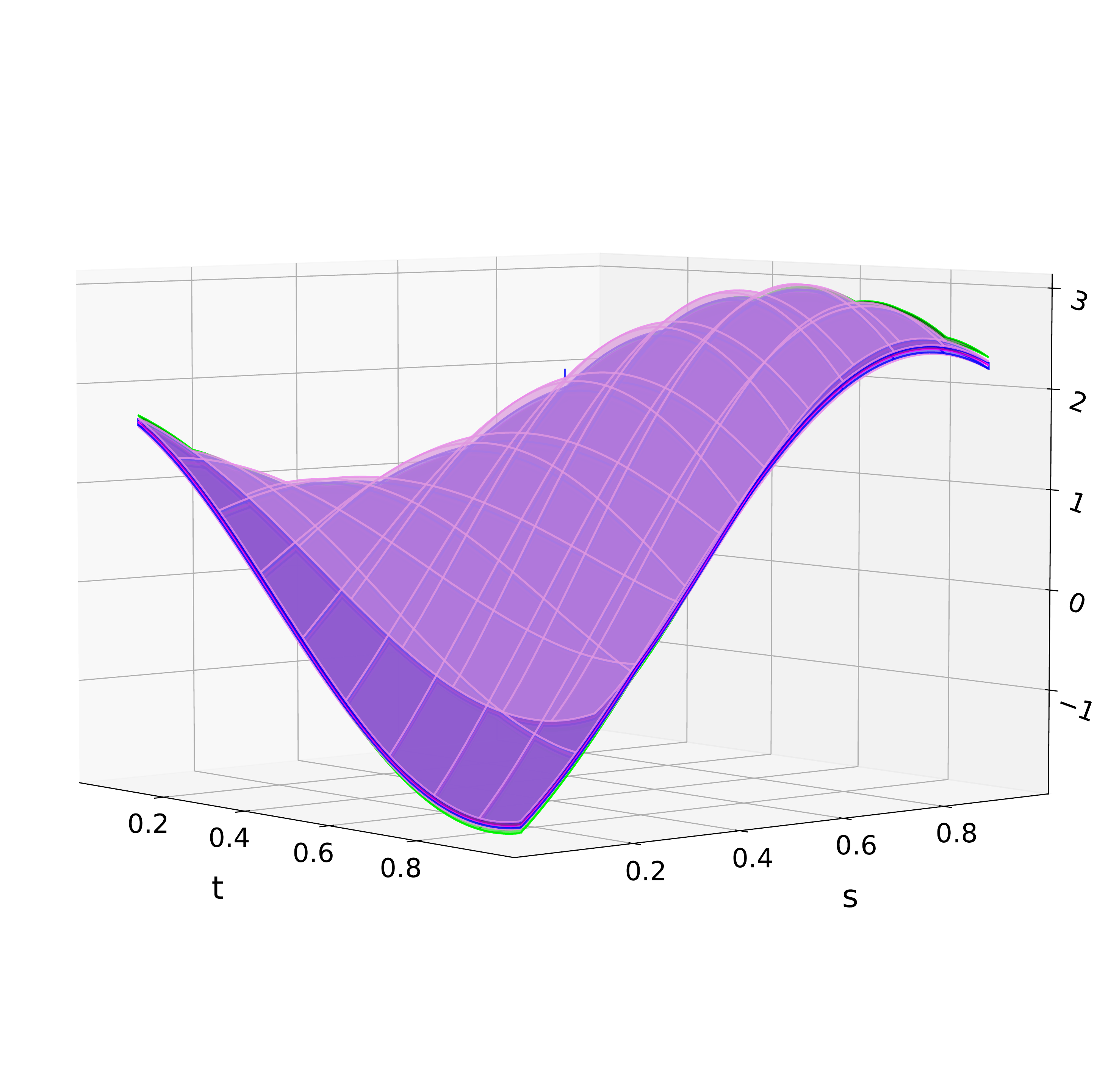} 
&  \includegraphics[scale=0.18]{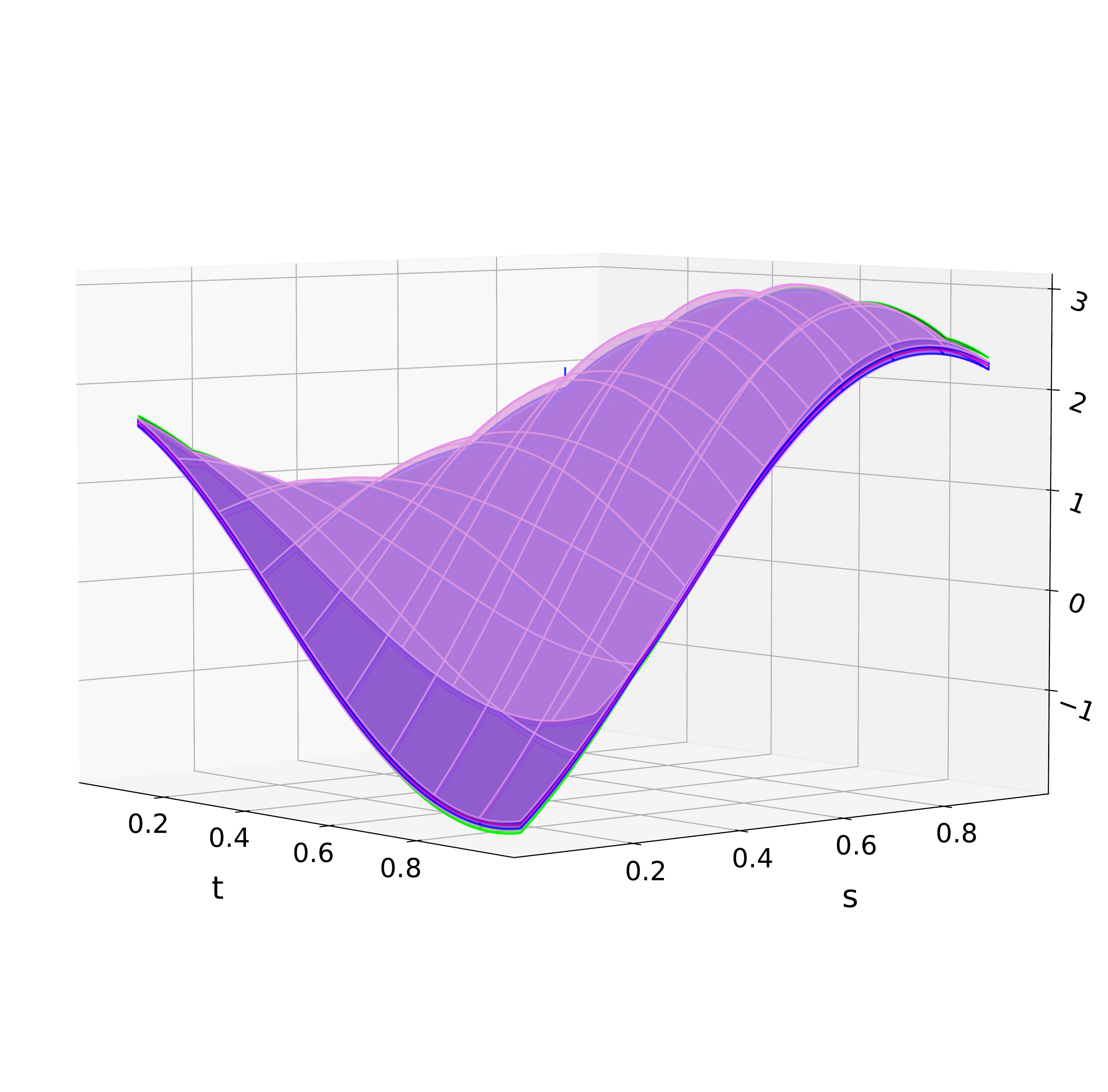} 
\\[-4ex]
$C_{1, 12}$ &  
 \includegraphics[scale=0.18]{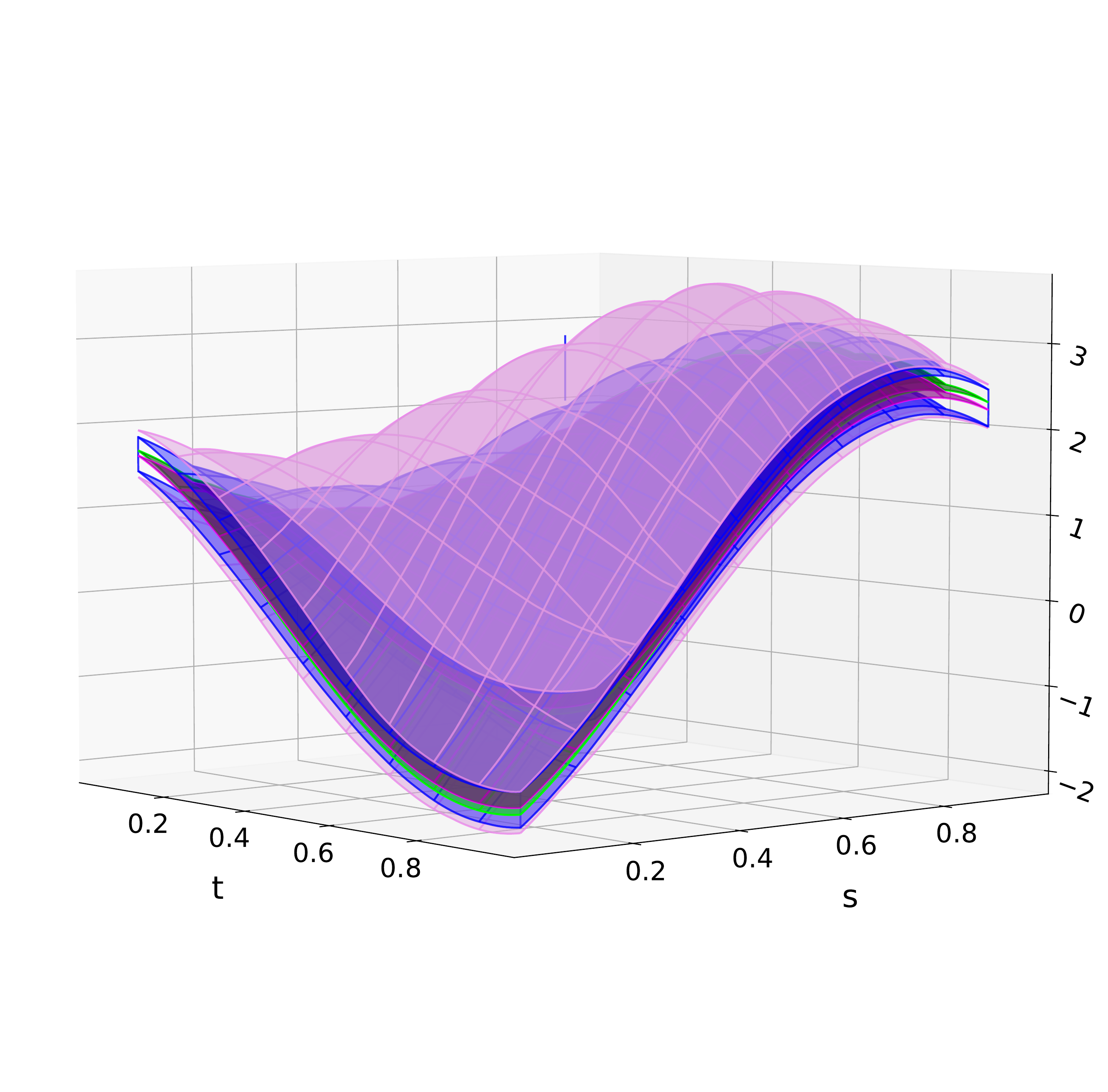}  & 
 \includegraphics[scale=0.18]{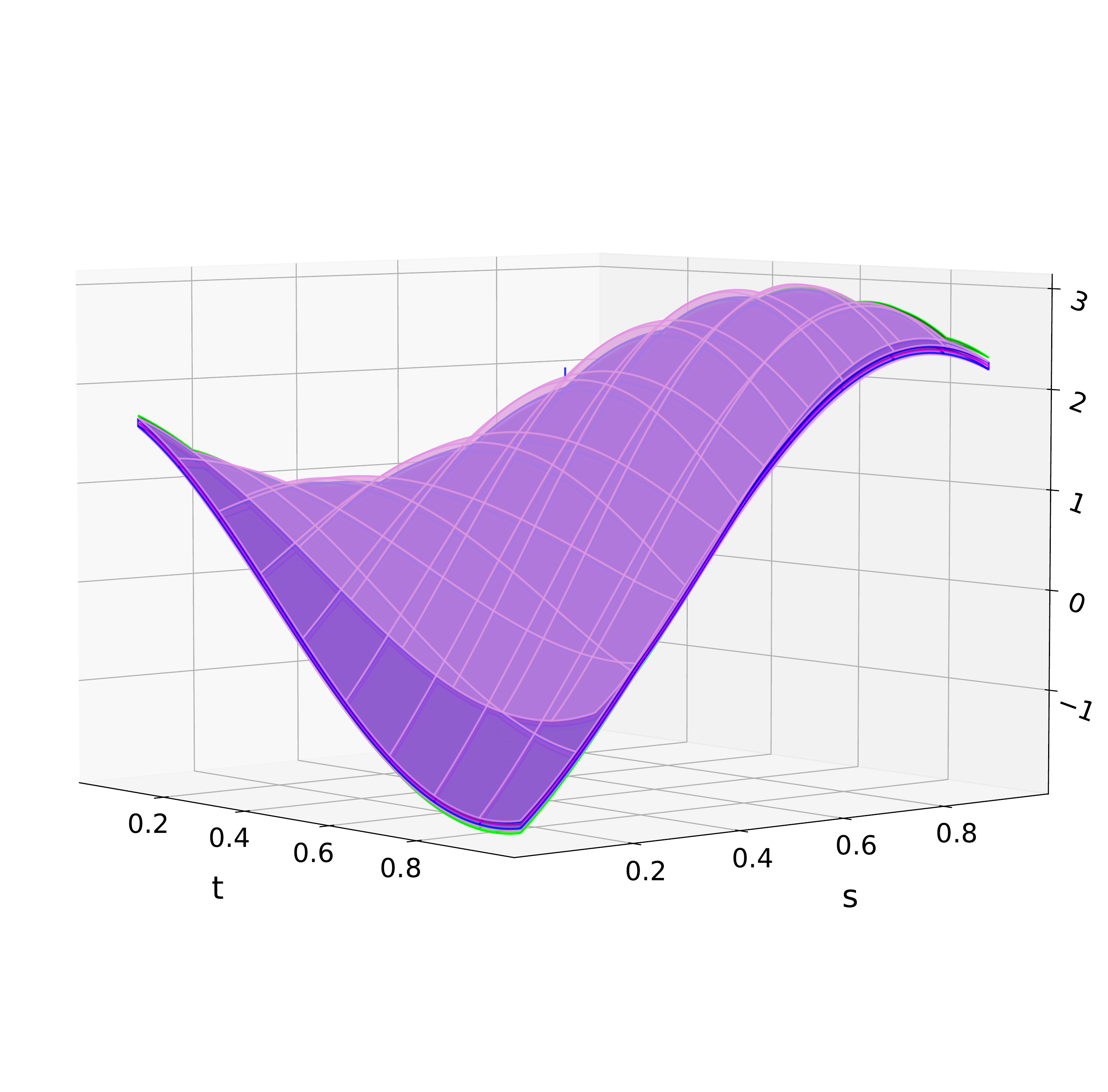} 
  \\[-4ex]
   
$C_{2, 8}$ &
 \includegraphics[scale=0.18]{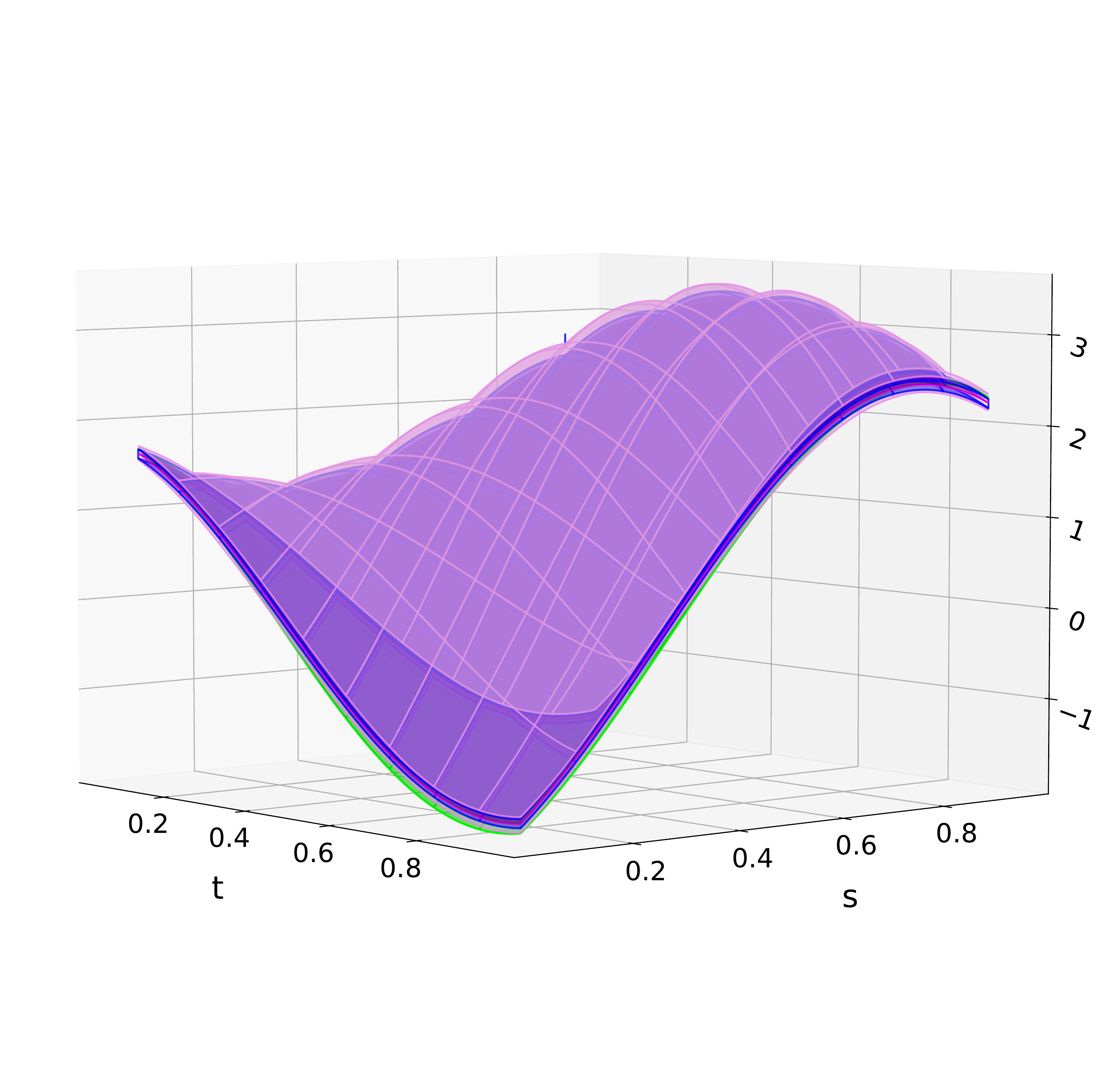}  & 
 \includegraphics[scale=0.18]{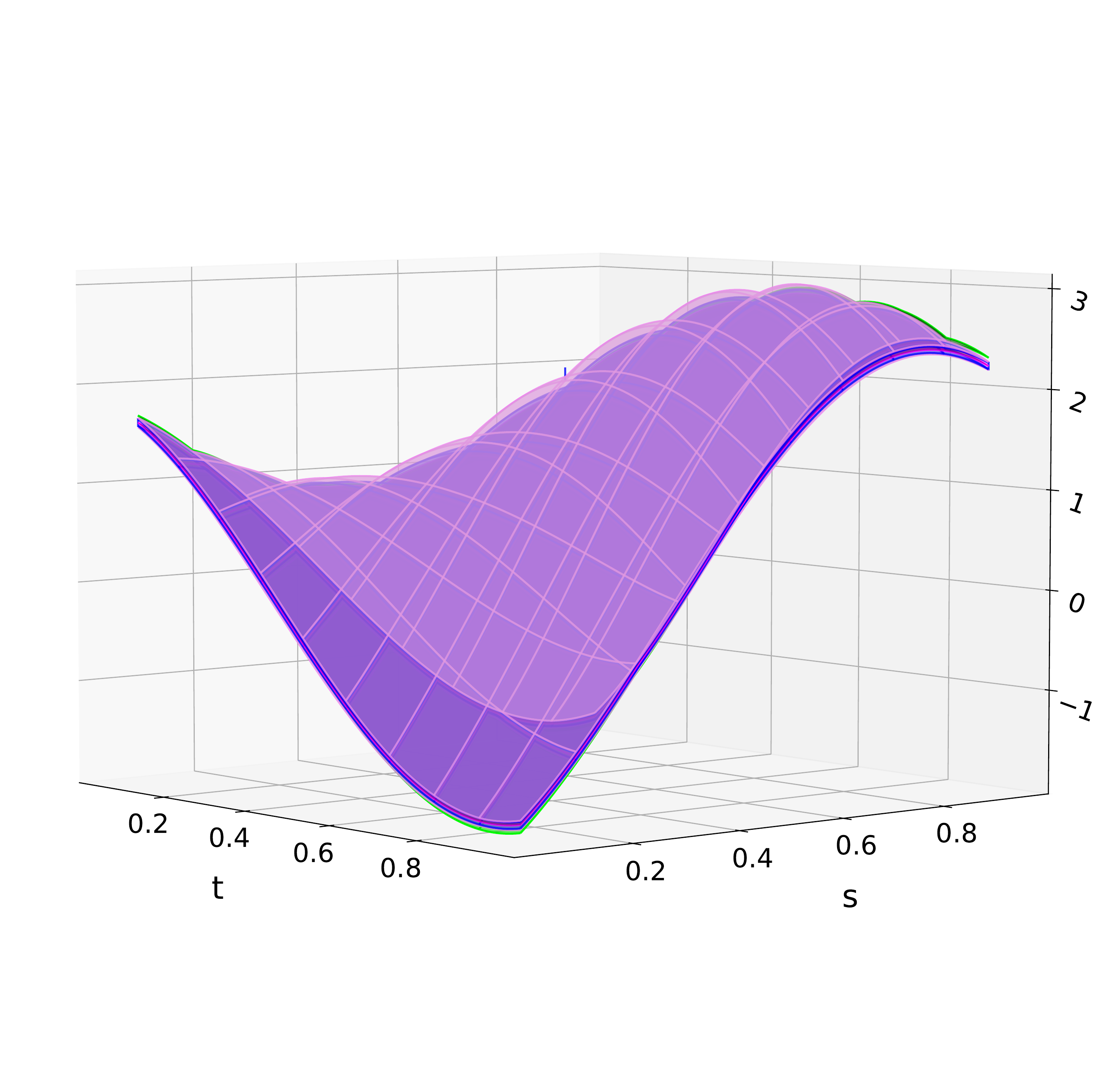} 
 \\[-4ex]
    
$C_{3, 1.2}$ &
 \includegraphics[scale=0.18]{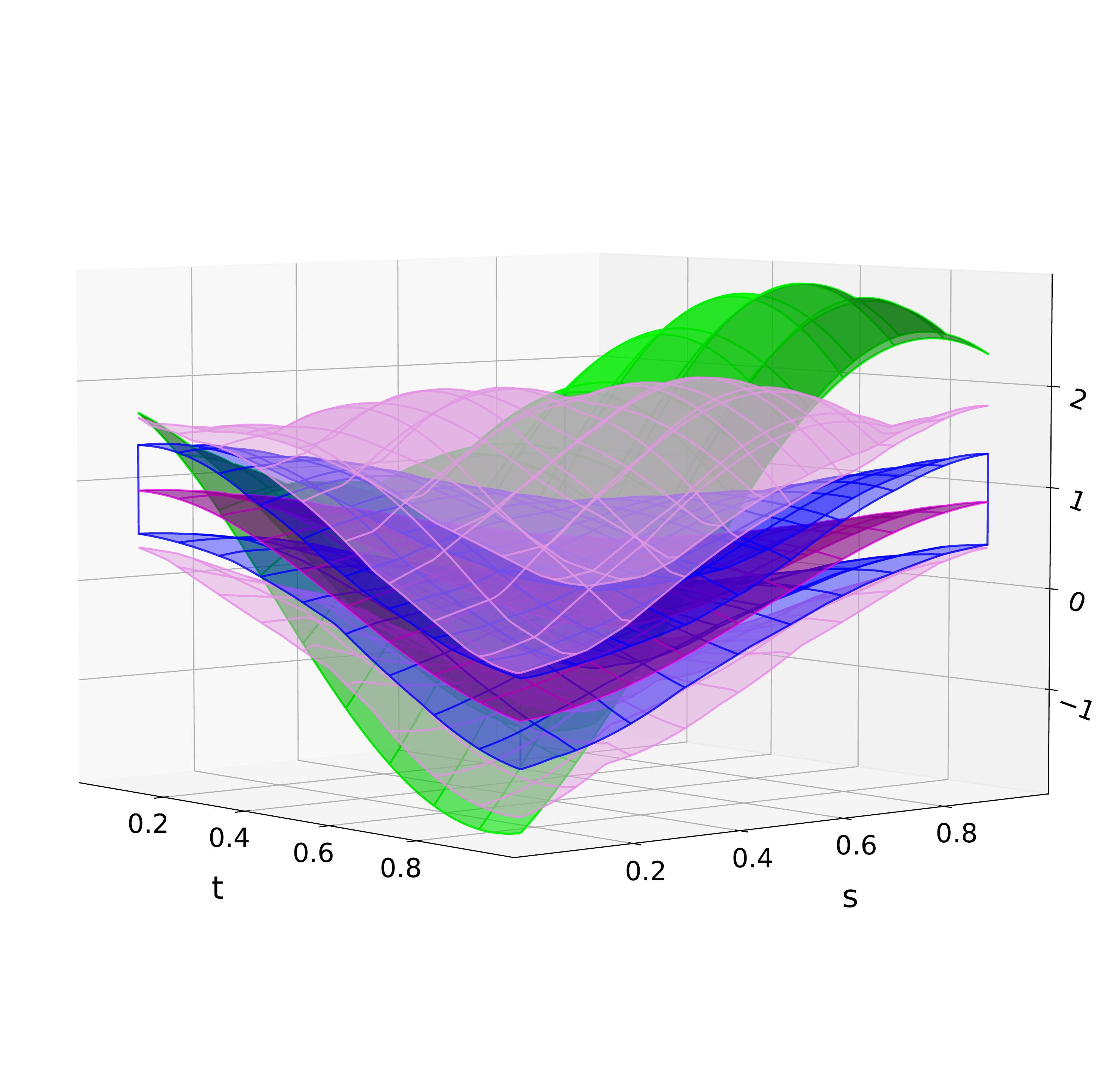}  & 
 \includegraphics[scale=0.18]{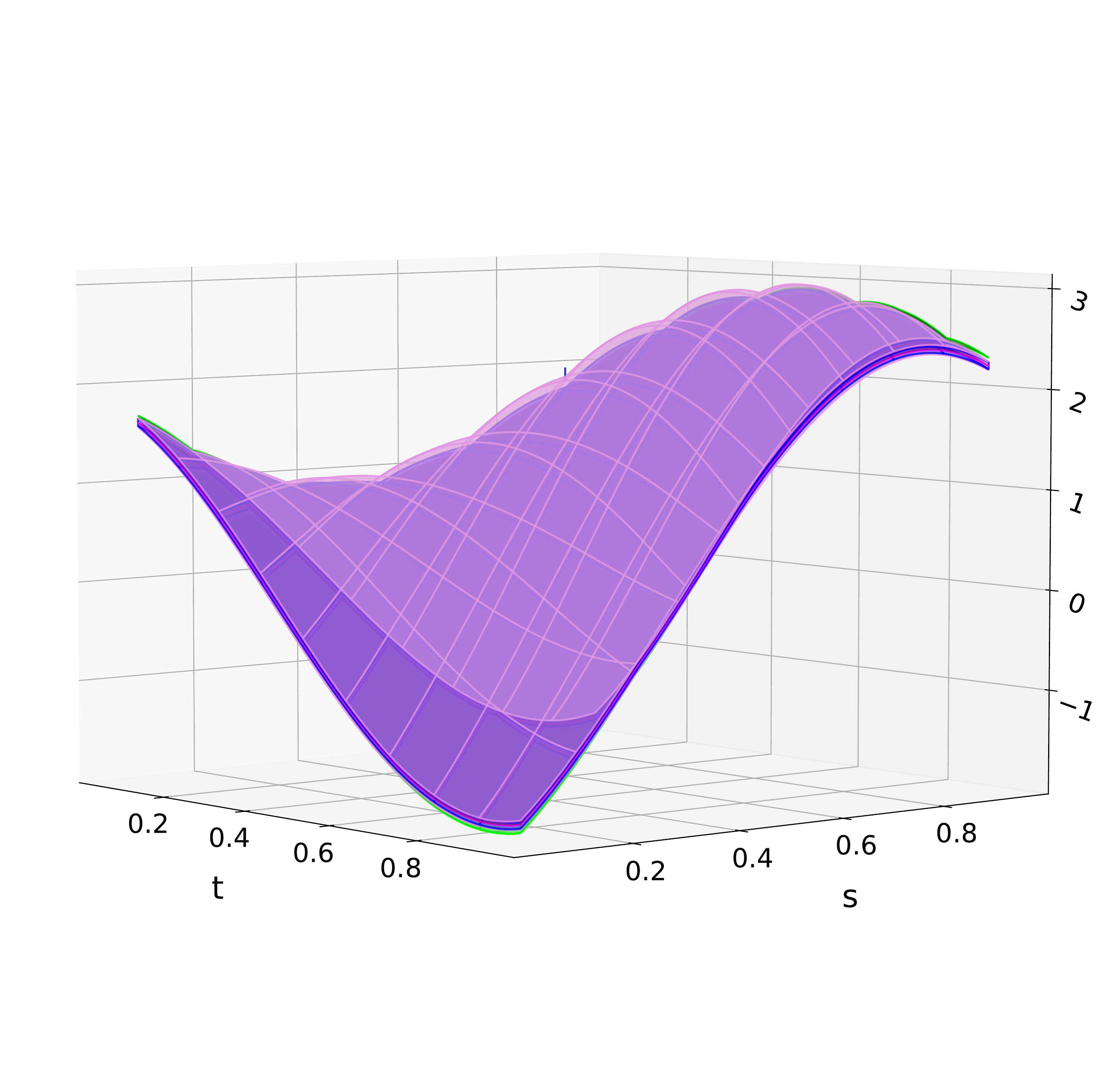}
\end{tabular}
\caption{\small \label{fig:wgamma-Upsilon1-YM}  Surface boxplot of the estimators for $\upsilon_0$ under \textbf{Model 2} with  $\Upsilon_0=\Upsilon_{0,1}$. 
The true function is shown in green, while the purple surface is the central 
surface of the $n_R = 1000$ estimates $\wup$. Columns correspond to estimation 
methods, while rows to $C_0$ and   to some of the three contamination  settings.}
\end{center} 
\end{figure}

The effect on the classical estimators of the quadratic operator is observed in Figures \ref{fig:wgamma-Upsilon0-YM} and \ref{fig:wgamma-Upsilon1-YM}, where it is clear the enlargement of the central 50\% region  under $C_{1, 12}$ and the damaging effect of contamination  $C_{2, 8}$ when $\Upsilon_0=0$, since the true surface is mostly beyond the area delimiting the non-atypical ones for values of $t$ and $s$ between 0.3 and 0.7. Under $C_{3,1.2}$, when the true model is a linear one, the whiskers are completely distorted even when the true surface lies within them. When considering quadratic terms in the model, this last contamination has an extreme effect, since the true surface  crosses the limits of the surfaces plot. Note that an estimate of the trimmed total variance $\var_{\trim}$ may be obtained as the difference between the MISE and the squared bias. Under $C_0$, the BIAS$_{\trim}$ equals 0.037 for both estimators and  the square root of $\var_{\trim}$ equals 0.0257 for the classical procedure and 0.0286 for the robust one, so the bias is much larger than the variability. In contrast, under \textbf{Model 1} when considering $\Upsilon_0=\Upsilon_{0,2}$ which varies within a similar range to that of the quadratic kernel under \textbf{Model 2}, the obtained estimate of the total variability ($\sqrt{\var_{\trim}}$) of $\wUps$ equal 2.2147 and 2.6251 for the least squares and $MM-$procedure, respectively and the trimmed biases are more than five times smaller, since they equal 0.3848 and 0.3818, respectively. The difference arising produces a distorting effect on the surface plots obtained under \textbf{Model 2} and $C_0$, which do not allow to see clearly the whiskers and central region. Furthermore, due to the stability of the robust procedure, the same behaviour arises for the $MM-$estimate under the considered contaminations. % and the plots   have a slightly green shape. 

\clearpage
\section{Tecator Data}{\label{sec:tecator}}

The Tecator data set was analysed, among others, in \citet{ferraty2006nonparametric}, \citet{aneiros2006semi}, \citet{yao2010functional}, \citet{shang2014bayesian}, \citet{huang2015sieve} and \citet{boente2020robust}
and it is available in the package \texttt{fda.usc} \citep{febrero2012statistical}, see also  \url{http://lib.stat.cmu.edu/datasets/tecator}. These data contain measurements taken on samples from finely chopped meat with different percentages of fat, protein and moisture content. Each observation consists of a spectrometric curve  that corresponds to the absorbance measured on an equally spaced grid of 100 wavelengths between 850 and 1050 nm. The contents of fat protein and moisture were also recorded through analytic chemistry methods. 

The goal of the analysis is to predict the fat content ($y$) using some characteristics of the spectrometric curve.

\citet{huang2015sieve} include also  the variables water and protein contents to predict the fat content and compared several models  in terms of their predictive properties. As a characteristic of the spectrometric curve, they used its second derivative which enters,  as the functional covariate in the model linearly, while the other two variables appear either through an additive non-parametric component or a varying coefficient model.

In this section, as in \citet{yao2005functional}, we use a linear and a quadratic model to predict the fat content from the first derivative $X$  of the spectrometric curve. The robust $MM-$estimators were calculated using the same $\rho-$functions as in our simulation study, and we choose $p=4$ principal directions which explain  more than 97\% of the total variability.

\begin{figure}[ht!]
 \begin{center}
 \small
 \renewcommand{\arraystretch}{0.2}
 \newcolumntype{M}{>{\centering\arraybackslash}m{\dimexpr.05\linewidth-1\tabcolsep}}
   \newcolumntype{G}{>{\centering\arraybackslash}m{\dimexpr.33\linewidth-1\tabcolsep}}
%\begin{tabular}{MGG}
\begin{tabular}{ cc}
Functional linear model &  Functional quadratic model  \\ 
   \includegraphics[scale=0.35]{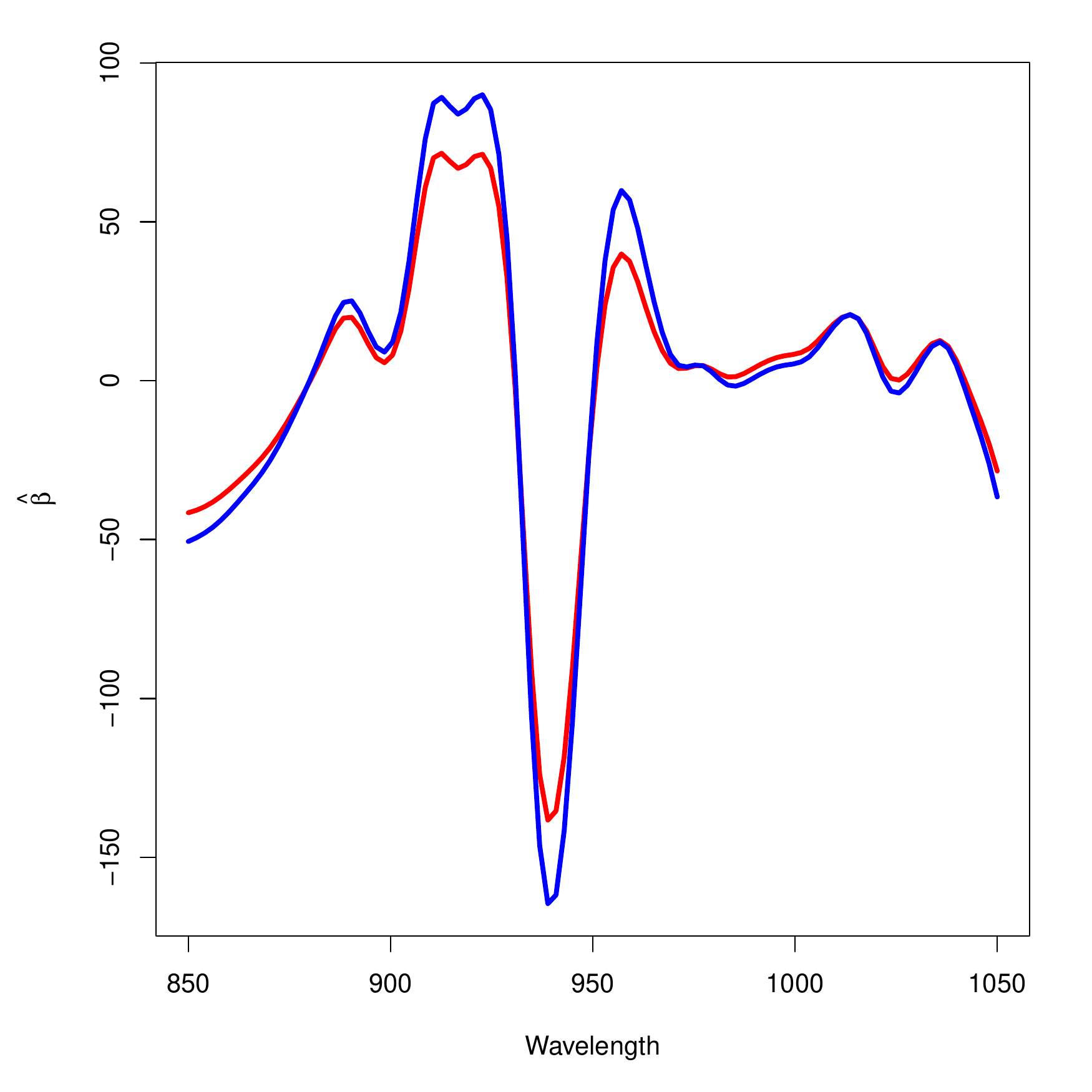} 
&  \includegraphics[scale=0.35]{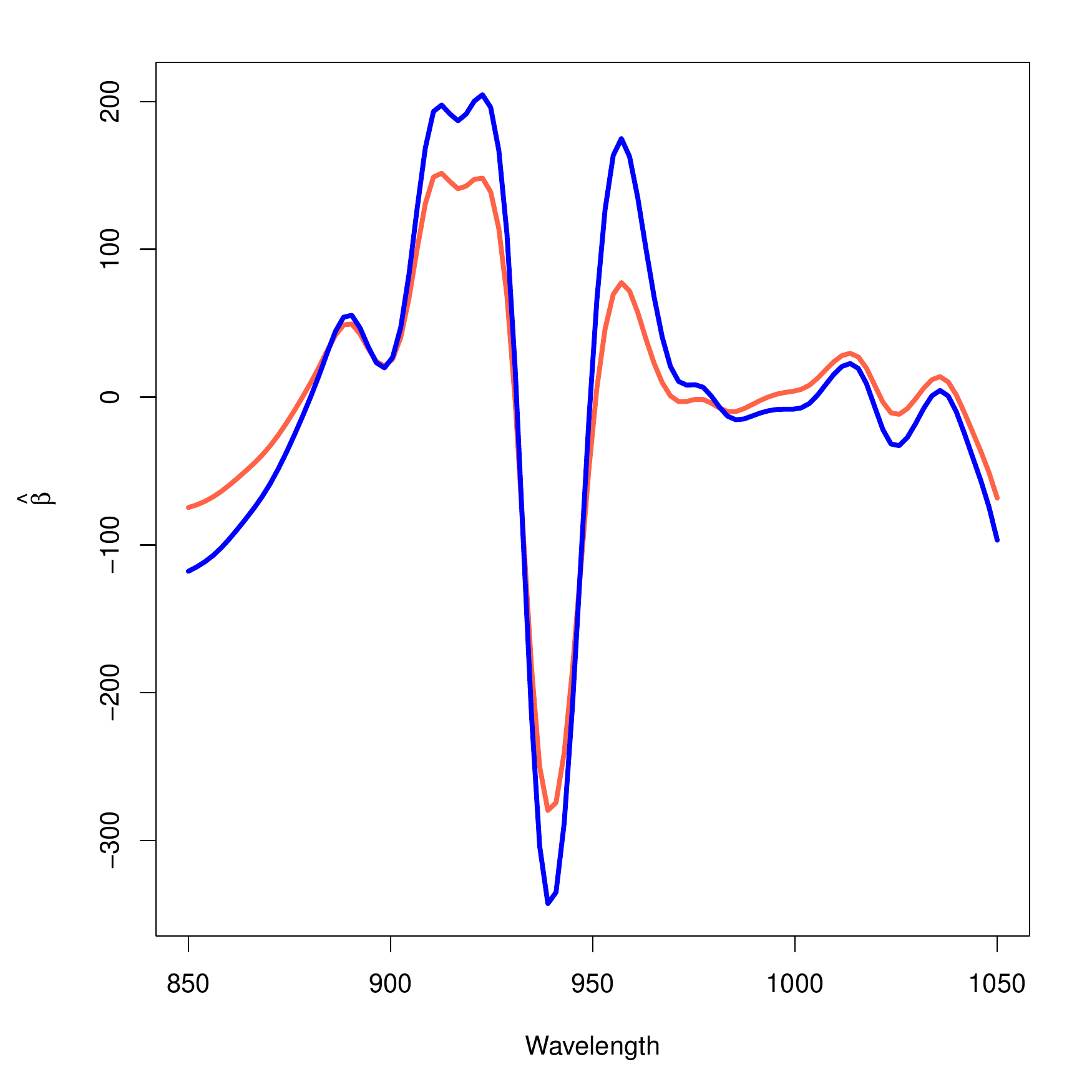} 
 
\end{tabular}
\caption{\small \label{fig:tecator-betas} Estimates of $\beta$ using a least squares (in solid red line) or an $MM-$estimator (in solid blue line), when using a functional linear model (left panel) or a quadratic one (right model).}
\end{center} 
\end{figure}

\begin{figure}[ht!]
 \begin{center}
 \small
 \renewcommand{\arraystretch}{0.2}
 \newcolumntype{M}{>{\centering\arraybackslash}m{\dimexpr.05\linewidth-1\tabcolsep}}
   \newcolumntype{G}{>{\centering\arraybackslash}m{\dimexpr.33\linewidth-1\tabcolsep}}
%\begin{tabular}{MGG}
\begin{tabular}{ cc}
$\wup_{\ls}$ &  $\wup_{\eme\eme}$  \\ 
   \includegraphics[scale=0.35]{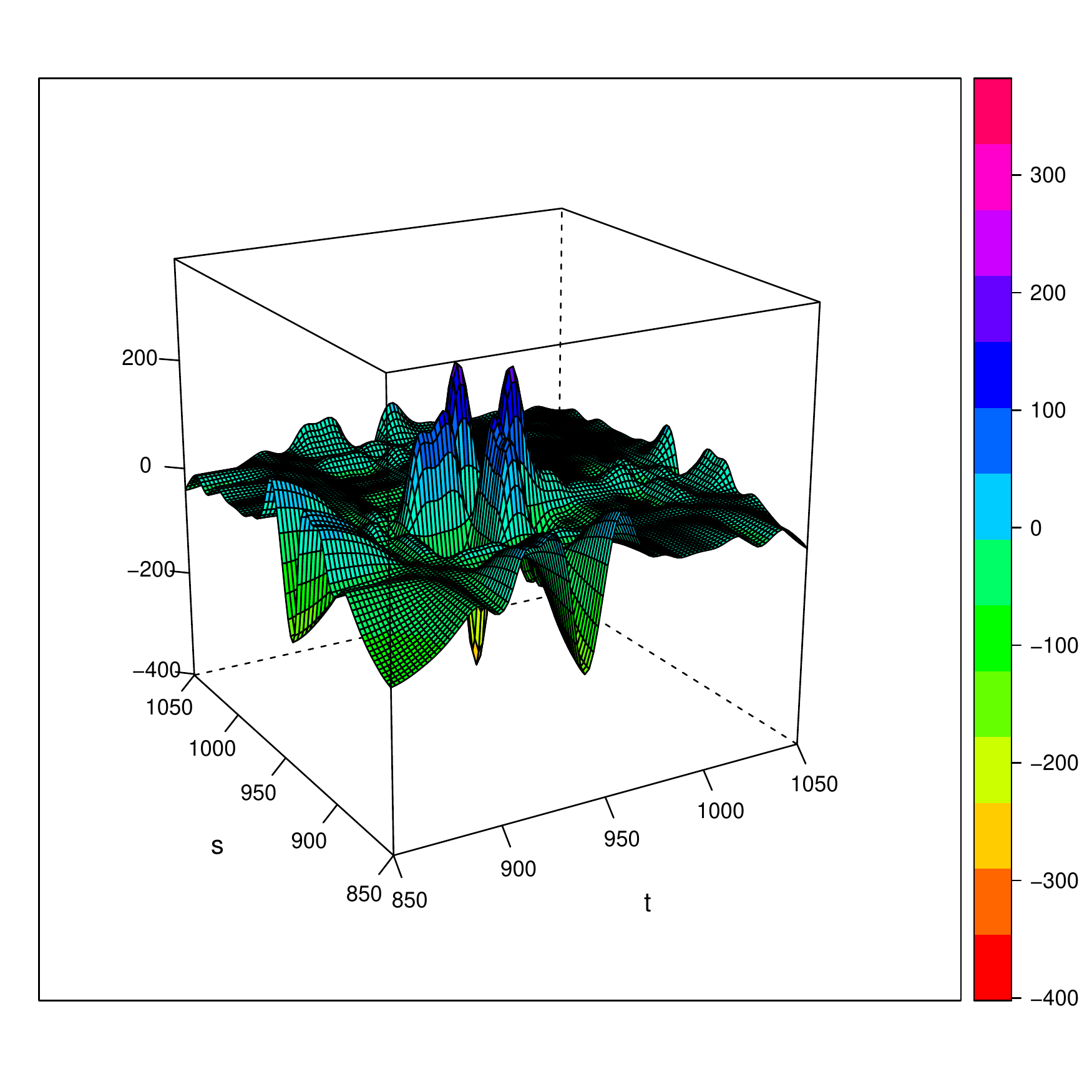} 
&  \includegraphics[scale=0.35]{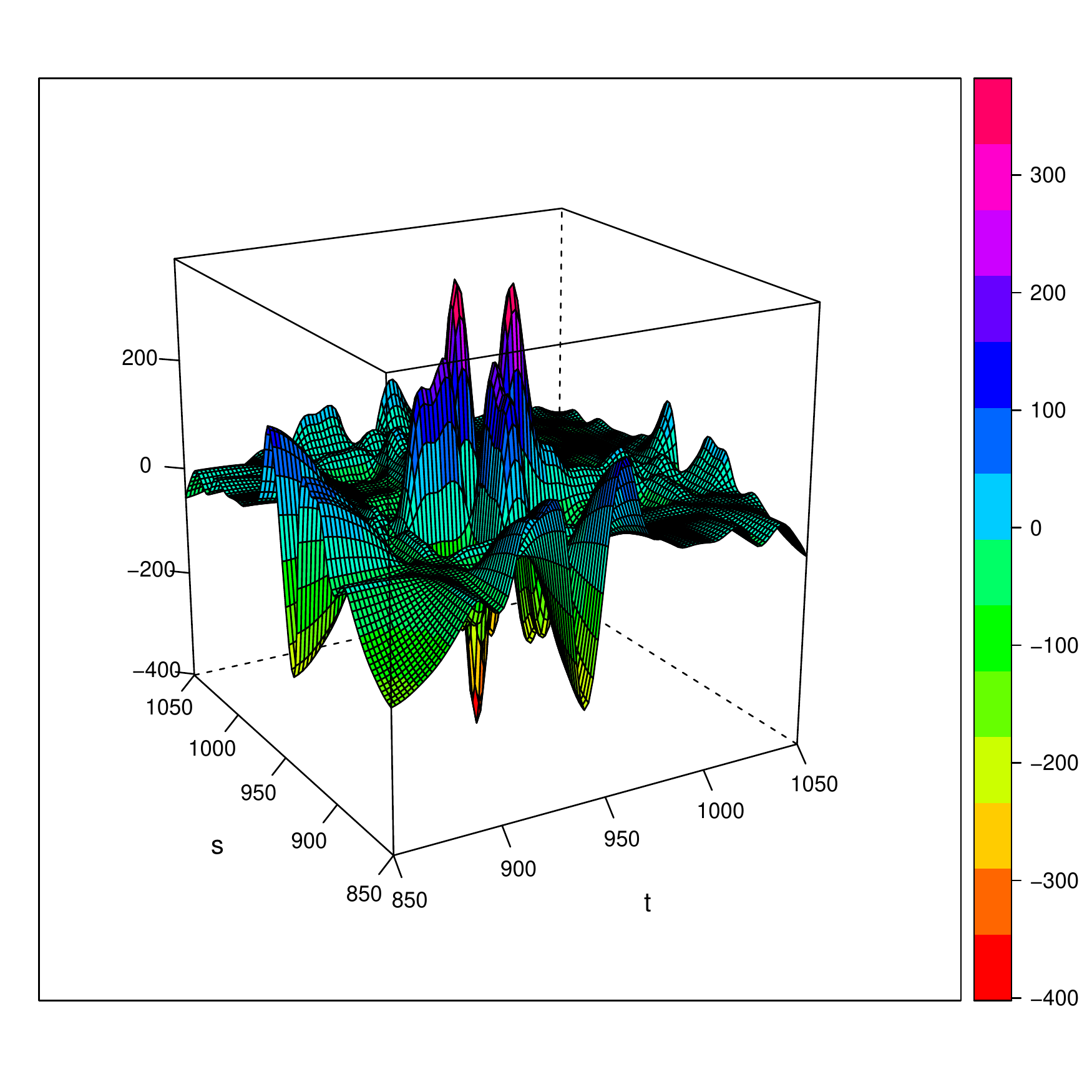} 
 \end{tabular}
\caption{\small \label{fig:tecator-gammas} Estimates of $\upsilon$ using a least squares, $\wup_{\ls}$, (left panel) or an $MM-$estimator,  $\wup_{\eme\eme}$, (right panel), when using a functional   quadratic model.}
\end{center} 
\end{figure}

The red and blue lines in the left and right panels of Figure \ref{fig:tecator-betas} show the estimates  $\wbeta$ 
obtained using the classical ($\wbeta_{\ls}$) and robust estimators ($\wbeta_{\eme\eme}$) when fitting a linear or a quadratic model, respectively. Note that the robust estimates take larger absolute values for wavelengths varying between 900 and 980 nm. It is also worth mentioning that the shape of the estimates obtained when fitting a linear or a quadratic model is quite similar, even when they vary in their range which is enlarged when a quadratic model is assumed.

\begin{figure}[ht!]
 \begin{center}
 \small
 \renewcommand{\arraystretch}{0.2}
 \newcolumntype{M}{>{\centering\arraybackslash}m{\dimexpr.05\linewidth-1\tabcolsep}}
   \newcolumntype{G}{>{\centering\arraybackslash}m{\dimexpr.33\linewidth-1\tabcolsep}}
%\begin{tabular}{MGG}
\begin{tabular}{ cc}
Least squares &  $MM-$estimators  \\ 
\multicolumn{2}{c}{Functional Linear Model}\\
   \includegraphics[scale=0.35]{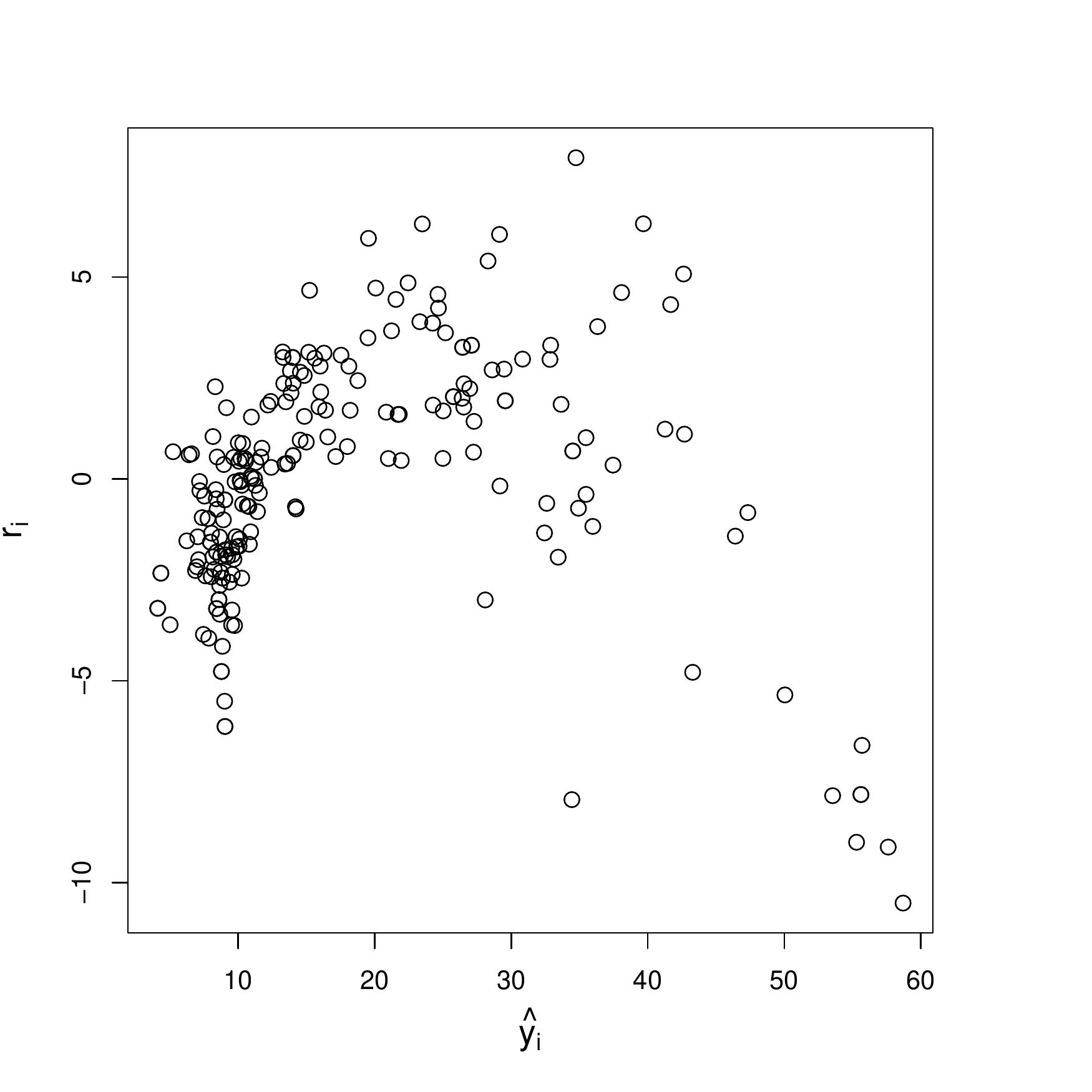} 
&  \includegraphics[scale=0.35]{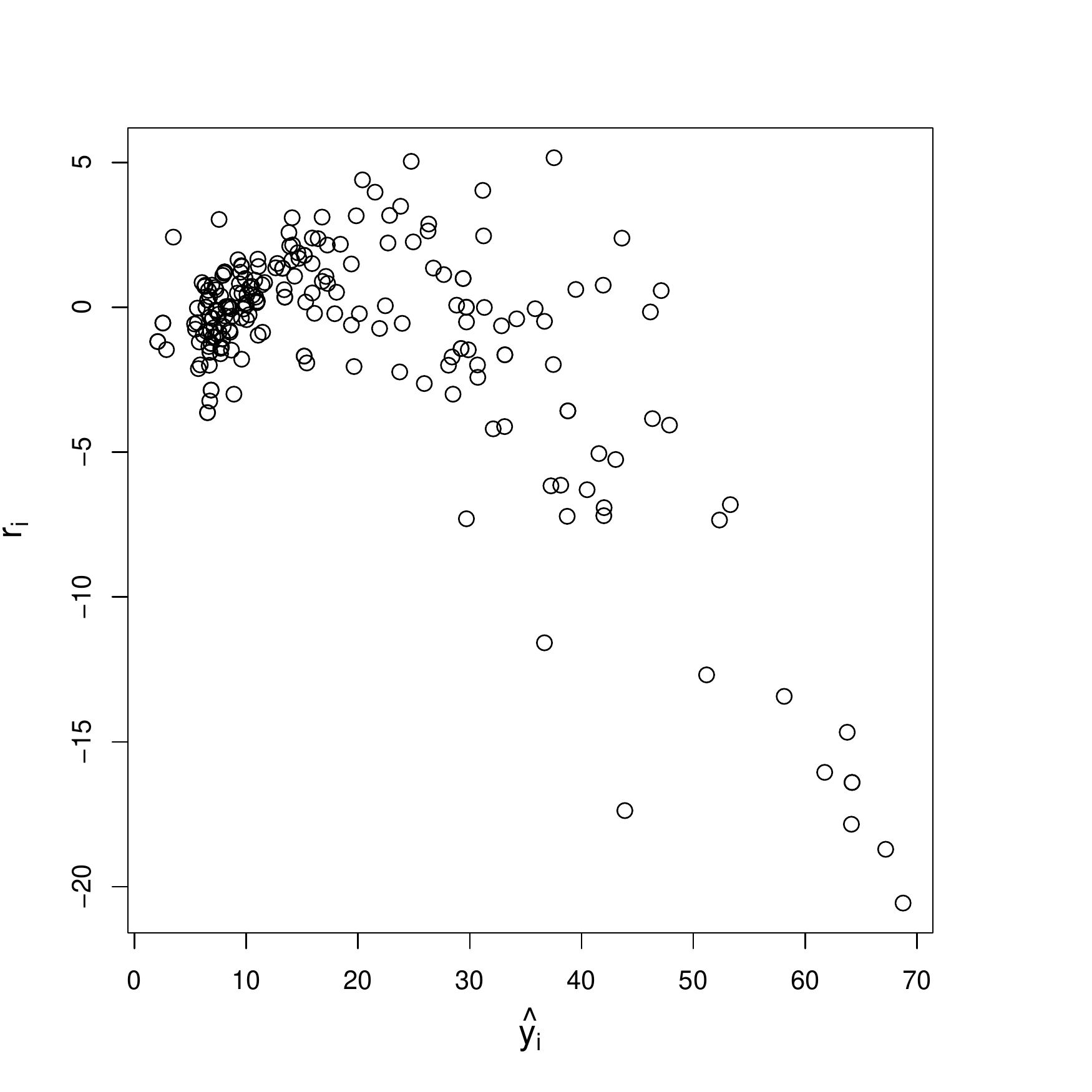} \\
\multicolumn{2}{c}{Functional Quadratic  Model}\\
   \includegraphics[scale=0.35]{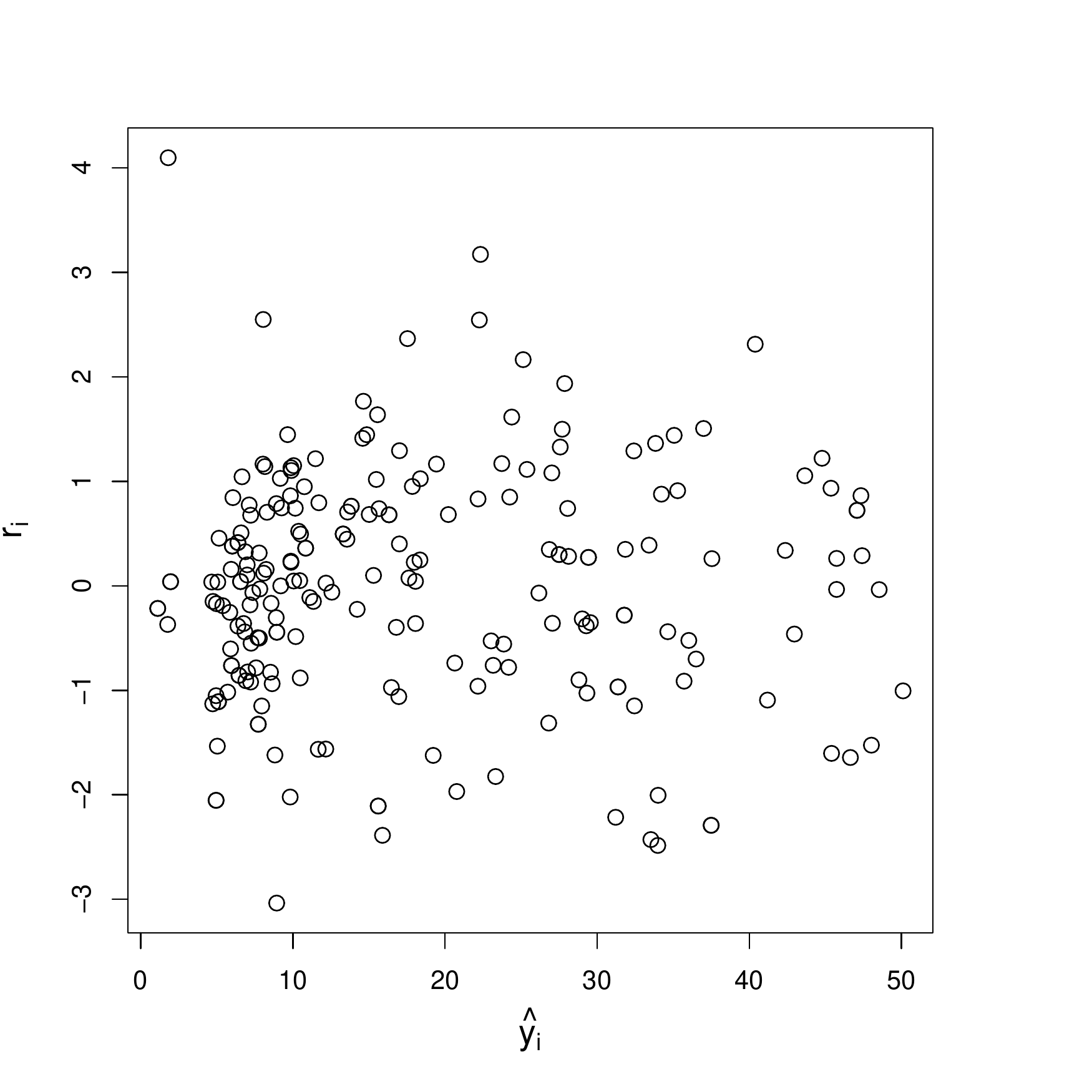} 
&  \includegraphics[scale=0.35]{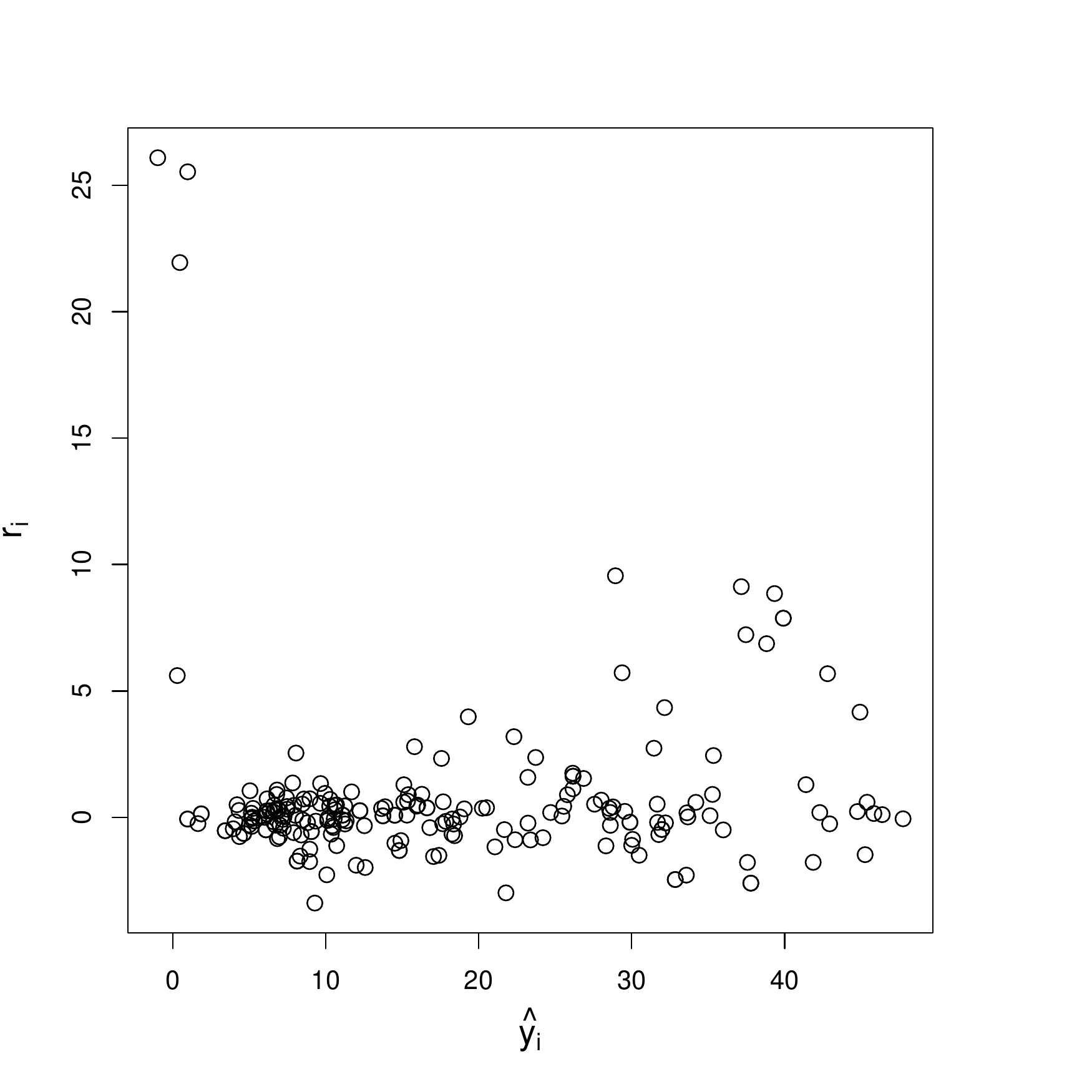} 
 \end{tabular}
\caption{\small \label{fig:tecator-residuals} Residuals from a classical (left panels) or robust (right panels) fit, when using a functional linear (first row) or   quadratic model (second row).}
\end{center} 
\end{figure}

\begin{figure}[ht!]
 \begin{center}
 \small
 \begin{tabular}{ cc}
 (a) & (b)\\
    \includegraphics[scale=0.35]{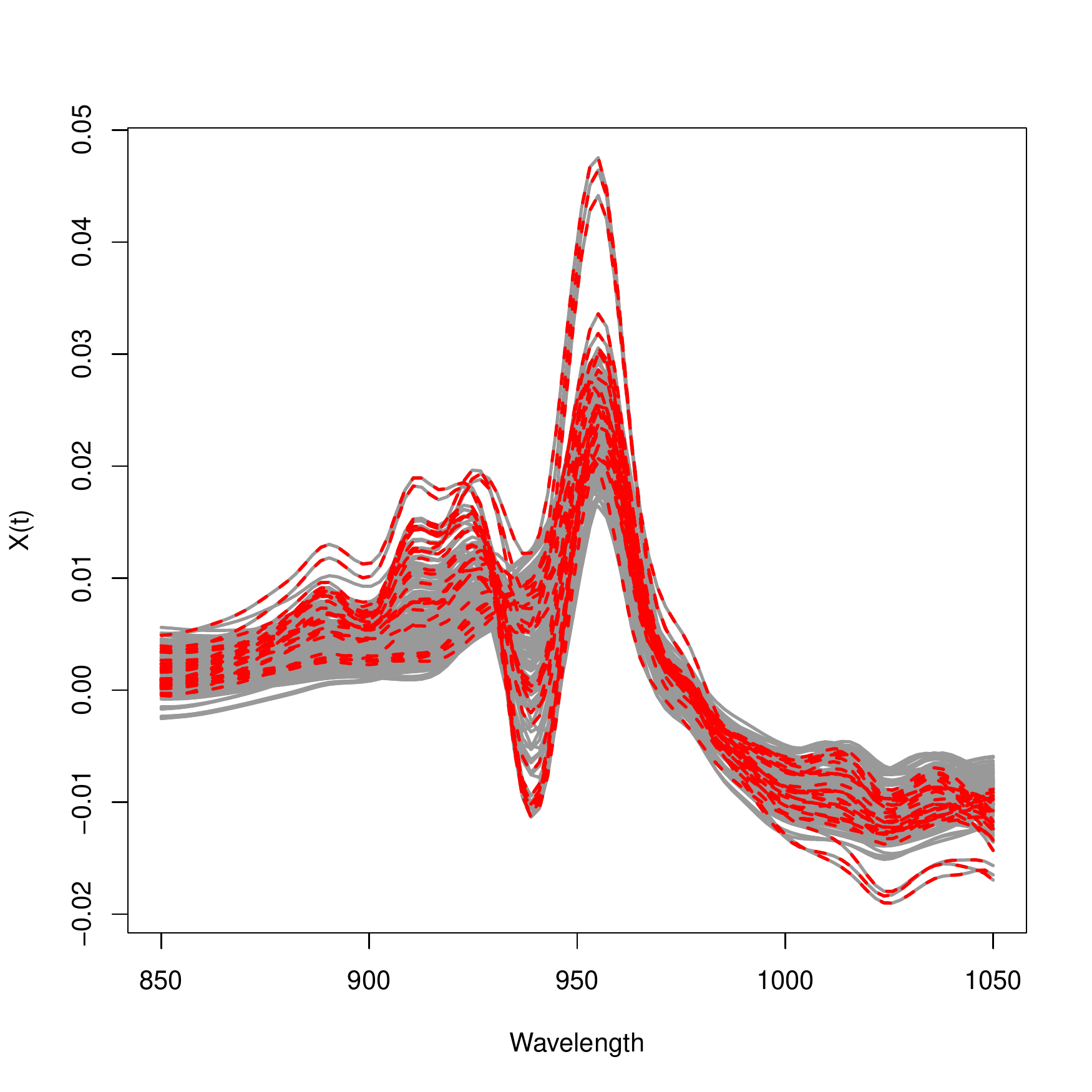} & 
    \includegraphics[scale=0.35]{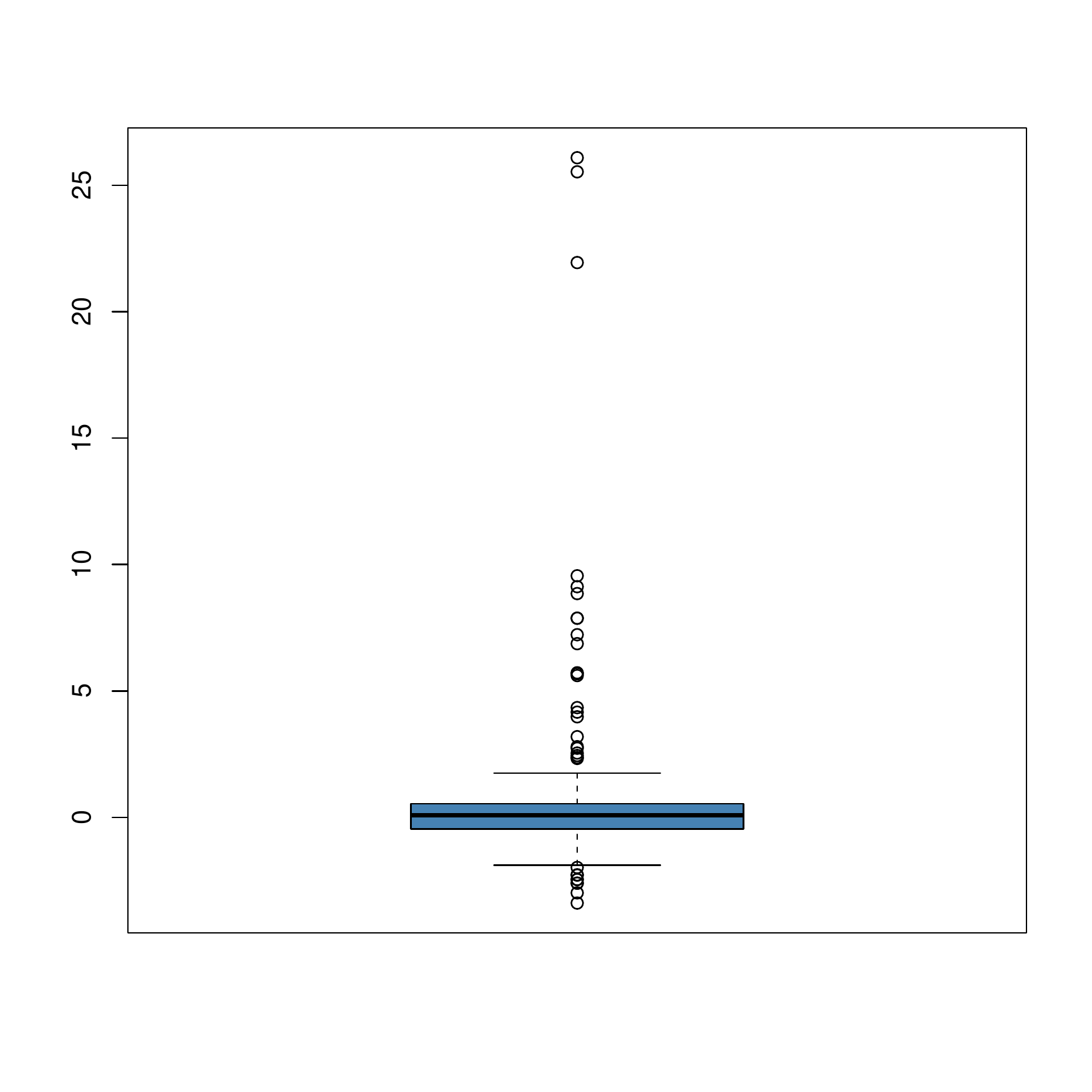}
 \end{tabular} 
\caption{\small \label{fig:tecator-outs} (a) Trajectories corresponding to the first derivative of the absorbance, in red we identify the covariates used in the the functional   quadratic model for which  the residuals are labelled as outliers by the boxplot. (b) Boxplot of the residuals from the robust quadratic fit. }
\end{center} 
\end{figure}

Figure \ref{fig:tecator-gammas} presents the plot of the quadratic kernel estimates $\wup_{\ls}$ and $\wup_{\eme\eme}$ obtained by each method when fitting a functional quadratic model. This Figure also reveals that the classical estimator presents a similar shape than the robust one but taking values in a smaller range, in particular, for the range of wavelengths close to 950 nm. The residual plots which are displayed in Figure \ref{fig:tecator-residuals} also show that the functional linear model does not seem to provide a reasonable fit neither for the classical nor for the robust method. Besides, when looking at the residuals from the robust quadratic fit, some atypical residuals are revealed. The boxplot of these residuals is given in the right panel of    Figure \ref{fig:tecator-outs} and identifies 32 observations as potential outliers, the corresponding covariates are displayed in red dashed lines in the left panel of Figure \ref{fig:tecator-outs}.

\begin{figure}[ht!]
 \begin{center}
 \small

    \includegraphics[scale=0.35]{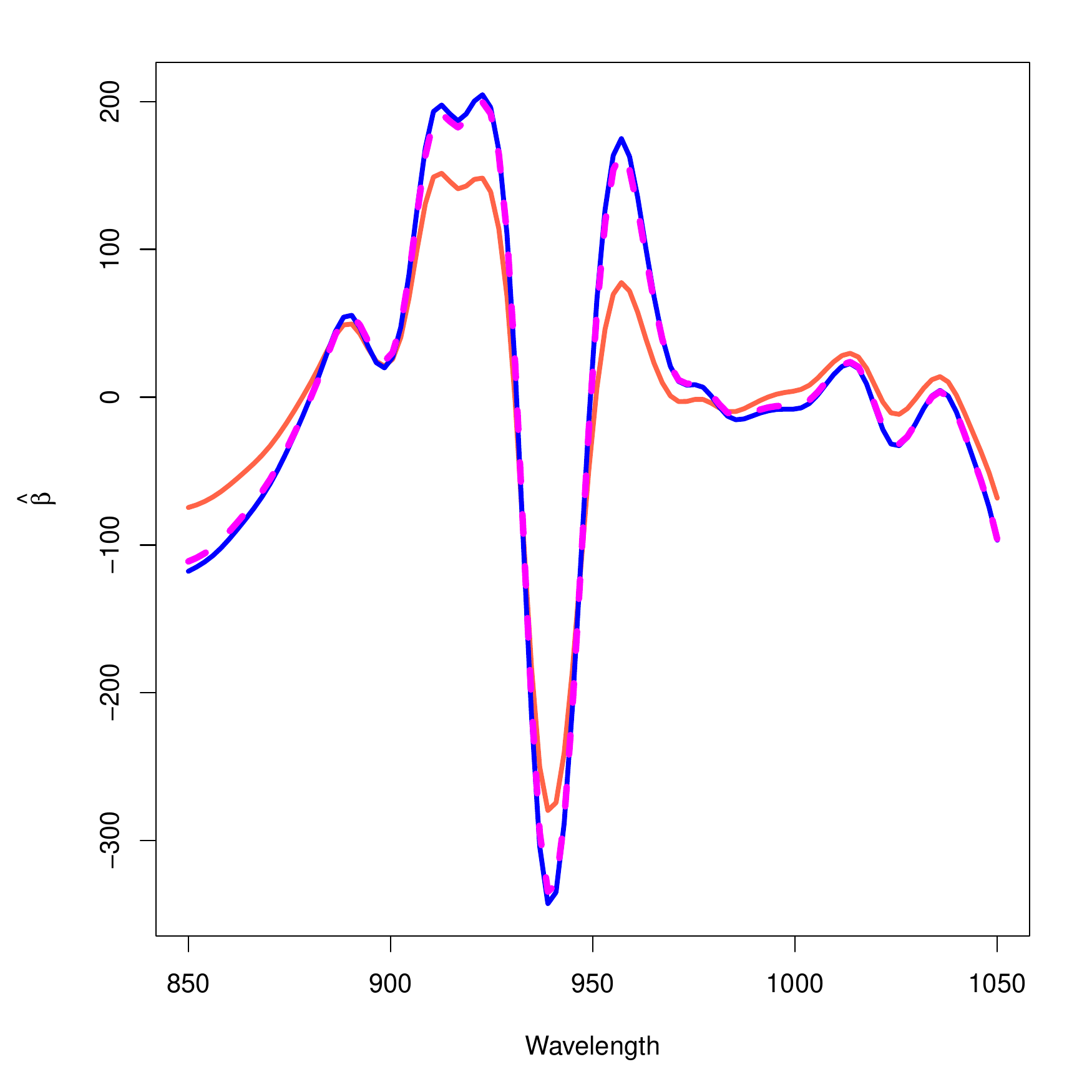} 
 
\caption{\small \label{fig:tecator-betas-SO} Estimates of $\beta$ using a least squares (in solid red line) or an $MM-$estimator (in solid blue line), when using a functional  quadratic model. The dashed pink line corresponds to the classical estimators computed after eliminating the potential atypical observations.}
\end{center} 
\end{figure}

\begin{figure}[ht!]
 \begin{center}
 \small
 \renewcommand{\arraystretch}{0.2}
 \newcolumntype{M}{>{\centering\arraybackslash}m{\dimexpr.05\linewidth-1\tabcolsep}}
   \newcolumntype{G}{>{\centering\arraybackslash}m{\dimexpr.33\linewidth-1\tabcolsep}}
%\begin{tabular}{MGG}
\begin{tabular}{ ccc}
$\wup_{\ls}$ &  $\wup_{\eme\eme}$  & $\wup_{\ls}^{(-\out)}$\\ 
   \includegraphics[scale=0.3 ]{GAMMA-CL.pdf} 
&  \includegraphics[scale=0.3 ]{GAMMA-ROB.pdf} 
& \includegraphics[scale=0.3 ]{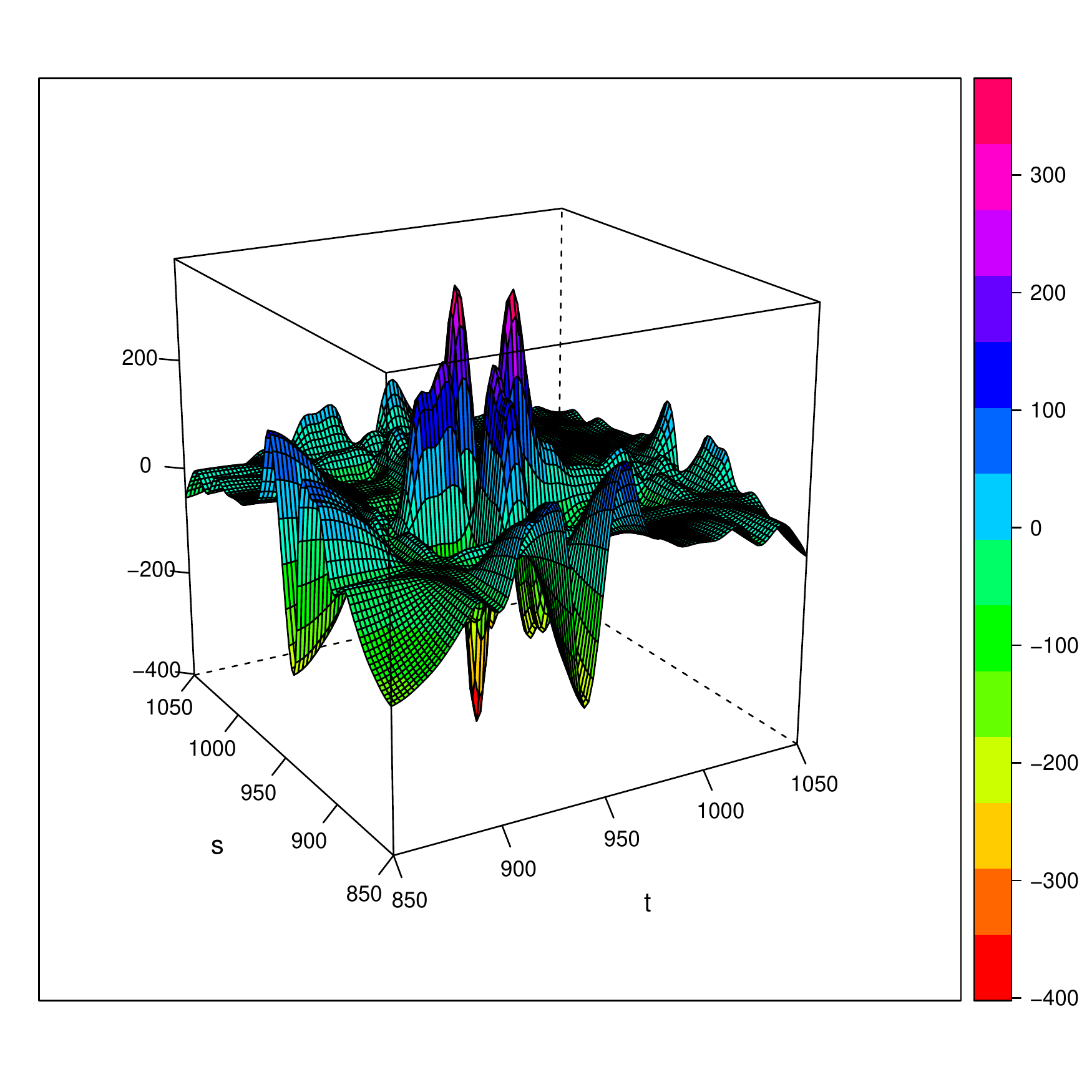} 
 \end{tabular}
\caption{\small \label{fig:tecator-gammas-SO} Estimates of $\upsilon$ using a least squares approach (left panel),   $MM-$method  (middle panel) or the classical procedure after eliminating the potential atypical observations (right panel).}
\end{center} 
\end{figure}

\begin{figure}[ht!]
 \begin{center}
 \small
 \renewcommand{\arraystretch}{0.2}
 \newcolumntype{M}{>{\centering\arraybackslash}m{\dimexpr.05\linewidth-1\tabcolsep}}
   \newcolumntype{G}{>{\centering\arraybackslash}m{\dimexpr.33\linewidth-1\tabcolsep}}
%\begin{tabular}{MGG}
\begin{tabular}{ cc}
$\wD$ &  $\wD^{(-\out)}$\\ 
   \includegraphics[scale=0.35 ]{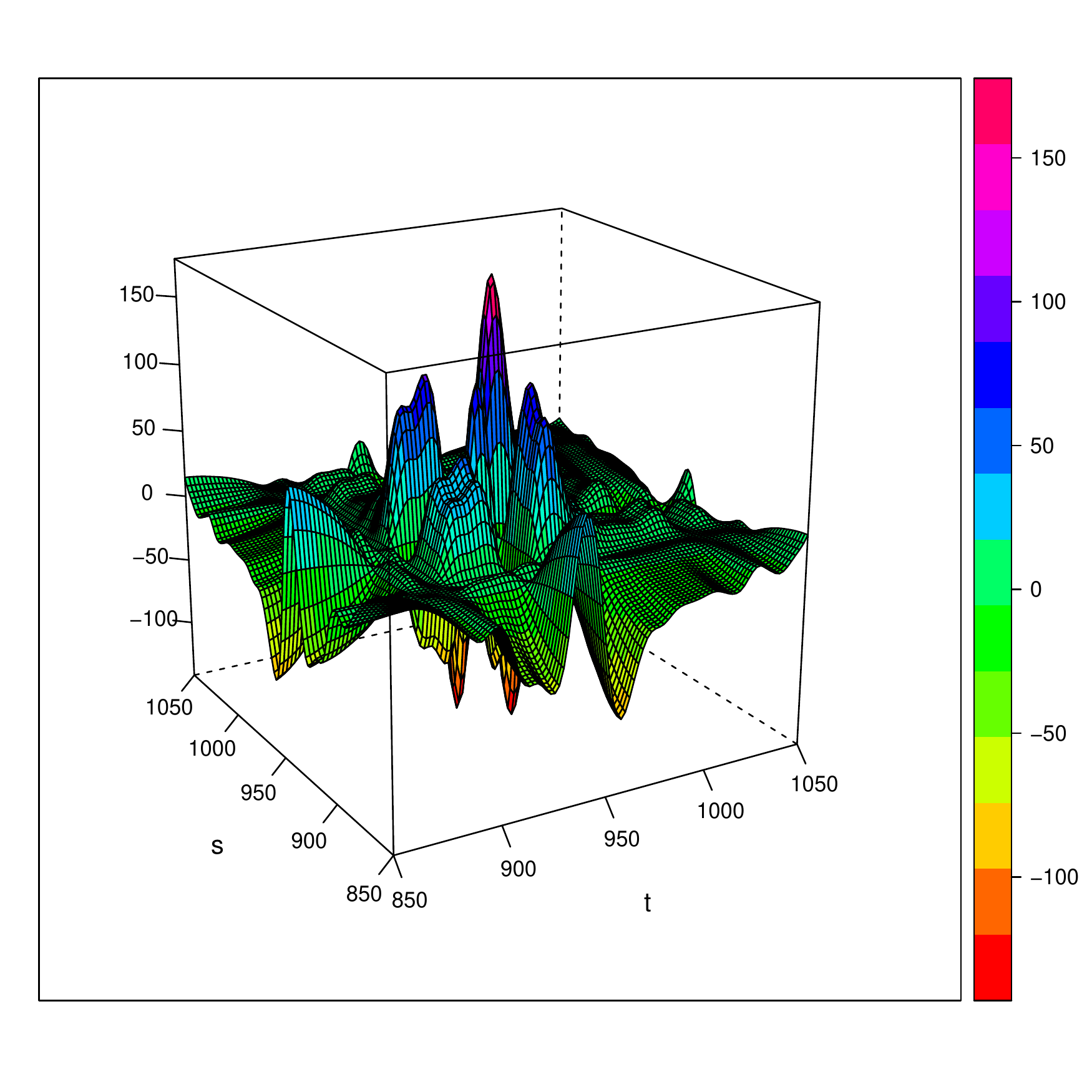} 
&  \includegraphics[scale=0.35 ]{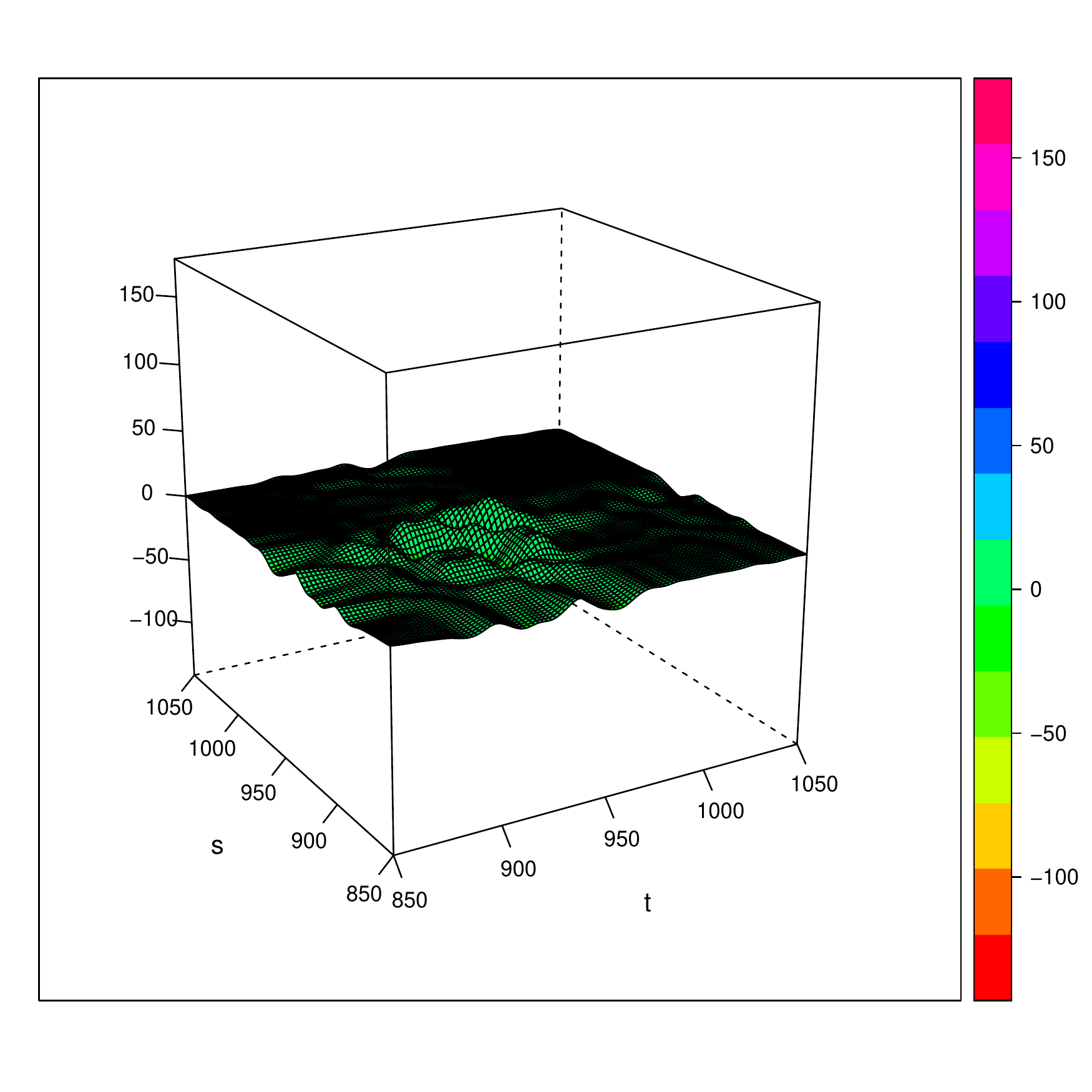}  
 \end{tabular}
\caption{\small \label{fig:tecator-DIF-gammas} Plot of the surfaces $\wD(s,t)=\wup_{\ls}(s,t)-\wup_{\eme\eme}(s,t)$ and $\wD^{(-\out)}(s,t)=\wup_{\ls}^{(-\out)}(s,t)-\wup_{\eme\eme}(s,t)$.}
\end{center} 
\end{figure}

We fitted again a quadratic model using the classical procedure  after eliminating the potential atypical observations. Figures \ref{fig:tecator-betas-SO} and \ref{fig:tecator-gammas-SO} display the obtained estimators together with the classical and robust estimates computed with all the data. In  Figure  \ref{fig:tecator-betas-SO} the classical and robust estimators of $\beta$ with  all the data are displayed in red and blue solid lines, respectively, while the least squares estimate  computed on the ``cleaned'' data set is presented in  a  dashed pink line. Note that for both the linear coefficient and the quadratic kernel, the shape of the classical estimators computed without the suspected atypical observations, resembles that of the robust ones. To visualize more clearly the  similarity between the surfaces related to the quadratic kernel estimators, we present in Figure \ref{fig:tecator-DIF-gammas} the surfaces $\wD =\wup_{\ls}-\wup_{\eme\eme}$ and $\wD^{(-\out)}=\wup_{\ls}^{(-\out)} -\wup_{\eme\eme}$. This figure highlights  the differences between the classical estimator computed with the whole data set and the robust one. Both when estimating the linear regression function or the quadratic operator,   the classical estimators computed without the detected potential outliers are very close to the robust ones, that is, the robust estimator behaves similarly to the classical one if one were able to manually remove suspected outliers.

\section{Final comments}{\label{sec:concl}}
In this paper, we propose robust estimators based on robust principal component analysis for functional quadratic regression models. Our estimators are robust against outliers in the response variable  and also in the functional explanatory variables. An extensive simulation study shows that our proposed estimators have good robustness and finite--sample statistical properties.    For finite--dimensional processes, where only   the coefficients of the regression and quadratic parameters over the linear spaces induced by the   eigenfunctions of the covariance operator are identifiable, the proposed procedure is Fisher--consistent. In such a situation, our requirements are closely related to those considered  in \citet{kalogridis2019robust}. However, we did not restrict our attention to this situation and we also derived Fisher--consistency when $X$ is an infinite--dimensional   process with $\mbox{ker}(\Gamma)=\{0\}$, under   smoothness conditions of the eigenfunctions of $\Gamma$.

We apply our method to a real data set and confirm that the robust $MM-$estimators remain reliable even when the data set contains atypical observations in the functional explanatory variables. Moreover, the residuals obtained from the robust fit provide a natural way to identify potential atypical observations.

%%%%%%%%%%%%%%%%%%%%%%%%%%%%%%%%%%%
%ACKNOWLEDGEMENT
%%%%%%%%%%%%%%%%%%%%%%%%%%%%%%

{\small \section*{Acknowledgements.}
This research was partially supported by Universidad de Buenos Aires [Grant 20020170100022\textsc{ba}]  and  \textsc{anpcyt} [Grant \textsc{pict} 2018-00740] at Argentina (Graciela Boente), the Ministerio de Ciencia e Innovaci\'on, Spain (MCIN/AEI/10.13039/501100011033) [Grant PID2020-116587GB-I00] (Graciela Boente). }
 
 \section{Appendix}{\label{sec:app}}
 
\noi \textsc{Proof of Proposition \ref{prop:FCKL}.} Note that  the Fisher--consistency of  $\mu_{\rob}$ and $\phi_{\rob,j}$  at $P_X$, i.e., $\mu_{\rob}(P_X)=\mu$ and $\phi_{\rob,j}(P_X)=\phi_j$, entail that $ \bx_p(P_X)= \bx_p$ and $\bz_p(P_X)=\bz_p$ where $\bx_p=(\langle X-\mu, \phi_1\rangle, \dots,\langle X-\mu, \phi_p\rangle)\trasp$ and $\bz_p=\vech(\{( x_{j}\, x_{  \ell}\})_{1 \le j\le \ell\le p}$. Moreover, if $p>q$ we have that $\langle X-\mu, \phi_j\rangle=0$, for any $j>q$, so that $\bx_p=(\bx_q\trasp,0,\dots,0)\trasp$ and similarly for $\bz_p$. 
Hence, for any $\bb_p=(b_1,\dots, b_p)=(\bb_q\trasp, \bb_{p-q}\trasp)\trasp\in \real^p$ and any finite-dimensional candidate $\beta_{\bb}=\sum_{j=1}^p b_j \phi_{\rob,j}(P_X)$ we have that $\langle X-\mu, \beta_{\bb}\rangle = \bb_q\trasp  \bx_q$. Similarly,  for any self--adjoint Hilbert--Schmidt operator with finite range such that
$$\Upsilon= \sum_{1\le j, \ell\le p}    v_{j\ell}  \phi_{\rob,j}(P_X)\otimes \phi_{\rob,\ell}(P_X)\,,  $$
we have that $\langle X-\mu,\Upsilon X\rangle= \bu_q\trasp \bz_q  $, where $\bu_p=\vech(\{(2-\uno_{j=\ell}) v_{j\ell}\})_{1 \le j\le \ell\le p}$. 
Thus, for any $(a, \beta, \Upsilon)\in \itC_p$, we have that
$$L(\alpha,\beta,\Upsilon,\mu_{\rob}(P_X),  \sigma)=\esp\rho_{1} \left ( \frac{  y-\alpha - \bb_q\trasp \bx_q- \bu_q\trasp \bz_q  }{\sigma} \right )\,.$$

Define  $\bb_0=(
\langle \beta_0, \phi_1\rangle, \dots, \langle \beta_0, \phi_q\rangle)\trasp$ and $v_{0,jj}=\langle \phi_j, \Upsilon_{0} \phi_j\rangle$ and $v_{0,j\ell}=v_{0,\ell\, j}= \langle \phi_j, \Upsilon_{0} \phi_\ell\rangle$, for $1\le j,\ell\le q$ and  $\bu_0=\vech(\{(2-\uno_{j=\ell}) v_{0,j\ell}\})_{1 \le j\le \ell\le q}$. Then,
\begin{equation}
\label{eq:ELE0}
L(\alpha_0,\beta_{0,q},\Upsilon_{0,q},\mu, \sigma)=\esp\rho_{1} \left ( \frac{  y - \alpha_0-\bb_0\trasp \bx_q - \bu_0\trasp \bz_q   }{\sigma} \right )=\esp\left( \rho_1\left(\epsilon \frac{\sigma_0}{\sigma}\right) \right)\,.
\end{equation}
 
Lemma 3.1 of \citet{yohai1987high} together with assumption \ref{ass:rho}  and the fact that $\widetilde{\epsilon}=\epsilon \sigma_0/\sigma$ satisfy assumption \ref{ass:densidad}, imply that for all $a \ne 0$, 
\begin{equation}
\esp  \left[ \rho_1 \left( \epsilon \frac{\sigma_0}{\sigma} -a \right) \right] > 
\esp  \left[ \rho_1 \left( \epsilon \frac{\sigma_0}{\sigma}  \right) \right]\,.
\label{eq:cotaesp}
\end{equation} 
Thus, taking conditional expectation, we obtain that for any $(a, \beta, \Upsilon)\in \itC_p$,  
\begin{align*}
L(\alpha,\beta,\Upsilon,\mu_{\rob}(P_X), \sigma)&=  \esp\rho_{1} \left ( \frac{  y - \alpha-\bb_q\trasp \bx_q - \bu_q\trasp \bz_q   }{\sigma} \right )\\
& =\esp\left\{\esp\left[\rho_{1} \left (  \epsilon\frac{\sigma_0}{\sigma} - \frac{(\alpha-\alpha_0)+ (\bb_q-\bb_0)\trasp \bx_q + (\bu_q-\bu_0)\trasp \bz_q}   {\sigma}  \right )|X\right]\right\}\\
&\ge  L(\alpha_0,\beta_{0,q},\Upsilon_{0,q}, \sigma)\,,
\end{align*}
where the last inequality is strict if  \ref{ass:probaX} holds and  $(\alpha,  \bb_q,\bu_q)\ne ( \alpha_0, \bb_0, \bu_0) $. Therefore, if we denote as $\bb_{0,p}= (\bb_0\trasp, \bcero_{p-q}\trasp)$ and $\bu_{0,p}=\vech(\{(2-\uno_{j=\ell}) v_{0,p,j\ell}\})_{1 \le j\le \ell\le p}$, where   $v_{0,p,j\ell}=v_{0,  j\ell}$, for $1\le j,\ell\le q$ and   $0$ otherwise, we have that the vector $(\alpha_0,\bb_{0,p},\bu_{0,p})$  is a solution of \eqref{eq:funfinitos}, meaning that $\alpha(P)=\alpha_0$, $\pi(\beta(P), \itH_q)=\beta_{0,q}$ and $\pi(\Upsilon(P), \itF_q)=\Upsilon_{0,q}$.

If in addition  \ref{ass:probaX} holds   and $p=q$, we have that $(\alpha_0,\bb_0,\bu_0)$ is the unique solution of \eqref{eq:funfinitos}. Indeed, given $(\alpha, \beta, \Upsilon)\in \itC_q$, $(\alpha, \beta, \Upsilon)\ne 0$ let
  $$\itA_0 \, = \, \left\{ X  : \Phi(X )=\alpha-\alpha_0+\langle X-\mu,\beta-\beta_{0,q} \rangle+ \langle X-\mu,(\Upsilon- \Upsilon_{0,q}) (X -\mu)\rangle = 0 \right\} \,,$$
   and $a(X)= \Phi(X)/\sigma$. Then, if, for any set $\itA$,  $\indica_{\itA}(x)$ equals 1 whenever $x\in \itA$ and $0$ otherwise, we have  that
$$
L(\alpha,\beta,\Upsilon,\mu_{\rob}(P_X), \sigma)=\esp \rho_1\left(\epsilon  \frac{\sigma_0}{\sigma} -\frac{ \Phi(X)}{\sigma}\right)= \esp\left\{  \rho_1\left(\epsilon \frac{\sigma_0}{\sigma}\right) \indica_{\itA_0}(X)\right\} \, + \,
\esp\left\{\esp\left[ \rho_1\left(\epsilon \frac{\sigma_0}{\sigma} -a(X)\right) |X\right]\indica_{\itA_0^{c}}(X)\right\} \,.
$$
Using (\ref{eq:cotaesp}) we get that, for any $X\notin \itA_0$,  
\begin{eqnarray*}
\esp\left[ \rho_1\left(\epsilon \frac{\sigma_0}{\sigma} -a(X)\right) |X= X_0 \right]&=& \esp\left[ \rho_1\left(\epsilon \frac{\sigma_0}{\sigma} -a(X_0 )\right) |X= X_0 \right]\\
&=& \esp\left[ \rho_1\left(\epsilon \frac{\sigma_0}{\sigma} -a(X_0 )\right)\right]>\esp  \left[ \rho \left( \epsilon \frac{\sigma_0}{\sigma}  \right) \right]\,,
\end{eqnarray*}
where the last equality follows from the fact that the errors are independent of the covariates.   Thus, taking into account that  \ref{ass:probaX} implies that  $\prob(\itA_0^{c})>0$, we obtain  
\begin{align*}
L(\alpha,\beta,\Upsilon,\mu_{\rob}(P_X), \sigma)&=  \esp\left\{  \rho_1\left(\epsilon \frac{\sigma_0}{\sigma}\right) \indica_{\itA_0}(X)\right\}  \, + \,
\esp\left\{\esp\left[ \rho_1\left(\epsilon \frac{\sigma_0}{\sigma} -a(X)\right) |X\right]\indica_{\itA_0^{c}}(X)\right\} 
\\
&> \esp\left\{  \rho_1\left(\epsilon \frac{\sigma_0}{\sigma}\right) \indica_{\itA_0}(X)\right\}\, + \,
\esp\left\{\esp\left[ \rho_1\left(\epsilon \frac{\sigma_0}{\sigma} \right) \right]\indica_{\itA_0^{c}}(X)\right\} = \esp\left( \rho_1\left(\epsilon \frac{\sigma_0}{\sigma}\right) \right) \,, 
\end{align*}
which together with \eqref{eq:ELE0} and the fact that $\mu_{\rob}(P_X)=\mu$, concludes the proof. \qed

\vskip0.2in

\noi \textsc{Proof of Proposition \ref{prop:Lphaciainfty}.} Note that as mentioned in Section \ref{sec:FCfinito}, under \ref{ass:densidad}    and   \ref{ass:rho}, for any
 $a\in \real$, $\beta\in L^2(0,1)$ and   $\Upsilon \in \itF$, we have that $L(\alpha,\beta,\Upsilon, \sigma_0)\ge L(\alpha_0,\beta_0,\Upsilon_0, \sigma_0)$. Thus for any $p\in \natu$,  $ L(\alpha_0,\beta_0,\Upsilon_0,  \mu,\sigma_0)\le L(\alpha_p(P), \beta_p(P), \Upsilon_p(P), \mu,\sigma_0)$. Hence, 
 $$L(\alpha_0,\beta_0,\Upsilon_0,  \mu,\sigma_0) \le \liminf_{p\to \infty} L(\alpha_p(P), \beta_p(P), \Upsilon_p(P), \mu,\sigma_0)\,.$$
On the other hand, taking into account that $\beta_{0,p}\in \itH_p$ and $\Upsilon_{0,p} \in \itF_p$, we get
\begin{equation}
\label{eq:cotaELE}
L(\alpha_p(P), \beta_p(P), \Upsilon_p(P), \mu,\sigma_0) = \argmin_{a\in \real,\beta\in \itH_p, \Upsilon\in \itF_p}  L(a,\beta,\Upsilon, \mu,\sigma_0)\le  L(\alpha_0,\beta_{0,p},\Upsilon_{0,p}, \mu,\sigma_0)\,.
\end{equation}
Using that $\|\beta_{0,p}-\beta_0\|_{L^2(0,1)}\to 0$ and $\|\Upsilon_{0,p}-\Upsilon_0\|_{\itF}\to 0$, the Cauchy-Schwartz inequality, the fact that  $\rho_1$ is a bounded continuous function and the  Bounded Convergence Theorem, we get that $ L(\alpha_0,\beta_{0,p},\Upsilon_{0,p}, \mu,\sigma_0) \to  L(\alpha_0,\beta_{0 },\Upsilon_{0}, \mu,\sigma_0)$, which together with \eqref{eq:cotaELE} leads to
$$\limsup_{p\to \infty} L(\alpha_p(P), \beta_p(P), \Upsilon_p(P), \mu,\sigma_0)  \le  L(\alpha_0,\beta_{0},\Upsilon_{0}, \mu,\sigma_0)\,,$$
concluding the proof. \qed

\vskip0.2in

\noi \textsc{Proof of Proposition \ref{prop:FCinfty}.} The proof uses similar arguments to those considered in the proof of Theorem 3.1 and Lemma S.1.4 in \citet{boente2020robust} but adapted to the present situation of a  quadratic model  and $L^2-$distances.

 Let us denote as $\itC=\real\times L^2(0,1)\times \itF$, $\itC^{\star}=\itC\cap\real\times \itW^{1,2}\times \itW_2^{1,2}$, $\theta=(a, \beta, \Upsilon)$ and  $\theta_0=(\alpha_0, \beta_0, \Upsilon_0)$. We will begin showing that, for any $\epsilon>0$,
\begin{equation}
\label{eq:infLAeps}
 \inf_{(a,\beta,\upsilon)\in \itA_\epsilon} L(a,\beta,\Upsilon, \mu, \sigma_0) > L(\alpha_0,\beta_0, \Upsilon_0,\mu, \sigma_0)\,,
\end{equation}
where $\upsilon$ the kernel related to the Hilbert--Schmidt operator $\Upsilon$ and $\itA_{\epsilon}=\{(a,\beta,\upsilon)\in \itC^{\star}: |a-\alpha_0|+  \|\beta-\beta_0\|_{\itW^{1,2} }+\|\upsilon-\upsilon_0\|_{\itW_2^{1,2}}\le M\,, \, d(\theta, \theta_0) \ge \epsilon\}$ with  $ d(\theta, \theta_0)=|a-\alpha_0|+\|\beta -\beta_0\|_{L^2(0,1)}+\|\upsilon -\upsilon_0\|_{L^2(\itT)}$ and $ \itT=(0,1)\times(0,1)$. From now on, we identify the quadratic operator with its kernel. 

Let $(a_k, \beta_k, \upsilon_k)\in \itA_{\epsilon}$ be such that $L_k= L(a_k, \beta_k, \Upsilon_k, \mu, \sigma_0) \to \inf_{(a,\beta,\upsilon)\in \itA_\epsilon} L(a,\beta,\Upsilon, \mu, \sigma_0) $. 

Recall that from the Rellich--Kondrachov Theorem,  $\itW^{1,2}$ and $\itW_2^{1,2}$ are compactly embedded in $ L^2(0,1)$ and $ L^2(\itT)$, respectively, so   using that $|a_k-\alpha_0|+  \|\beta_k-\beta_0\|_{\itW^{1,2} }+\|\upsilon_k-\upsilon_0\|_{\itW_2^{1,2}}\le M$, for all $k\ge 1$,   we have that there exists a subsequence $ {k_j}$ such that $a_{k_j}-\alpha_0 \to a_0$, $\beta_{k_j}-\beta_0\to b_0$ in $L^2(0,1)$ and  $\upsilon_{k_j}-\upsilon_0\to w_0$ in $L^2(\itT)$. 

Denote as  $\wtalfa_0=\alpha_0+a_0$, $\wtbeta_0=\beta_0+b_0$ and $\wtup_0=\upsilon_0+w_0$. Therefore, using that $(a_k, \beta_k, \upsilon_k)\in \itA_{\epsilon}$, we get that   $d(\wttheta_0, \theta_0) \ge \epsilon$ where $\wttheta_0=(\wtalfa_0, \wtbeta_0, \wtup_0)$. Taking into account that     $\rho_1$ is a bounded continuous function, from  the Bounded Convergence Theorem and the Cauchy--Schwartz inequality, we get that $L_{k_j}\to  L(\wtalfa_0, \wtbeta_0, \wtup_0, \mu, \sigma_0) $. Therefore, we obtain that  $ \inf_{(a,\beta,\upsilon)\in \itA_\epsilon} L(a,\beta,\Upsilon, \mu, \sigma_0) =L(\wtalfa_0, \wtbeta_0, \wtup_0, \mu, \sigma_0) $. Using that $(\alpha_0,\beta_0,\Upsilon_0)$ is the unique minimizer of $L(\alpha,\beta,\Upsilon, \mu, \sigma)$ for any $\sigma>0$, since \ref{ass:probaX1} holds, we obtain that  $L(\wtalfa_0, \wtbeta_0, \wtup_0, \mu, \sigma_0) > L(\alpha_0 , \beta_0 , \upsilon_0 , \mu, \sigma_0)$ which concludes the proof of \eqref{eq:infLAeps}.

 The proof will be completed if we  show that there exists $M>0$ such that,    
 \begin{equation}
 \label{eq:limsupbetap}
 \limsup_{p\to \infty}|\alpha_p(P)-\alpha_0|+\|\beta_p(P)-\beta_0\|_{\itW^{1,2} }+\|\upsilon_p(P)-\upsilon_0\|_{\itW_2^{1,2}}\le M\,.
 \end{equation}
% The proof follows the same steps as in those used in the proof of Lemma S.1.4 in \citet{boente2020robust}. 
 
 Given $\delta>0$, define $K_\delta$ such that for any $K\ge K_\delta$,
\begin{equation}
\label{eq:XK} 
\prob(\|X\|+ \|X\|^2\ge K)<\delta\,.
\end{equation}
Fix   $\theta=(a,\beta,\upsilon) \in \itC^{\star} $, $\theta\ne 0$, such that $|a|+  \|\beta\|_{\itW^{1,2} }+\|\upsilon\|_{\itW_2^{1,2}}\le 1$ and let $\varphi_{\theta}>0$ be a continuity point of the distribution of $|a+\langle X, \beta \rangle +\langle X, \Upsilon X\rangle|$ such that
\begin{equation}
\label{eq:cotaprobc}
\prob \left(|a+\langle X, \beta \rangle +\langle X, \Upsilon X\rangle|< \varphi_{\theta}\right)< c \,.
\end{equation}
Then, if $a^{\star}\in \real$, $\beta^{\star}  \in L^2(0,1)$ and  $\upsilon^{\star}  \in L^2(\itT)$ are such that 
$$  \Delta(\theta^{\star}, \theta)=\max\left(|a^{\star}-a|,\|\beta^{\star}-\beta\|_{L^2(0,1)}, \|\upsilon^{\star}-\upsilon\|_{L^2(\itT)}\right)< \vartheta_{\theta}\,,$$ 
where $\vartheta_{\theta}= \varphi_{\theta}/(2(K+1))$, we have that
\begin{align*}
\prob \left(|a^{\star}+\langle X, \beta^{\star} \rangle +\langle X, \Upsilon^{\star} X\rangle| \ge \frac{ \varphi_{\theta}}2 \right)  & \ge  A(\theta)\,,
\end{align*}
where $ A(\theta)=   \prob \left(|a +\langle X, \beta  \rangle +\langle X, \Upsilon  X\rangle|\ge \varphi_{\theta}\right)- \prob\left(\vartheta_{\theta} (1+\|X\|+\|X\|^2) \ge  { \varphi_{\theta}}/2 \right)$.
Hence, noting that from \eqref{eq:XK} and \eqref{eq:cotaprobc}, $A(\theta)>  1-c-\delta$, we conclude that
\begin{equation}
\label{eq:infimodelta}
\inf_{ \Delta(\theta^{\star}, \theta)< \vartheta_{\theta}}\prob \left(|a^{\star}+\langle X, \beta^{\star} \rangle +\langle X, \Upsilon^{\star} X\rangle|  \ge  \frac{ \varphi_{\theta}}2 \right)
\ge  A(\theta)>  1-c-\delta\,.
\end{equation}
Taking into account that $\itV=[-1,1]\times \{\beta\in L^2(0,1): \|\beta\|_{\itW^{1,2} }\le 1\}\times \{\upsilon\in L^2(\itT):  \|\upsilon\|_{\itW_2^{1,2}}\le 1\}$ is compact with the topology in $\real \times  L^2(0,1)\times L^2(\itT)$, we can take a finite sub--cover from the covering of $\itV$ given by $\{ \itB(\theta, \vartheta_{\theta})\}_{\theta\in \itV}$, where $\itB(\theta, \rho)$ stands for the open ball with center $\theta$ and radius $\rho$, that is, $\itB(\theta, \rho)=\{(u, f, w)\in \real \times  L^2(0,1)\times L^2(\itT): \max\left(|u-a|,\|f-\beta\|_{L^2(0,1)}, \|w-\upsilon\|_{L^2(\itT)}\right)<\rho \}$. The   compactness of $\itV$ entails that there exist $\theta_j=(a_j,\beta_j, \upsilon_j)\in \itV$, $1\le j\le s$,  such that $\itV \subset \cup_{j=1}^s \itB(\theta_j,\vartheta_j )$ with $\vartheta_j=\vartheta_{\theta_j}$. Therefore, from \eqref{eq:infimodelta}, we obtain that
$$
\min_{1\le j\le s}\inf_{\Delta(\theta, \theta_j) < \vartheta_j}\prob \left(|a +\langle X, \beta  \rangle +\langle X, \Upsilon  X\rangle|> \frac{\varphi_j}{2}\right)> 1-c-\delta\,, 
$$
with $\varphi_j=\varphi_{\theta_j}$, meaning that for any $(a,\beta,\upsilon)\in  \itV$, there exist $1\le j\le s$ such that 
\begin{equation}
\label{eq:infimoenC1}
 \prob \left(|a +\langle X, \beta  \rangle +\langle X, \Upsilon  X\rangle|> \frac{\varphi_j}{2}\right)> 1-c-\delta\,.
\end{equation}
Taking into account that from Proposition \ref{prop:Lphaciainfty}, 
$$\lim_{p\to \infty}  L(\alpha_p(P), \beta_p(P), \Upsilon_p(P), \mu,\sigma_0)  = L(\alpha_0,\beta_0,\Upsilon_0,\mu, \sigma_0) =b_{\rho_1}\,,$$
and that $c< 1- b_{\rho_1}$, we have that there exists $p_0\in \natu$ such that for each $p\ge p_0$, 
\begin{equation*}
\label{eq:cotaLpbrho}
L(\alpha_p(P), \beta_p(P), \Upsilon_p(P), \mu,\sigma_0) \le b_{\rho_1}+\frac{\xi}2\,,
\end{equation*}
where $\xi < 1-c-b_{\rho_1}$. 

In order to derive \eqref{eq:limsupbetap}, it will be enough to show that there exist $M>0$  such that,
$$\inf_{(a, \beta, \upsilon)\in\itD(\theta_0, M)} L(a,\beta,\Upsilon, \mu, \sigma_0) \ge b_{\rho_1}+\xi\;,$$ 
where $\itD(\theta_0, M)=\{\theta=(a, \beta, \upsilon)\in \itC^{\star}: |a-\alpha_0|+  \|\beta-\beta_0\|_{\itW^{1,2} }+\|\upsilon-\upsilon_0\|_{\itW_2^{1,2}} > M\}$. Denote as $R(u)=\esp\rho\left( \epsilon  - u/\sigma_0\right)$.  First note that  the independence between the errors and covariates entails that
\begin{align*}
L(a,\beta ,\Upsilon, \mu, \sigma_0) & =\esp\rho\left(\epsilon -\frac{ (a-\alpha_0)+\langle X, \beta-\beta_0 \rangle + \langle X, (\Upsilon-\Upsilon_0)X \rangle}{\sigma_0}\right)\\
& = \esp R\left((a-\alpha_0)+\langle X, \beta-\beta_0 \rangle + \langle X, (\Upsilon-\Upsilon_0)X \rangle \right)\,.
\end{align*}
Using that  $\lim_{|u|\to +\infty} R(u)=1$, we get that for any $\delta>0$, there exists  $u_0$ such that, for any $u$ such that $|u|\ge u_0$,  
\begin{equation}
\label{eq:epsM}
 R(u) >1-\delta\,.
\end{equation}
Choose $M> 2 \; u_0/ \min_{1\le j\le s}(\varphi_j)$, where $\varphi_j$ is given in \eqref{eq:infimoenC1} and   let $(a_k,\beta_k, \upsilon_k)\in \real\times \itW^{1,2}\times \itW_2^{1,2}$ be such that $\nu_k=|a_k-a_0|+\|\beta_k-\beta_0\|_{\itW^{1,2}}+\|\upsilon_k-\upsilon_0\|_{\itW_2^{1,2}}>M$ and 
$$\lim_{k\to \infty} L(a_k,\beta_k,\upsilon_k,\mu, \sigma_0) = \inf_{(a, \beta, \upsilon)\in\itD(\theta_0, M)} L(a,\beta,\upsilon, \mu, \sigma_0)\,.$$ 
Denote as  $\wta_k= (a_k-\alpha_0)/\nu_k$, $\wtbeta_k=(\beta_k-\beta_0)/\nu_k$ and $\wtup_k=(\upsilon_k-\upsilon_0)/\nu_k$, then $(\wta_k,\wtbeta_k,\wtup_k)\in \itV$, thus using \eqref{eq:infimoenC1}, we obtain  that there exists $1\le j=j(k)\le s$ such that 
$$
 \prob \left(|\wta_k+\langle X, \wtbeta_k \rangle +\langle X, \wtUpsilon_k X\rangle|> \frac{\varphi_j}{2}\right)> 1-c-\delta\,. 
$$
Using that  $\nu_k>M>2\, u_0/\phi_j$ and denoting as $u_k(X)=\nu_k (\wta_k+\langle X, \wtbeta_k \rangle +\langle X, \wtUpsilon_k X\rangle)$, we obtain that  $|u_k(X)|>u_0$ whenever $|\wta_k+\langle X, \wtbeta_k \rangle +\langle X, \wtUpsilon_k X\rangle|> {\varphi_j}/2$, which together with \eqref{eq:epsM} leads to 
\begin{eqnarray*} 
L(a_k,\beta_k,\Upsilon_k,\mu, \sigma_0)&=&  \esp R\left( a_k+ \langle X, \beta_k\rangle + \langle X, \Upsilon_k X\rangle  \right)=   \esp R\left(u_k(X) \right)\\
& \ge & \esp\left\{ R\left(u_k(X)\right)\indica_{\left|\wta_k+\langle X, \wtbeta_k \rangle +\langle X, \wtUpsilon_k X\rangle \right|> {\varphi_j}/2} \right\} \\
& > & (1-\delta)\, \prob\left(\left|\wta_k+\langle X, \wtbeta_k \rangle +\langle X, \wtUpsilon_k X\rangle\right|>\frac{\varphi_j}2\right) \\
&>& (1-c-\delta)\,(1-\delta)\,,
\end{eqnarray*}
where the last inequality follows from  \eqref{eq:infimoenC1}. Therefore,
$$\inf_{(a, \beta, \upsilon)\in\itD(\theta_0, M)} L(a,\beta,\Upsilon, \mu, \sigma_0)\ge (1-c-\delta)(1-\delta)\,.$$
The proof follows now easily noting that $\lim_{\delta\to 0} (1-c-\delta)(1-\delta)=1-c>b_{\rho_1}+\xi$, so we can choose $\delta$ and consequently $M$ such that   
$$\inf_{(a, \beta, \upsilon)\in\itD(\theta_0, M)} L(a,\beta,\Upsilon, \mu, \sigma_0) >b_{\rho_1}+\xi > L(\alpha_p(P), \beta_p(P), \Upsilon_p(P), \mu,\sigma_0) \,,$$
which shows that $|\alpha_p(P)-\alpha_0|+\|\beta_p(P)-\beta_0\|_{\itW^{1,2} }+\|\upsilon_p(P)-\upsilon_0\|_{\itW_2^{1,2}}\le M$, for any $p\ge p_0$, concluding the proof. \qed
 
%%%%%%%%%%%%%%%%%%%%%%%%%%%%%%%%%%%
 %REFERENCES
%%%%%%%%%%%%%%%%%%%%%%%%%%%%%%
\small
\nocite{*}
\bibliographystyle{apalike}
\bibliography{referencias2}

\end{document}